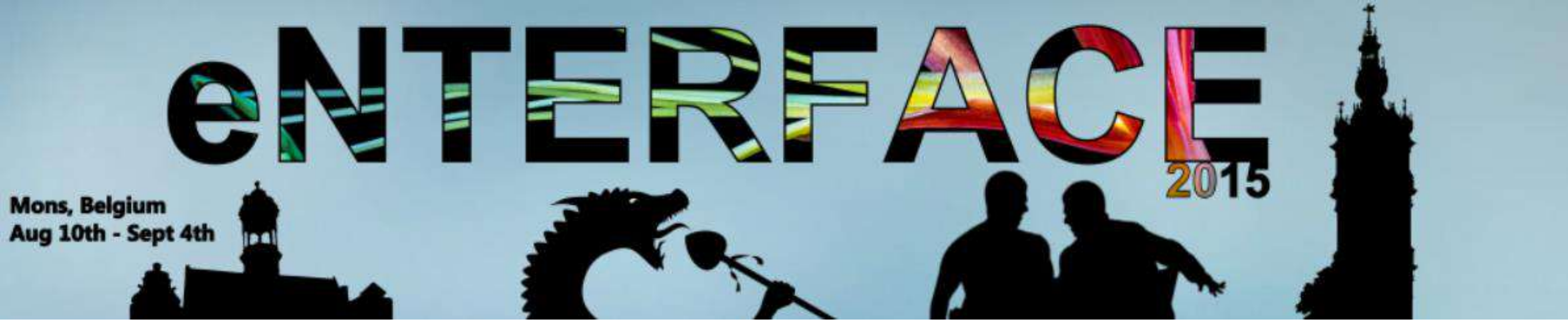

# Proceedings of eNTERFACE'15

## The 11th Summer Workshop on Multimodal Interfaces

August 10th - September 4th, 2015
Numediart Institute, University of Mons

Mons, Belgium

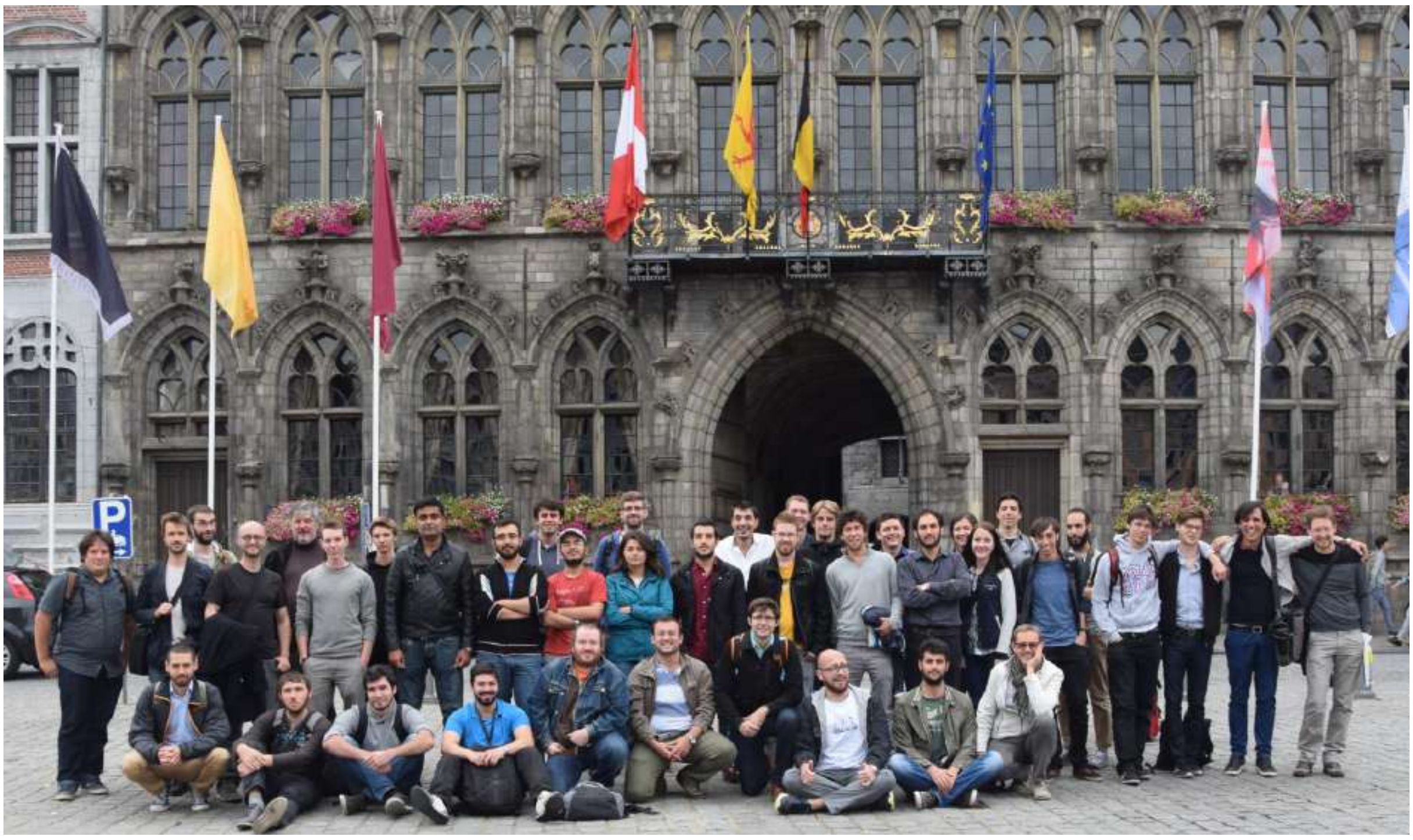

numediart
INSTITUT DE RECHERCHE POUR LES TECHNOLOGIES CREATIVES DE L'UMONS
UMONS
Université de Mons
fnrs
FREEDOM TO RESEARCH
MONS 2015

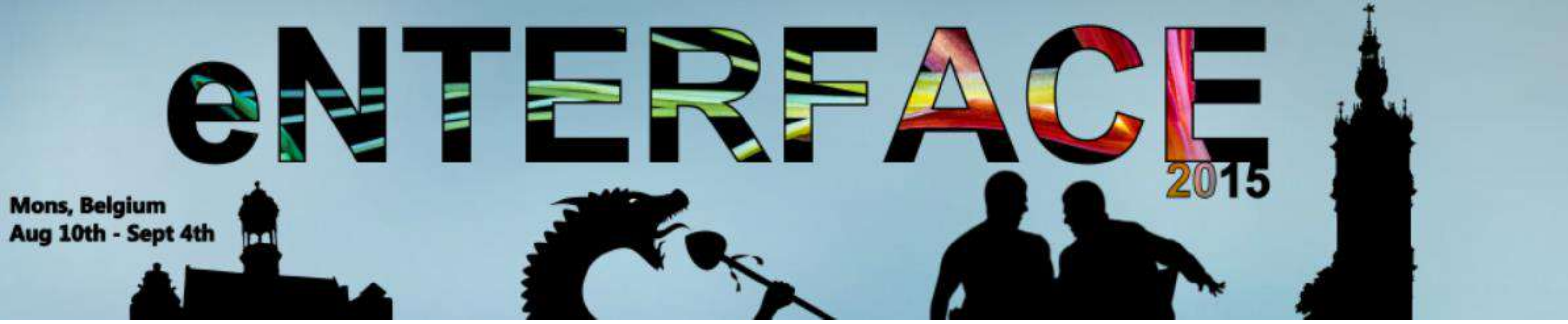

# Overview

## Preface

## Project #1:

### Prototyping a New Audio-Visual Instrument Based on Extraction of High-Level Features on Full-Body Motion

*Jo□elle Tilmanne, Nicolas d'Alessandro, Petr Barborka, Furkan Bayansar, Francisco Bernard, Rebecca Fiebrink, Alexis Heloir, Edgar Hemery, Sohaib Laraba, Alexis Moinet, Fabrizio Nunnari[5], Thierry Ravet, Loïc Reboursière, Alvaro Sarasua, Mickaël Tits, Noé Tits, and François Zajéga*

## Project #2:

### Analysis of the qualities of human movement in individual action

*Paolo Alborno, Ksenia Kolykhalova, Emma Frid, Damiano Malafronte, and Lisanne Huis in't Veld*

## Project #3:

### EASA : Environment Aware Social Agent

*Hüseyin Cakmak, Kevin El Haddad, Nicolas Riche, Julien Leroy, Pierre Marighetto, Bekir Berker Türker, Hossein Khaki, Roberto Pulisci, Emer Gilmartin, Fasih Haider, Kübra Cengiz, Martin Sulir, Ilaria Torre, Shabbir Marzban, Ramazan Yazıcı, Furkan Burak Bâgcı, Vedat Gazi Kılı, Hilal Sezer, Sena Büsra Yenge.*

## Project #4:

### MOMMA: Museum MOtion & Mood MApping

*Charles-Alexandre Delestage, Sylvie Leleu-Merviel, Muriel Meyer-Chemenska, Daniel Schmitt, Willy Yvart*

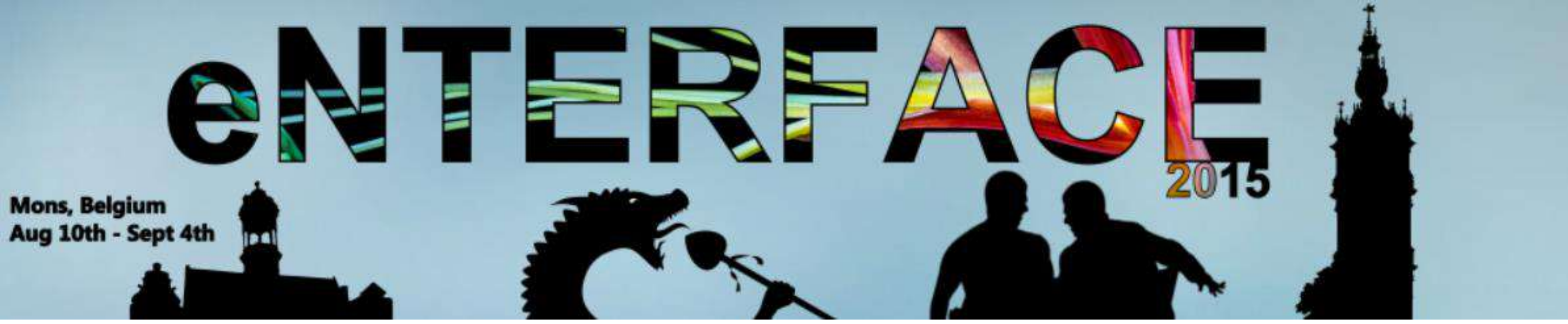

## Project #5:

VideoSketcher: Innovative Query Modes for Searching Videos through Sketches, Motion and Sound

*Stéphane Dupont, Ozan Can Altiok, Aysegül Bumin, Ceren Dikmen, Ivan Giangreco, Silvan Heller, Emre Külah, Gueorgui Pironkov, Luca Rossetto, Yusuf Sahillioglu, Heiko Schuldt, Omar Seddati, Yusuf Setinkaya, Metin Sezgin, Claudiu Tanase, Emre Toyan, Sean Wood, Doguhan Yeke*

## Project #6:

Enter the ROBiGAME: Serious game for stroke patients with upper limbs rehabilitation

*Françcois Rocca, Pierre-Henri De Deken, Alessandra Bandrabur, and Matei Mancas*

## Project #7:

BigDatArt: Browsing and using big data in a creative way

*Fabien Grisard, Axel Jean-Caurant, Vincent Courboulay, and Matei Mancas*

## Project #8:

Automatic detection of planar surfaces from uv maps

*Radhwan Ben Madhkour, Ambroise Moreau*

numediart
INSTITUT DE RECHERCHE POUR LES TECHNOLOGIES CREATIVES DE L'UMONS
UMONS
Université de Mons
fnrs
FREEDOM TO RESEARCH
MONS 2015

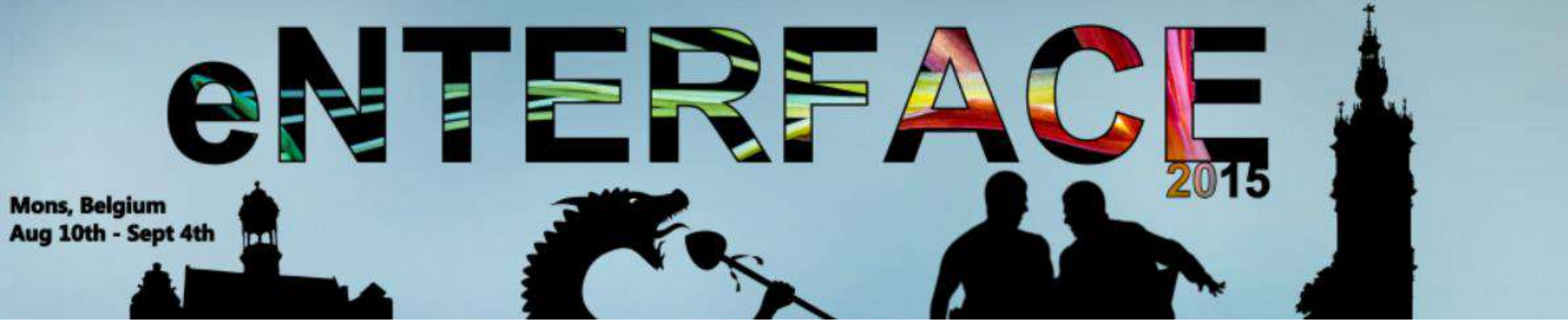

# Preface

The 11th Summer Workshop on Multimodal Interfaces eNTERFACE'15 was hosted by the Numediart Institute of Creative Technologies of the University of Mons from August 10th to September 2015. During the four weeks, students and researchers from all over the world came together in the Numediart Institute of the University of Mons to work on eight selected projects structured around "multimodal interfaces".

The eNTERFACE workshops aim at establishing a tradition of collaborative, localized research and development work by gathering, in a single place, a team of leading professionals in multimodal human-machine interfaces together with students (both graduate and undergraduate), to work on a prespecified list of challenges, for four complete weeks. In this respect, it is an innovative and intensive collaboration scheme, designed to allow researchers to integrate their software tools, deploy demonstrators, collect novel databases, and work side by side with a great number of experts. It brings together dozens of researchers for a whole month, subsequently it is the largest workshop on multimodal interfaces.

The eNTERFACE was initiated by the FP6 Network of Excellence SIMILAR. It was organized by Faculté Polytechnique de Mons (Belgium) in 2005, University of Zagreb (Croatia) in 2006, Bogaziçi University (Turkey) in 2007, CNRS-LIMSI (France) in 2008, University of Genova (Italy) in 2009, University of Amsterdam (The Netherlands) in 2010, University of West Bohemia (Czech Republic) in 2011, Metz Supélec (France) in 2012, New University of Lisbon (Portugal) in 2013, and University of Basque Country (Spain) in 2014.

eNTERFACE'15 returned to Mons in 2015, exactly ten years after the first workshop held in the same place. This years was also a special year for the city of Mons which was European Cultural Capital.

We would really like to thank the entire local organizing committee which made this event a great success both from a professional and personal point of view.

And finally, we would like to thank to all the participants to this workshop for coming in Mons!

Matei Mancas and Christian Frisson

*Publication chairs*

numediart
INSTITUT DE RECHERCHE POUR LES TECHNOLOGIES CREATIVES DE L'UMONS
UMONS
Université de Mons
fnrs
FREEDOM TO RESEARCH
MONS 2015

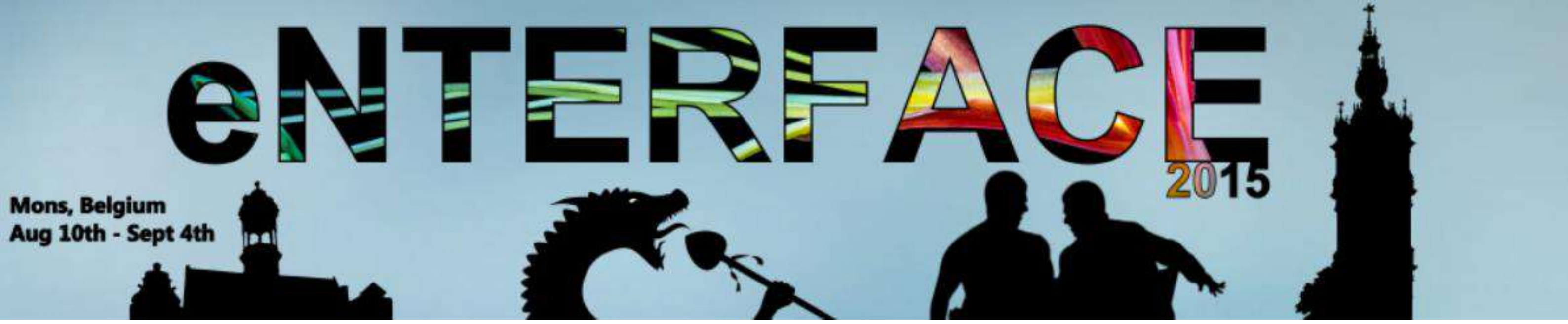

numediart
INSTITUT DE RECHERCHE POUR LES TECHNOLOGIES CREATIVES DE L'UMONS

UMONS
Université de Mons

fnrs
FREEDOM TO RESEARCH

MONS 2015

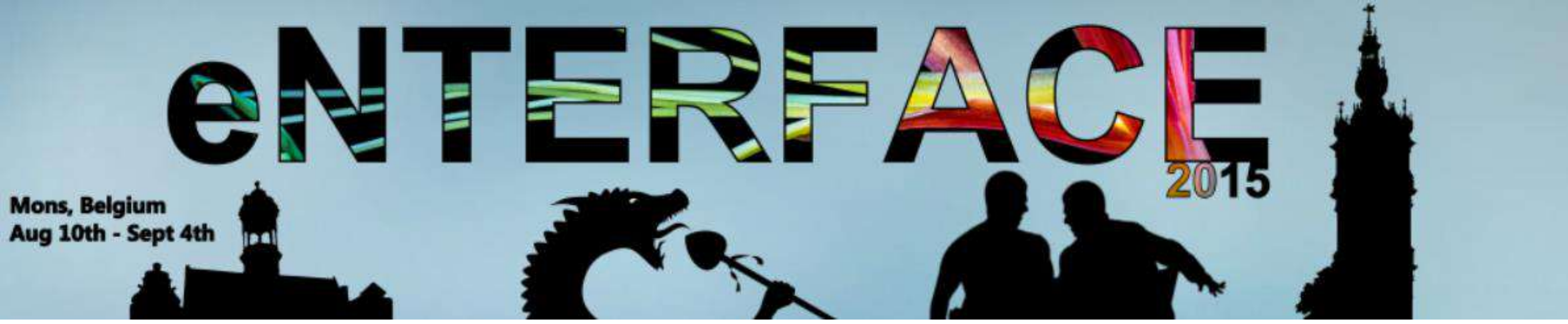

## Project #1:

# Prototyping a New Audio-Visual Instrument Based on Extraction of High-Level Features on Full-Body Motion

*Jo�elle Tilmanne, Nicolas d'Alessandro, Petr Barborka, Furkan Bayansar, Francisco Bernard, Rebecca Fiebrink, Alexis Heloir, Edgar Hemery, Sohaib Laraba, Alexis Moinet, Fabrizio Nunnari[5], Thierry Ravet, Loïc Reboursière, Alvaro Sarasua, Mickaël Tits, Noé Tits, and François Zajéga*

# Prototyping a New Audio-Visual Instrument Based on Extraction of High-Level Features on Full-Body Motion

Joëlle Tilmanne[1], Nicolas d'Alessandro[1], Petr Barborka[2], Furkan Bayansar[3], Francisco Bernardo[4], Rebecca Fiebrink[4], Alexis Heloir[8,5], Edgar Hemery[6], Sohaib Laraba[1], Alexis Moinet[1], Fabrizio Nunnari[5], Thierry Ravet[1], Loïc Reboursière[1], Alvaro Sarasua[7], Mickaël Tits[1], Noé Tits[1], and François Zajéga

[1] Numediart Institute, University of Mons, Belgium
[2] Cybernetics, University of West Bohemia, Czech Republic
[3] Middle East Technical University, Ankara, Turkey
[4] Goldsmiths University, London, United Kingdom
[5] German Research Centre for Artificial Intelligence, Germany
[6] MINES ParisTech, Paris, France
[7] MTG, Universitat Pompeu Fabra, Spain
[8] LAMIH UMR CNRS/UVHC 8201, Valenciennes, France

**Abstract.** Skeletal data acquisition generates a huge amount of high-dimensionality data. In many fields where motion capture techniques are now used, practitioners would greatly benefit from high-level representations of these motion sequences. However meaningful motion data dimensionality reduction is not a trivial task and the selection of the best set of features will largely depend on the considered use case, hence enhancing the need for a fast customization and prototyping tool. In this work, we present a prototyping tool for motion representation and interaction design based on the MotionMachine framework, as well as use cases of new audio-visual instruments that use full-body motion to drive sound and visuals. These instruments have been developed using the proposed prototyping environment. The development of these instruments is a proof of concept and demonstrates the potential of an elaborate choice of higher-level feature extraction techniques in improving the human/computer interaction and leading to more expressive experiences.

## 1 Introduction

### 1.1 Context

Over the last ten years, an important amount of motion capture techniques have emerged. However most of these techniques – such as inertial suits or optical markers tracking – did remain expensive and cumbersome. More recently, the democratization of depth cameras like the Microsoft Kinect has considerably changed the scope of markerless mocap research. Indeed the massive dissemination of these sensors gave many new research groups the chance to jump in the field and provide various resources (databases, software, results) to the scientific community [1]. This technological breakthrough has brought motion capture into new application domains, like health [2], education or arts [3].

Skeletal data acquisition generates a huge amount of high-dimensionality data. Such raw data is neither very readable, nor very reusable. The design of interesting mapping strategies for gesture-controlled interaction is often hampered by these considerations and very basic gestures are usually selected for such interaction because of the lack of existing or readily-available higher-level motion features. In many fields where motion capture techniques are now used, practitioners would greatly benefit from high-level representations of these motion sequences. Such features could be understood by sight or highly related to the studied phenomenon (expertise, style, etc.), so that they could be mapped with sonification, visualization or used in machine learning. In literature, we can find e.g. top-down description of full-body motion [4], geometrical descriptors of the 3D skeleton [5] or application of dimension-reduction techniques to extract higher-level aspects of gestures. Our perspective on this situation is that there is no generic solution to motion high-level feature extraction. The optimal solution will be highly dependent on the use case. There is therefore a need for a very intuitive prototyping platform, where a set of feature extraction modules can quickly be plugged in and their efficiency evaluated in real-time, both from offline and online skeletal data. The Numediart institute has developed such a tool for real-time prototyping of motion data processing, called MotionMachine [6]. It is used as an independent feature extraction layer, taking OSC data from various sensors and streaming OSC features in real-time. It is a C++ library and it currently has bindings with openFrameworks for the visualization module.

Some of our previous works on gesture recognition, mapping and synthesis [7] have demonstrated the need for high level motion feature extraction in order to reduce the dimensionality of the motion data and hence reduce the complexity and improve the performances of motion recognition models, especially when the training data is scarce. These approaches drew the lines of a combined optimization of both motion representation and motion recognition models.

### 1.2 Related Works

Several tools exist to manipulate and visualize motion capture data with the goal of creating motion-enabled applications. The MOCAP toolbox [8] is among these tools, giving access to many high-level processing functions, but it is only available for Matlab and therefore not very suitable for real-time, iterative and interactive testing and performance. When it comes to performance-driven tools, we find software like the RAMToolkit [9] with the opposite issue: the tool is not generic and rather very specific to a single use case (Kinect-captured dancers or specific mocap markers placement). In the field of mapping software for new interface design, the complexity of motion capture data is still not properly faced. For instance, tools like Max or LibMapper [10] lack of serious mocap toolkits with high-level feature extraction mechanisms. Most of the time, such abstraction of raw data is achieved and fine-tuned for a specific use case.

### 1.3 Outline of the Paper

In this paper, we will first present the MotionMachine framework, a new C++ environment for the rapid prototyping of motion capture based interaction (see Section 2). In Section 3, we further develop our approach towards motion feature extraction within MotionMachine and give an overview of motion feature extraction categories and processors that have been implemented in this project. Section 4 highlights the different mapping and synthesis use cases that have been developed as proofs of concept. Finally we drive some conclusions on the project and highlight key perspectives that will lead to further research in Section 5.

## 2 MotionMachine Framework

MotionMachine is an open source C++ library that enables the rapid prototyping of motion features, their extraction on standardized motion capture data structures coming from typical motion capture file formats and live OSC streams and their selection so as to represent motion in the considered use case. It has first been introduced in [6]. In this Section, we first give an overview of the main structure of MotionMachine and then focus on what has been refined in this project: *timed containers*, i.e. the underlying data-encoding backbone enabling for both offline batch and real-time processing.

### 2.1 General Structure

The overall data flow used in MotionMachine is presented in Figure 1. The library is built from two independent modules: one for data representation and feature extraction (built on the top of the *Armadillo* C++ library [11]), the other to take care of 2D and 3D scenes visualization and general user interaction aspects (built on the top of the *openFrameworks* C++ library [13]).

Four important core features are available in the MotionMachine framework:

1. *Skeletal Model Independent Motion Data*: Nowadays the number of available motion capture sensors is significantly growing. Most of these devices provide skeletal data, i.e. changes in the position and/or orientation of 3D joints and/or segments. For the moment, most of the achieved motion processing is model-specific and the corresponding know-how is not directly transferable to another model, though the underlying morphology is similar. In MotionMachine, we have developed a model-independent formalism for storing and accessing such skeletal data through APIs. The back-end is based on timed containers, which are further explained in the next Section. The front-end gives an easy-to-understand access to a few nested data structures [6].

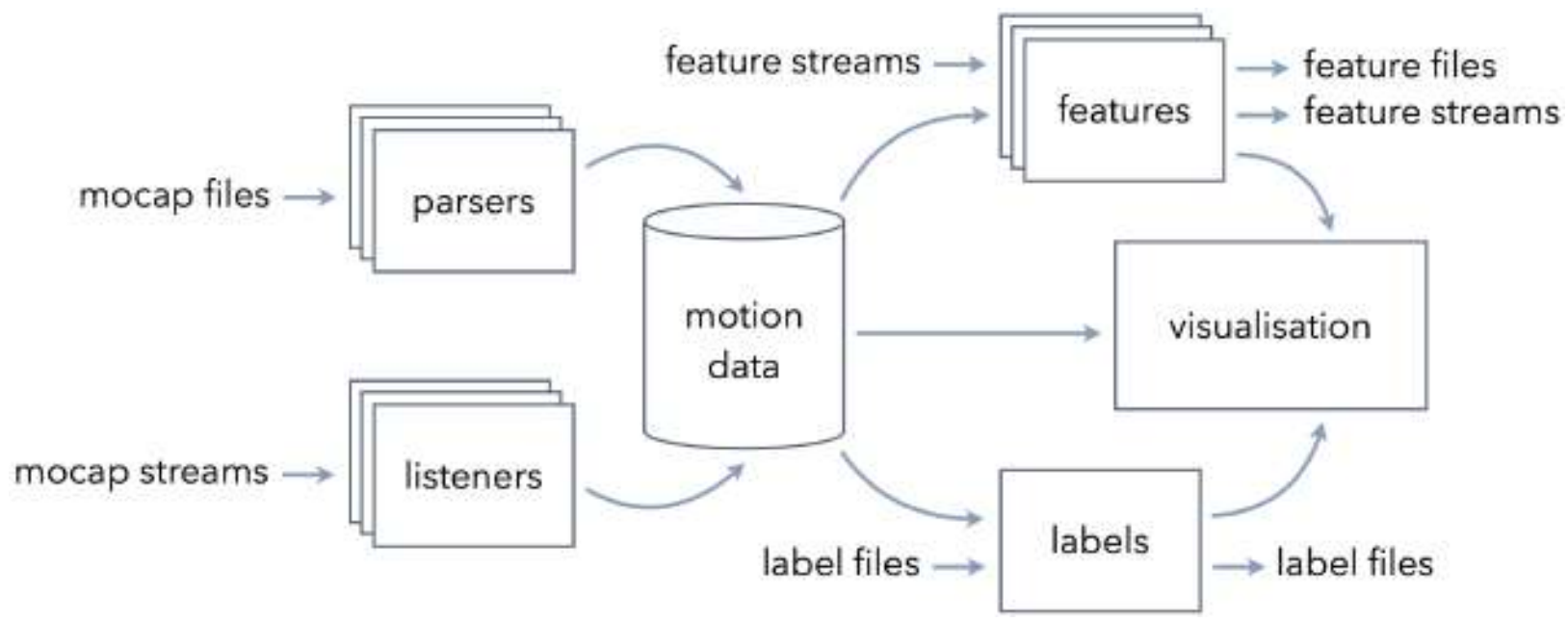

**Fig. 1.** Overall data flow used in MotionMachine: modular structure to process mocap data files and/or streams into features files and streams, labels files and visualization.

2. *Collections of Motion Feature Extractors*: MotionMachine is built around the idea that developers can write custom code to be inserted in the motion capture data processing pipeline, while still preserving the intuitiveness and efficiency of the overall environment. The design principle underlying the available collection of feature extractors is essentially container-driven and based on the idea that offline batch and windowed real-time processing should both be available by default for any built-in or third-party feature. We give an in-depth description of this idea in Subsection 2.2.

3. *Interactive 2D/3D Scene View*: In MotionMachine, we wanted to improve the affordance of motion capture data processing by solving several visualization issues and bring the user faster to his/her valuable work. Such improvements were achieved by balancing the apparent complexity of the environment. As a result, the library comes with an integrated 2D/3D scene viewer for displaying mocap data on screen and interacting with the contents [6]. On the one hand, many visualization aspects are automatically determined from data types, included the so-called "degraded" modes where not all the information is available. On the other hand, 3D and 2D time lines are automatically synchronized (embedded app time handlers), helping to observe available motion capture data from different viewpoints.

4. *Annotation Layer*: In MotionMachine, we have integrated a lightweight annotation scheme. It allows the programmatic and UI-based insertion of *Labels* alongside the motion capture data. It means that the time tag of these *Labels* can be automatically derived from signal properties in the feature extraction code or added manually by the user. Moreover these *Labels* get properly rendered in the 2D view and can then be rearranged manually. *Labels* can also be imported from and exported to label text files (*.lab* extension).

### 2.2 Timed Containers

Time management can be a serious issue in motion capture based interactive applications: the application has its own life cycle and time line, input files may have their own frame rate, time stamps or nothing, input streams may have their own frame rate, time stamps or nothing, and any piece of data might reveal opportunities to be processed as one batch (offline), frame by frame or within a sliding window. Moreover, as MotionMachine is intended to be open-ended, i.e. with third-party feature extractors, we aim at optimizing the modularity of the code base, e.g. not duplicate code for offline vs. real-time scenarios.

In this project, we have decided to foster the role of underlying data containers in the process of framing the way time and skeletal structures are handled in MotionMachine, as well as how the feature extractors should be developed. Our *timed containers* aim at providing all the logic that enables the processing of motion capture data files and streams. The two main behaviors are:

1. *Hierarchical structure of data is transparent*: Motion capture data and more generally 3D animation data are described by using a hierarchical structure. On one hand, the position and the orientation of a 6 DOF limb can be encoded with absolute values in a global world coordinate system. On the other hand these data can be computed in local coordinate system linked to a parent object, e.g. foot orientation is given relatively to the data of the leg. Following the aimed application, local or global values are the most appropriate description. The timed containers implement this kinematic chain and allow the user to choose between global or local motion descriptors.

2. *Time management within data is transparent*: In order to enable the processing of data extracted from different sources, MotionMachine provides a time modeling that respects two main constraints. Firstly the timed container accepts data with a fixed frame rate or frames with time stamps, if they have not been recorded with a unique temporally stable device. Secondly the containers are able to store data from a real-time streams and data from offline prerecorded files. To do that, the container can switch between a ring buffer mode and a regular buffer mode, i.e. with a directly indexed-timed relation as showed in Figure 2. Series of APIs are provided to the developer so as to access the data at a given time, no matter the underlying time metric.

## 3 Motion Feature Extraction

One very transversal aspect of this project was to grow an amount of motion feature extractors that can be applied on various kinds of input data (files or streams). In this Section, we give an overview of what we aimed at keeping common among all the implemented features and then we give a list of feature categories and implementations that have been achieved.

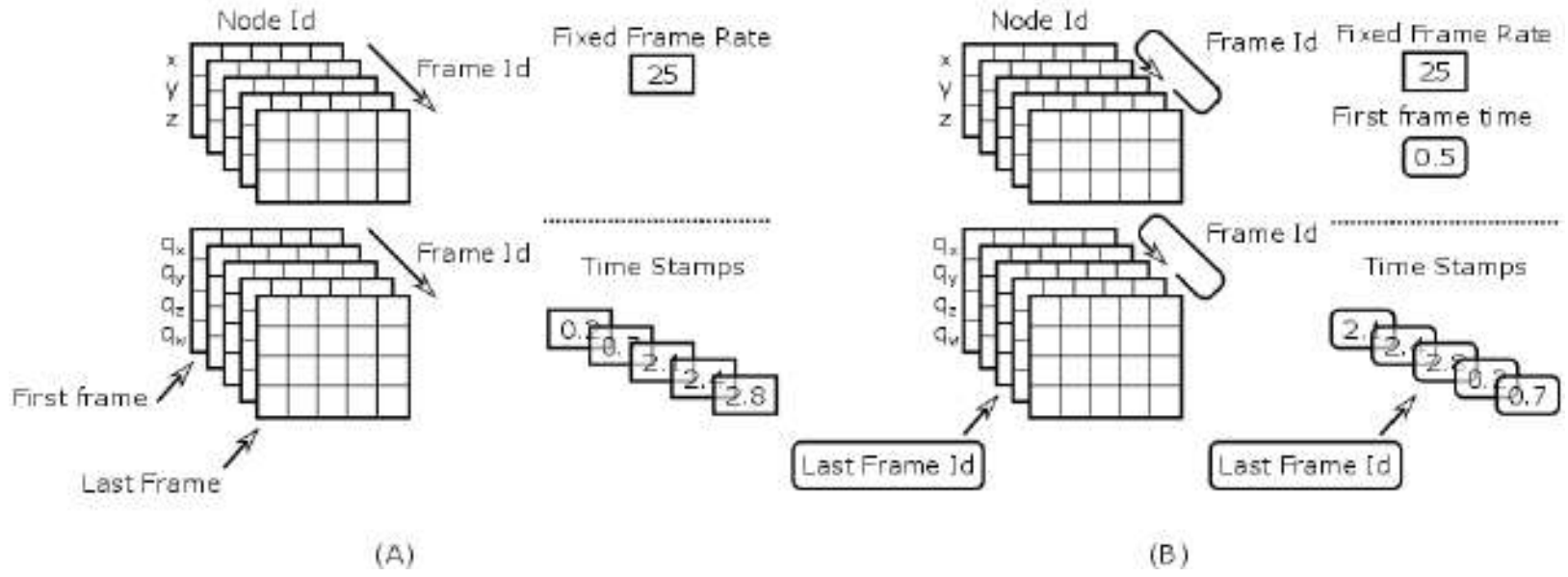

**Fig. 2.** Representation of the four modes to store the motion capture data in memory buffer. Position of each node is encoded by 3 Cartesian coordinates; rotation of each bone is encoded with 4 components of a unit quaternion. (A) Example of a track working in an offline mode. (B) Example of a track in real-time mode. In each case, the time can be modeled with a fixed frame rate or a time stamp vector. All time-related parameters are updated to insure a transparent management for the developer.

We acknowledged this idea that feature extractor developers should not be "forced" to follow a DSP chain format. Indeed the variability within motion capture based use cases is quite significant and we do not want to load MotionMachine with a complex and cumbersome patching environment, nor preventing developers to design their software following their scenario-driven constrains.

The one constraint that we put onto the development of motion feature extractors within MotionMachine is to take timed containers as inputs and outputs arguments and work with their APIs for accessing structural and time-handling information. This design choice also has another implication: it makes a lot of sense to work with Armadillo for the vector and matrix operations vs. transferring data into another format then back. This choice is motivated by the ability of Armadillo to do parallel computation on arrays using CPU optimization.

Ten different feature categories have been implemented and are presented here under. Six are high level motion feature categories, for each of which several feature extraction functions have been implemented inside the MotionMachine framework. Three are gesture recognition algorithms, either fully implemented within MotionMachine (relation features and DTW), based on external libraries (HMM-based gestures recognition based on the HTK library), or communicating with external modules (e.g. Wekinator). The last feature category consists in parsing data extracted offline in the OpenSIM software.

### 3.1 Balance

The balance/weight category represents the idea of stability, balance and equilibrium, which can be perceived when looking at a skeleton with a given posture. It encapsulates concepts such as center of mass (CoM), floor contact points and base of support, in both continuous and binary ways: continuous features give an evolving metrics about the studied property, binary features represent a state.

To compute these high level balance features, some intermediate features and geometrical operations are required and presented further in this Subsection.

### Center of Mass (CoM)

*Description* The center of mass (CoM) of a body is the point at which the whole mass may be considered as concentrated [15]. The center of mass of the whole body can be computed as the weighed average of the center of mass of each body segment in 3D space. The "center of mass" feature is tridimensional since it is computed as the 3D-coordinates ($xyz$ position) of the center of mass of either one of the body segments or of the whole body.

*Computation* The position of the center of mass of a body depends on the distribution of the mass in the considered body. For the human body, several tables can be found in the literature and give an estimation of the position of the center of mass for each of the human body segments under the form of a ratio between the extremities of the considered segment. We have based our implementation of the center of mass computation on de Leva's work [16].

For each body segment, the COM is computed as the ratio between the position of both extremities of the considered segment, using the ratios found in [16]. The global center of mass is computed as a weighted sum of the position of the center of mass of each one of the body segments, the weight being defined by the relative weight of each segment of the body as presented in de Leva's paper.

$$CoM_{global} = \sum_{i=1}^{N} w_i * CoM_i \qquad (1)$$

with $N$= number of body segment considered, $w_i$ the weighted mass of body segment $i$ and $CoM_i$ the center of mass of segment $i$.

*Visualization* A visualization of the center of mass of each segment as well as of the global center of mass has been implemented. The center of mass of each body segment is represented by a green sphere whose center is located at the $xyz$ coordinates of the center of mass of the considered segment and whose diameter is proportional to the weighted mass of the segment. The global center of mass of the body is represented by a red sphere, as illustrated in Figure 3.

### Floor Contact Points

*Description* When considering motion sequences performed on a standard planar ground, the equilibrium or balance of the body is highly dependent on the contact points of the body with the ground. This feature calculates which joints of the body can be considered as in contact with the ground.

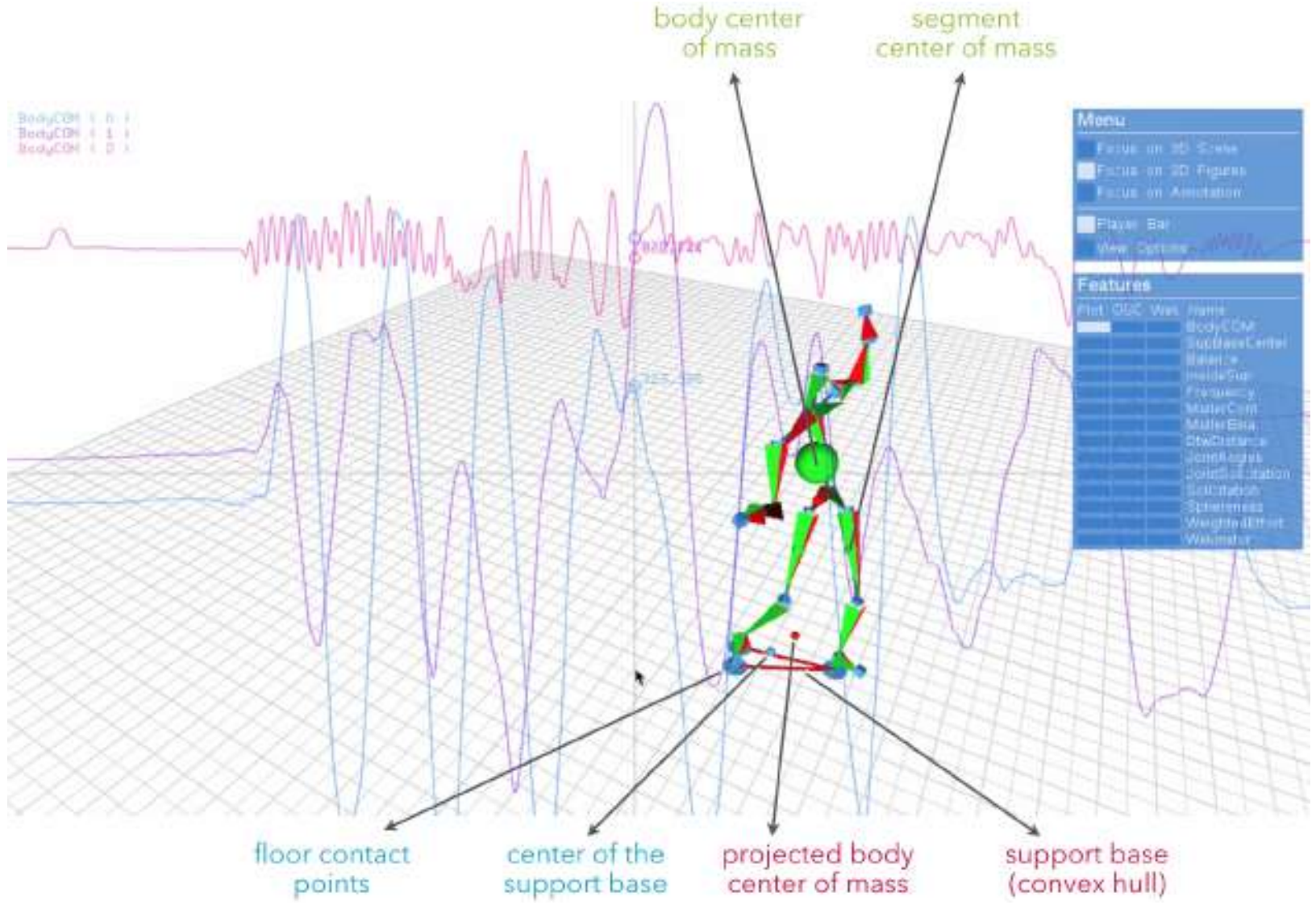

**Fig. 3.** Illustration of the Balance feature elements: body center of mass (with $xyz$ position in 2D), segment center of mass, floor contact points, support base (convex hull), center of support base and projected body center of mass.

*Computation* For each joint, this feature calculates if it is touching the ground or not. The binary decision is based on the height of each joint, using a threshold which can be adjusted for each joint, since the physical position of the mocap markers on the actor's body will change the height of the marker when the considered segment is actually touching the ground. By default, the threshold is set to 5cm, except for foot and ankle, for which it is computed as the mean height of the first ten frames of the motion sequence + $1.5cm$).

*Visualization* Each floor contact point is displayed as a blue circle on the ground, using for its center position the x and y position of the corresponding joint.

**Convex Hull of Support Base**

*Description* The convex hull of a set of points is the smallest convex envelope that contains all of the points in the set. This geometrical feature is used in our case for instance to compute the contact base area based on the set of contact points with the ground. It can also be used to compute the area corresponding to the space occupation of the body on the ground at any given time, or to compute the area covered by a subject during a given motion sequence by computing the convex hull of the positions of the center of the body over time.

*Computation* The computation of the convex hull is and iterative process. From a set of points in a plane (defined by $xy$ positions), it computes the set of points defining the convex hull containing all the points. Our implementation of the convex hull computation is based on [17]. From the defined convex hull, typical metrics can be determined such as contact base perimeter or area.

*Visualization* The implemented convex hull visualization consists in drawing the lines between the points defining the convex envelope, so as to show the area.

**Binary Balance Value**

*Description* In MotionMachine, the discrete value of the balance feature is named $isInsideCOMSupport$. This feature calculates if the projection of the body center of mass on the ground is inside the support base or not. The feature can take one of three values at each frame: 1 if the $CoM_{global}$ is inside the support base, 0 if it is outside, and $-2$ if there is no support base (for instance when the subject is jumping). This feature can be seen as a ternary balance feature since it represents the equilibrium of the posture at each frame.

*Computation* This discrete balance feature is computed by testing if the projection of the $CoM_{global}$ on the ground is inside the convex hull defined by the support base. This test of being inside is achieved by testing if the projected point is on the right of each of the successive segments of the hull.

*Visualization* The projection of the global center of mass on the ground is represented as a green sphere if the $CoM_{global}$ is inside the support base, a red sphere if it is outside, and an orange one if there is no support base.

**Continuous Balance Value**

*Description* This continuous representation of the balance of the body is computed as the distance between the center of the support base and the projection of the body center of mass on the ground. When used in combination with the discrete value of the balance ($isInsideCoMSupport$), it illustrates how far from the equilibrium the current body posture is. The higher the feature is, the poorer the balance is. Another implementation that should be considered in the future would be to compute the distance from the border of the support base area (inside or outside) rather than the distance from the center of that area.

*Computation* The center of the support base is computed as the mean of the support joints positions. The continuous balance feature is then computed as the distance between the center of support base and the projection of the $CoM_{global}$ on the ground. Small distance means that the $CoM_{global}$ and the center of the support based are more aligned vertically, large distance the opposite.

*Visualization* The visualization of the continuous balance value has been done in the 2D view. Indeed peaks and valleys in the graph would inform quite clearly about moments where the body was particularly (un)balanced.

### 3.2 Rhythm and Periodicity

This Section describes our work in implementing a first set of periodicity-related features. It goes from low-level analysis to fundamental frequency estimation.

#### Auto-correlation

*Description* For a given signal, its auto-correlation is the cross-correlation of the signal with itself. The cross-correlation between two signals of length $N$ measures their correlation for all possible delays $0, \ldots, N-1$. In the case of a pseudo-periodic signal, the auto-correlation would present a peak for each delay that is a multiple of the fundamental period, thus providing a way to detect whether a signal is periodic and to measure its period, if any. This feature takes as input a 1D signal whose values correspond to a temporal sequence of either node positions, angles or any feature described in Section 3.

*Computation* The input signal of length $L$ is split into blocks, or *analysis windows*, of length $N$ with a shift of $H$ samples between the first samples of two successive blocks. For each analysis window $x_h(n)$ – with $h = 0, 1, \ldots \frac{L-N}{H}$ – its auto-correlation $R_h(m)$ is computed as:

$$R_h(m) = \mathcal{F}^{-1}(|\mathcal{F}(x_h(n)\, w(n))|^2) \tag{2}$$

where $w(n)$ is a *weighting window* and $\mathcal{F}$ and $\mathcal{F}^{-1}$ represent the *Discrete Fourier Transform* (DFT) and its *Inverse* transform (IDFT) respectively. The DFT and IDFT are computed over $2N$ points but only the first $N+1$ points are stored in the final result as $R_H(m)$ is symmetrical. Optionally, the average value of $x_h(n)$ can be removed before computing the DFT. The Equation becomes:

$$R_h(m) = \mathcal{F}^{-1}(|\mathcal{F}((x_h(n) - \bar{x}_h(n))\, w(n))|^2) \tag{3}$$

This latter approach gives better results, especially when applied to node positions with a non-zero long-term average (typically, the vertical position of the head). Finally, the auto-correlation vectors $R_h(m)$ are stored in an array $R$ with $\frac{L-N}{H}$ columns and $N+1$ lines.

*Visualization* The auto-correlation array $R$ can be displayed as a 2D curve changing over time. Shortly, the $h^{\text{th}}$ column $R_h(m)$ can be shown on screen for all time stamps corresponding to frame indices $f$ such that: $h \times H \leq f < (h+1) \times H$.

#### Power Spectral Density

*Description* The Power Spectral Density (PSD) is a measure of how the power of a given signal is distributed across the frequency domain. Random signals tend to have a continuous PSD and periodic signals exhibit some peaks.

*Computation* The computation is similar to that of the auto-correlation. An input signal is split into analysis windows $x_h(n)$ and a PSD $S_h(m)$ is computed for each $x_h(n)$ and stored in an array $S$. The computation is made by means of the squared absolute value of the DFT of $x_h(n)$ over $N$ points:

$$S_h(m) = |\mathcal{F}(x_h(n)\, w(n))|^2 \tag{4}$$

Note that a weighting window $w(n)$ is applied before computing the DFT. Since the result is symmetrical, we store only $N/2 + 1$ points in $S$. Also note that the average value $\bar{x}_h(n)$ can be subtracted before computation:

$$S_h(m) = |\mathcal{F}((x_h(n) - \bar{x}_h(n))\, w(n))|^2 \tag{5}$$

*Visualization* Likewise, the visualization for $S$ is similar to $R$, except that values in $S$ are always positive and better represented on a logarithmic scale (dB).

**Periodicity from Peak**

*Computation* This feature takes as input the array resulting from one of the two periodicity-related features here above (Auto-correlation or PSD). For each frame of the input, it looks for a peak in a region that corresponds to acceptable values of periodicity (e.g. a human being cannot jump repeatedly at 100Hz but 2Hz would be reasonable). A peak is defined as a value in a frame that is larger than its $2K$ direct neighbors and above a given threshold. Results using the auto-correlation function were much better and smoother than with the PSD.

*Visualization* As presented in Figure 4, the periodicity can be visualized as a dot superimposed on the peak detected in the graph of the auto-correlation/PSD function, accompanied by an overlay text with the value of the periodicity (or the indication "no period" when no peak is found in the look-up region).

### 3.3 Ergonomics

Ergonomic features are related to bio-mechanical aspects of the movement. They are based on musculo-skeletal modeling of the body, aiming to describe the quality of movement in terms of comfort, robustness, or load.

**Postural Load**

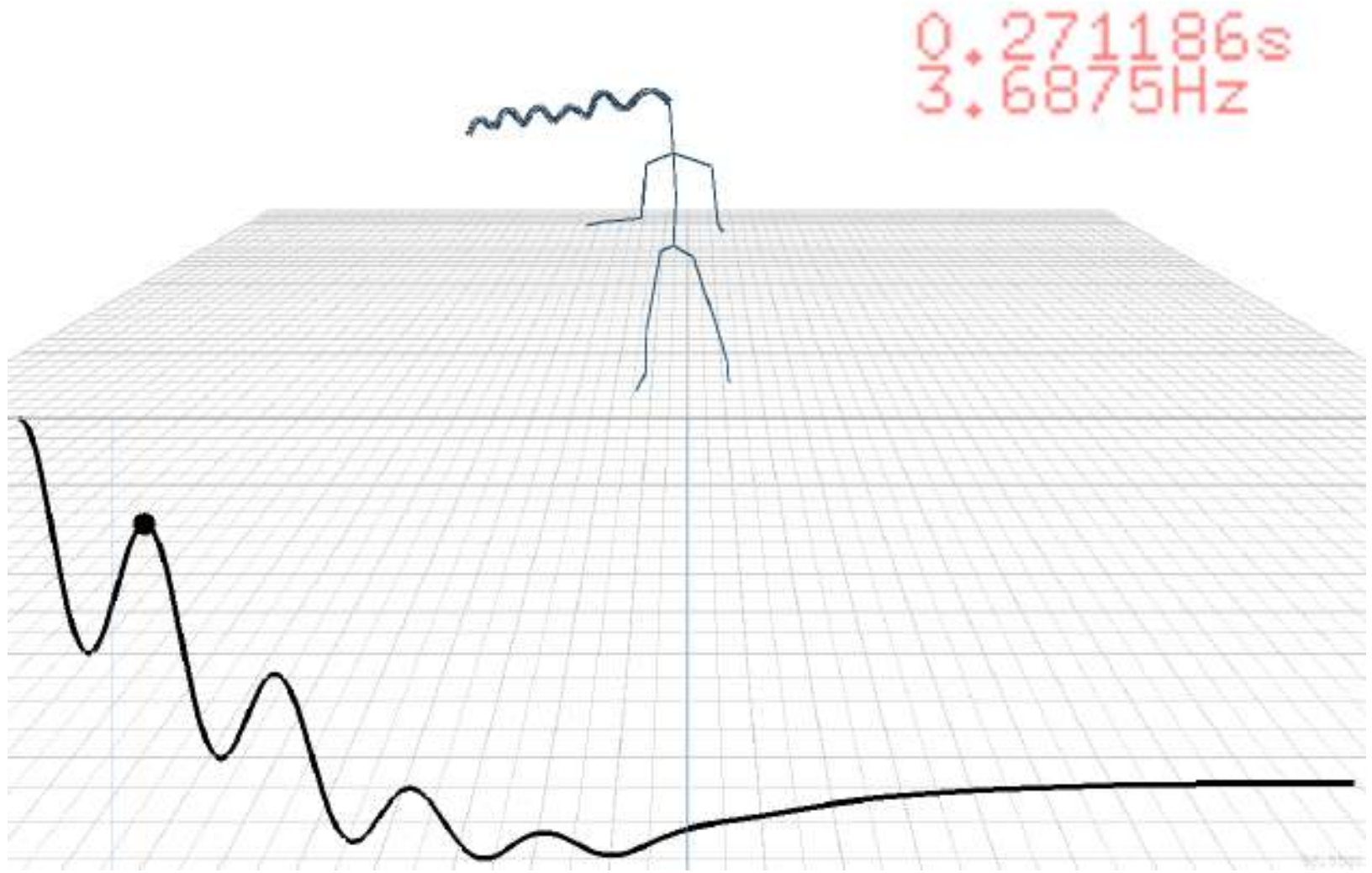

**Fig. 4.** The recorded movement is that of a body hopping repeatedly, as we can see in the trajectory of the head (shown attached to the head on graph). The auto-correlation function (black curve) of the Z position of the head presents a peak (black dot) at a position corresponding to the periodicity of the hops (0.27 s or $3.7Hz$ in this case).

*Description* The postural load of a body joint is an indication of the perceived stressfulness of a joint, according to its orientation, taking into account each degree of freedom (DoF) of the joint. The postural load of the whole body provides information on the comfort of a posture, and on the risk of injury. The postural load allows assessment of the comfort of a posture, and is widely used in industrial ergonomics, but it can also be used in performance assessment.

*Computation* The computation of the postural load requires mocap data containing joints orientations. The algorithm defined below is inspired from Andreoni et al. (2009) [22]. For each DoF of a joint, a relation exists between the perceived discomfort and the orientation of the joint. Tables of average perceived discomforts according to the orientation of a joint on a DoF have been set in the ergonomics literature to report these relations in [20] and [21].

The load, or stress of a joint $S_j$ can be defined as the sum of independent perceived discomforts associated with each DOF of the joint $S_{ij}$. In our case, the perceived discomforts $S_{ij}$ were spline-interpolated from the tables, as proposed in [22], in order to have continued values:

$$S_{ij} = f_{ij}(\theta_{ij}) \tag{6}$$

where $\theta_{ij}$ is the orientation of the joint j on the axis i, and the function $f_{ij}$ represents its relation based on spline functions and the aforementioned tables:

$$S_j = \sum_{i=1}^{m_j} S_{ij} \tag{7}$$

where $m_j$ is the number of DoF of the joint j. The postural load results from the sum of each joint stress, as described in the previous Equation:

$$PosturalLoad = \sum_{j=1}^{n} S_j \tag{8}$$

Andreoni et al. (2009) [22] proposed to define the postural load as a weighted sum of each joint stress, where the weight for each joint is proportional to the mass of the corresponding distal body district. However, as the perceived discomforts, i.e. feelings of pain, are supposed to inherently depend on the mass of the distal body district, we omitted these weights in the current study.

*Visualization* Figure 5 shows an example of the postural load analysis on a contemporary dance performance. In the 3D scene, the dancer is rendered at the most constraining pose. The 2D curve shows the computed postural load, and the peak corresponding to the maximal postural load. The colored sliders in the canvas view on the left display the stresses of each joint. In the case of this specific pose, the most stressed joints are the hips and the shoulders.

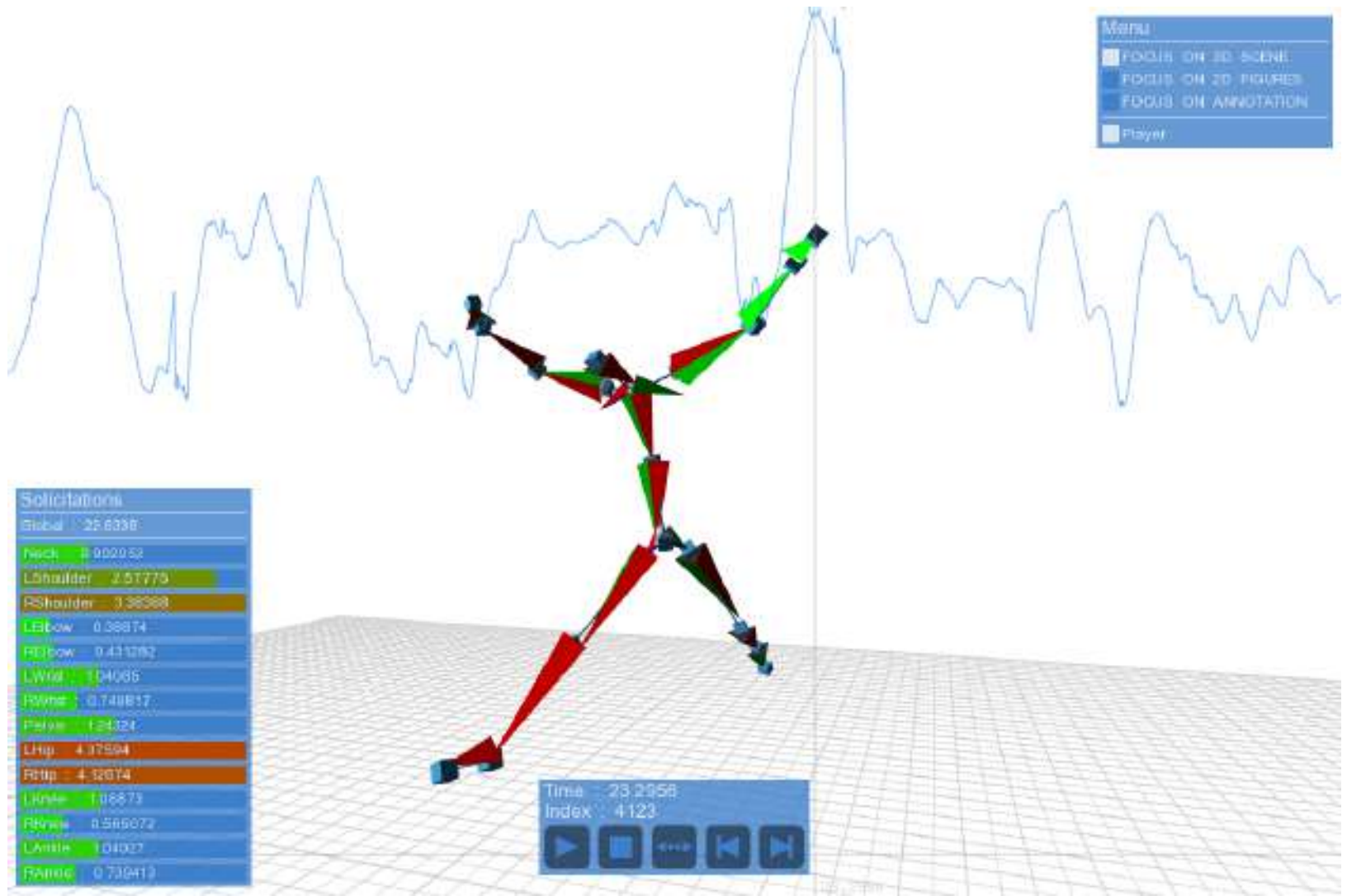

**Fig. 5.** An illustration of the postural load visualization in MotionMachine.

**Sphereness**

*Description* The sphereness is a novel exploratory feature inspired from a concept of Taijiquan. In the philosophy of Taijiquan, the body and the movements are represented through mental images, to enhance the transmission of the sensations of the art of Taijiquan from the teacher to the pupil. The sphere of the body is one of these images. The body is surrounded by an imaginary sphere, and controls the sphere with its movements. This image fosters the idea of moving the body as a whole rather than individuated segments. Empirical principles of Taijiquan rule the way the body interacts with the sphere: the body must fit the shape of the sphere, and the sphere must not be too wide nor too small. These empirical concepts can be linked to more concrete physical principles. The spherical shape of the body can be related to robustness, or to notions of balance, and the restrictions on the sphere expansion may be linked to Hill's muscle model (the force-length relationship), explaining that the muscles have less power when they are too expanded or too crouched. The sphereness feature tends to describe these relations between the body and the sphere.

*Computation* The sphereness is computed through three steps:

1. the body center of mass (CoM) is computed and is defined as the center of the sphere. See Equation 1 of Section 3.1) for its computation details;

2. the euclidean distance ($d_J$ is calculated between the each one of the five end-effectors of the body (head, hands and feet) and the body CoM:

$$d_J = |\overrightarrow{p_J} - \overrightarrow{p_{COM}}| \tag{9}$$

   for $J = \{head,\ left\ hand,\ right\ hand,\ left\ foot,\ right\ foot\}$;

3. the mean and the standard deviation of these distances are processed:

$$r = \frac{\sum\limits_{J=1}^{5} d_J}{5} \tag{10}$$

$$\sigma = \sqrt{\frac{\sum\limits_{J=1}^{5} (d_j - r)^2}{5}} \tag{11}$$

The radius $r$ of the sphere is defined as the mean of these distances, and indicates if the body is expanded or crouched. The standard deviation $\sigma$ indicates the deviation of the body from the spherical shape, i.e. its *sphereness*.

*Visualization* Figure 6 shows an example of sphereness visualization on a Taijiquan exercise. The pose shown on the left is a rest pose, and the pose shown on the right is the basic Taijiquan posture, also called the "Tree posture". The 3D scene shows two spheres. The big one is the aforementioned sphere, whose radius is the mean of the five distances of the five end-effectors to the CoM. The radius of the small sphere is the radius of the first one minus the deviation. In the 2D scene, the upper curve displays the radius of the sphere, and the lower curve displays the deviation. We can observe that the Taijiquan posture is as expanded as the rest pose, but is much closer to a spherical shape.

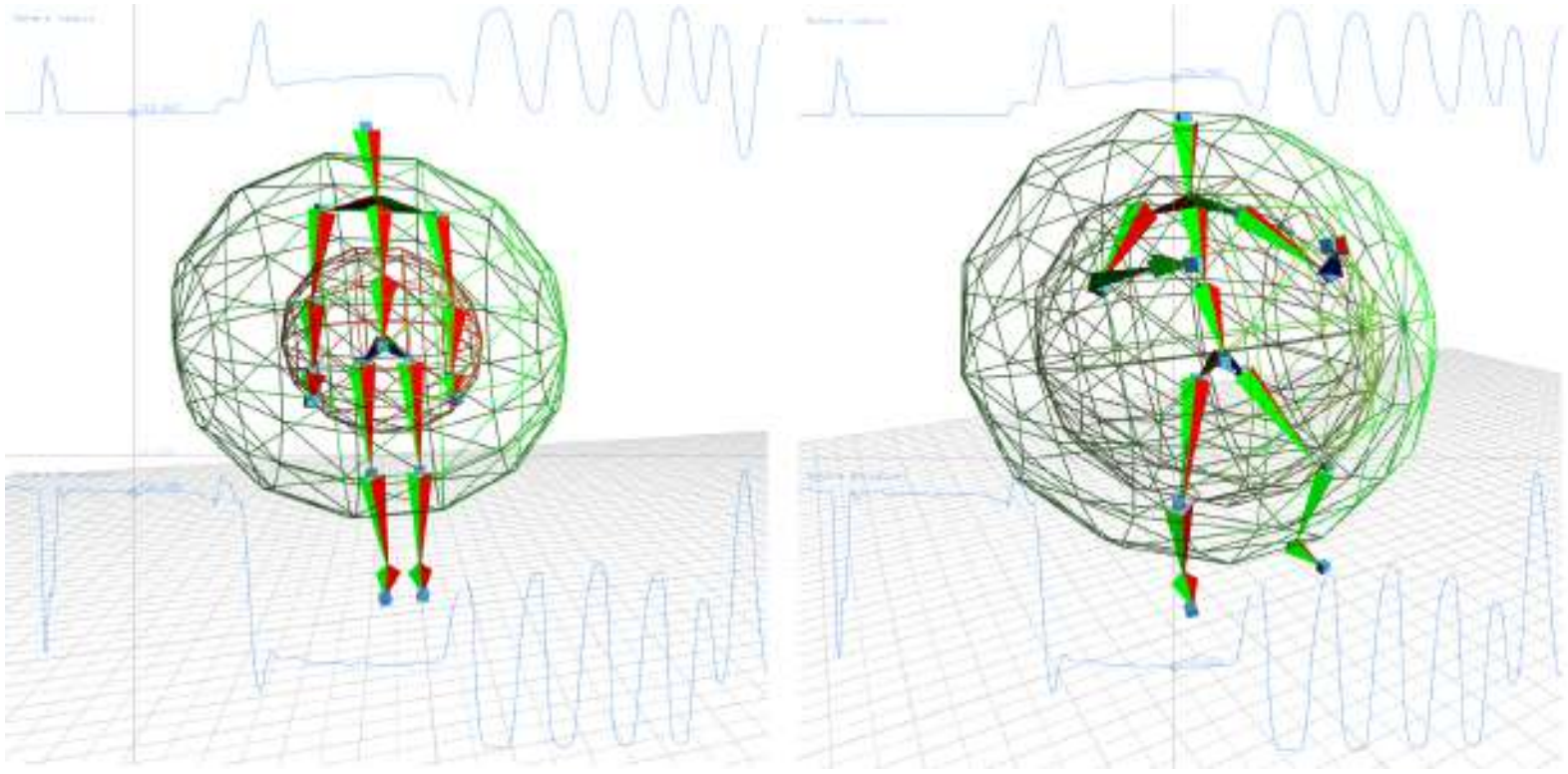

**Fig. 6.** An illustration of the sphereness visualization in MotionMachine.

### 3.4 Effort (Laban)

Effort features, derived from the Laban Movement Analysis (LMA) [4], describe the quality of the motion in various dimensions: dynamics, shape, space, expression, etc. Different categorizations can be found in literature but we focused on three of them for the implementation: weight, time and space. Several effort features take the kinetic energy feature as their starting point.

#### Kinetic Energy

*Computation* We compute an estimation of the kinetic energy $E^k$ of a body part $k$ (we can have one value per skeletal joint) as the square of its velocity:

$$E^k(t_i) = v^k(t_i)^2 \tag{12}$$

with $v^k(t_i)$, the velocity of the body part $k$ at time $t_i$. The kinetic energy of the whole body is determined as a weighted sum for all the body parts:

$$E(t) = \sum_{k \in K} \alpha_k . E^k(t_i) \tag{13}$$

with $\alpha_k$, the the normalized weight associated to body part $k$ [16]. Figure 7 shows the whole kinetic energy of the body during one performance, and kinetic energies corresponding to all the joints of a given skeletal model.

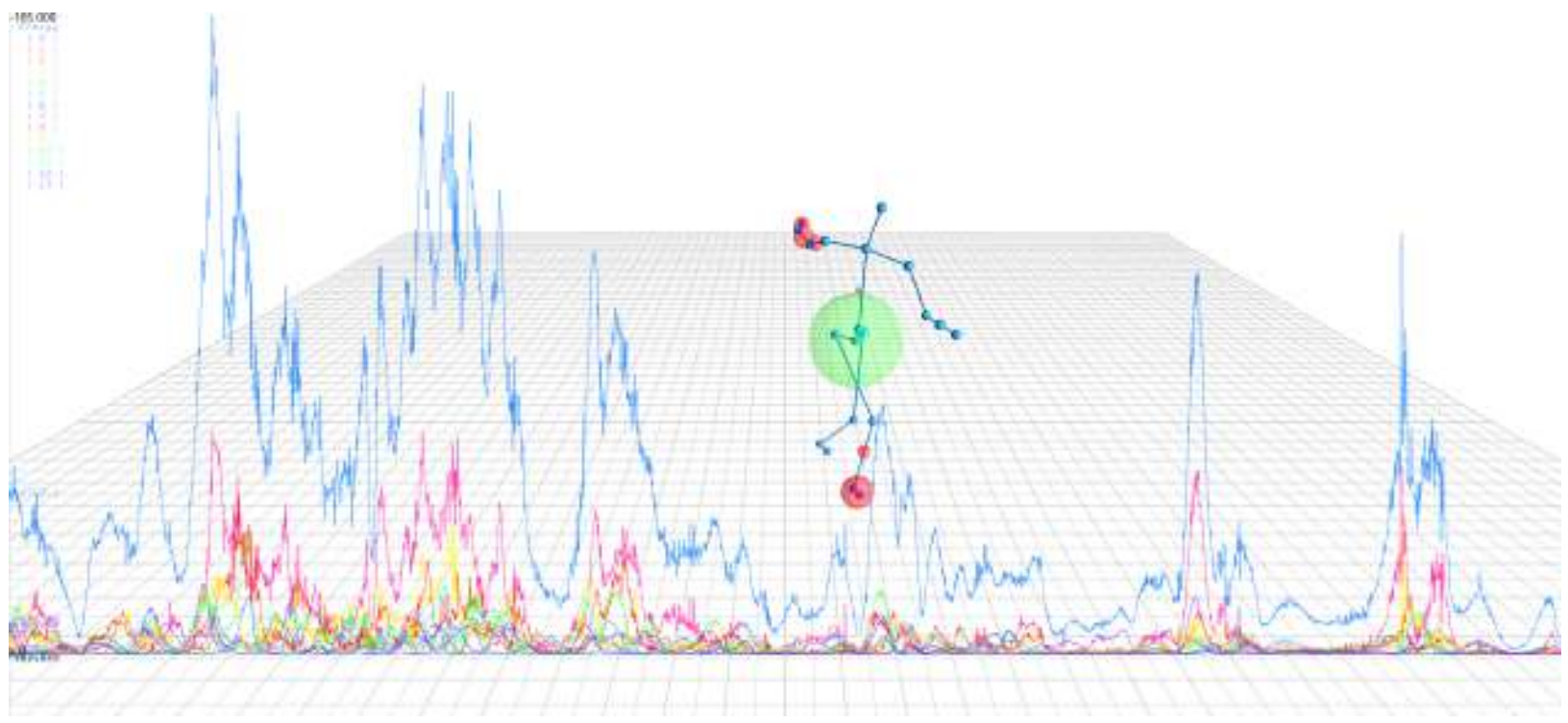

**Fig. 7.** Kinetic energy: Blue: Whole body. Other colors: different body parts.

### Weight Effort

*Description* The weight effort refers to physical properties of the motion. In LMA, the two opposite weights for a motion can be *strong* or *light*.

*Computation* The weight effort is estimated by computing the maximum of kinetic energy over a defined time interval [24], as shown in Equation below:

$$WeightEffort(T) = \max_{i=1:T}(E(t_i)) \tag{14}$$

with $T$, the size of the analysis window. Figure 8 shows the weight effort values computed for different window sizes: $0.1s$, $0.5s$, $1s$ and $5s$.

### Time Effort

*Description* The time effort represents the sense of abruptness and change over time. A motion having a high time effort is defined as *sudden* or *urgent*. A motion having less time effort is *sustained* or *steady*.

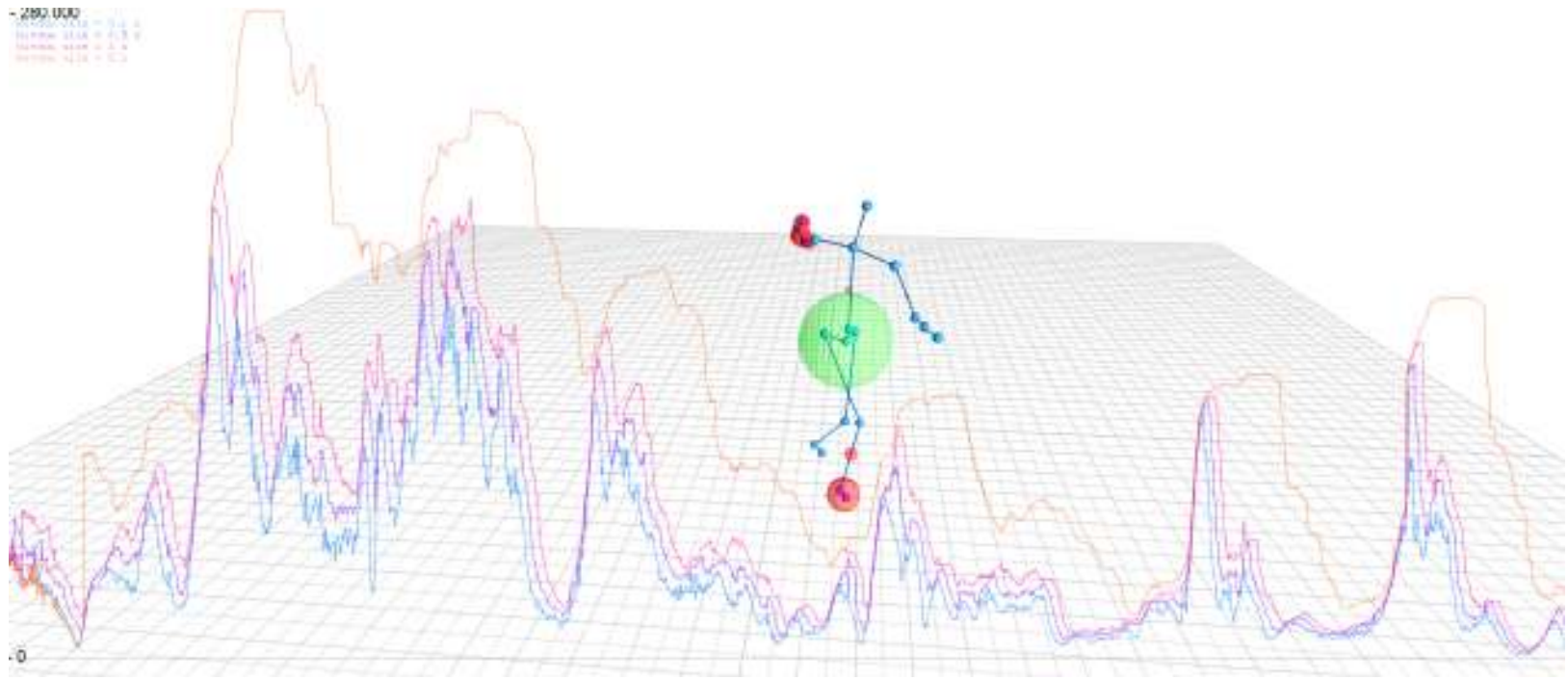

**Fig. 8.** Weight Effort: blue: T=0.1s, purple: T=0.5s, pink: T=1s, orange: T=5s.

*Computation* The time effort is estimated by computing the weighted sum of the accelerations over time for a given body part (can be skeletal joint):

$$TimeEffort^k(T) = 1/T \sum_{i=1}^{T} a^k.(t_i) \tag{15}$$

where $a^k(t_i)$ is the acceleration of the body part $k$ at time $t_i$. The time effort for the whole body is computed as a weighted sum of time efforts of body parts:

$$TimeEffort(T) = \sum_{k \in K} \alpha_k.TimeEffort^k(T) \tag{16}$$

Figure 9 shows the time effort values computed, for the different body parts and the whole body, for different window sizes: $0.1s$, $0.5s$, $1s$ and $5s$.

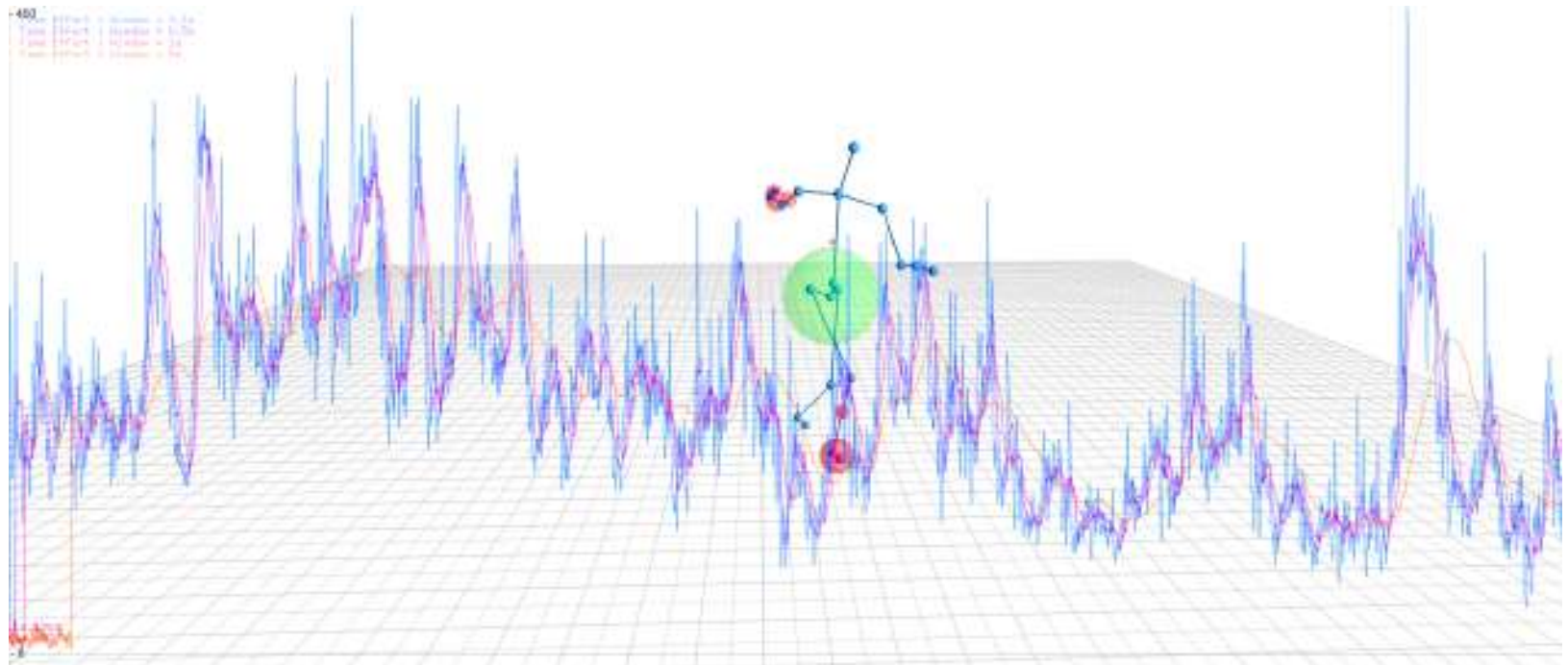

**Fig. 9.** Time Effort: blue: T=0.1s, purple: T=0.5s, pink: T=1s, orange: T=5s.

### Space Effort

*Description* The space effort defines the directness of the movement. By computing the space effort we can say that a movement is considered as *direct* (focused and toward a particular spot) or *indirect* (multi-focused and flexible).

*Computation* For one body part $k$, we compute the space effort as follows:

$$SpaceEffort^k(T) = \sum_{i=2}^{T} \frac{||x^k(t_i) - x^k(t_i - 1)||}{||x^k(T) - x^k(1)||} \tag{17}$$

Similarly to other metrics, the space effort for the whole body is computed as a weighted sum of space efforts on the different body parts:

$$SpaceEffort(T) = \sum_{k \in K} \alpha_k.SpaceEffort^k(T) \tag{18}$$

Figure 10 shows the space effort computed, for the different body parts and the whole body, for different window sizes: $0.1s$, $0.5s$, $1s$ and $5s$.

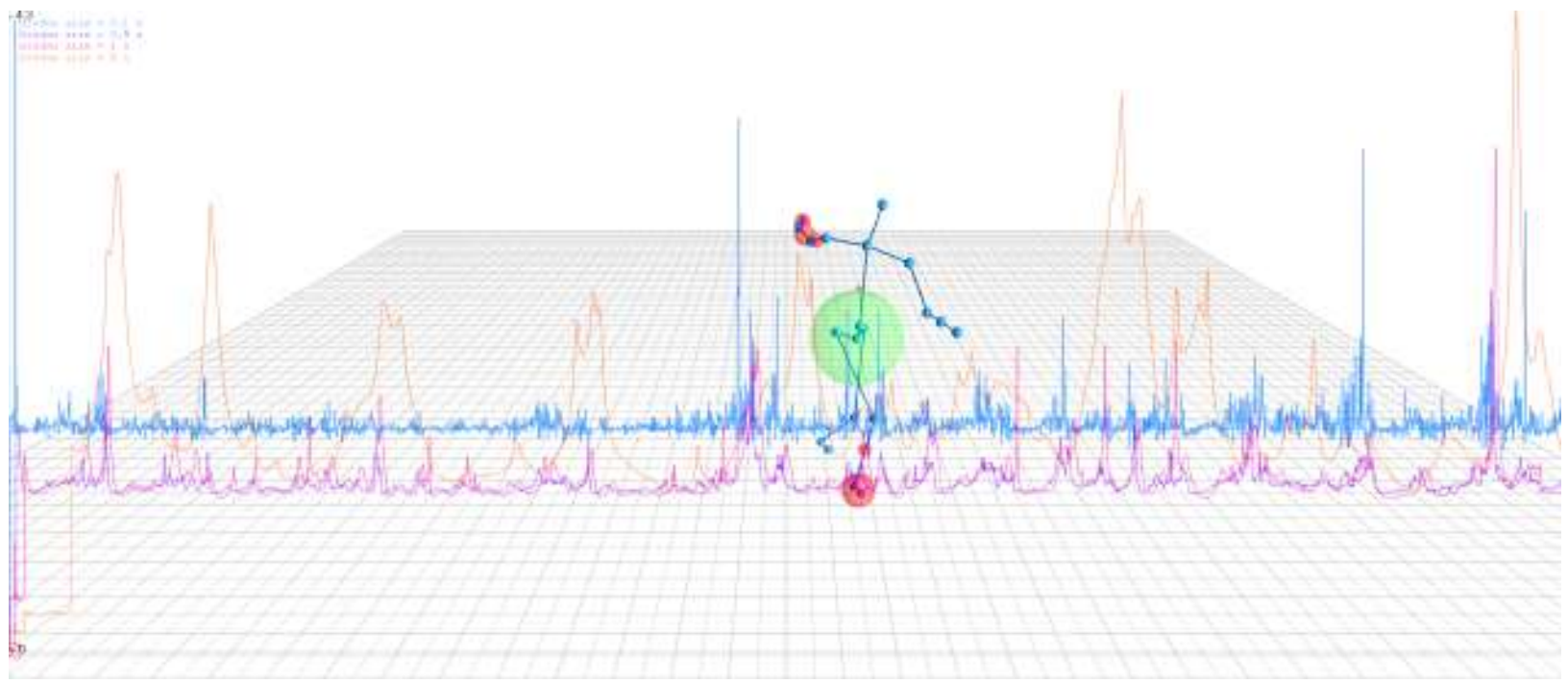

**Fig. 10.** Space Effort: blue: T=0.1s, purple: T=0.5s, pink: T=1s, orange: T=5s.

## 3.5 Space (Laban)

Space features of Laban Movement Analysis [4] describe the relation between the body and the environment. In the scope of this research, two features were developed: the covered distance, and the covered area.

### Covered Distance

*Description* This feature refers to the accumulated distance that is covered on the ground over a period of time by a given body part or the whole body.

*Computation* The covered distance can be processed on any node of the body, though the pelvis is used by default, in order to analyze global displacement of the body on the scene. The body CoM could also be used. The node is first projected on the ground. The cumulative sum is then processed on the lengths of the segments linking two successive positions of the projected node:

$$\overrightarrow{p_i} = (x_i, y_i)$$

$$d_i = |\overrightarrow{p_i} - \overrightarrow{p_{i-1}}| \tag{19}$$

$$D_i = \sum_{j=1}^{i} d_i$$

where $\overrightarrow{p_i}$ is the projection on the ground of the node, $d_i$ the distance between two successive frames, and $D_i$ the accumulated distance at frame i.

*Visualization* Figure 11 shows a visualization of the Laban space component in MotionMachine. In the 3D scene, the trace of the pelvis on the ground is displayed in gray, and its convex hull (covered area), in red. In the 2D scene, the upper curve is the covered distance, the lower curve the covered area.

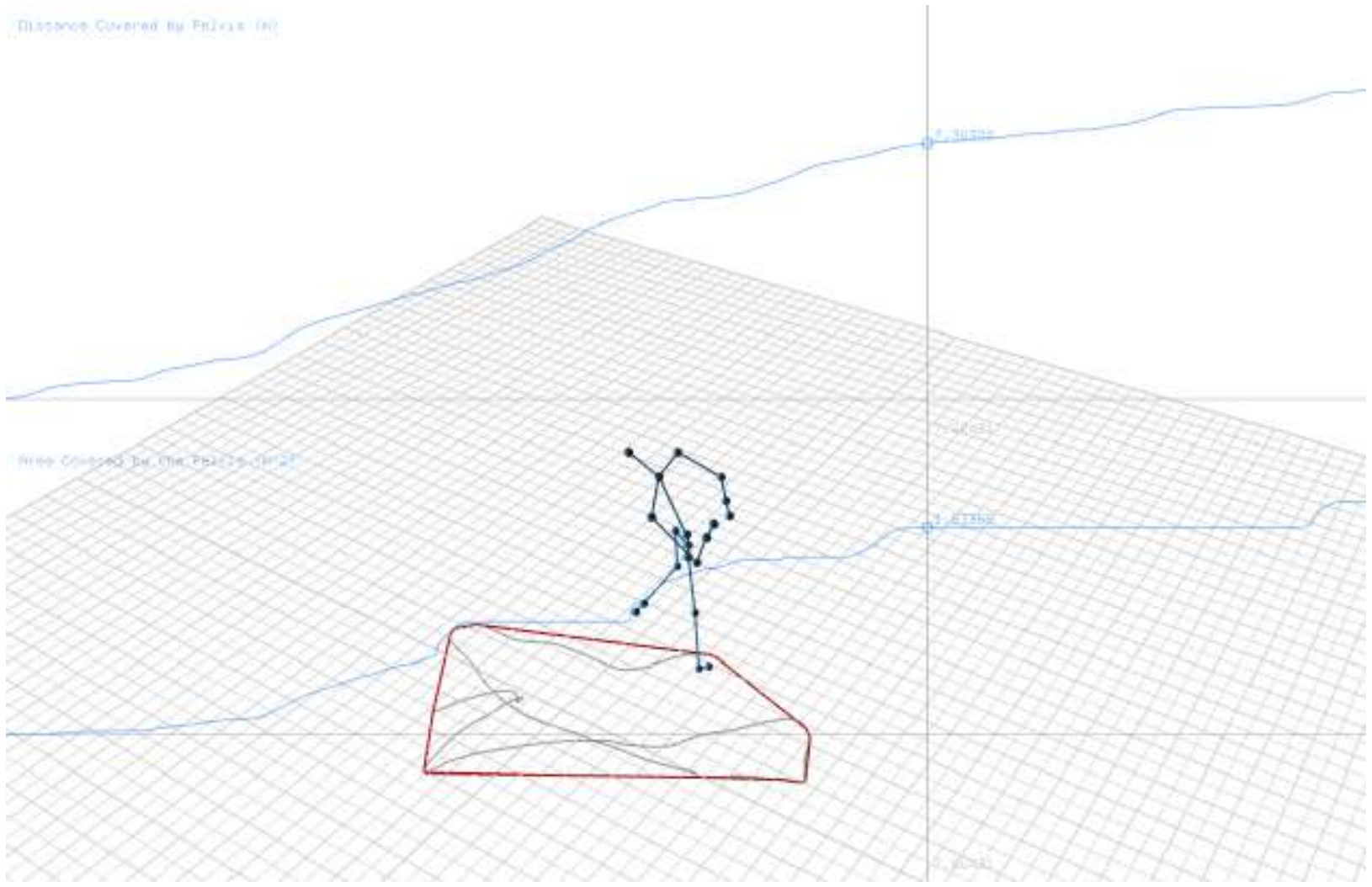

**Fig. 11.** Laban space component visualization in MotionMachine.

**Covered Area**

*Description* This feature refers to the area that is covered on the ground over a period of time. It describes how widely the scene is explored by the body.

*Computation* The covered area is based on the convex hull, presented in Section 3.1. The convex hull is calculated on the trace of the pelvis projected on the ground. The area is then computed on the points of the resulting convex hull, with an algorithm derived from Findley's polygon area algorithm [18].

*Visualization* The covered area is represented in Figure 11, by the lower curve in the 2D scene, and the red convex hull on the ground in the 3D scene.

### 3.6 Smoothness/Jerk

The jerk is a kinematic descriptor corresponding to the rate of change of the movement acceleration. Low values of jerk correspond to smooth movements, while high values appear for complex, discontinuous movements.

#### Jerk Vector

*Description* For each frame, a jerk vector can be determined. It corresponds to the evaluation of the rate of change in each of the $xyz$ coordinates of each joint.

*Computation* For one joint $k$, it can be computed as:

$$\mathbf{j}^k(t_i) = \frac{x^k(t_{i+2}) - 2 \cdot x^k(t_{i+1}) + 2 \cdot x^k(t_{i-1}) - x^k(t_{i-2})}{2 \cdot \delta t^3} \tag{20}$$

#### Jerk Magnitude

*Description* From the jerk computed on each joint and each dimension – therefore a jerk vector –, we will preferably compute a scalar value for each joint. This value represents the overall jerkiness of a given joint.

*Computation* The jerk magnitude is the length of the jerk vector:

$$j^k t_i = \sqrt{\mathbf{j}_x^k(t_i)^2 + \mathbf{j}_y^k(t_i)^2 + \mathbf{j}_z^k(t_i)^2} \tag{21}$$

#### Flow Effort

*Description* The flow effort describes the continuity of the movement over time. Low values describe fluid movement, high values for restrained movement [19].

*Computation* For one joint $k$ and a time window of length $T$, it is computed as:

$$Flow^k(T) = \frac{1}{T}\sum_{i=1}^{T} j^k(t_i) \tag{22}$$

For a set of joints $\mathbb{K}$, we just aggregate the flow efforts of all joints $k \in \mathbb{K}$:

$$Flow(T) = \sum_{k\in\mathbb{K}} \alpha_k \cdot Flow^k(T) \tag{23}$$

### 3.7 HMM Decoding with HTK

*Description* In previous works, we detailed how it is possible to implement a gesture decoding system based on Hidden Markov Models (HMMs) [23]. The HMMs represent the gesture with a succession of states. At each state, local statistics of the observations apply and both local statistics and state transition probabilities are determined by training. Hence it is possible to identify the most likely gesture corresponding to new observations and compute an approximation of the current state, which reflects the progression in the executed gesture.

*Computation* The HMMs used for gesture recognition are trained using HTK (Hidden Markov Model Toolkit [14]). Each input data frame contains the body skeleton pose data. We can choose to describe this pose with the position of skeleton nodes or the orientation of the bones. The gesture recognition is performed by using a Viterbi algorithm. We used an implementation proposed by the *MLPack* library [12]. The Viterbi algorithm is applied on the whole temporal window of data stored in a MotionMachine Track data type.

### 3.8 DTW/Muller Gesture Recognition

This method of gesture recognition has been initially proposed by Müller [5] who decomposes a movement into a set of relational characteristics between joints of a skeleton, reflecting simple human body motions. Binary features are then converted from one value per frame to one value per segment. Forty features are extracted and can be interpreted as a summary of the mocap data under the form of characteristics understandable from an anatomical point of view.

For each gesture to recognize, we hence know which characteristics are active/inactive and their evolution over time. A motion template is defined by computing the mean of the relational features based on several executions of the same gesture. This motion template hence highlights the zones where the relational features differ between several occurrences of the same gesture (in black in the Figure 13) and the zones where they are the same (in white in the Figure 13), which can be considered as a definition of the gesture to be recognized. Zones in orange in Figure 13 are considered are zones of uncertainty.

The last step consists in measuring the similarity between the template $X$ of size $K$ and the motion to be compared, $Y$ of size $L$ (with $L > K$) or a continuous movement sequence. A distance between the motions is measured based on features outside of the uncertainty zone of the motion template.

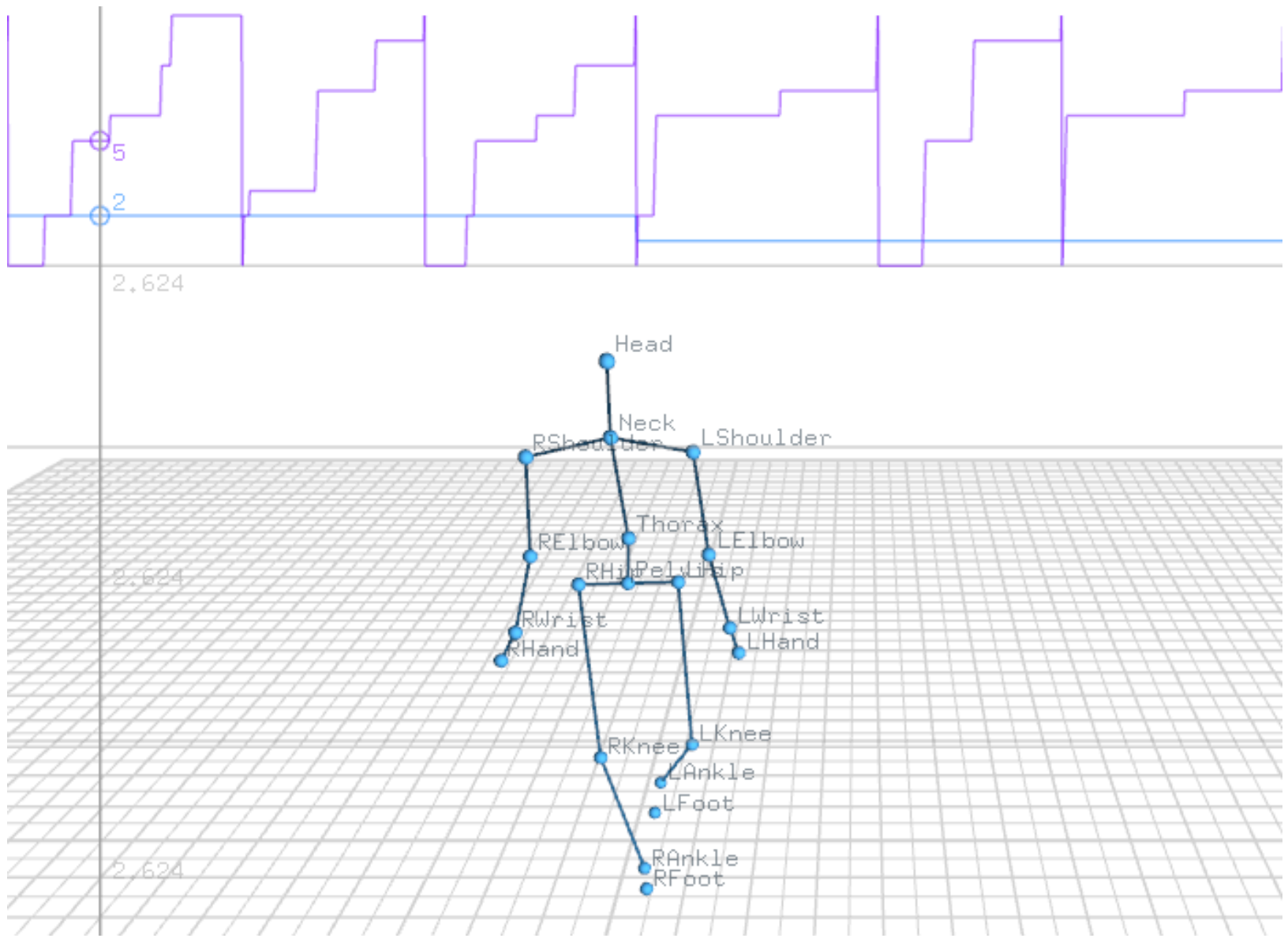

**Fig. 12.** HMM-based gesture decoding of a dance step sample. Four HMMs for four different dance steps were trained with HTK. Each gesture was modeled with 10 states. On the top of the Figure, we can see the decoded gesture and the most likely current state. The cursor is pointing at a frame decoded as gesture 2 and state 5.

$$\begin{cases} D(k,l) = c(k,l) + min(D_{k-1,l-1}, D_{k-2,l-2}, D_{k-1,l-2}), k \in [3:K], l \in [3:L] \\ D(k,l) = c(k,l) + min(D_{k-1,l-1}, D_{k-1,l-2}), k \in [1:2], l \in [3:L] \\ D(k,l) = c(k,l) + min(D_{k-1,l-1}, D_{k-2,l-1}), k \in [3:K], l \in [1:2] \end{cases} \tag{24}$$

where $c(k,l)$ is the normalized Manhattan distance:

$$c(k,l) = \frac{1}{I(k)} \sum_{i \in I(k)} |X(k)_i - Y(l)_i| \tag{25}$$

$I(k)$ is a vector containing the indices $i$ of the features outside the uncertainty zone. Figure 14 shows the distance between the motion template and a full dance motion sequence (in blue) and intervals recognized as corresponding to the motion template (highlighted in green).

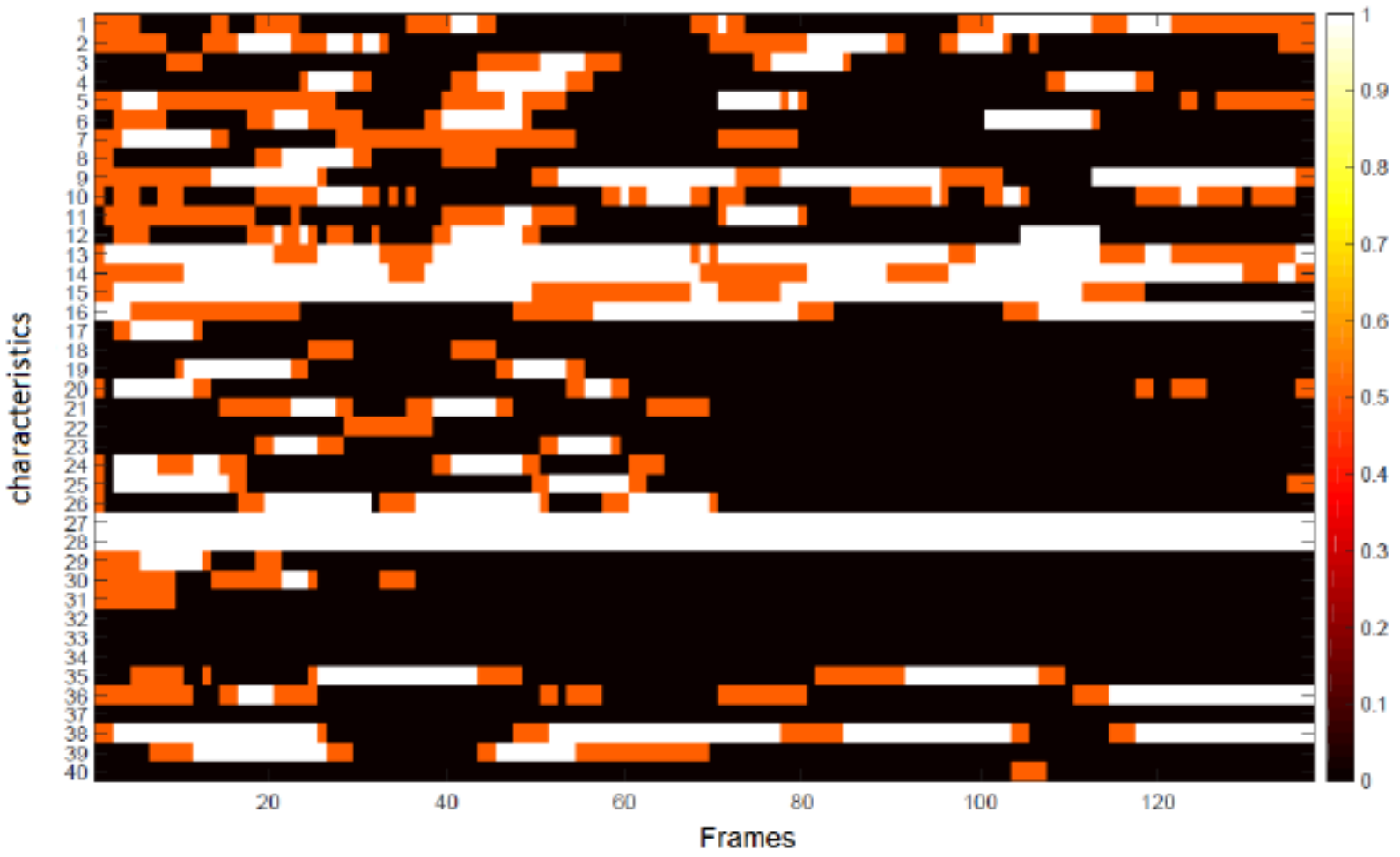

**Fig. 13.** Example of the motion template of a dance sequence.

### 3.9 Wekinator

Wekinator is a standalone, user-facing software tool for applying supervised machine learning algorithms interactively to real-time applications. It was originally developed by Fiebrink [25] to support system builders in analyzing data from real-time audio and sensors, and in creating real-time systems that are data-controlled. One of the most common applications of Wekinator to date has been the creation of new digital musical instruments, in which the actions of a musician are captured using bespoke sensing interfaces (on-body sensors, sensors within tangible objects, etc.) and mapped to control over sound synthesis.

Wekinator provides supervised learning algorithms for classification (applying a category label to incoming data), regression (mapping incoming data to a real number) and temporal modeling (identifying temporal patterns within a stream of data). At this time, Wekinator's algorithms for classification include k-nearest neighbors, support vector machines, decision trees, AdaBoost and Naive Bayes; regression algorithms include multilayer perceptron neural networks and linear and polynomial regression; and temporal modeling is achieved using dynamic time warping. These are general-purpose algorithms with wide applicability to diverse types of sensor inputs and diverse modeling goals.

Whereas many popular machine learning toolkits (e.g., Weka) support users in modeling a fixed data set (e.g., comparing many algorithms or feature representations on the same data set in order to choose the best configuration), Wekinator supports an *interactive* approach in which users are able and encouraged to change the training data set in order to alter model performance. For instance, a user who wishes to build a gesture classifier for a bespoke gesture vocabulary can provide an initial training set consisting of a few examples of each gesture type, train a model on the data set, and evaluate the model by testing it on new gestures in real-time. If the model makes mistakes on a particular gesture, the user can augment the old training set with new examples of that gesture, and retrain the model with a reasonable expectation that is performance on the gesture will improve. The training data set effectively becomes a user interface through which users can communicate their intentions for the trained model. This type of interactive editing of the training set is appropriate in scenarios where the user is capable of generating or curating appropriate new training examples. It can enable efficient learning on small data sets, and it leaves flexibility for users to explore many variations of modeling goals as they discover what can accurately be modeled. For instance, a user building a bespoke gesture classifier may iteratively increase or decrease the number of gesture classes until he or she is satisfied with both the degree of model accuracy and the granularity of classification [26]. Such human-in-the-loop approaches to building machine learning systems, also called *interactive machine learning*, are an active topic of research within human-computer interaction.

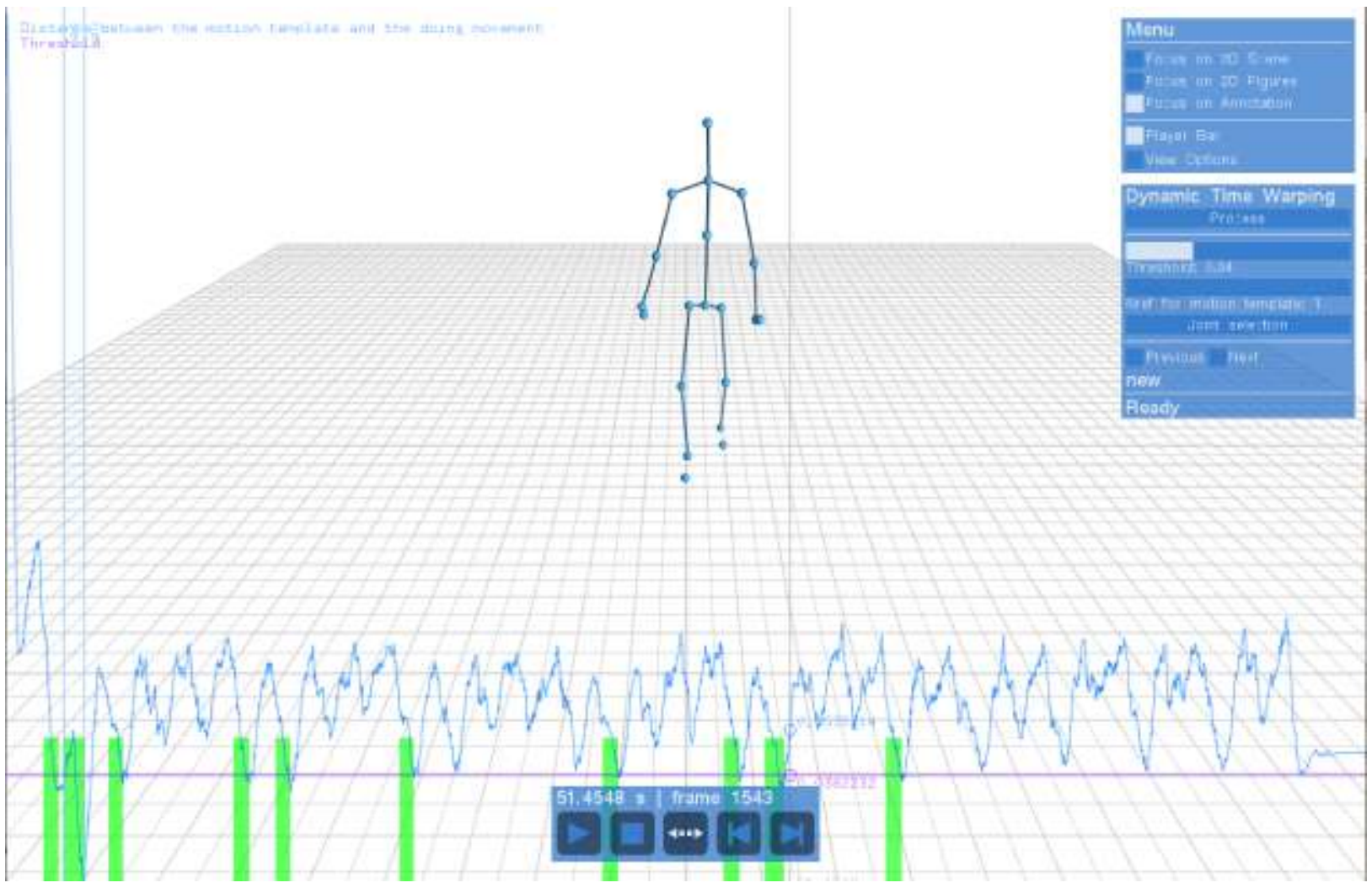

**Fig. 14.** DTW-based gesture recognition in MotionMachine. The similarity score between the considered motion template (forward dance step) and a dance motion sequence is represented as a blue line, the threshold is represented as a purple line and the time intervals recognized as similar to the motion template are shown in green.

Wekinator is a standalone application that does not perform feature extraction, sound synthesis, or animation itself. Rather, it receives streams of feature vectors (e.g., raw or processed data from sensors) from any other application via Open Sound Control (OSC) protocol [27]. OSC is a lightweight, UDP-based communication protocol for sending data between applications, which may reside on the same or different physical machines.) When Wekinator has one or more trained models, it also sends out model outputs to any other application using OSC. For example, a simple webcam-based controller could be constructed with the following pipeline: a C++ program using OpenCV extracts 10 computer vision features from a webcam and sends them to Wekinator via OSC; Wekinator passes these features as inputs to a set of 3 neural networks, each of which computes a real-valued output; these outputs are sent via OSC to the Max/MSP sound synthesis environment where they drive 3 sound synthesis parameters.

**Integrating Wekinator into MotionMachine** Wekinator's on-the-fly, interactive machine learning capabilities were integrated with the MotionMachine framework by implementing user interface control for machine learning within the MotionMachine interface, and by implementing mechanisms for Wekinator and MotionMachine to pass data and machine learning results to each other in batch and real-time modes. This architecture, shown in Figure 15, provides a light-weight coupling between MotionMachine and Wekinator; the two applications are run in parallel and communicate using OSC messages (which are sent via UDP) for real-time control and the local file system for large data transfers.

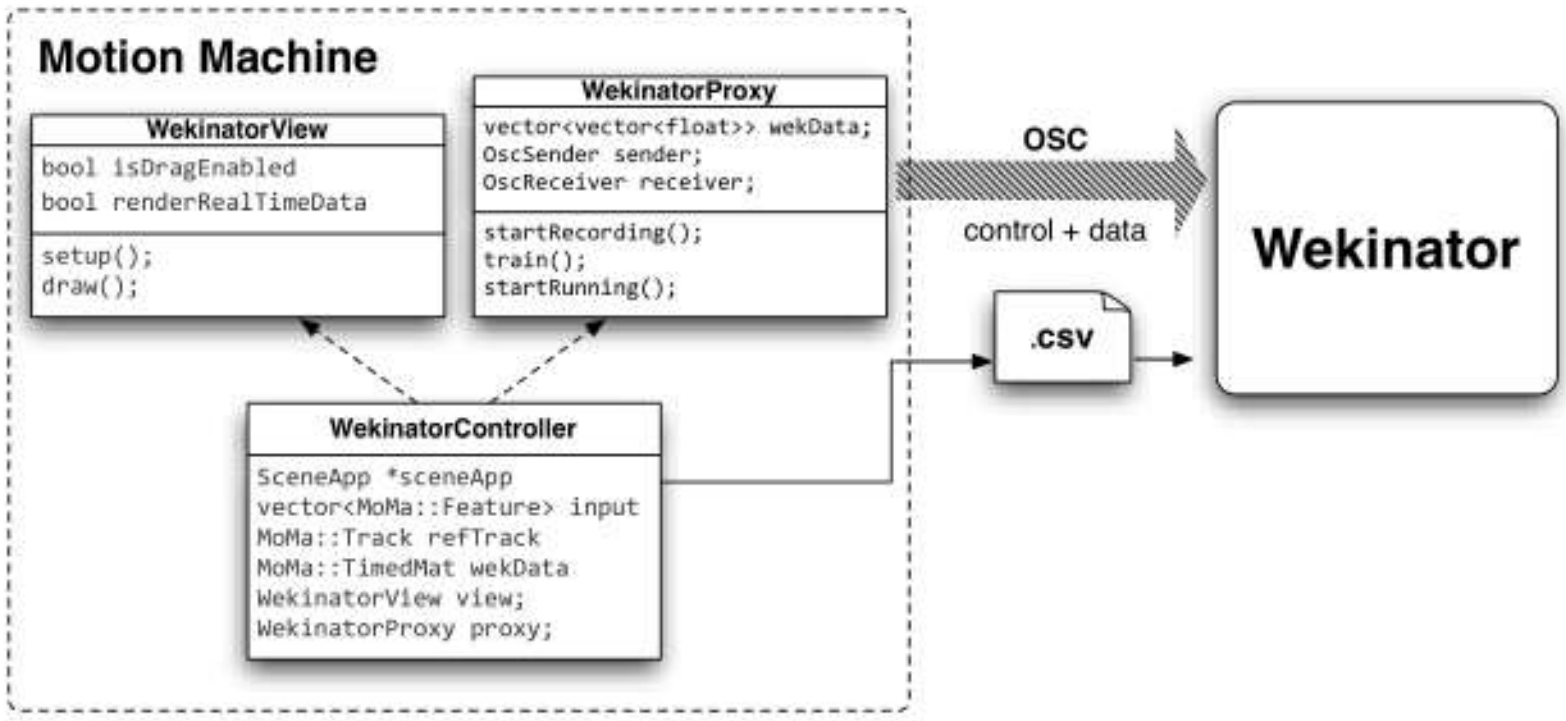

**Fig. 15.** MoMa - Wekinator Integration architecture

The new user interface elements for interacting with Wekinator within MotionMachine are shown in Figure 16. Once a Track object has been filled with mocap data in MotionMachine, a user can select any sub-sequence of the stream and send it to Wekinator as a set of labeled training examples. The labels for these examples are supplied using the integrated annotation system of MotionMachine. Training can be triggered from MotionMachine too.

Once Wekinator's model(s) have been trained, they can be run in two modes. To run in *batch* mode, the user first employs the mouse to select a sub-sequence (see Figure 16, "Batch selection" label in red) of MotionMachine data, then clicks on "Send selected data for class". The sub-sequence is sent to Wekinator as a batch, and Wekinator computes model outputs for each example in this sub-sequence and returns the batch of outputs to Motion Machine where they are plotted on screen (Figure 16). To run in *real-time* mode, the user can scrub or play back the MotionMachine sequence in real-time. Each new data point is sent from MotionMachine to Wekinator using OSC in real-time, and Wekinator's models compute their output values for this next data point. These output values may be simultaneously sent to any other environment to drive sonification, animation, or visualization from the real-time data. The MotionMachine user interface also allows the user to refine model behaviors by adding additional training examples and re-train the model(s). The user can also delete models.

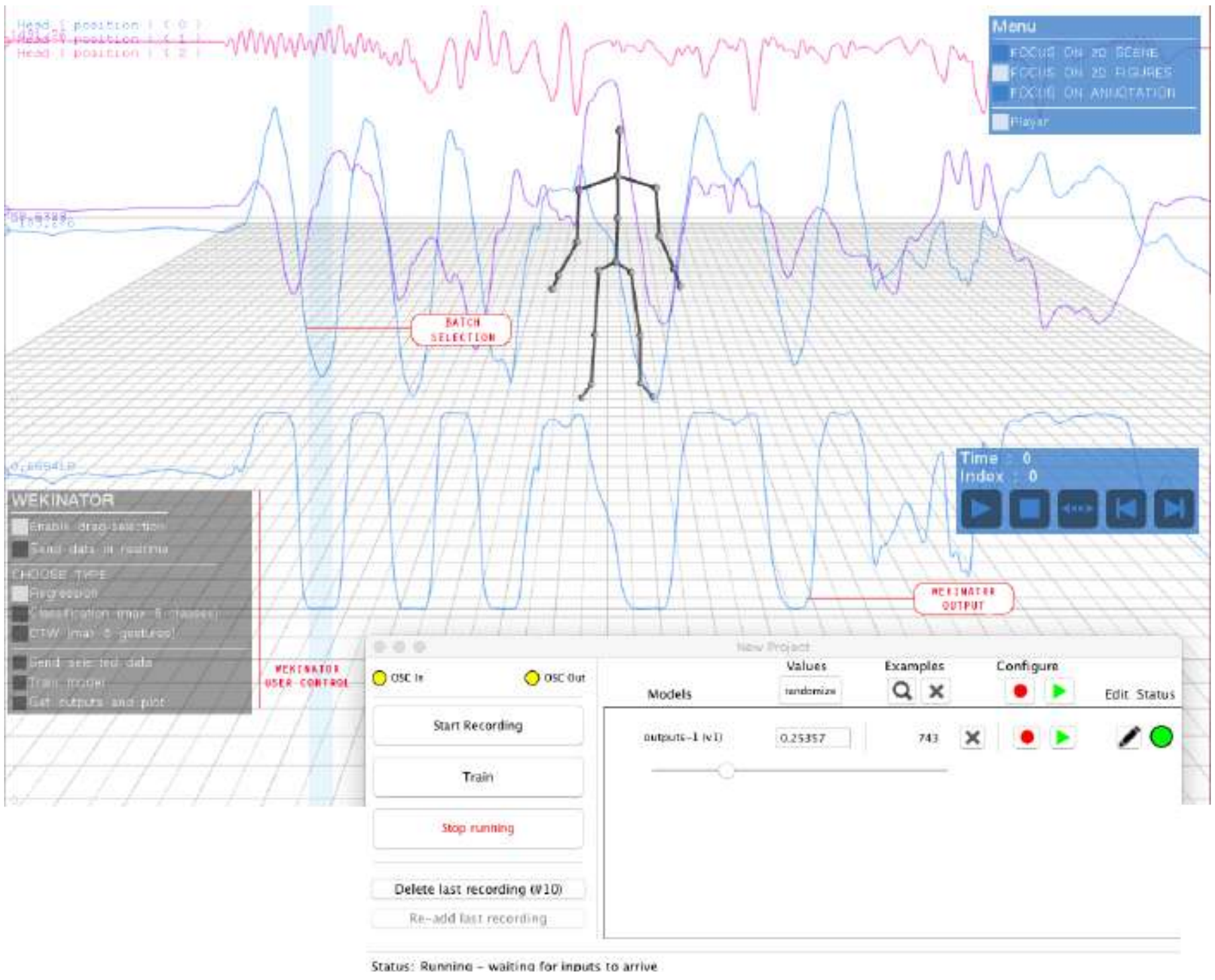

**Fig. 16.** Integration with Wekinator

The integration of Wekinator with MotionMachine was accomplished through a self-contained structure, and it's design is inspired by the Proxy pattern and Model-View-Controller design patterns (ref?). That structure is comprised of three classes (Controller, View and Proxy), implemented in Motion Machine's Core library. The Proxy class implements all the communication protocol with Wekinator over OSC through a UDP connection. This protocol, which is wrapped in public methods of the class, implements operations such as sending atomic data and bundles, setting the classifiers, and starting and stopping the essential Wekinator operations (recording, training and running). The *WekinatorView* class implements all the necessary UI controls for the user to interact with and trigger commands that will action the protocol exchange with Wekinator. The user is presented controls that enable the selection of data interactively through drag operation, buttons to select the learning algorithms types (between Regression, Classification and DTW), selecting and sending data to classes. The *WekinatorController* class contains the previous two classes as components, integrating them with the track and feature data from MotionMachine's general scene, and manipulating the proxy according to the user control events.

This loose coupling between Wekinator and MotionMachine will enable both software projects to evolve independently without concern for breaking dependencies, which is critical given that the two code bases are managed by organizationally and geographically distinct teams. Furthermore, because MotionMachine is coupled to Wekinator only through its *WekinatorController* class, the implementation of this class could be easily changed to encapsulate native C++ machine learning code, or to reference other machine learning frameworks. Likewise, Wekinator's mechanisms for exchanging data and control with an external application now allow it to serve as a machine learning back-end for other user interfaces, whether alternative user interfaces within MotionMachine or other user-facing systems such as the RepoVizz online data repository. Another similar proxy implementation has been created with [28], a cloud based system for storage and visualization of synchronous multi-modal, time-aligned data.

### 3.10 Parsing OpenSIM Data

A better knowledge of the various muscle activation patterns which occur while subjects perform expressive movement or use gesture-based interface has the potential to help interaction designers and choreographers improving the ergonomics of their creations [29]. It is now possible to infer such muscle activation patterns from a set of captured kinematic trajectories using bio-medical simulation techniques. OpenSim[9] offers an implementation of these techniques.

By analyzing motion captured data, OpenSim outputs the simulated time-varying values for the toque moment exerted at each joint and muscle activation data. The torque moment exerted at a joint is a measure of the quantity of torsion resulting from the weight of the body and the dynamic of the movement. The magnitude of the torque is strictly related to the quantity of the effort which is put by the muscles connected to the joint.

The muscle activation is a measure of the energy consumed by a subject to contract the muscle. It can be used to predict muscle strain and fatigue. The software developed so far focuses on the visualization of the torque moment at each joint. The visualization of muscle activation is planned for future work.

OpenSim and MotionMachine address two different tasks: bio-mechanical simulation vs. visualization of motion capture data. Motion Machine is tuned to displays kinematic information of entities along trajectories described using 3 degrees of freedom in an absolute Cartesian space. As such, a conversion of the data format provided by OpenSim is needed (See Figure 17).

[9] https://simtk.org/home/opensim

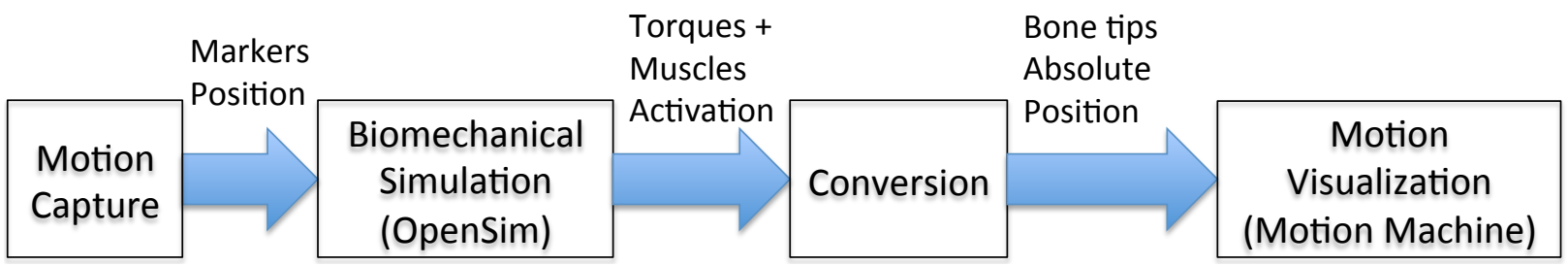

**Fig. 17.** The conversion process from OpenSim to MotionMachine.

The conversion software developed for this module performs the conversion through two main steps. First, it converts the motion model of OpenSim (set of interconnected joints) into a hierarchical bone structure. Second, it converts the motion information of OpenSim, which is expressed as rotation of joints around specific axes, into absolute movement of bone tips in the Cartesian space.

One of the most challenging parts of the development consisted of parsing the skeletal description system used by OpenSim. OpenSim associates each joint to a set of *coordinates*. Each coordinate is then associated to a rotational or a translational transformation. Each rotation simulates one of the degrees of freedom of a joint. The translations are used to improve the simulation of some joints (such as the knees) where two connected bones shift on each other while rotating. To comply with the needs of Motion Machine, the conversion process approximate the translation to a fixed averaged value and accumulates the different rotational components into a single quaternion associated to a bone.

Finally, the software performs some minor tasks, such as the alignment of the coordinate systems and the creation of a file describing the mapping between the coordinate names and the associated joint, which is later needed to represent the joint moments as visual cues to the joints.

## 4 Mapping and Synthesis

In this Section, we give an overview of the audiovisual use cases that have been developed during the project. It covers four scenarios grouped into two categories: visualization and sonification works. The extend of the use cases illustrates the versatility and intuitiveness of the environment that we have developed.

### 4.1 Visualization

**Bloom** MotionMachine already had a UI that displays the skeleton and the curves of the features. A new paradigm was to be found to propose something interesting and aside of the scientific approach. Based on former studies (harmonic plants and followers[10]), the idea of seeds floating in a volume around the skeleton quickly arouse. Each seed is placed randomly in the space and moves slowly, generally towards the center of the space, where the skeleton tends to be. When a seed enters the attraction radius of a skeleton's node, it starts to accelerate towards it (pull phase). When it is close enough, the node pushes the seed away (push phase). By a careful tuning of the pull and push radius, the seed is never stuck around a node, and keeps moving around the whole skeleton.

[10] http://lab.frankiezafe.org

The model has many advantages:

- autonomous: seeds are not depending on a skeleton;
- rich paradigm: seeds can grow into many forms, perfect for OOP;
- fast to setup: points moving in a 3d space is easy to monitor.

*BNode structure* The approach taken from the start was to work outside of the main MotionMachine environment, that already contains a set of methods to draw. A *ofxMoMa* add-on has been built to make all the important classes accessible in a standard openFrameworks project. To make nodes information accessible to the seeds, we have implemented a *BNode* structure containing:

- position (*ofVec3f*);
- orientation (*ofQuaternion*);
- relation to parent *BNode* (if some);
- relations to *BNode* children;
- features - for each of them:
  - a Boolean if the feature is enabled;
  - one or several float(s) for the feature value(s).

*BNodes* are updated at each update by a specialized class called *BNodeManager*. The manager extracts the relevant info from the track and other objects (*TimedMat* mainly) and updates the *BNode*. All the *BNodes* are stored in a pivot structure shared upon all the objects, called *SeedExchange*. Therefore, when a Seed is created, it receives a pointer to the *SeedExchange* structure and is able to read the *BNodes* parameters. So we make them accessible through the nodes.

*Normalization* Normalization is a key point when you want to use data to control other data. Typically, the possible range is scaled between 0 and 1. The feature *Postural Load* (cf. Subsection 3.3) was one of the first to be ready. It was computed on the node's level, making it a perfect candidate to feed the seeds. The main issue was that it was not normalised. A generic *FeatureNormaliser*, separated from the other features availble, has been developped for that purpose. The results of the normaliser are available in the *BNodes*.

*Attaching the seed* The first problem was where to stop the seed? Some 3d geometry has been required to solve this issue. At each frame, the seed knows which nodes influence it. Using the same information, an attach point can be processed. The idea was to attach it as close as possible from its actual position, meaning that if it is above a bone (a segment between 2 *BNode*), the attachment must be located between them. The procedure deals with 2 different cases:

- seed is outside the bone: the attach point equal to the closest *BNode*;
- seed is above the bone: the attach point between start and end of the bone.

The second problem was how and when to stop the seed? Deciding when to attach the seed it is a not trivial. A parameter satisfaction in each seed is a container that rises or shrinks. This parameter is bounded by an arbitrary minimum and maximum. As it rises or shrinks continuously and gradually, it can be used as a dampener and delay the actions by waiting for it to reach one of the bounds. Because we wanted the satisfaction to behave differently depending on the kind of seed, we needed a way to describe how the seed reacts to the feature and how much of satisfaction is added or removed at each update.

*Affinity* The way satisfaction evolves is managed by this class. An affinity instance represents the reaction to a value. It can be attached to any kind of normalized value. It has few parameters and a main method called *process()*:

- minimum and maximum input value, each of them between 0 and 1, allowing to narrow the range of response;
- low and high response, representing the normalised output when value is high (at maximum or above) and when it is low (minimum or below), between -1 and 1;
- increase and decrease, two multipliers applied on low and high responses, that represent the amount of satisfaction to add, both being above 0 but not bounded to 1;
- power allowing to curve the response range and increase the response at the bounds;
- and other minor parameters.

This concept is crucial. It allows to define at a high level and in a centralised way the reactions of the seeds in constant interaction with nodes.

*Life cycle*

- Rooting: When satisfaction reaches the maximum for the first time, the attachment process happens. At that moment, its behaviour changes radically. It stops moving freely and stays close to the skeleton and follows the evolution feature of closest BNode;
- Grow and decay: The satisfaction is used to decide when to add a new part to the plant (growing process) or when to remove a part (decay process). This part of the life cycle does not happen in the Seed class and is managed in the sub-classes.
- Death: If the feature is perceived as too negative by the seed, it will eventually die and detach from the skeleton. The death process has not been finished in the current version of the code, due to time constraints.

*Weed and salpetre* Two sub-classes of Seed have been implemented: *Weed* (green leaves) and *Salpetre* (a mycelium and white corollas). Each of these sub-classes have a different *Affinity* configuration related to *Postural Load* feature. *Weed* likes the *Postural Load* to be high. Meaning that, when the *Postural Load* of the closest *BNode* is high, the satisfaction of the plant increases. *Salpetre* has the opposite reaction. When the *Postural Load* is low, its satisfaction increases.

As they both feed on the same feature, the importance of the *Affinity* configuration is clearly highlighted. Due to the feature process, *Salpetre* tends to appear at the extremities of the skeleton, where it is generally low, and weed concentrates around the hips where it is generally high.

*Graphics* The graphical objects include a damping effect, motion being as important as pure shape part in the aesthetics. We have implemented a base class that groups all required parameters such as positions, color and a simple physical model. This base object is called *Position*. The *Salpetre*'s mycelium is a growing tree of positions, rendered as transparent lines. The *Weed*'s leaves require a more complex object. They both use *Branch* objects, that inherits from *Position*. *Branches* render 4 points that describe the corners of a quad. UV's coordinates can be set on each of these points, enabling texturing.

**Distributed Effort** When analyzing a huge amount of high-dimensionality data, scientists need visualization techniques which can present the data in a way that foster exploration. In traditional exploratory data analysis, scientists rely of static charts in order to have visual cues about the nature of the data. Such visualization techniques are inappropriate for data like real-time streams of sensor data. In the biomechanical simulation context, the data is a set of animation curves reporting the value of torque moments applied at each joint of the skeleton, as well as information about the activation of muscles.

The software developed in this module focuses on the visualization of torque moments at joints. The information about a torque moment is characterized by a spatial information (on which joint the moment is exerted), a temporal information (at which time the moment is exerted), and a scalar information which is the magnitude of the moment.

The visualization technique represents the amount of torque exerted on each joint as a 2D circular semi-transparent disk displayed using the billboard technique over each joint. The magnitude of the torque moment modulates both the transparency and the size of the disk. The modulation of the disk area guarantees for the perception of the magnitude. The modulation of the transparency helps the perception of localized effort when the cue of several moments overlap in the screen. Figure 18 shows a screenshot.

This visualization technique helps the data analyzer to quickly spot concentrations and peaks of muscular effort in all areas of the body at a glance. At its maximum extent, the disk has a maximum fixed size and no transparency (full opacity). The maximum extent of the disk is reached when the magnitude reaches, or overtake, a value of three times the standard deviation of the signal over the whole animation. The magnitude is linearly mapped to the transparency factor (in the range $[0, 1]$) and to the area of the disk. As the magnitude of the moment decreases, the transparency is brought to 1 and the disk area to 0. Both effects bring the disk to invisibility.

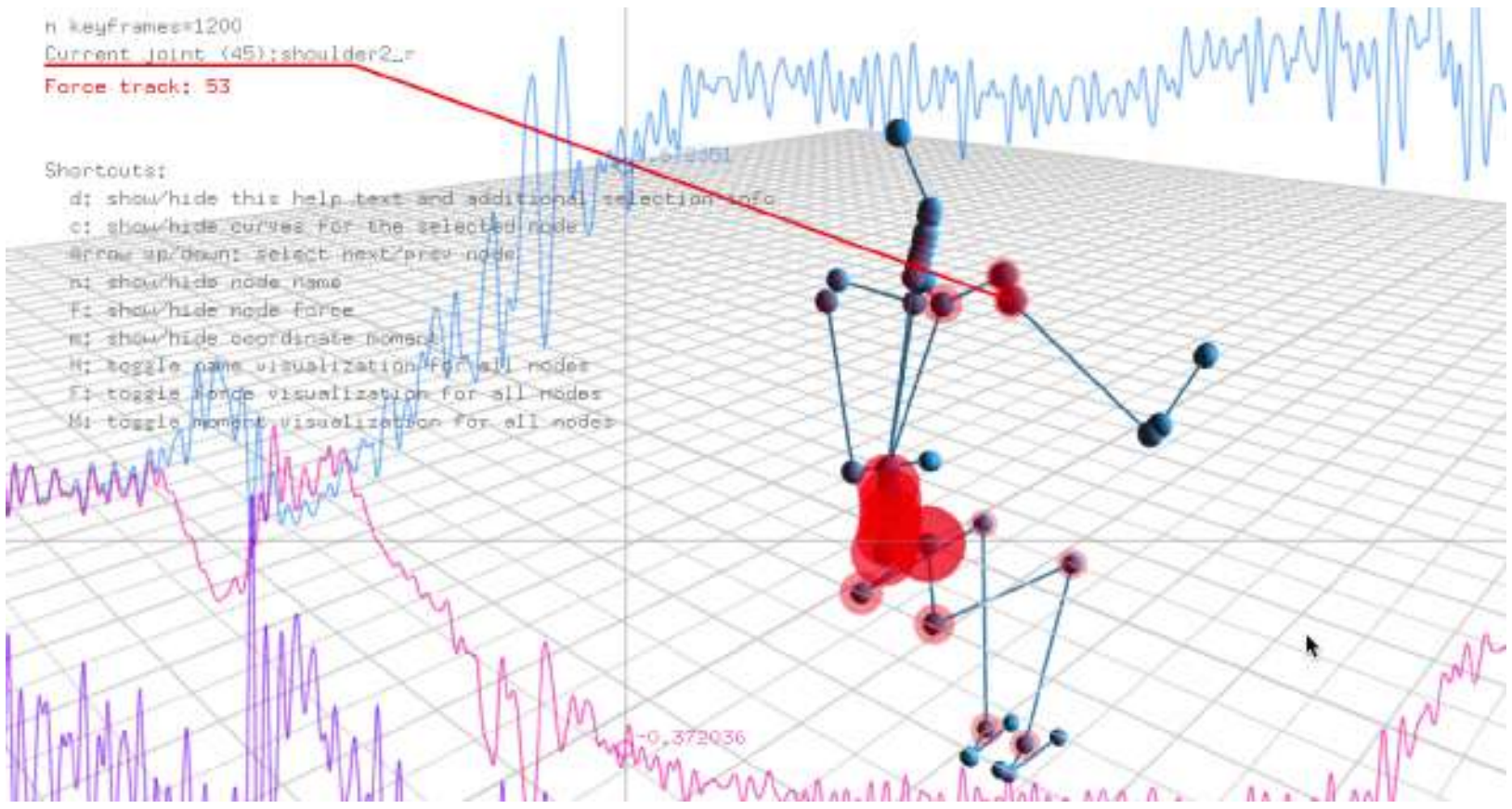

**Fig. 18.** A screen shot of the torque moments visualization in Motion Machine.

The implementation allows for an interactive exploration of the information. It is possible to cycle through the joints and (de)activate the visualization of the visual cue (the disk) as well as the whole animation curve for that joint. It is also possible to toggle the visualization of the disks for all joints at once. However, in order to avoid cluttering the screen with too much information, the animation curve is visualized only for the currently selected joint. As a commodity, it is also possible to save the list of joints for which the moment torque cue is visualized, and re-load it in further analysis sessions.

### 4.2 Sonification

**CataRT** Sonification of high-level features extracted from gestural data opens the way for an infinity of possibilities in terms of mapping strategies and choice of sound synthesis engines. However one can recognize two main trends in the use of sonification: as an end result in the field of art installations or performances and as a feedback in the even broader field of human-computer interaction. In our case, we allowed ourselves the freedom to go in the artistic direction, which at first seems tentative but leaves us a wide choice and few guidelines.

Depending on the type of gestures to sonify, one can favour one type of feature and one type of sound category. As there is no science behind, all the mappings between the gesture and the sound will inevitably be subjective and highly dependent on the material that is being used. Hence, by thinking in terms of what is available and what makes sense, one can start making connections between sounds and gestures.

We decided to work mostly with Taijiquan gestures – recorded previously in the interest of working with the *Sphereness* feature (see Section 3.3). These recordings where our best candidates for sonification as they are smooth, slow and sometimes reveal symmetrical patterns. These attributes are convenient as they are revealed in the sonification more clearly than noisy and quirky input signals. As a criteria for success, we aim at making the link between the gesture and the sound obvious and plausible for the observer/listener.

From this gestural data, we have tried several sound synthesis engines. One that drew our attention the most is the the corpus-based concatenative sound synthesis engine CaraRT provided as a library for Max/MSP by the IMTR team of the Ircam [30]. The principle of Corpus-based sound synthesis is the following: one creates a corpus of source sounds, picked up from sound samples of different durations. The engine segments the sound samples into units, computes some audio features such as the fundamental frequency, energy, periodicity and the first-order autocorrelation coefficient that describes the spectral tilt [30]. The CataRT GUI displays a two dimensional matrix in which one can place the sound units according to any combination of the four sound features mentioned above. Once two features have been assigned to the x and y axis one can see the sound units scattered as coloured circles in the matrix (see figure 19).

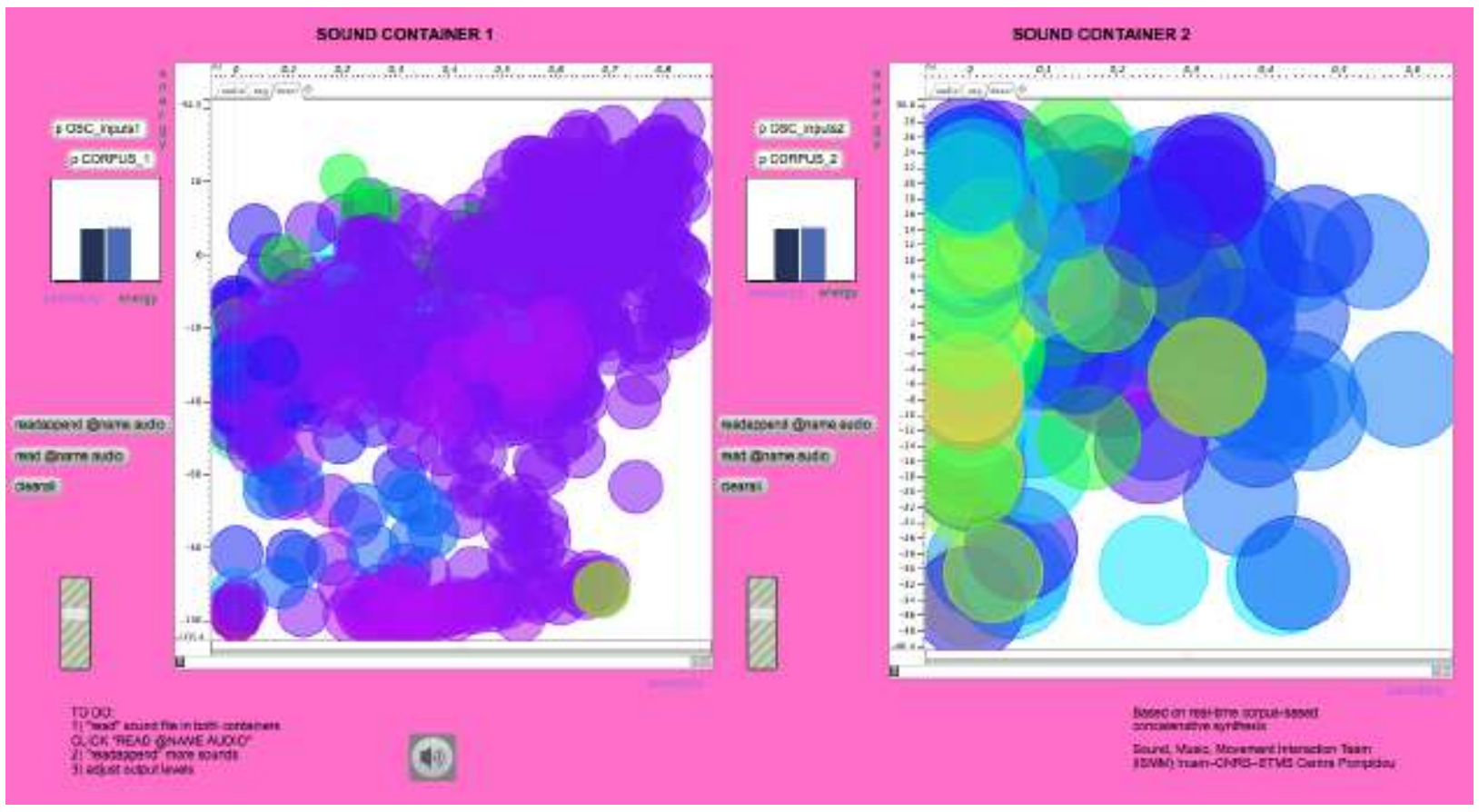

**Fig. 19.** CataRT sound containers for Motion Machine

CatarRT allows to retrieve these sound units by navigating in the sound containers. A nearest neighbor algorithm finds the closest circle that is on one trajectory and plays back the corresponding sound unit. The most basic tool to navigate in these containers is the mouse. Hence, the $xy$ coordinates of the cursor are mapped onto the $xy$ axis of the matrix. A basic gestural control using the body position can be easily implemented: the $xy$ position of the body center of mass can be directly mapped to the $xy$ axis of the matrix.

However, a more meaningful approach would be to map gesture features and sound features which have higher level connections to our perception. Hence, we have chosen the $x$-axis of the sound matrix to represent the periodicity dimension and the $y$-axis the energy dimension. In terms of gesture features, we use the *Periodicity* feature (see Section 3.2) to control the sound periodicity and the Sphereness feature (Section 3.3) to control the energy of the sound. The periodic patterns of the gestures have a straight correlation with the periodicity of the sound. The periodicity of the sound signal ranges from periodic signals such as pure sine waves to aperiodic signals which are often percussive and noisy. Finally, the $y$-axis of the sound matrix, attributed to the energy of the signal is controlled by the *Sphereness* feature.

The phenomenon is as described here: as the body limbs contract themselves in a smaller spherical volume, the sonic energy gets fainter and on the contrary, as the limbs open themselves and expands in the physical space, the sonic energy gets larger. This mapping and the resulting sonification yet sounds quite mysterious, grainy but quiet and have a certain smoothness thanks to the fluidity of the Tai chi gestures. The sounds following closely the sequence of movements, stimulate both visual and sonic modalities into a unified experience after several seconds of observation and listening. Highly dependent on the material used, namely the sound samples and the gestural data, the outcome is encouraging in pursuing more explorations of gesture to sound mappings using this approach.

**Blotar** Blotar is a physical modeling synthesizer that is part of PeRColate, an open-source distribution containing a set of synthesis and signal processing algorithms for Max/MSP [31] based off the Synthesis Toolkit [32]. We used a Blotar in a Max/MSP patch bound to Wekinator, which was itself bound to our application, all through OSC, to build a realtime sonification setup that could provide meaningful aural feedback for the body motion data (see Figure 20).

We used Wekinator to learn from the fluctuations of the *Sphereness* feature extracted from the data (see Section 3.3). Our graphical user interface allows the user to select variable sized segments from the plotting and send it to different classifiers in Wekinator. For instance, by selecting the peaks and troughs in the plot, the user is selecting data that corresponds to the extremes of the range of movement that is achieved between arms-closed and arms-wide-open stances. We fed this data to Wekinator as training samples for nine regression models, which were mapped to nine parameters of the Blotar synthesizer. A previous study of the control space of Blotar had revealed its immense control and sound spaces, and we chose to parameterise it to sound thin and fluty, or metallic and more energetic. We achieved an interesting result with regression. The troughs in the Sphereness plot were mapped to the fluty sound parameters, while the peaks where mapped to the distorted sound parameters. The final result could be experienced as a gradual morphing between these two different sounds.

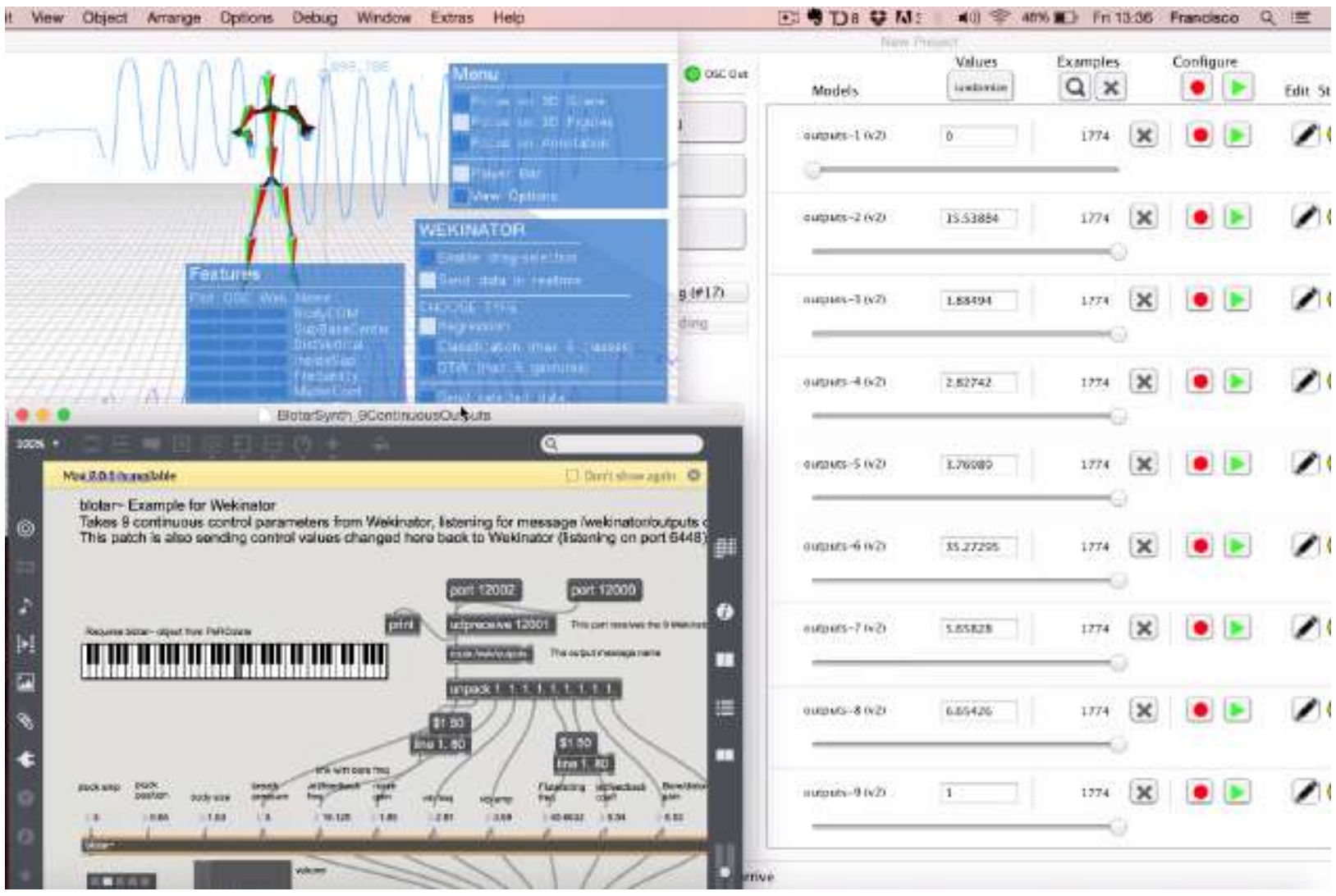

**Fig. 20.** Sphereness - Blotar - Wekinator setup

## 5 Conclusion

In this work, we have adressed the issue of fast prototyping of new audiovisual instruments as well as the implementation of new features and the interconnection with external modules using the MotionMachine framework as a common backbone. This project enabled us to validate and further refine the core of the framework as well as to further expand its capabilities by implementing new features (balance, weight, periodicity, etc.), enabling the parsing of data coming from OpenSim and its visualization, and implementing the communication of the MotionMachine framework with external modules such as Wekinator. The whole framework and the approach of using it for fast prototyping was finally tested and validated on specific use cases, each one corresponding to different constraints (realtime or offline, analysis purposes vs interactive audiovisual instrument, etc.). The validation of the framework by a group of people outside of the original development team enabled us to release a stable version of the framework and to disseminate it in the open source community.

Further work will include the addition of new motion analysis features in the platform, as well as its use and validation in longer term projects and use cases.

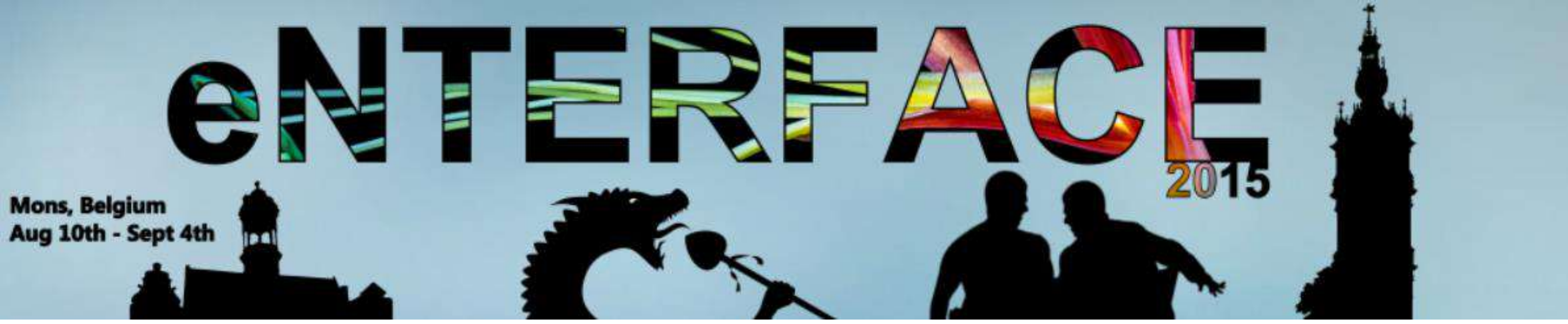

**Project #2:**

# Analysis of the qualities of human movement in individual action

*Paolo Alborno, Ksenia Kolykhalova, Emma Frid, Damiano Malafronte, and Lisanne Huis in't Veld*

# Chapter 1
# Analysis of the qualities of human movement in individual action

Paolo Alborno[1], Ksenia Kolykhalova[1], Emma Frid[2], Damiano Malafronte[1], and Lisanne Huis in 't Veld[3]

**Abstract**

The project was organized as a preliminary study for Use Case #1 of the Horizon 2020 Research Project "Dance in the Dark" (H2020 ICT Project n.645553 - http://dance.dibris.unige.it).

The main objective of the DANCE project is to study and develop novel techniques and algorithms for the automated measuring of non-verbal bodily expression and the emotional qualities conveyed by human movement, in order to enable the perception of nonverbal artistic whole-body experiences to visual impaired people. In the framework of the eNTERFACE '15 Workshop we investigated methods for analyzing human movements in terms of expressive qualities. When analyzing an individual action we were mainly concentrating on the quality of motion and on elements suggesting different emotions. We developed a system to automatically extract several movement features and transfer them to the auditory domain through interactive sonification.We performed an experiment with 26 participants and collected a dataset made of video and audio recordings plus accelerometer data. Finally, we performed a perception study through questionnaires, in order to evaluate and validate the system.

As real time application of our system we developed a game named "Move in the Dark", which has been presented in the Mundaneum Museum of Mons, Belgium and Festival della Scienza, Genova, Italy (27 November 2015).

University of Genova, Italy
e-mail: {paolo.alborno,ksenia.kolykhalova,damiano.malafronte}dibris.unige.it
KTH Royal Institute of Technology, Sweden
e-mail: {emmafrid}@kth.se
Maastricht University, Netherlands
e-mail: {e.huisintveld}@maastrichtuniversity.nl

## 1.1 Introduction

The proposed work is presented as part of the EU Project DANCE - Dance in the Dark (H2020 ICT Project n.645553 - http://dance.dibris.unige.it). The DANCE project aims to innovate state of art on recognition of emotions expressed by the movements of the human body, given human movement as social language. We propose to develop technologies of sensory substitution enabling the perception of dance performance, and in general of bodily movement, through interactive sonifications.

In previous studies it has been experimentally shown that movement qualities may communicate social relations and intentions, such as: emotional states [5], affiliation [12], cultural background [16], dominance [18], agreement [3], group cohesion[9], empathy[18].

Several approaches dedicated to automatically detecting emotional states from body movement were recently developed. In [1] the authors proposed to detect emotional states from low-level features of hand movements, such as maximum distance of the hand from the body, its average speed and acceleration. In [6], expressive qualities of the movement such as amount of motion, contraction and directness indexes as well as velocity and acceleration were used to classify four emotions (anger, sadness, joy, and pleasure) with dynamic time warping classifiers. Also Gross and colleagues in [7] analysed expressive quality (Effort-Shape model) and kinematics (range of motion and the velocity) of emotional hand gestures such as knocking: six different emotions were investigated.

Despite the increased interest in modelling emotional behaviour, analysis of expressive qualities of human full-body movement is rather under-investigated with respect, e.g., to facial expressions or prosody in speech.

Many previous studies, indeed, focus on emotion recognition based on audio and facial expression data. Therefore our project is in the direction of further investigation of expressive movement qualities of full body movements. The main focus was on single performers and the quality of the respective motion, rather than on the type and direction of the movements.

DANCE Use Case #1 served as a foundation for experiments' set up on sensory substitution by means of associating appropriate interactive sonifications to the extracted movement features. In our experiment, we used sonification to transform parameters of human movement qualities into sound in order to investigate the perception of users.

The remaining part of this paper is structured as follows. The next section describes the particular case of Use Case #1 of DANCE Project and section 3 presents movements features: their definition, computation and extraction.

After that, the following section presents the system architecture. Section 5 presents the perceptive studies for testing and validation of work of the system. Section 6 describes the interactive demonstration - game “Move in the Dark”, implemented in the framework of the workshop.

## 1.2 Experimental Scenario

A user who is temporarily deprived of vision is learning to recognize (her own as well as others’) movement qualities only by using the auditory channel. Sensory deprivation is a mean to amplify user’s capabilities and sensibility in recognizing individual movement qualities by means of the auditory modality. The structure of the described use case can be summarized in two phases:

Phase 1

The user (normal sighted and blindfolded) learns how the qualities of gesture are translated into sound, and comprehends how to exploit such an inner vision, induced by sound, of the movement features (e.g., energy, fluidity, rigidity, impulsivity and so on).

Phase 2

The user, who in Phase 1 has been familiarized with the mechanisms of interactive sonification of her own movements, is able to recognize movement qualities of another person just by listening the interactive sonification.

## 1.3 Movement Features

Starting from the literature from experimental psychology and HCI, e.g., [19], [14], [2], [4], humanistic theories and the arts, e.g. [10], [11] and from meetings with dance experts, a collection of expressive movement qualities was defined to be considered in the DANCE project, including smooth, light/heavy, fluid, impulsive, sudden/sustained, symmetric, contracted/expanded, energetic and synchronised.

In our experiment three main movement features were considered and extracted based on the data from accelerometers: Energy, Fluidity and Impulsivity.

To extract information about the user motion, we used four Nexus S commercial Android smartphones. Each Nexus S are equipped with InvenSense

MPU-6050 three-axis gyroscope and accelerometer. The accelerometer produces data in units of acceleration (distance over time2), and the gyroscope produces data in units of rotational velocity (rotation distance over time). The scale of sensibility of each of the MPU-6050 is set to +/- 2g. The smartphones were tied to the arms and legs of the user for a total of four accelerometers.

### *1.3.1 Energy*

The quantity of energy spent by the user is computed from the total amount of displacement of the tracked joints. We define the Weighted Energy Index - $WEI$(f) as the weighted sum of the kinetic energy. Given the three dimensional velocities of the $i-th$ joint at time frame $f$, the total velocity of the joint at time frame $f$ is defined by:

$$v_i(f) = \sqrt{(v_1^2(f) + v_2^2(f) + v_3^2(f)} \tag{1.1}$$

The energy produced by the single joint is computed by:

$$J_i(f) = 1/2 \cdot m \cdot v_i^2 \tag{1.2}$$

With $N = number$ of accelerometers, i.e. 4, and $m = 1$ for each joint.
The maximum value between the joints energies at each frame is selected:

$$M(f) = max[J_1, ..., J_N] \tag{1.3}$$

$$k = argmax[J_1, ..., J_N] \tag{1.4}$$

And the total amount of energy is:

$$WEI(f) = M(f) + \sum_{i \neq k} C_i(f), \tag{1.5}$$

Where $C$ is the single contribute, $i = 1...N$, $i \neq k$ and $T$ is a threshold value in [0,1] used to limit the $WEI$ index.

$$C_i(f) = \frac{J_i(f)}{M(f)} \cdot \frac{(1-T)}{(N-1)} \tag{1.6}$$

### *1.3.2 Fluidity*

We define a fluid movement as an uninterrupted movement characterized by low jerkiness and without abrupt variations in speed. Fluid movements are sometimes associated with slow and sluggish movements, in contrast with large and energetic body movement.

To measure fluidity we defined a Fluidity Index $FI$ based on inverse of the jerkiness of the movement produced by the joints.

$$FI = \frac{1}{(\dot{a_1}^2 + \dot{a_2}^2 + \dot{a_3}^2)} \tag{1.7}$$

### *1.3.3 Impulsivity*

Impulsivity can be defined as a temporal perturbation of a regime motion [8]. Impulsivity lacks of premeditation i.e. it is performed without a significant preparation phase, quickly and with a high energy change [13], [15].

We defined an impulsivity index $II$ based on the inverse of the fluidity and abrupt changes of energy :

$$II = \frac{1}{FI} \cdot \frac{WEI(f)}{WEI(f-m)} \tag{1.8}$$

where $m$ is fixed and arbitrary. The ratio between the values of the WEI index at frame f and frame f-m is used to evaluate whether the energy evolution contains quick changes or peaks, and it is proportional the impulsivity of the movement.

## 1.4 System Architecture

EyesWeb.

The EyesWeb open platform (`http://www.infomus.org/eyesweb_ita.php`), designed at Casa Paganini - InfoMus research centre of University of Genoa, is a development and prototyping software environment for both research purposes and interactive applications. It supports multimodal analysis, real-time processing of non verbal expressive gestures, research on synchronization, coordination, and entrainment in dance and music performance. We used the EyesWeb platform as one of the core modules for developing the interactive demos and applications planned in the project. The system has been implemented in terms of a collection of software modules for the EyesWeb XMI platform. Extraction of features for analysis of affective behavior has been implemented as software modules part of the EyesWeb DANCE Library.

Max/MSP.

The software Max/MSP was used for interactive sonification. EyesWeb and Max/MSP were connected via the Open Sound Control (OSC) protocol [20].

Sonification was used to enable participants to understand specific qualities of their movement in real-time simply by listening to the synthesized sound generated by their own movement. We developed three synthesis models based on a granular synthesis: one for energy, one for fluidity and one for impulsivity. The granular synthesis technique was opted since this method allows for creation of a wide variety of complex sound textures. The granular synthesis method can be seen as a form of additive synthesis, in which the sound results from the additive combination of many smaller sonic grains [17]. A *grain* is a short sonic event with an amplitude envelope with the shape of a quasi-Gaussian bell curve, typically of the length 1-50 milliseconds [17]. We developed a flexible granular synthesis engine in which the sound buffer could be easily altered in order to produce many different types of sound textures.

The first two sound models (Energy, Fluidity) used the same sound buffer; a short recording of a section of string instruments. The third sound model (Impulsivity) used a buffer of a sound reminding of an explosion. Motion features were sent from the motion analysis in EyesWeb and mapped to the following parameters in the granular synthesis engine:

1. Energy
   - velocity mapped to maximum grain amplitude and overall amplitude of the outputted continuous stream of sound
   - grains of randomized length between 1 and 20 ms
   - inter-grain delay set to 12 ms; producing a continuous stream sound
2. Fluidity
   - velocity mapped to inter-grain time and amplitude as for Energy
   - fluidity mapped to grain envelope slope, inter-grain delay (randomized to 0-200 ms),
   - grains of randomized length between 10 and 190 ms.
3. Impulsivity:
   - mapped to triggering of single grains (randomized length between 1-50 ms) of sound buffer with sharp envelope.

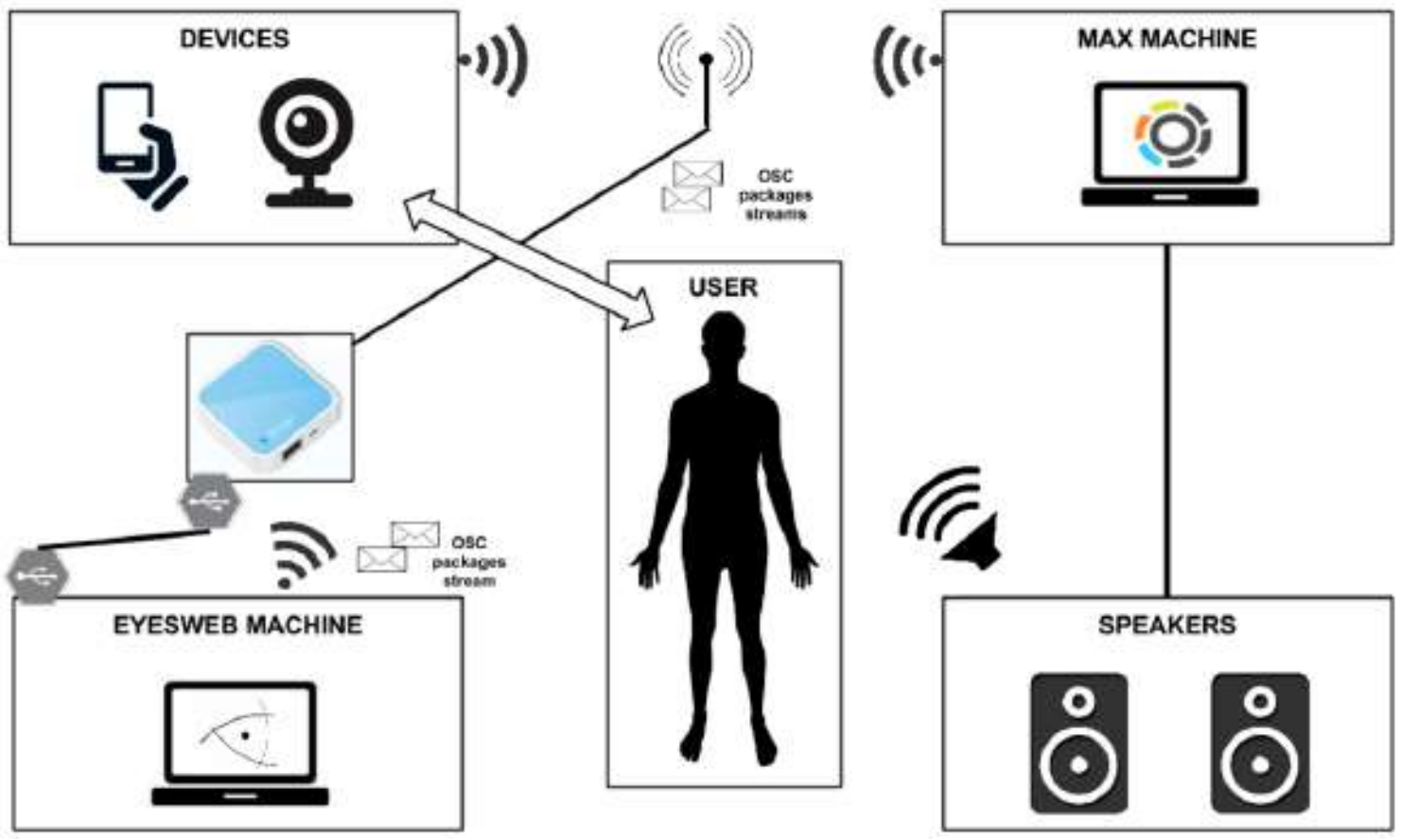

**Fig. 1.1** System Architecture

## 1.5 Evaluation

In the first part of our experiment, participants explored how our system mapped their movements into sound and got acquainted with the sonifications. Then participants were asked to recognize the movement qualities only by listening to pre-recorded audio-files that were created using the same sonifications. In this way, participants were in “blind” conditions and we investigated the user’s capabilities and sensibilities in recognizing individual movement qualities only by means of the auditory modality.

### *1.5.1 Participants*

Two groups of participants were asked to provide a rating of the system. The first group of participants included 16 people (13 M, 3 F; different nationality; mean age 29 years, STD=5.67). The second group of participants consisted of 9 people (6 M, 3 F; mean age 26 years with STD=4.06).

### *1.5.2 Materials and method*

Experiment consists of two phases. The first group participated in both phase 1 (“Learning phase”) and 2 (“Validation of the system”). The second group participated only in phase 2. The perceptual evaluation was performed in order to compare the ability of trained and not trained users on understanding which kind of movement feature represented by a particular sonification and if the sound is intuitive enough to be perceived without prior knowledge. The two phases of the experiment can be described as follows:

Phase 1

“Learning phase”: participants were asked to move freely. Four Android smartphones were attached with armbands to their forearms and calfs. Participants had one minute for exploration of each movement feature (energy, fluidity/rigidity, impulsivity), i.e. they can freely move and listen to the sound produced by the system. After trial of the system, participants provided ratings, describing their personal opinion and experience they had intearcting with our system. The questionnaire using a six-points Likert scale is shown in Table 1.1.

The second group of evaluators were introduced as a control group; they did not participate in the first phase of experiment, i.e. the learning phase. Instead, they immediately proceeded to the questionnaire for evaluation of the short audio segments. We investigated the ability of not trained participants, to recognize which movement quality mapped into sound without prior knowledge.

Phase 2

“Validation of the system”: we constructed a questionnaire with short audio recordings that had been generated from actual movements, using our system. Both groups of participants (Trained and Control) provided answers to the questionnaire, trying to understand which movement feature sonification they hear, no video feedback was provided.

In the questionnaire with short audio segments we used two types of questions with respect to two different sonification models that we developed and used in experiment (fluidity and impulsivity). The sonification model of fluidity is based on the energy, therefore first type of questions asked our participants to chose two out of four conditions: ”low energy or high energy” and ”fluid or rigid” (where rigid is a way to express ”not fluid” in this experiment). Second type of questions concerned about impulsivity sonifications allowed to chose between ”impulsive or not impulsive”. A diagram of the experimental setup is shown in Figure 1.2. Apart from the interview data from both groups, we have collected video, audio and accelerometers data of the first group.

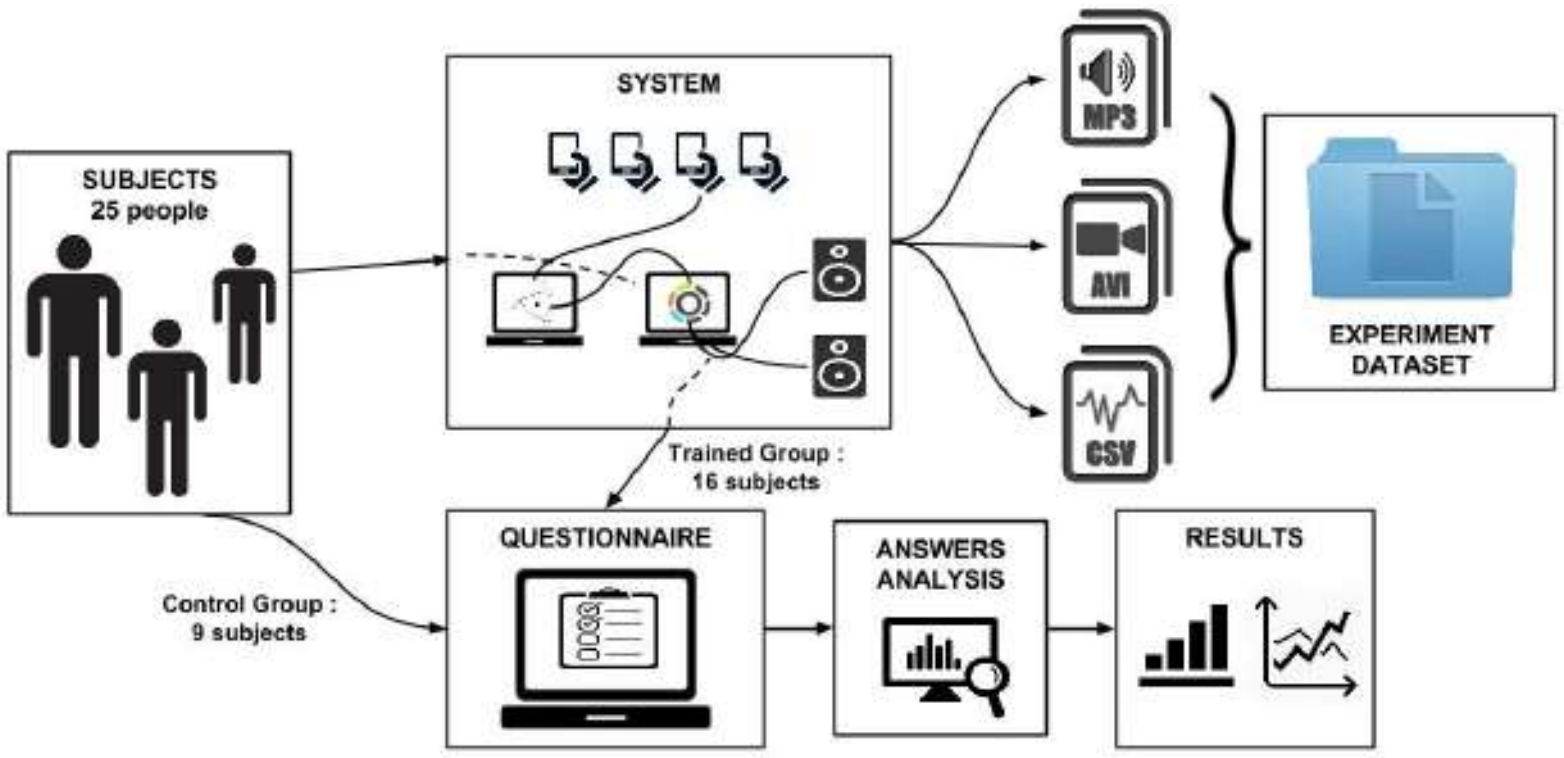

**Fig. 1.2** Diagram of Experimental setup

## *1.5.3 Results*

According to the collected feedback from the first group of participants, we were able to draw the following conclusions: Sound was responsive to the movement (mean 4 with STD 0.73) and easy to manipulate (mean 3.81 with STD 0.655). Learning of the system was interesting (mean 4.19 with STD 0.655) and intuitive (mean 4.06 with STD 0.929). Participants stated that it was easy to manipulate the sound corresponding to the following movement features: fluidity (mean 3.63 with STD 0.957), energy (mean 4.31 with STD 0.793) and that they could distinguish between rigidity and impulsivity sonifications (mean 3.63 with STD 1.147). According to respondents, they were at their own pace wearing the system equipment meaning that they were comfortable enough (mean 3.75 with STD 1).

In Table 1.1 detailed information of the feedback from participants is provided.

Next, we checked whether the prior knowledge of the system inuenced the recognition of the audio sonifications.

First, we calculated the proportion of the correct answers, when participants could recognize the movement quality from short audio segments. Table 1.2 displays the percentage of correct recognitions by the two groups (Trained and Control) of participants for each movement feature.

An ANOVA was conducted independently for each analysed movement feature. The results of the analysis between groups are the following:

Low Energy (F=0.566, p=0.459); High Energy (F=0.566, p=0.459), which showed no significant difference between trained and control groups.

For analysis of the movement feature Fluidity/Rigidity, between the groups, a four one-way ANOVAS has been conducted, comparing all four conditions: rigid low energy (F= 10.015, p=0.004), rigid high energy (F=0.007, p=0.933), fluid low energy (F=1.659, p=0.210), fluid high energy (F=0.131,

**Table 1.1** Feedback about the system. (6-points Likert scale from 0 - not at all to 5 - very much)

| Question | Min | Max | Mean | STD |
|---|---|---|---|---|
| How responsive was the sound to the actions you performed? | 3 | 5 | 4.00 | 0.730 |
| How well could you manipulate the sounds? | 3 | 5 | 3.81 | 0.655 |
| How compelling was your sense of moving around in the system? | 3 | 5 | 3.81 | 0.655 |
| Did you at your own pace wearing the system equipment? | 2 | 5 | 3.75 | 1.000 |
| I thought learning how to use the system was interesting | 3 | 5 | 4.19 | 0.655 |
| I thought learning how to use the system was intuitive | 2 | 5 | 4.06 | 0.929 |
| I thought learning how to use the system was confusing | 1 | 4 | 2.00 | 1.155 |
| I thought the sounds were pleasing | 1 | 5 | 3.13 | 1.258 |
| It was easy for me to manipulate the sounds related to fluidity | 2 | 5 | 3.63 | 0.957 |
| I had to put too much strength into creating an impulse | 1 | 4 | 2.44 | 1.153 |
| It was easy for me to differentiate rigid and impulsive movements | 2 | 5 | 3.63 | 1.147 |
| It was easy for me to manipulate the sounds related to energy | 3 | 5 | 4.31 | 0.793 |
| I could notice the difference between using one, two or more limbs during the energy experiment | 1 | 5 | 2.88 | 1.310 |

p=0.720). According to results we found that trained group of participants are significantly better in recognizing rigid low energy movements with respect to Control group (that had no prior knowledge about system and did not hear any sonifications prior interview).

Analysis of the ability to recognize the movement quality Impulsivity did not show a significant difference between the groups: No Impulse (F=1.341, p=0.259) and Impulse (F=0.001, p=0.978).

Therefore, we were able to conclude that the chosen sound models were intuitive for both groups, with and without prior knowledge.

## 1.6 Demo Game "Move in the Dark"

As application of our experimental system, we developed interactive multiuser and multimodal game named "Move in the Dark".

**Table 1.2** Proportion of correct answers in recognition of the movement qualities from auditory stimuli (first group of participants - "Trained", second group of participant - "Control")

| **Group** | **Feature** | **Varirity** | **Mean** | **STD. Err** | **95% Confidence Interval** | |
|---|---|---|---|---|---|---|
| | | | | | Lower Bound | Upper Bound |
| Trained | Rigid | Low | 0.547 | 0.051 | 0.441 | 0.652 |
| | | High | 0.797 | 0.061 | 0.670 | 0.923 |
| | Fluid | Low | 0.828 | 0.049 | 0.726 | 0.930 |
| | | High | 0.375 | 0.069 | 0.232 | 0.518 |
| | Impulse | No | 0.937 | 0.0360 | 0.860 | 1.0144 |
| | | Yes | 0.890 | 0.0393 | 0.806 | 0.974 |
| | Energy | Low | 0.750 | 0.00 | 0.750 | 0.750 |
| | | High | 0.9375 | 0.036 | 0.860 | 1.014 |
| Control | Rigid | Low | 0.278 | 0.068 | 0.137 | 0.419 |
| | | High | 0.806 | 0.081 | 0.637 | 0.974 |
| | Fluid | Low | 0.722 | 0.066 | 0.586 | 0.858 |
| | | High | 0.333 | 0.092 | 0.143 | 0.524 |
| | Energy | Low | 0.722 | 0.050 | 0.606 | 0.837 |
| | | High | 0.9167 | 0.041 | 0.8206 | 1.0128 |
| | Impulse | No | 0.861 | 0.0605 | 0.721 | 1.0007 |
| | | Yes | 0.888 | 0.0291 | 0.829 | 0.950 |

Two users compete with each other, blindfolded. Listening to the sound around them, they have to start moving accordingly to what they hear. The better the user is able to translate her body gestures, the more points the user is gaining in order to be the winner. The type of sound is randomly chosen between sonifications of fluid and impulsive movement. Each sound is synthetically generated by the system and the volume is controlled by a random value representing the energy. The diagram in Figure 1.3 describes the game step by step.

The Demo game has been designed with a use of the EyesWeb Platform, Max Software, Graphical Editor, EyesWeb Designer and accelerometers of 4 Android Smartphones.

A Move in the Dark session consists of three parts:

Prepare.

At the beginning of the game two users can specify their names and how long they want to play. Before starting the game, the user should be blindfolded.

Hear and move.

The system will now start to produce sonifications related to movement qualities as fluid, rigid and impulsive. The user interprets the sounds and translate them into movements. Meanwhile the players are moving, graphs related to each considered quality are displayed into the game interface for the audience.

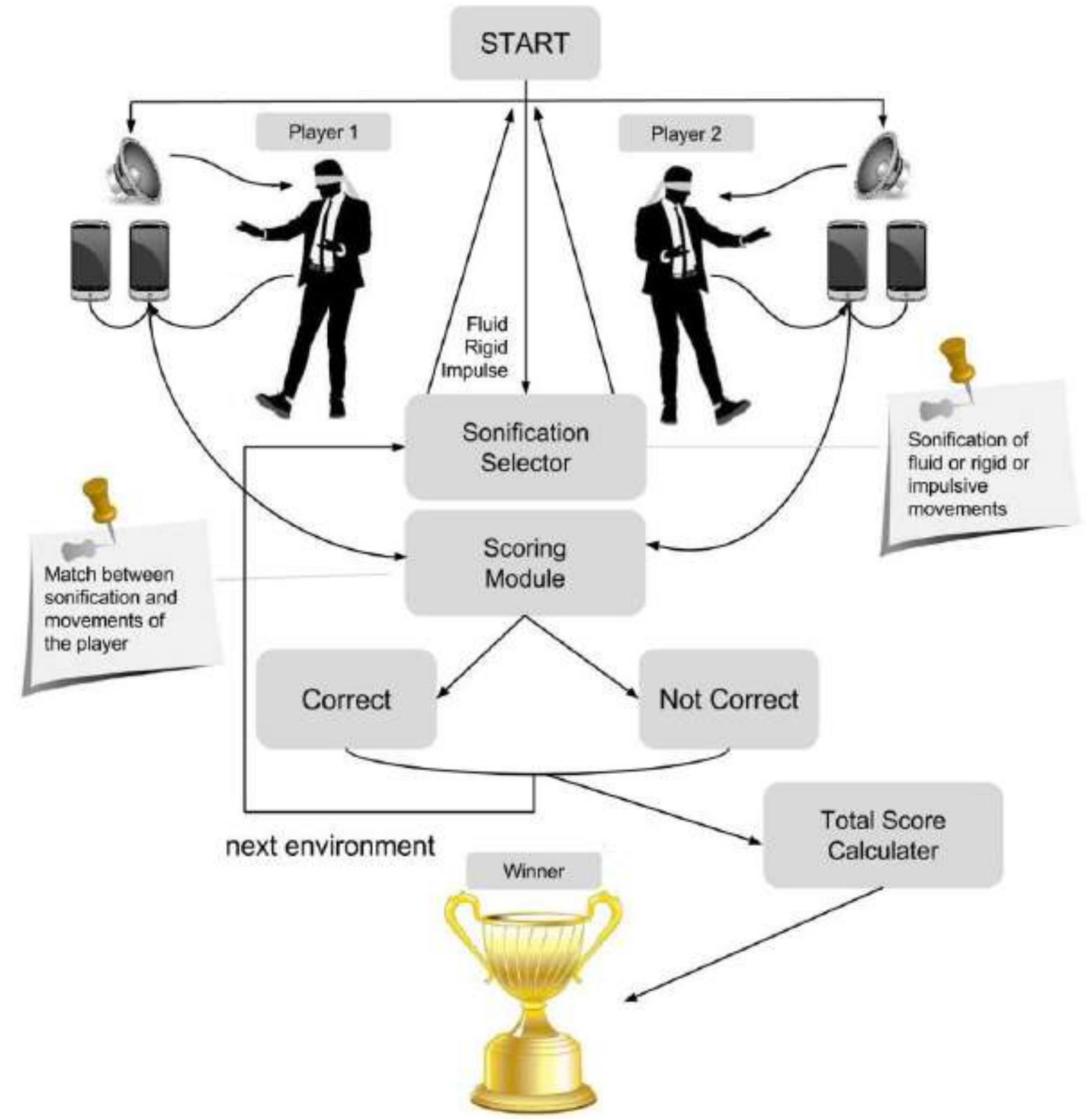

**Fig. 1.3** Game "Move in the Dark". The diagram of the game structure

Check your scores.

Depending on how much each player has been able to correctly translate the sonification into the related feature, a winner will be determined by the system. At the end of the game, the acquired scores are displayed.

The "Move in the Dark" game has been presented in the Mundaneum Museum of Mons, Belgium and Festival della Scienza, Genova, Italy (27 November 2015).

## 1.7 Conclusions and future work

The project at the eNTERFACE Workshop 2015 was carried out in collaboration between three institutions: University of Genova (Italy), KTH Royal Institute of Technology (Sweden) and Maastricht University (Netherlands) at the premises of the Numediart Lab in the University of Mons, Belgium. It is considered as a preliminary study for Use Case #1 of H2020 DANCE Project.

We acquired a deeper understanding of the computational models and theories meaningful to achieve the goals of the DANCE Project, as well as the paths for investigation in the following work related to extraction of high level movement qualities from the sensors data.

With the stated objectives in mind, we developed a working system able to describe a number of movements qualities and translate them into the auditory domain, that was successfully perceived by the participants who took part in the experimental study.

In the future we are planning to define new algorithms related to additional features detection and adapt the implemented techniques to be effectually used with video, Kinect and MoCap data.

## 1.8 Acknowledgements

This work was partially supported by the EU H2020 ICT DANCE Project no.645553.

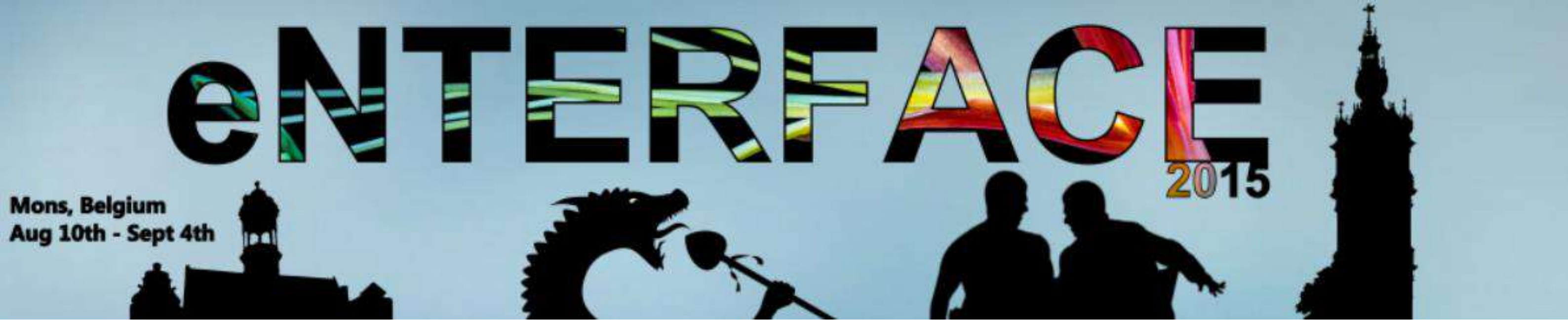

## Project #3:

# EASA : Environment Aware Social Agent

*Hüseyin Cakmak, Kevin El Haddad, Nicolas Riche, Julien Leroy, Pierre Marighetto, Bekir Berker Türker, Hossein Khaki, Roberto Pulisci, Emer Gilmartin, Fasih Haider, Kübra Cengiz, Martin Sulir, Ilaria Torre, Shabbir Marzban, Ramazan Yazıcı, Furkan Burak Bâgcı, Vedat Gazi Kılı, Hilal Sezer, Sena Büsra Yenge.*

# EASA : Environment Aware Social Agent

Hüseyin Çakmak, Kevin El Haddad, Nicolas Riche, Julien Leroy, Pierre Marighetto, Bekir Berker Türker, Hossein Khaki, Roberto Pulisci, Emer Gilmartin, Fasih Haider, Kübra Cengiz, Martin Sulir, Ilaria Torre, Shabbir Marzban, Ramazan Yazıcı, Furkan Burak Bağcı, Vedat Gazi Kılı, Hilal Sezer, Sena Büşra Yenge.

## Abstract

Enhancing human-machine interaction by adding emotions to the machine‘s way of expression is one of the main topics of current research. This would improve the interaction, on the assumption that the more human-like the machine behavior is, the more comfortable the interaction with it will be. This project aims to create an environment-aware emotional avatar. This avatar will be placed in an experimental framework which is described as follows: Participants will be interacting with each other in a limited room in which the avatar will be present as a spectator. The avatar should be able to recognize when different scenario events occur and react in an affective way. It will react through different affective expressions and utterances. This report summarizes the advances made towards the building of such a system during the eNTERFACE'15 Workshop.

Hüseyin Çakmak · Kevin El Haddad · Nicolas Riche · Julien Leroy · Pierre Marighetto
University of Mons, Belgium e-mail: huseyin.cakmak@umons.ac.be

Bekir Berker Türker · Hossein Khaki · Shabbir Marzban
Koç University, Turkey

Emer Gilmartin · Fasih Haider
Trinity College Dublin, Ireland

Kübra Cengiz
Istanbul Technical University (ITU), Turkey

Furkan Burak Bağcı · Vedat Gazi Kılı · Hilal Sezer · Sena Büşra Yenge
Turgut zal University, Turkey

Ramazan Yazıcı
Izmir Technical Institute, Turkey

Ilaria Torre
Plymouth University, UK

Martin Sulir
Technical University of Košice, Slovakia

## 1 Introduction

This project aims at building a social agent aware of its environment and able to express emotion-like audiovisual signals. This is of course a project which cannot be finalized in one month. This is why the project has been designed in different work packages that are mostly independent from each other. This allows to build sub-teams that are responsible of specific parts of the project without having to constantly discuss design choices with other teams. The work packages that have been defined are explained below :

WP0 is the **DATA COLLECTION** package and is one of the most important tasks in this project. The database consists of audio, video and 3D recordings of the participants as well as annotations of what happened and when during each session. This information will be fed to WP1 and WP2.

WP1 is the **ATTENTION** package. In this package, the team in charge has to predict where the avatar should be looking at a given time and given what is happening in the environment. Although the processes and algorithms behind it are very complex, the simple objective of this package is to output the Euler angles that define the head posture of the avatar at any time.

WP2 is the **AFFECT RECOGNITION** package. The aim here is to be able to recognize what is happening in the scene in terms of emotions of the participants and if any unexpected events are happening.

WP3 is the **DECISION** package. It takes the outputs of WP1 and WP2 as input. This part of the system is responsible for making the decision of what the avatar is supposed to do. Many things are possible here, from mimicking other participants to building complex artificial intelligence. In this work, this part was simply reproducing the affects that are recognized from the scene.

WP4 is the **SYNTHESIS** package. In this package, we build the audio and animation data corresponding to the decisions made in WP3.

WP5 is the **RENDERING** package. This package takes care of rendering the final animation based on the audio and visual trajectories built in WP4.

The organization of this report follows exactly the order of the work packages described above. Each work package is described in one section and the report finishes with a conclusion section.

## 2 Work Package 0 : Database Collection

We recorded a multi-modal database that includes videos of the participants, audio track of each participant as well as from a microphone in the middle of the experiment room and the stimuli videos. As a post-processing step, affect bursts from each participant have been annotated. All the modalities have been synchronized. Behavioral features have been extracted from the Kinect data of the participant and are also synchronized and available in the database. Finally, a Matlab toolbox has been coded to ease the visualization of the data available with examples of usage.

The emotional expressions that are considered in this work are amusement (smile, laughter and amused speech), disgust and surprise – which can be positive or negative. These emotions are triggered either by the content of the visioned videos or by the unexpected intervention of an external person.

### *2.1 Setup*

Figure 1 shows a schematic of the experimental setup that has been used. Each of the participants was provided with a personal wireless microphone (Sennheiser HSP4) so that they would not be disturbed by the microphone in their interactions with others. A multi-directional microphone (Rode Podecaster) was also placed in the middle of the room to record the ambiance sound. All the four microphones were linked to an external sound card (RME Fireface UCX). Recording all the audio on a single external sound card allows to have all of them synchronized. Two computers were used in the setup: one for audio recordings (yellow on the figure) and one for Kinect recordings. Indeed, a powerful computer was needed to record the Kinect data, as well as a lot of disk space, since the Kinect captures around 100 Mb of data per second. Therefore, we saved the data on an external hard drive through a USB 3.0 interface for speed requirements (pink on the figure). For the stimuli, the setup included a wide screen television and two high quality loud speakers. The television was controlled by the same computer as the sound recordings and the loudspeakers were connected to the external sound card.

### *2.2 Stimuli*

The stimuli consisted of a set of videos that were found on the web. Two different sets of 7 videos were used during the recordings, but one given group of participants was shown only one set; each set was around 13 minutes long. Each set of videos contained material which were expected to elicit the four different emotional states

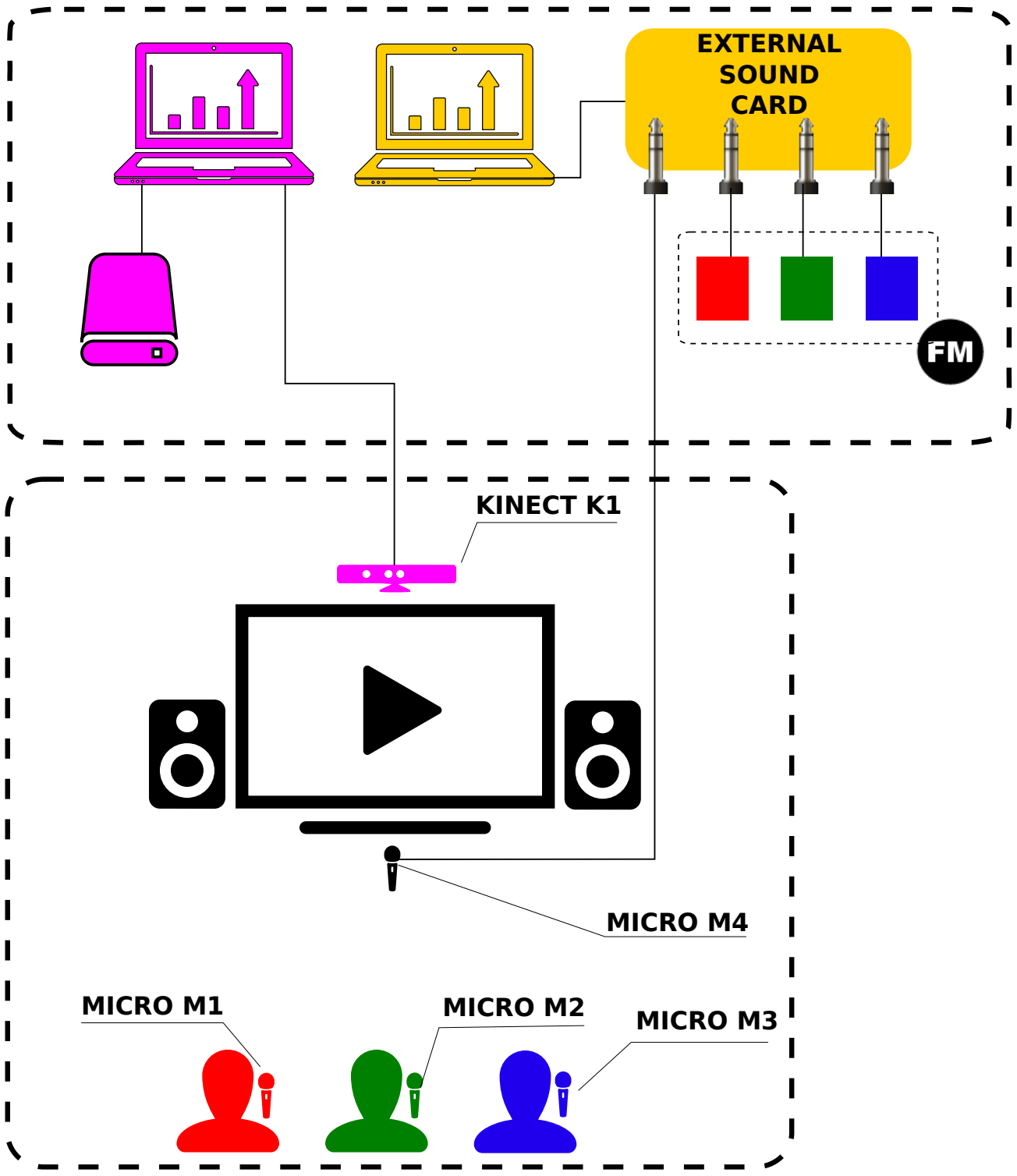

**Fig. 1** Schematic of the experimental setup

that we wanted to trigger.

## *2.3 Synchronization*

Before the beginning of the first video, while recording, two clap signals were played with exactly two seconds between them. This was done in order to be able to synchronize the Kinect data with the microphones data after the recordings. Indeed, as the microphones and Kinect were recorded on separate computers, a synchronization between the two was necessary. As the Kinect records audio as well as other types of data (all being synchronized), we can use the audio of the Kinect as an anchor to synchronize the audio from the microphones with the audio from the Kinect and therefore have all the data synchronized.

## 2.4 Annotations

Annotations of the different Social Communicative Events (SCE) event and external intervention have been done using the video annotation software ELAN. Table 1 gives the different tiers that have been used in the annotation scheme.

**Table 1** List of tiers that have been used in annotations

| TIER NAME | DESCRIPTION |
|---|---|
| amuse | When there is amusement which is visible even if it is not audible |
| laugh | When there is laughter which is audible even if it is not visible |
| surprise | When the participant is surprised |
| disgust | When the participant is disgusted |
| misc | Unclassified detected reaction |

Table 2 shows the number of time each tier appears on each recordings.

**Table 2** Number of occurrence of each SCE per recording. (amu = amusement, lau = laughter, sur = surprise, dis = disgust, misc = miscellaneous)

| | amu | lau | sur | dis | misc |
|---|---|---|---|---|---|
| Rec 1 | 111 | 84 | 2 | 7 | 7 |
| Rec 2 | 46 | 10 | 3 | 171 | 2 |
| Rec 4 | 30 | 7 | 3 | 117 | 3 |
| Rec 7 | 65 | 46 | 6 | 220 | 20 |
| Rec 9 | 88 | 70 | 9 | 239 | 7 |
| Rec 10 | 20 | 11 | 0 | 44 | 4 |
| Total | 360 | 228 | 23 | 798 | 43 |

Table 3 shows the mean time of tier apparition on each recording.

**Table 3** Mean duration of each SCE per recording. (amu = amusement, lau = laughter, sur = surprise, dis = disgust, misc = miscellaneous)

| | amu | lau | sur | dis | misc |
|---|---|---|---|---|---|
| Rec 1 | 8.62 | 2.30 | 0.70 | 1.41 | 8.29 |
| Rec 2 | 3.92 | 2.58 | 1.42 | 0.91 | 2.82 |
| Rec 4 | 8.82 | 3.56 | 3.66 | 1.58 | 18.77 |
| Rec 7 | 4.12 | 1.96 | 0.99 | 1.01 | 4.20 |
| Rec 9 | 3.52 | 1.79 | 2.45 | 1.40 | 2.79 |
| Rec 10 | 4.23 | 5.17 | 0 | 3.55 | 3.49 |
| Total | 6.54 | 2.72 | 2.44 | 1.41 | 6.42 |

## 2.5 Visualization interface

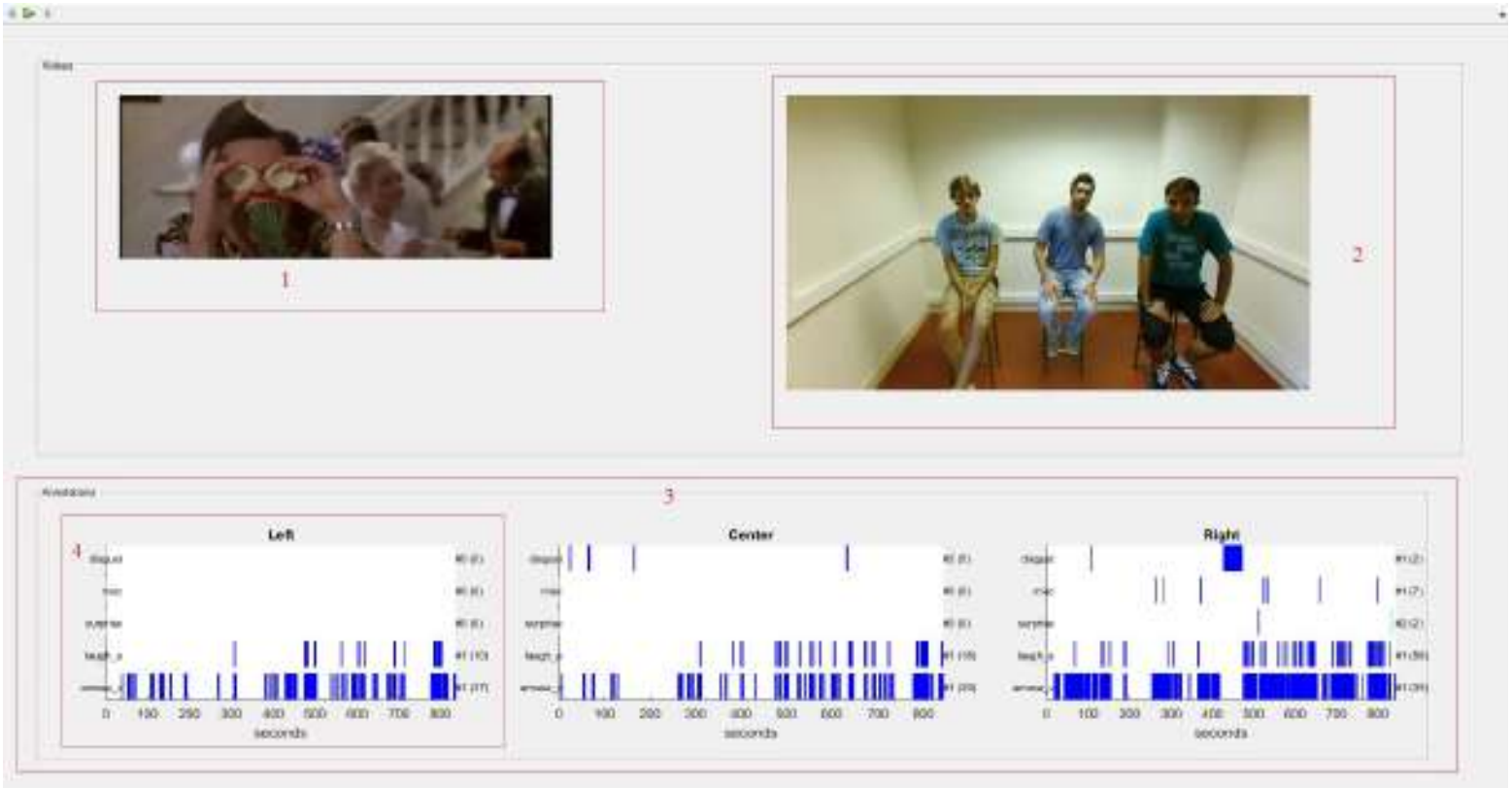

**Fig. 2** ToolBox screenshot

A ToolBox is proposed to visualize dataset videos and annotations. Fig. 2 shows an overview of this interface. The different sections are:

1. Stimulus video frame, corresponding to the current frame being showed in the recorded video section. A menu button allows to switch between the recordings and video sets.
2. Recorded video frame. Play/Pause button, as well as Previous Peak, Previous, Next and Next Peak buttons, allows to navigate through the video.
3. Annotations of the current recording. This section is split into $n$ parts (one per participant), with $n$ being the number of participants in the current video (2 or 3). Subsection position corresponds to participant position in video shown. Next point describes these subsections.
4. Participant corresponding annotation. Annotations were imported from EAF files using SALEM ToolBox [8].

## 2.6 Extracting Kinect Features

Three visual features data, RGB video and the audio streams are being extracted. Audio is used to synchronize other audio recording sources. RGB video is then multiplexed with synchronized scene microphone to be used in annotations. Lastly, the visual features data, used for the recognition of emotional expressions, are body features, facial features and facial animation parameters (FAPs). Body features consist of 25 full body joint positions and angles in the Cartesian coordinates system listed in [1]. Facial features are 8 quaternary flags providing different states of the face which are happy, engaged, left/right eye closed, mouth open/moved, wearing glasses, and looking away flags. Each flag has five possible values, 0 to 4, which indicate the detection result as unknown, no, maybe, and yes respectively. FAPs are defined based on muscle activation on human face in the framework of MPEG-4 standard. They form a set of high level descriptors of facial animations which are very useful in the analysis and synthesis studies [15].

In the aim of recognition system building, we have used only 6 joints (head, neck, shoulder left and right, spine shoulder, spine mid) from upper-body which are specially selected as less noisy and dominantly moving ones by our eyes inspection. Our modified frame level feature vector for body is as follows.

$$f_{body} = [J_v J_o \dot{J_o}]$$

Since 3D locations of the individuals joints are meaningless for classification purposes, we only include the first derivative of positions and indicate it as joint velocity, $J_v$. We also included joint orientations, $J_o$, and its first derivatives, $\dot{J_o}$, to introduce dynamic features as well. In total, we have 58 dimension feature vector per body, since one of the selected joint is Head and has no orientation as a joint point.

## 2.7 Contents of the Database

The contents of the database are given in Table 4 (all the elements are synchronized with each other).

**Table 4** Contents of the database

| | CONTENTS |
|---|---|
| 1 | Audio files for each recording session: one for each participant and one multimodal microphone in the middle of the room that gets all the different sounds. |
| 2 | RGB Video file of the participant |
| 3 | Stimuli Video |
| 4 | Per subject features extracted from the Kinect data |
| 5 | Annotations of the SCE and external interventions in the database in the ELAN format |
| 6 | A Matlab toolbox for facilitating the navigation in the database and its visualization |

## 3 Work Package 1 : Saliency Detection

When dealing with great number of stimuli, humans naturally tend to focus on a subset of them, discarding less ”interesting” signals. The purpose of this selection is to reduce the amount of information to process and identify as quickly as possible those parts of our environment that are key to our survival. Such mechanism is called human attention.
Regarding human vision, visual attention has been studied through many papers, such as [9]. The objective is to determine where humans tend to look on an image. The result is represented as a density map, with values between 0 and 1, showing the probability of a pixel to be seen by a person. This map is called the saliency map. Fig 3 shows an example of a saliency map.

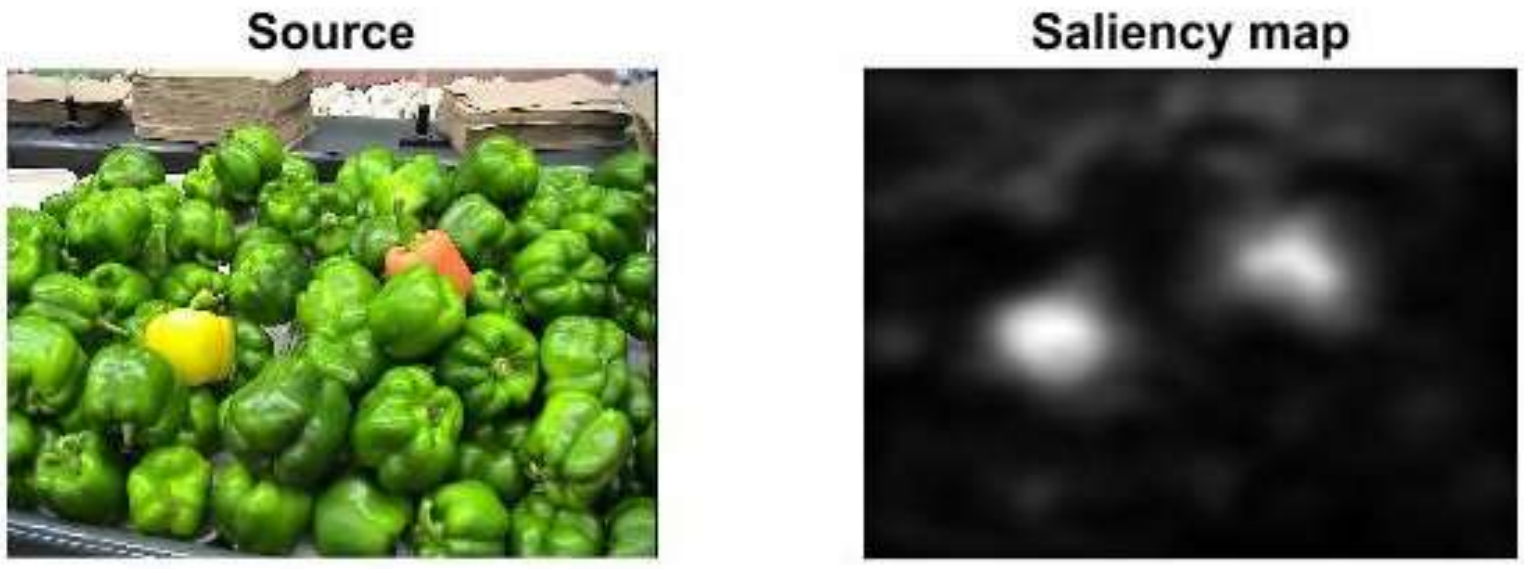

**Fig. 3** Example of saliency map.

This work package aims to detect areas that attract attention in the environment and determine an eye position for the avatar, using saliency models. As saliency models usually focus on one source of stimuli (such as audio, image or video), the objective here is to:

- Compute a saliency map on audio stimuli only.
- Compute a second map using video stimuli.
- Create a 3D representation of the scene, where a third saliency method can be used
- Project audio and video results onto the 3D scene representation, in order to merge the different maps, resulting in a final saliency map and thus, a gaze location for the avatar.

### 3.1 Audio stimuli

Audio saliency focus only on audio stimuli, without considering visual input. The objective here is to find temporally when the sound is more attractive to humans. In order to find it, we merge two different models:

- Schauerte model [17] is used to determine the regions of surprise on the global soundtrack.
- Mancas model [13] transform sound's spectrum, in order to use image rarity-based saliency model on it. Fig. 5 shows the different steps from the audio input to the result.

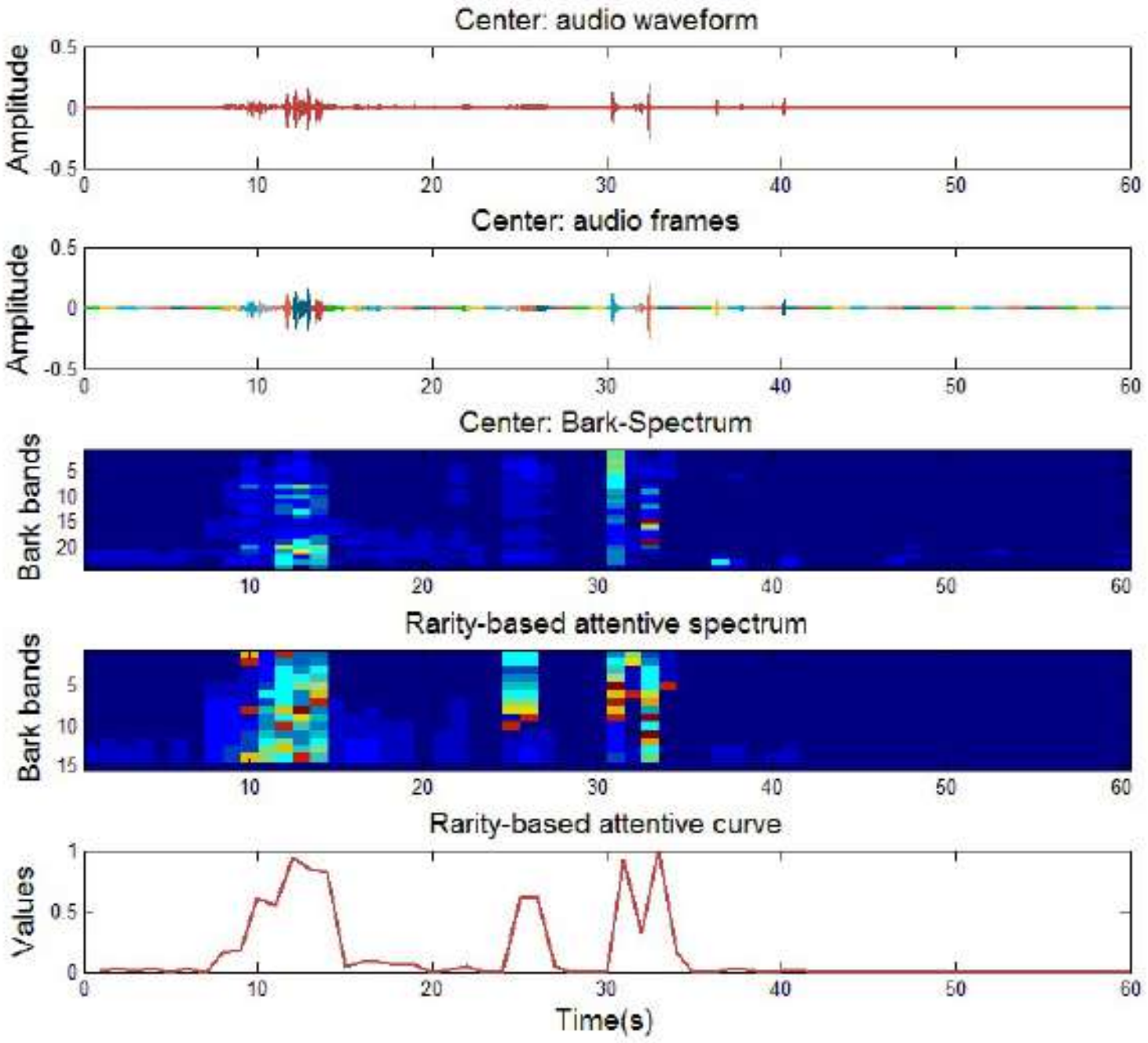

**Fig. 4** Mancas model overview. From top to bottom: (1) Audio input - (2) Audio input divided into frames, fitting video frames - (3) Bark spectrum - (4) Spectrum adapted to fit Mancas model - (5) Rarity-based model result.

These models are used for each soundtrack coming from participants' microphones. Thus, we obtain the "moments of attention" for every participant by mixing the results of this two models. Fig 5 shows, for 3 participants, the inputs, both models results and the final result.

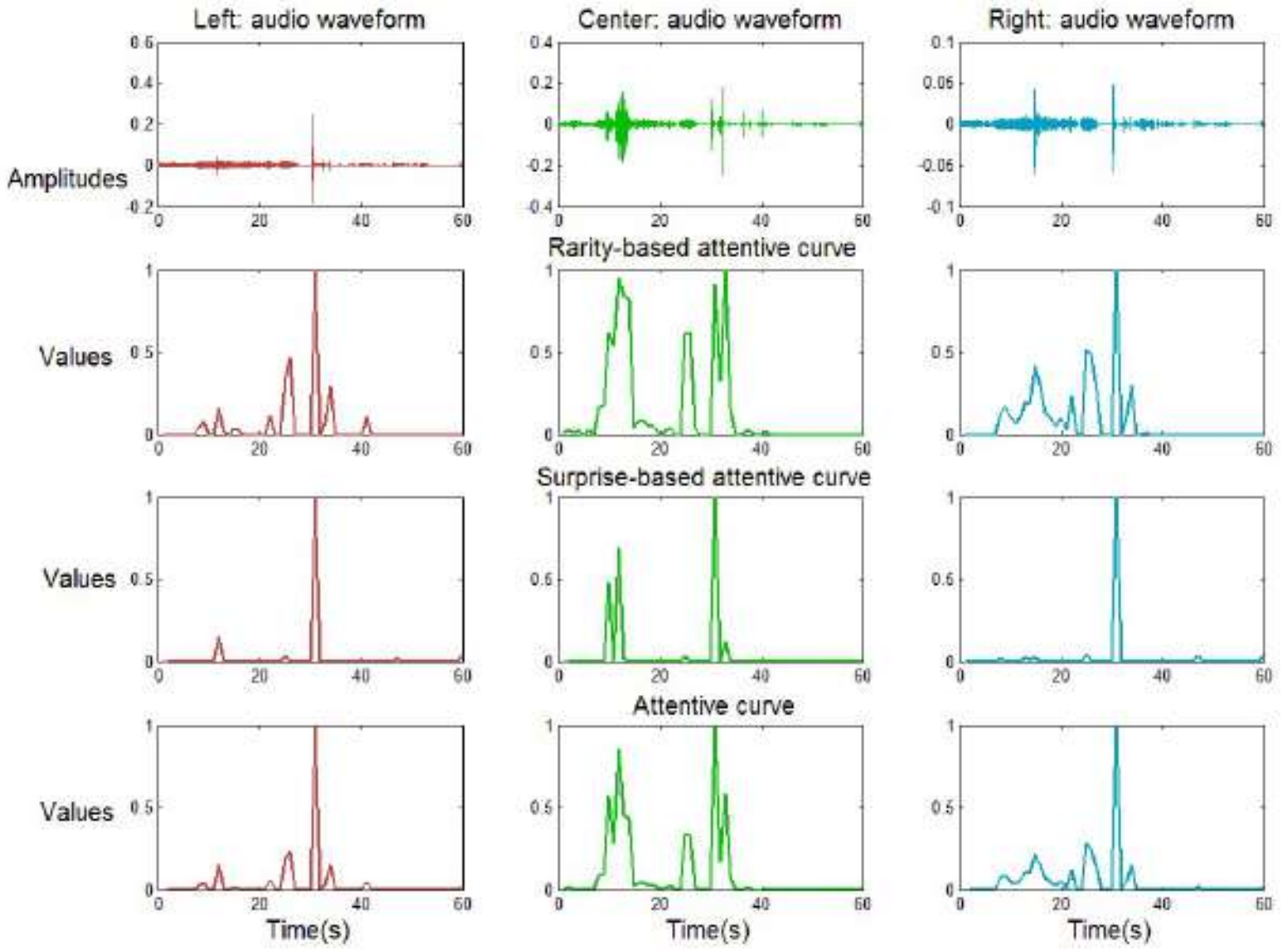

**Fig. 5** Audio saliency results. From top to bottom: (1) Audio input - (2) Mancas rarity-based model - (3) Schauerte surprise-based model - (4) Final result.

## *3.2 Hyperaptor: Saliency model on video stimuli*

Video inputs come from the recorded participants and the video stimuli. To obtain a saliency map, one for each video, we used a simplified version of Hyperaptor model [3]. Fig. 6 shows a schema of the version we used in this work.

The three last phase of the original Hyperaptor model are not used (temporal stabilization, high level priors and object-oriented mechanism). Fixed cameras for recordings is the reason for deleting the temporal stabilization. We choose not to use the object-oriented mechanism, which is a combination between saliency and superpixel segmentation, to avoid redundancy with the saliency model on the 3D scene, using supervoxels. Finally high level priors were not quite interesting in the context of this project.

As the next section will be focused on obtaining static information on a 3D projected version of the environment, the main objective of this model was to obtain dynamic saliency as good as possible.

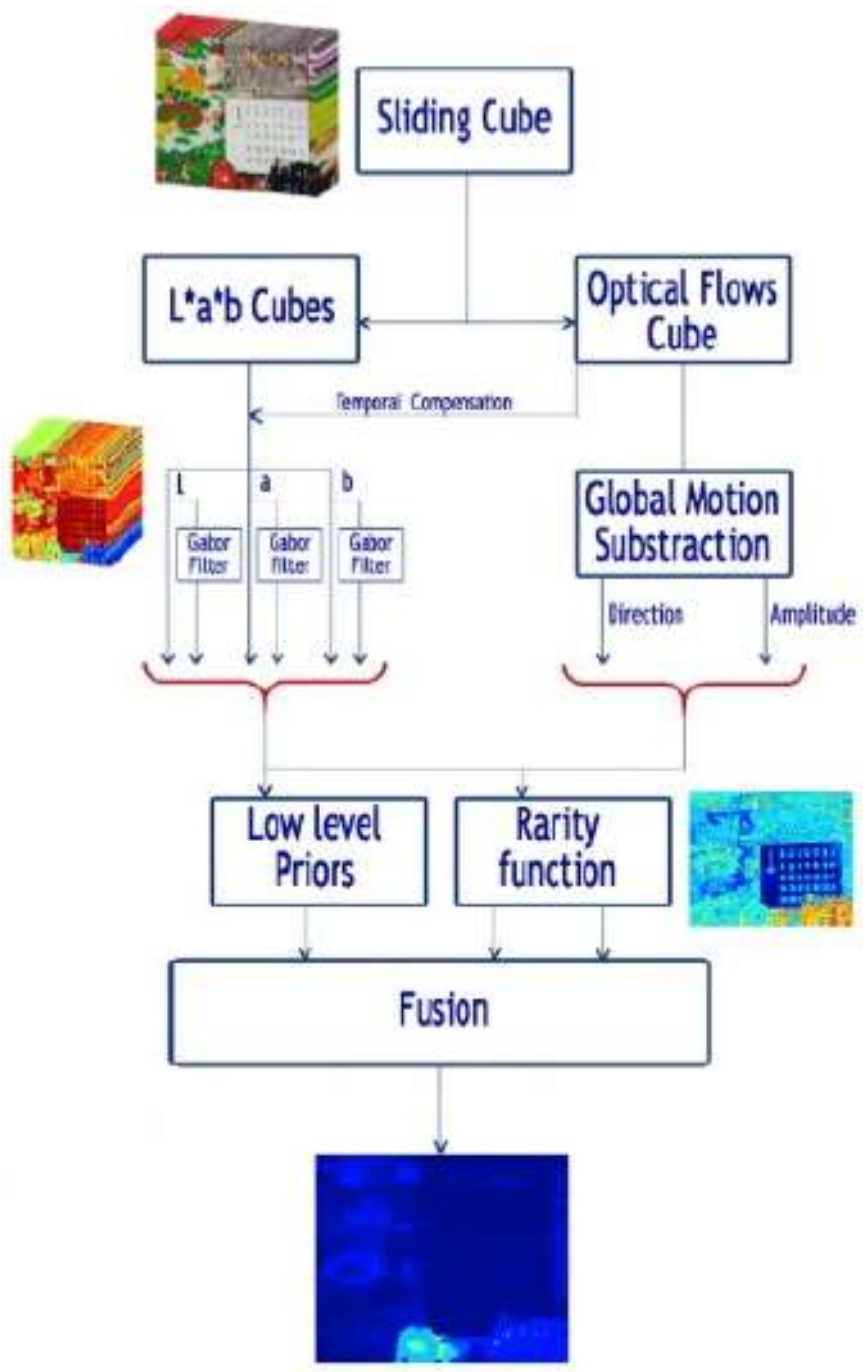

**Fig. 6** Schema of simplified Hyperaptor version used in this work.

## *3.3 SuperRare3D: Supervoxel 3D Saliency*

We used a novel object-oriented algorithm of bottom-up attention dedicated to analyze colored point clouds [11]. This model uses a rarity-based approach not based on superpixels as in [12] but on supervoxels. Supervoxels consist of an over-segmentation of a point cloud in regions with comparable sizes and characteristics (in terms of color and other 3D features). More details on supervoxels and the method used here are provided in the corresponding paper. Our approach has four major interests:

1. Supervoxels let us reduce the amount of processing and allow our method to work on organized or unorganized clouds. Thus, it can analyze point clouds or even fused point clouds coming from various sensors.
2. Supervoxels allow us to have an object-oriented approach in the 3D space.
3. Supervoxels multi-level decomposition allows us to maintain detection performance regardless of the size of the salient objects present in the data.
4. This approach provides a bottom-up 3D saliency map which is viewer-independent. It is possible to add viewer-dependent top-down information as a viewer-dependent centered Gaussian and depth information.

This algorithm can be divided into three major stages: (1) supervoxels decomposition, (2) supervoxel rarity-based saliency mechanism, (3) fusion. Fig. 7 shows the different steps of SuperRare3D (SR3D).

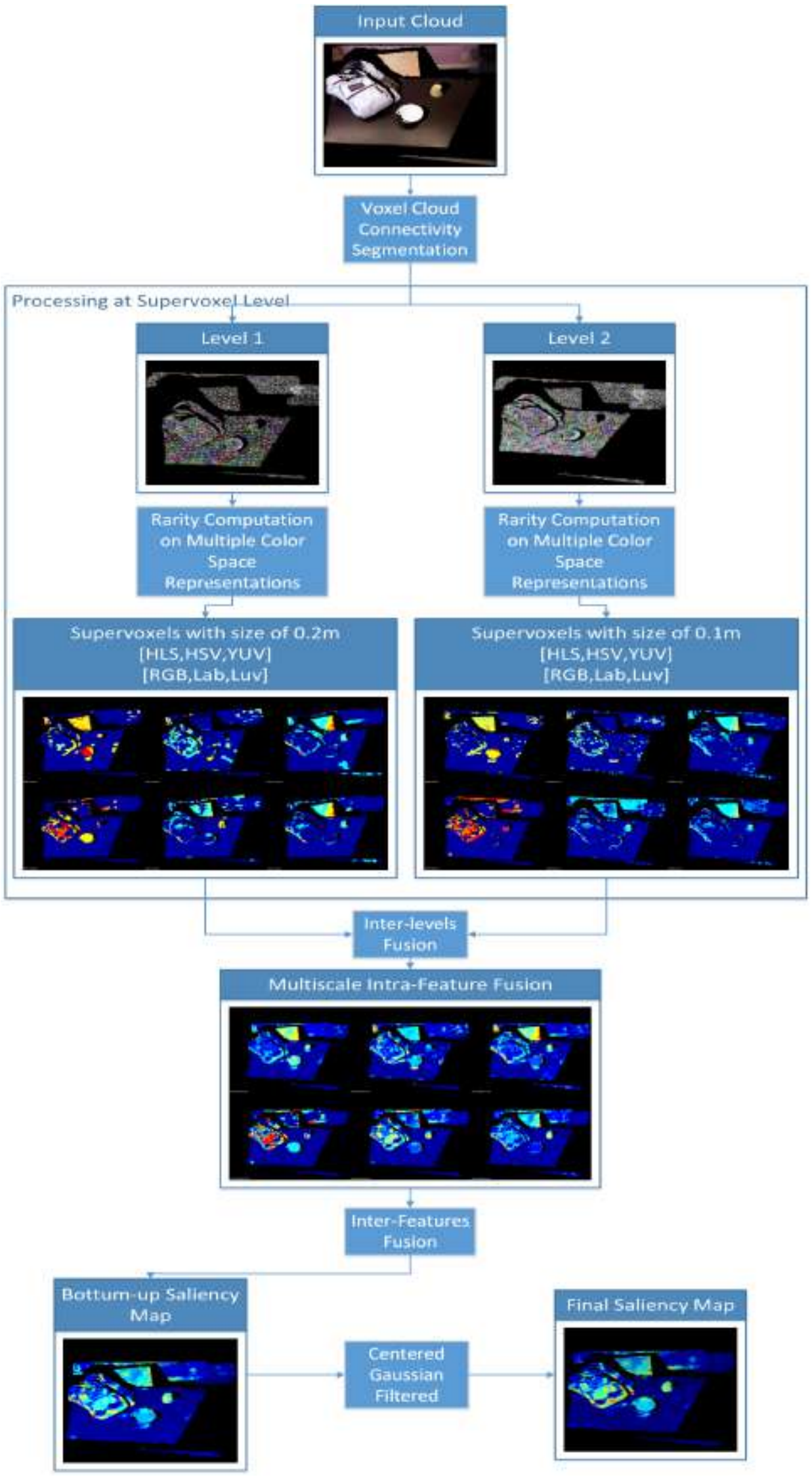

**Fig. 7** Our method is divided in 3 major steps: (1) multiscale supervoxels decomposition, (2) color rarity applied on multiple color spaces, (3) inter-level and inter-feature fusion. A top-down centered gaussian can be use to simulate the human centric preference [10].

### *3.4 Fusion*

Multimodal aspect of our environment comes from the mix of different salient source of information. In our experimental framework, 3 sources of different dimensions were joined together, such as:

1. 3D saliency source, coming from the analysis of the environment with our 3D saliency model, dealing with point clouds given by an RGBD camera. Thus, SR3D will create a 3D saliency map linked to the 3D environment computed.
2. 2D saliency source, coming from the analysis with Hyperaptor model on experiment's recorded video. 2D saliency maps will then be created for each frame of the video.
3. 1D saliency source, coming from the analysis of audio sources represented by different microphones set in the experimental environment. From those signals, an audio saliency map is extracted. This map represents, for each microphone, the moment in time when the sound is the most salient.

In view of the differences in dimensions from the present sources, the first step of our fusion process was to ”project” them in the same 3D world. To achieve this ”projection” and then fusion, a same 3D world coordinate system should be defined. For this, a 3D reconstruction of the experimentation environment was made and served as a reference. Each source was then calibrated and ”projected” on this reference environment, allowing a spatial fusion. Regarding the 2D and 3D flows, these were obtained using RGBD cameras, data once represented as point clouds can simply be registered on the world reference. Audio streams from multiple microphones were installed on the participants, these are defined by the experimental setting as static and will be represented in the space by the 3D position of the subjects. The audio information will be shown as a 3D Gaussian function of the intensity of the extracted audio saliency mechanism centered on the 3D position of the corresponding microphone. The Gaussian representation is achieved by the convolution of the points in the vicinity of the center with a Gaussian function.

Each of these sources have now their projection on a 3D representation, more specifically a point cloud. A fusion process will then be made between the different point clouds. This process involves several steps:

- First, we will generate a dynamic 3D saliency map fusing the map obtained with SR3D and the map from an OCTREE based change detector.
- Second, the 3D dynamic saliency map will be fused with the video saliency map to create a global visual attention map including all visual sources.
- Third, a 3D audio-visual saliency map is obtained by merging the audio map with the previous global visual map.

Fig. 8 shows the different step of the fusion used in the process.

Several algorithms fusions will be used. For the creation of 3D dynamic saliency map, as confidence in the change detection system is high, we will use a weighting of SR3D saliency map with binary output of the detector. The fusion will be effected as:

$$S_{DynamicSR3D} = S_{SR3D} + S_{SR3D} x B_{octree} \quad (1)$$

with $S_{SR3D}$, SR3D saliency map and $B_{octree}$, the binary map from the octree change detector. The moving areas will therefore have a doubled saliency value compared to the static value, giving more importance to the dynamic aspect. In the case of other fusions, a non-linear combination is used. This fusion is similar to the one employed in our rarity mechanisms.

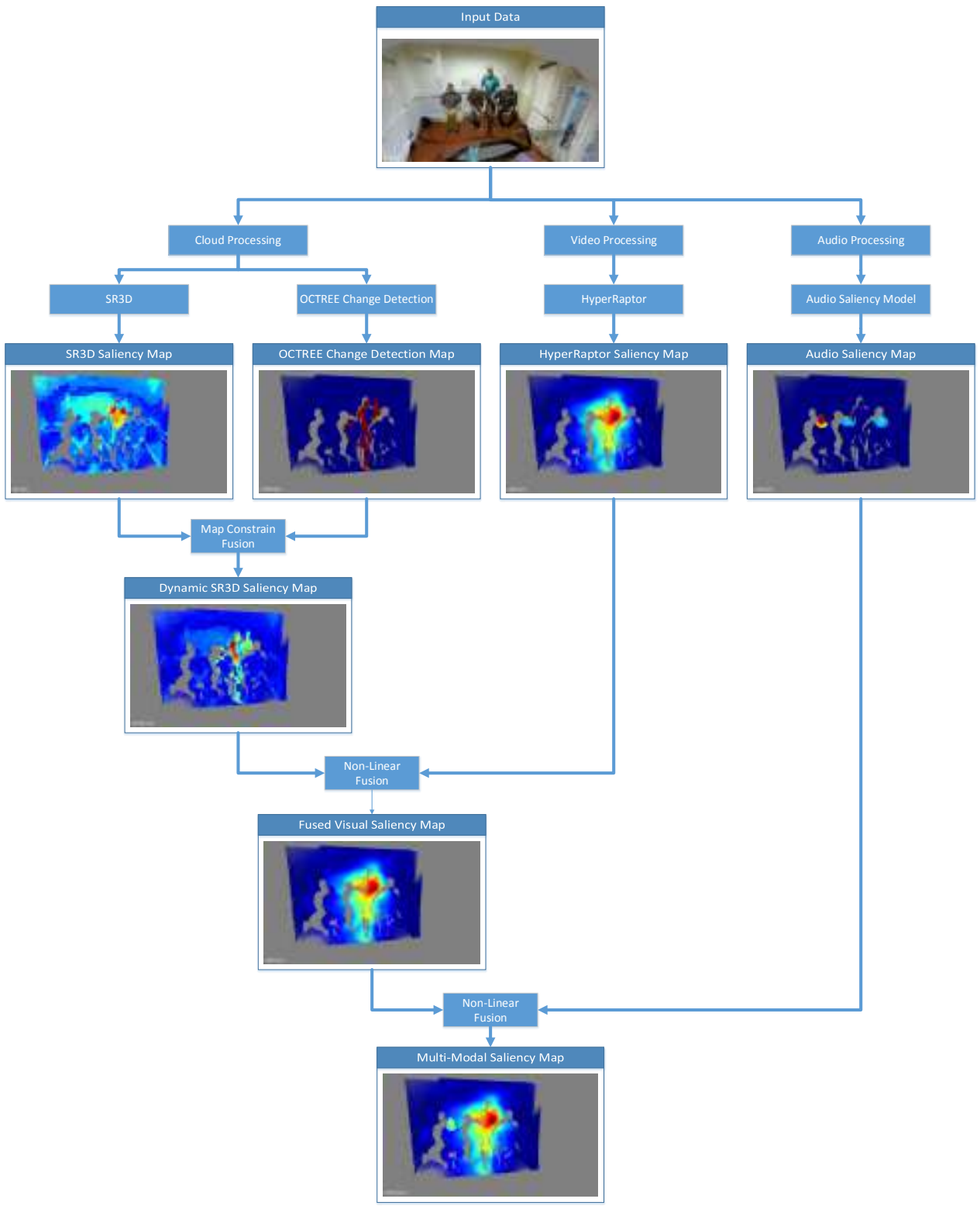

**Fig. 8** Schema representing the different step of the multimodal saliency fusion.

# 4 Work Package 2 : Affect Recognition

## *4.1 Handling the missing values*

Facial features and FAPs are related to body features. It means that if the body feature type is missed, the other two feature types will not be accessible anymore. Fortunately, missed values of body features were negligible and we just pick the available time stamps. For the facial features, when they are zero, detection result is unknown. To eliminate the zero entries, first we removed the frames with consecutive zeros that are more than 10 frames long and then we applied a median filter with the length of 21 to substitute the zero values. Since the facial features are quaternary flags and the length of the median filter is a odd number, median filter replaces the zero values with the median of the window which is a value between 1 to 3.

For the facial animation parameters, zero values represent missed values too but the ratio of them over the FAPs dimensions were different. Hence, initially we calculate the ratio of missed values over the all participants and then remove the dimension with high missed values ratio. The mean and maximum of the ratio of missed values over the different participants are presented in table 5. Based on table 5, we removed the first, second and tenth features of FAPs sets. We then applied linear interpolation for the rest of the missed values of each dimension.

## *4.2 Correlation matrix*

We performed two basic analysis on our dataset. First we analyzed the correlation between facial features and the emotional expression, particularly amusement, disgust, and idle. The correlation matrix is depicted in Figure 9. The first 8 rows are the correlation between the 8 facial features and the last two rows are the correlation of each facial features with amused versus idle and disgust versus idle respectively. We consider 0 for idle and 1 for disgust or amused frame. Based on Figure 9, MouthOpen and LookingAway flags have positive correlation with the amused and disgust. Happy flag is highly correlated with the amused but not with the disgust.

## *4.3 Feature Summarization*

As a summarization method of Kinect features, we used statistical functionals [14] on temporal segmented data instances. In other words, a segmentation process is done by sliding a window on frame level features of annotated data. This resulted in series of segments which have the same length in time. We then calculate statistics on frame level features for each of the segment to capture temporal behavior in a fixed length feature. This produces a data instance in our feature domain.

**Table 5** The mean and maximum of the ratio of missed values over the different participants

| feature name | mean of the ratio of missed values | max of the ratio of missed values |
|---|---|---|
| 1. JawOpen | 31.72% | 92.84% |
| 2. LipPucker | 75.08% | 97.57% |
| 3. JawSlideRight | 6.10% | 15.48% |
| 4. LipStretcherRight | 12.03% | 37.45% |
| 5. LipStretcherLeft | 9.65% | 25.51% |
| 6. LipCornerPullerLeft | 7.27% | 20.49% |
| 7. LipCornerPullerRight | 7.15% | 19.27% |
| 8. LipCornerDepressorLeft | 7.84% | 19.96% |
| 9. LipCornerDepressorRight | 7.25% | 17.23% |
| 10. LeftcheekPuff | 15.56% | 79.65% |
| 11. RightcheekPuff | 12.93% | 31.48% |
| 12. LefteyeClosed | 7.92% | 16.75% |
| 13. RighteyeClosed | 13.34% | 25.62% |
| 14. RighteyebrowLowerer | 6.10% | 15.48% |
| 15. LefteyebrowLowerer | 6.10% | 15.48% |
| 16. LowerlipDepressorLeft | 11.94% | 30.94% |
| 17. LowerlipDepressorRight | 9.52% | 21.98% |
| 18. Pitch | 6.10% | 15.48% |
| 19. Yaw | 6.10% | 15.48% |
| 20. Roll | 6.10% | 15.48% |

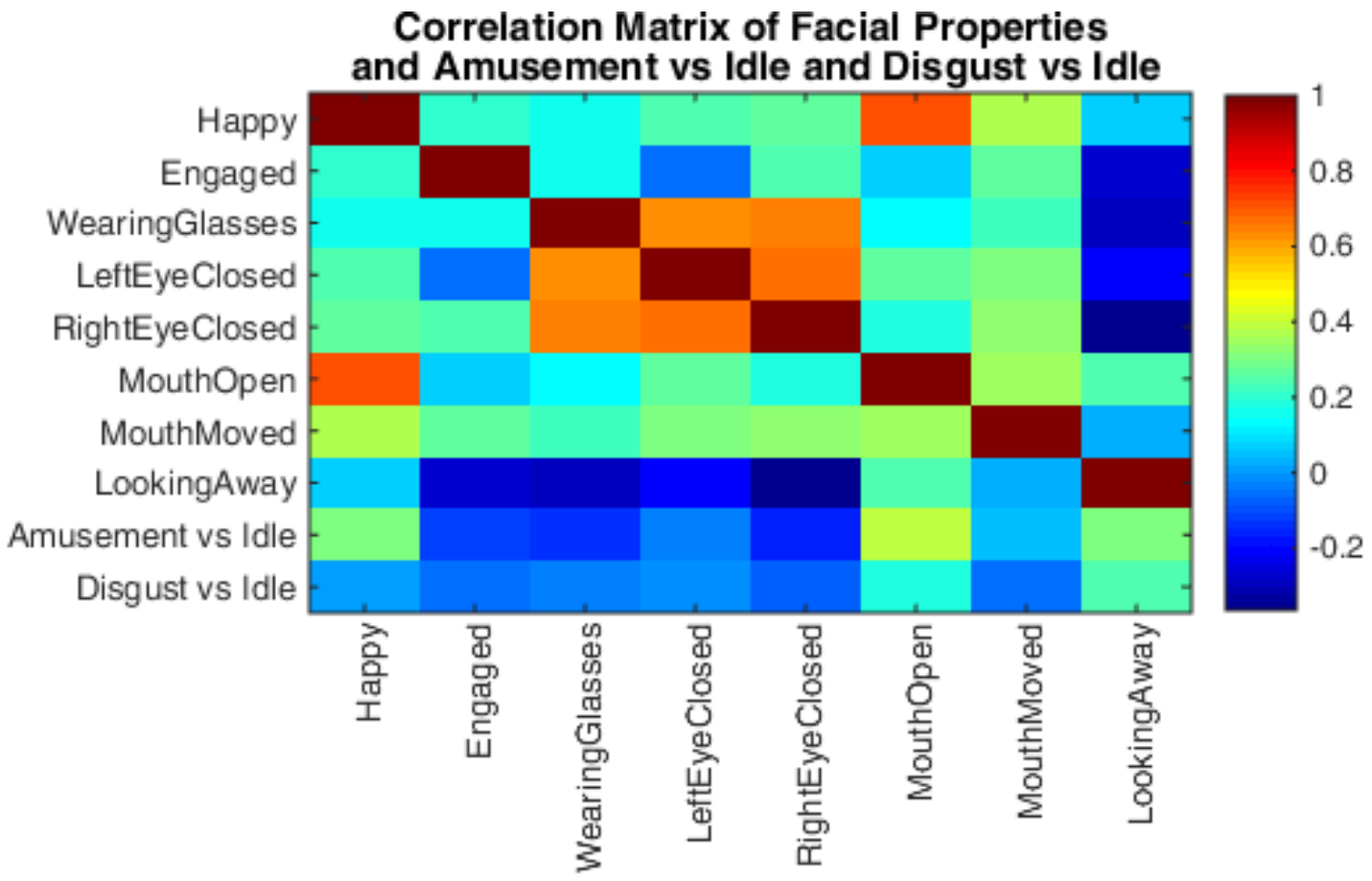

**Fig. 9** The mean and maximum of the ratio of missed values over the different participants

**Table 6** The number of segments for each participant and overall

| Speaker ID | 1 | 2 | 3 | 4 | 5 | 6 | 7 | 8 | 9 | 10 | 11 | overall |
|---|---|---|---|---|---|---|---|---|---|---|---|---|
| # Idle segments | 323 | 342 | 336 | 276 | 274 | 324 | 328 | 312 | 336 | 420 | 428 | 3699 |
| # amused segments | 353 | 415 | 974 | 60 | 102 | 23 | 267 | 870 | 282 | 132 | 230 | 3708 |
| # disgust segments | 19 | 0 | 89 | 10 | 2 | 0 | 7 | 182 | 64 | 37 | 43 | 453 |

The statistics are mean, standard deviation, skewness, kurtosis, range, min, max, first quantile, third quantile, median quantile, inter-quantile ranges. Since we have 11 statistics to compute for each frame level feature dimension, we end up having segment level feature vector with 11 times of frame level feature size.

Lastly, we have feature scaling and PCA steps just before feeding the data to classifier. In feature scaling, we basically scale the features to zero mean unit variance. In PCA we are reducing feature dimension while preserving 90% of the variance.

## 4.4 *Affect recognition result and discussion*

After feature summarization, in total, we have 7860 segments out of four selected recordings. Among those 7860, 3699 were idle, 3708 segments were amused and 453 segments were disgust. We performed the affect recognition on the first four recordings. The number of segments for each participant is gathered in Table 6.

**Table 7** Classification accuracy result for amused, disgust, and idle expression from body features, facial features, and FAPs.

| Speaker ID | 1 | 2 | 3 | 4 | 5 | 6 | 7 | 8 | 9 | 10 | 11 | overall |
|---|---|---|---|---|---|---|---|---|---|---|---|---|
| Body features | 49.64% | 57.86% | 42.32% | 58.67% | 49.74% | 32.28% | 54.82% | 40.18% | 53.67% | 56.54% | 49.07% | 48.33% |
| Facial features | 51.88% | 49.54% | 48.93% | 51.73% | 45.63% | 33.64% | 51.16% | 40.12% | 60.76% | 44.29% | 47.08% | 47.68% |
| FAPs | 46.39% | 37.38% | 41.17% | 48.84% | 22.22% | 31.12% | 45.02% | 34.38% | 52.49% | 40.24% | 32.24% | 39.48% |

In the recognition system, we build classifiers based on the visual features to distinguish between the amused, disgust, and idle. Classifiers are trained by using SVM [5] with radial basis function kernel. In training and testing scheme, we have 3 fold inner cross validation for exhaustive grid search of cost and gamma while the outer fold is for independent test operations.

Our experiments are organized in an eleven fold and the data of each participant are in a fold. We train and adjust the SVM parameters with the features of ten clips and test on the other one. Hence the analysis is person independent. To overcome the unbalance problem, we repeat the samples of smaller classes randomly in training. Hence the chance rate is 33.33% for a three class classification problem. The overall and each fold classification results for the three different features are depicted in Table 7. The overall results show that all three modality work better than the chance rate. Particularly, the body and facial features perform near 50% accuracy. Since the amused and disgust are highly correlated with facial expression, unlike the results of Table 7, we expect higher accuracy from the FAPs channel. We think the reason is related to the noisy features captured by Kinect because of insufficient light while recording.

## 5 WP 3 : Decision Management

This package could not be developed in detail and with the relevant research it need in the scope of this first enterface EASA project. We simply used a mimicking policy that relies on the information retrieved from the recognition package. This means that based on the affects recognized on the participants, the most important one is taken as the behavior of the avatar. For example, if 2 of the 3 participants are laughing, then the decision package outputs that at that moment the avatar should laugh as well. Of course, this may not work in some cases but if we consider the scenario in which participants are watching videos and the avatar is one of them also watching the video, the constraints make that this simple mimicry process is expected to be most of the time relevant.

## 6 WP 4 : Synthesis

As previously mentioned, the avatar should be able to express 4 different types of SCE (Amusement, disgust, positive and negative surprise) on three different arousal levels each.

For this project, data were gathered from previously recorded databases. These data consist of audio-visual non-verbal expressions of the previously mentioned SCE. Laugh data were taken from the database presented in [2] to represent amusement. The other expressions were taken from another database in which the same person recorded in the previously mentioned database was asked to express disgust, positive and negative surprises for several times on three different levels each. For these expressions, the motion capture data were recorded using Natural Point's Optitrack system. This was done using 12 infra-red emitting cameras, 33 facial markers placed on an actor as well as 4 headband markers placed on his head. The facial expression coordinates were given by the facial markers while the head movements were calculated from the headband markers. Thus giving us a total of 105 features per frame recorded. The frame rate at which the cameras were recording was 100 fps. Along with the movement capture data, the audio was also captured at 44.1 kHz and stored in 16 bit PCM WAV files.

These databases were recorded in similar conditions.

The audio and movement capture data we are using, were previously synchronized together and processed. The post-processing following the recordings is out of the scope of this report.

Each of the SCE are manually separated and annotated in three different arousal levels. This annotation was made by three annotators.

For the purpose of this project the coordinates and head motion can be projected directly onto the avatar without the need of synthesis. Indeed, each SCE is represented at three levels of arousal. Therefore this level can be controlled by choosing among the ones we have for each emotion. Thus, we have control over the arousal level of the SCE, but, this control is of a discrete type (i.e., the arousal level cannot be controlled on continuous scale). A synthesis system would enable us to have a continuous control over the arousal level of each of these SCE. This is why, the eNTERFACE month was dedicated to conduct research to obtain an AV SCE synthesis system. Preliminary results of these researches are presented in this section.

The synthesis is done separately for the audio and visual data and then synchronized. Our approach is to use HMMs to model each of the modalities. Indeed, they proved useful for modeling non verbal utterances in previous work [7, 6]

An audio-visual laughter synthesis system has already been proposed in [4]. So, in what follows, the work presented concerns only the other SCEs (disgust, positive and negative surprise).

Due to the 1 month time constraint most of the work was focused on the audio cue. Although, data was gathered for both modalities. Table 8 shows the amount of data gathered per SCE and per level. Level 1 (L1) refers to the lowest level, Level 2 (L2) to the medium level and Level 3 (L3) to the highest level. This table is valid for both modalities since the data for one modality is available for the other as well.

**Table 8** Number of samples per SCE and per level

| | disgust | +surprise | -surprise |
|---|---|---|---|
| Level 1 | 40 | 37 | 34 |
| Level 2 | 25 | 7 | 11 |
| Level 3 | 19 | 34 | 39 |

## *6.1 Audio*

For this work and in order to model the SCEs, 5-state left-to-right HMMs were used. HMM models were trained for each emotion and each level separately. The features extracted were 25 MFCCs and the log value of the pitch with their derivatives and double derivatives. They were modeled in each state with a multivariate single Gaussian distribution. The implementation of this system was made using the available HMM-based Speech Synthesis System (HTS) [16]. In order to obtain intermediate levels to the 3 we already have, interpolation is used. After the models for each SCE and each level was trained, the hts-engine software was used to interpolate one SCE of a certain level to the same SCE of another level. The output signal is the result of a weighted interpolation between one model and another.

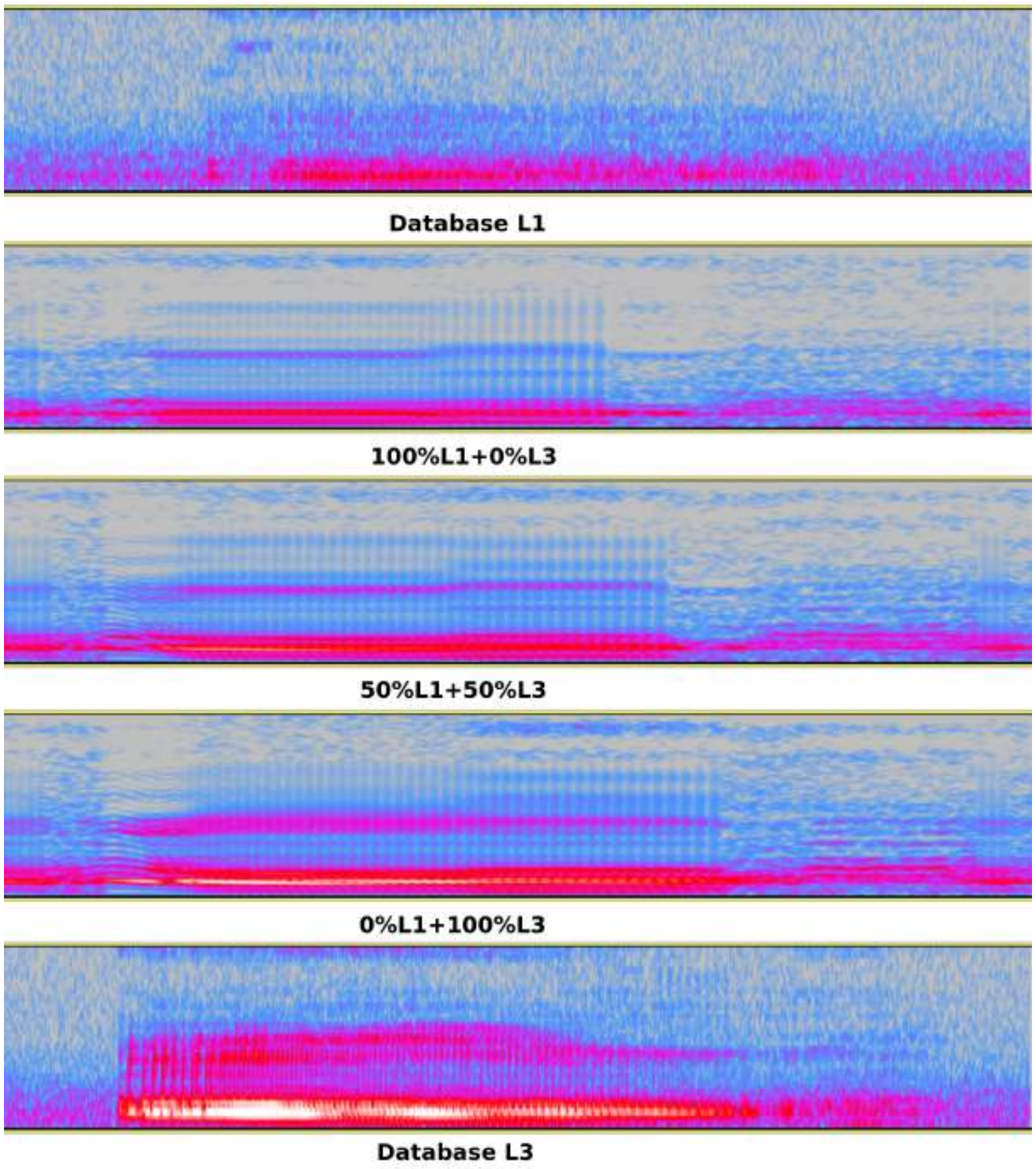

**Fig. 10** Representation in the spectral domain of examples SEC (of disgust in this example) taken from the database collected (top and bottom) and the interpolated synthesized SCEs between them

Fig 10 shows examples of SCEs from the database and synthesized SCEs. In this figure, the synthesized signals presented are the outcome of a weighted interpolation between L1 and L3 of the SCE disgust with weights as shown in Fig 10.

### *6.2 Visual*

Concerning the motion capture, we used the same approach as described in [4] for the HMM modeling. A PCA was applied to the 105 features. The four first principal components (PCs) were used to train the HMMs.

HMMs were trained for each SCE and each level. An interpolation technique was then used as previously described to obtain intermediate levels. After synthesis, an inverse PCA is applied to obtain the initial feature dimensions.

## 7 WP 5 : Rendering

The rendering is done using the open source 3D animation software called Blender. During the workshop, a real-time version of the avatar has been developed. It is able to receive data on a frame by frame basis.

In each frame 105 values are read :

- 3 head translations and 3 head rotation
- 3 translations for each of the 33 facial markers

The data is sent using the OSC protocol. The agent is also controllable in a wizard-of-oz manner by sending the content of a CSV file. This way predefined sequences may be played on the avatar.

A friendly user interface has been tested on an Android phone to send OSC messages to the agent. The system is currently lacking a computer-based interface.

## 8 Conclusions

In this project, 20 researchers from 6 countries and 10 different institution worked together to develop the different packages of this project. For each package, the advances that we were able to make are :

WP0 - DATA COLLECTION We were able to collect a database of social communicative events (SCE) that includes the audio of each participant of 6 sessions of about 13 minutes each. Many SCE were collected and annotated. A Matlab toolkit has also been developed to ease the navigation and the visualization of the database.

WP1 - ATTENTION Several existing attention algorithms, working on different sources (such as audio, video and 3D) were successfully applied in the scope of this project, mixed together and gave good results. As a result, the system is able to give the information of where the avatar should look at any moment.

WP2 - AFFECT RECOGNITION In the recognition package, we were able to build preliminary models and test techniques for recognition of SCE. The results are promising although there is still room for improvements. Features for each participant were extracted from the raw kinect data.

WP3 - DECISION The decision module is the less developed one among all the work packages and is only due to a lack of time within the scope of the project to explore such a large topic. The simplest decision protocol that we had defined beforehand was a mimicking behavior of the avatar. A simple python code to handle this was implemented.

WP4 - SYNTHESIS For the synthesis, preliminary models were built for disgust and surprise synthesis. To do this, a small database of these expressions were recorded for both audio and facial motion capture data. They were acted expressions. Existing laughter models were used for synthesis. Apart from the models, recorded original audiovisual sequences were also used. A simple sequence concatenation algorithm for audio and visual data was implemented.

WP5 - RENDERING The existing 3D controllable avatar was improved during this workshop and was made real-time capable using the game engine of blender. An OSC interface was implemented in python and its use was demonstrated through an android application sending OSC commands to the avatar in real time.

Several advances and implementations were made during the project. Many things are still to be done though. Perspective for the project include :

- Testing the whole pipeline in a concrete scenario and make the evaluation of the appropriateness of the behavior of the avatar, possibly through perception tests and objective comparison between the behavior of real participants and the avatar.
- Other scenarios than the TV watching one may definitely be recorded to enhance the existing database.
- The saliency detection could be improved by working on a better and more complex fusion of the different attention source (audio, video, 3D).
- The affect recognition would benefit from training on a larger dataset. Other machine learning methods and their comparison would also be a good perspective of improvement.
- The decision module could definitely be upgraded and many techniques of machine learning of dialogue systems may be tested.
- For the synthesis work, models could be improved by working on the specific characteristics of each affect considered and visual models that could be trained during the workshop due to a lack of time could be done as well.

- For the rendering part, the model developed is now real-time capable. Interfaces may be built on top of the existing system to ease the control. For example, a WOZ may be implemented for real-time human controlled interactions of the avatar.

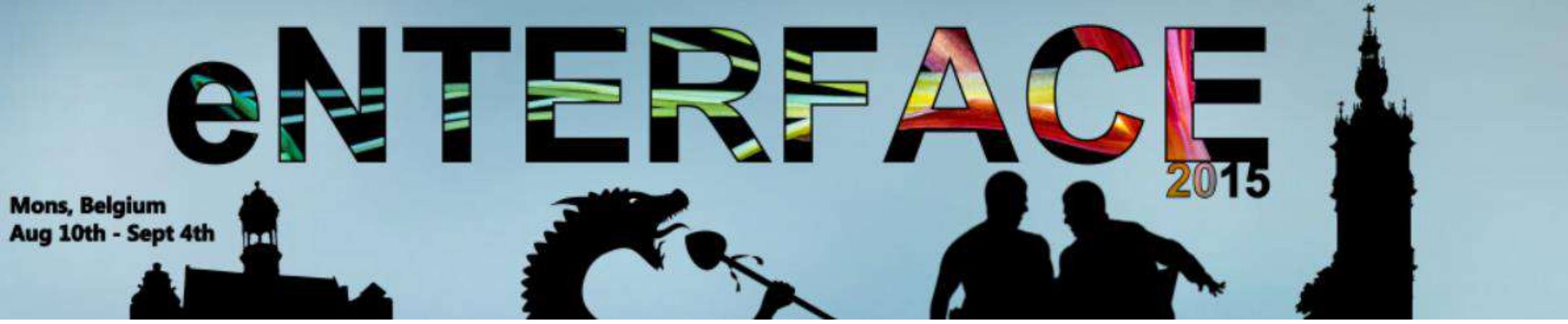

## Project #4:

# MOMMA: Museum MOtion & Mood MApping

*Charles-Alexandre Delestage, Sylvie Leleu-Merviel, Muriel Meyer-Chemenska, Daniel Schmitt, Willy Yvart*

# MOMMA: Museum MOtion & Mood MApping

Charles-Alexandre Delestage[1], Sylvie Leleu-Merviel[1], Muriel Meyer-Chemenska[2], Daniel Schmitt[1], Willy Yvart[1 3]

[1]University of Valenciennes and Hainaut-Cambrésis, DeVisu Laboratory, France
[2]Metapraxis, France
[3]University of Mons, TCTS Laboratory, Belgium

## Abstract

Technologies available today allow the experience of the visit to be documented precisely within a somewhat ecological framework of research museum curators and, more broadly, any individuals involved in the domain. However, they still need simple technical devices and easy-to-implement investigation methods in order to obtain precise, reliable, and easily interpretable results. The M4Museum project aims at testing a particular setup of mixed technologies as to assess their usefulness in analysing a visitor's experience. It is based on the intimate experience of visitors during their route in a natural and autonomous setting, as it synchronously records a video tracing their visual perception (visual field and position of the gaze), their emotional state, and the circuit travelled. It is also a case of application for SYM in a more challenging environment.

## 1. Introduction

Nowadays, the better understanding of the public, its motivations and expectations represents keys of the understanding of strengths, weaknesses and possible improvement of a given facility. To a great extent, this project was a kind of a feasibility study attached to a larger project. During this workshop we aimed at determining strengths and weaknesses of an exhibition by trying to characterize the visitor experience throughout his/her sense-making processes.

Indeed, sense-making processes are expected to be influenced by the visitors affects (emotions, mood, feelings, etc.) especially when it comes to being confronted a piece of art that is specially intended to produce feelings. In this case, affects can be seen as part of the first parameters regarding their potential impact on the personal experience. Thus, in this workshop, we decided to test the feasibility and the workability of a prototypical tool aimed at helping the researchers access that very private kind of data.

Making this access easier could lead, in our opinion, to new tools for assessment, analysis and decision making in a wide variety of areas. Although available existing technologies, we were unable to find any simple or integrative set-up capable of giving a precise, reliable and easy-to-implement measure with easily interpretable results.

The main goal within the MoMMa Project is to sketch a protocol to get the visitor's intimate experience during his/her visit in an ecological[1] and autonomous setting[2]. Trying to follow the adage saying that there is no better data than more data, we decided, in this study, to try to make a synchronous record of their visual perception with focalization spots and their localization in the museum, as well as their emotional state. This approach could create a viable common ground between qualitative and quantitative research.

## 2. User experience: a survey

[1]trying to be the least invasive
[2]trying to give to the visitor a certain freedom in his/her movement

### 2.1. Museal experience and sense-making processes

In museums, the 'course of experience' research programme [33] aims to identify, describe, and understand how visitors construct meaning and knowledge during a museum visit under natural and autonomous conditions (without a guide or teacher). The question concerning the museum experience is : what makes sense from the visitors's point of view during their visit?

Sense-making can be schematized this way: when confronted with reality, the person has to join up the perceived dots and his/her available reference points (past experience, knowledge, patterns, etc.) in order to create a coherent drawing, his/her reality, avoiding, if possible, confusion, non-sense, etc. This construction is trapped into the very limited spatio-temporal configuration, in other words, the exact 'drawing' is only available to the person at that moment and in the place where and when it happens.

That is to say 'Every visitor lives in an environment that is meaningful on a personal level ' [35]. However, we know that if this environment as it is perceived by the visitor can be reproduced (i.e. by presenting to the person a recording of his/her point of view), he or she may relive this experience, with, this time, adequate conditions to be able to describe to the researcher. This ability to describe what has been seen and felt as well as the thoughts that come to mind is based on revivification, namely the ability to re-experience (in quality) what has already been experienced [28]. Hence, in our experimental set-up we decided to rely on subjective re-situ interviews we present hereafter.

### 2.2. Mood as a central parameter in qualitative research

If it is a common ground to say that pieces of art are created in order to produce feelings or emotional experience, the study of the bystander's mood is sometimes neglected in many areas of science. However, since the early work of Herbert Simon [30] on the link between decision making processes and emotional experience, mood and affects in general have been, with highs and lows, at the very centre of the scientist's focus, at least in human sciences. Findings have shown the importance of this parameter[20]; nevertheless, in the literature, it is very hard to find any sufficiently stable or rigorous definitions or distinctions between different possible affects - according to Gross and Barrett : '*It is widely agreed that emotion refers to a collection of psychological states that include subjective experience, expressive behaviour (e.g., facial, bodily, verbal), and peripheral physiological responses (e.g., heart rate, respiration). It is also widely agreed that emotions are a central feature in any psychological model of the human mind. Beyond these two points of agreement, however, almost everything else seems to be subject to debate*' [13]. Quoting Dennett we could regret this epistemological lack as '*Rigorous arguments only work on well-defined materials* ' [8]. Using the disambiguation exposed by Yvart *et alli* [38] we decided to work with the affective concept of mood which refers to something we can attach to Heidegger's 'stimmung'and 'befindlichkeit'[14]. From a strictly emotional point of view it lasts longer and has milder but more pervasive effects on the person compared with emotions. But from a more general point of view it can be seen as a central parameter of attentional, dispositional and sense-making processes [38].

Following evidence between learning and felt happiness [10] and between enjoyable museal experience and reported knowledge acquisitions [28], we thus proposed to rely, as a work hypothesis, a strong link between sense-making processes and mood. In this way, by characterizing the contextualized emotional experience of a visitor during an exhibition, we think possible to characterize the efficiency of the exhibition according to its main objectives: bring something to the visitor and help him to create sense out of his/her experience. The information about mood is no substitute for re-situ interviews but can be seen as a valuable additional source of knowledge.

Be that as it may, we first need a tool to characterize moodal experience, then we need to contextualize this experience (where and when that moodal configuration occurred). Studies of emotions and affects managed to raise such an enthusiasm and an interest that it is now possible to talk about an industry of emotion [7].

## 3. Experimental setup

MOMMA directly follows Daniel Schmitt's work on trying to characterize the strengths and weaknesses of a museal exhibition [27] by taking a look at the sense-making processes in context. His main experiment consisted in the use of a POV[3] camera worn by the subject during his/her visit (with or without gaze capture). After that, the subject was invited to comment the exhibition and his/her experience relying on the upsurge effect of the POV video; this technique is called '*Subjective Re-situ Interview*'. In order to access verbalized affects, the researcher used a discrete VAS (Visual Analog Scale) represent for the hedonistic valence (from -3 to +3).

MOMMA was the opportunity to go one step further by introducing SYM, a new tool developed in order to help the user in expressing his/her mood at any moment of an experimental setup, adding the arousal information to the hedonistic valence one and replacing the discrete VAS by a continuous graphical space. Furthermore we could add latest generation eye-tracking glasses and localization techniques in order to contextualize the experience of the visitor in order to diagnose the exhibition in itself.

[3] Point of View.

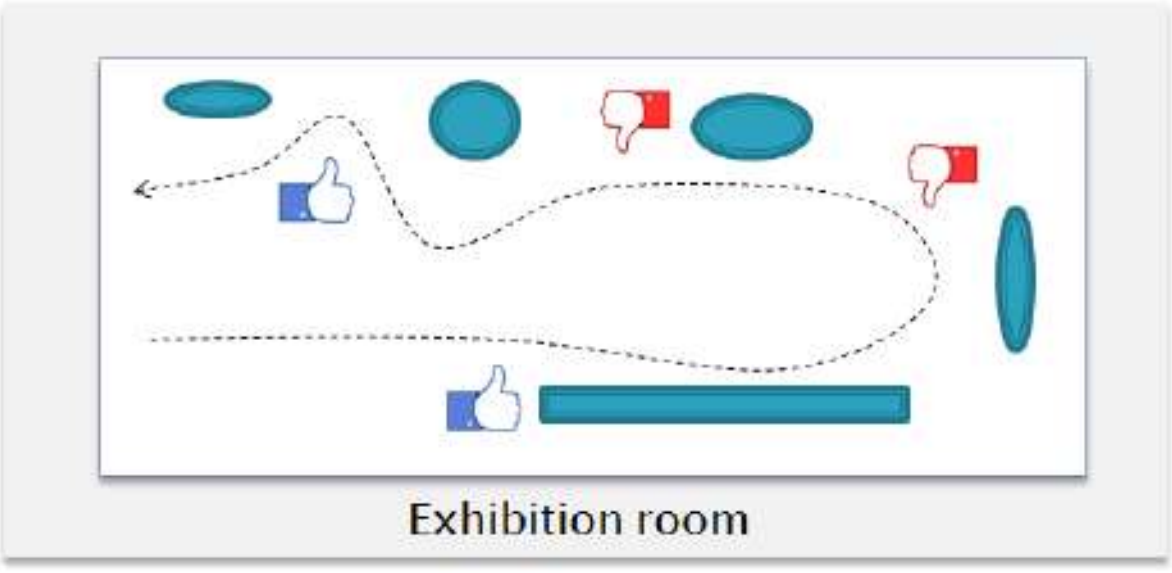

Figure 1. Aimed representation of the MoMMa project

The main purpose of the workshop was to obtain a prototypical tool that could lead, after many improvements, to the constitution of qualitative and quantitative global evaluations of an exhibition setup [see Figure 1].

## 3.1. Eye Tracking

Eye tracking is about keeping a trace of the eye movements of an user in his/her visual field [17]. This technique is generally used for a purpose of interactivity (i.e. touchless control of some devices) or for a purpose of diagnosis (i.e. to determine salient parts in a picture) [9] as has been studied for a long time [23].

Nowadays, there are two main ways to perform this kind of measurement: an 'invasive' but wearable way (the person wears equipment on the head) and a 'non-invasive' but static one (the equipment is set up at a defined place, the person does not wear any special gear but cannot move). Technically, as for calibration, the two methods give similar results, penalizing, in general, people wearing glasses - the measurement takes account of the movements of the eye with infrared light that is easily reflected by glasses.

Although still technically improving and being modified, these techniques still suffer from technical limits, but, according to Duchowski [9], eye tracking techniques are able to provide a '*quantitative measure of real-time overt attention*'. Our objective in this project is to try to marry these quantitative measures to more qualitative ones relying on affective states.

For ecological reasons (according to Bateson's meaning [5]), in other words, in order to minimize the influence on the subject, non-invasive methods seem to be the best option as we can expect fewer experimental biases. However, as our experiments were performed in a museum, we were not able to use non-invasive methods such as the FaceLab, as they require a fixed environment for the user (the eye-tracking measurement is an indirect one consisting in filming the retina). Fortunately, we were able to test one of the lastest Tobii Glasses. They consisted in headgear resembling glasses, with a full-HD front scene camera with 50 / 100Hz gaze sampling cameras. This way, we were finally able to obtain, a POV recording with focal measurement as an additional video layer.

## 3.2. Subjective Re-situ Interviews

The measurement of gaze and focal points seems critical for museum curators, as it gives them information about what people look and read - if an information plaque is read or not or if people look at the collections. To their mind this measurement is seen as a qualitative justification for expenses (should we pay for bigger plaques? etc.). However, from our standpoint, it only allows us to know what the visitors have been seeing, but not what they have been looking at attentively or retaining.

So, during this experiment, we only considered the eye-tracker as a tool to preserve a trace of the world as perceived by each visitor and as an anchorage point allowing further re-situ interviews. The visitors were equipped with glasses and after about 30 minutes of free visit, the eye tracker was removed and the visitors placed in front of a video screen, close to the site of the visit. The video recording of the visit with the gaze points was then projected on this screen, and the visitor was invited to comment on his/her experience of the visit.

During the subjective re-situ interview, the visitor spontaneously divided his/her actions into units which were significant and fully coherent from his/her own point of view. The subjective re-situ interview was then transcribed for further analysis. The aim was to identify what was taken into account by the visitor at each and every moment: what he/she looked at (with attention or intention) and did in addition to the expectations, preoccupations, and knowledge called upon, with the aim of documenting each fragment of the meaningful sequence and then reconstruct the visitor's experience.

For any piece of art shown to the public, visitor tried to isolate a fragment, a shape, a color, that raised expectations. These expectations are often identified as questions and tensions (why a shape, a color, a relationship...) that call for answers. So each visitor mobilizes his/her knowledge, but also his/her experiences as memories, images , dreams... anything that may help to link him/her to this fragment of reality, always in relation to the question he or she asks.

Empathy for example, is a possible form of resolution of the tension from their point of view. To feel the environment of a painting, its atmosphere may be sufficient to "understand" something. Empathy is a way to connect the visitor to the painting, to his/her reality of the painting. Then the visitor "knows" something and this knowledge allows him to make sense. In this case, empathy is an operative relationship, but we also found relationships built from memories, dreams or reminiscence.

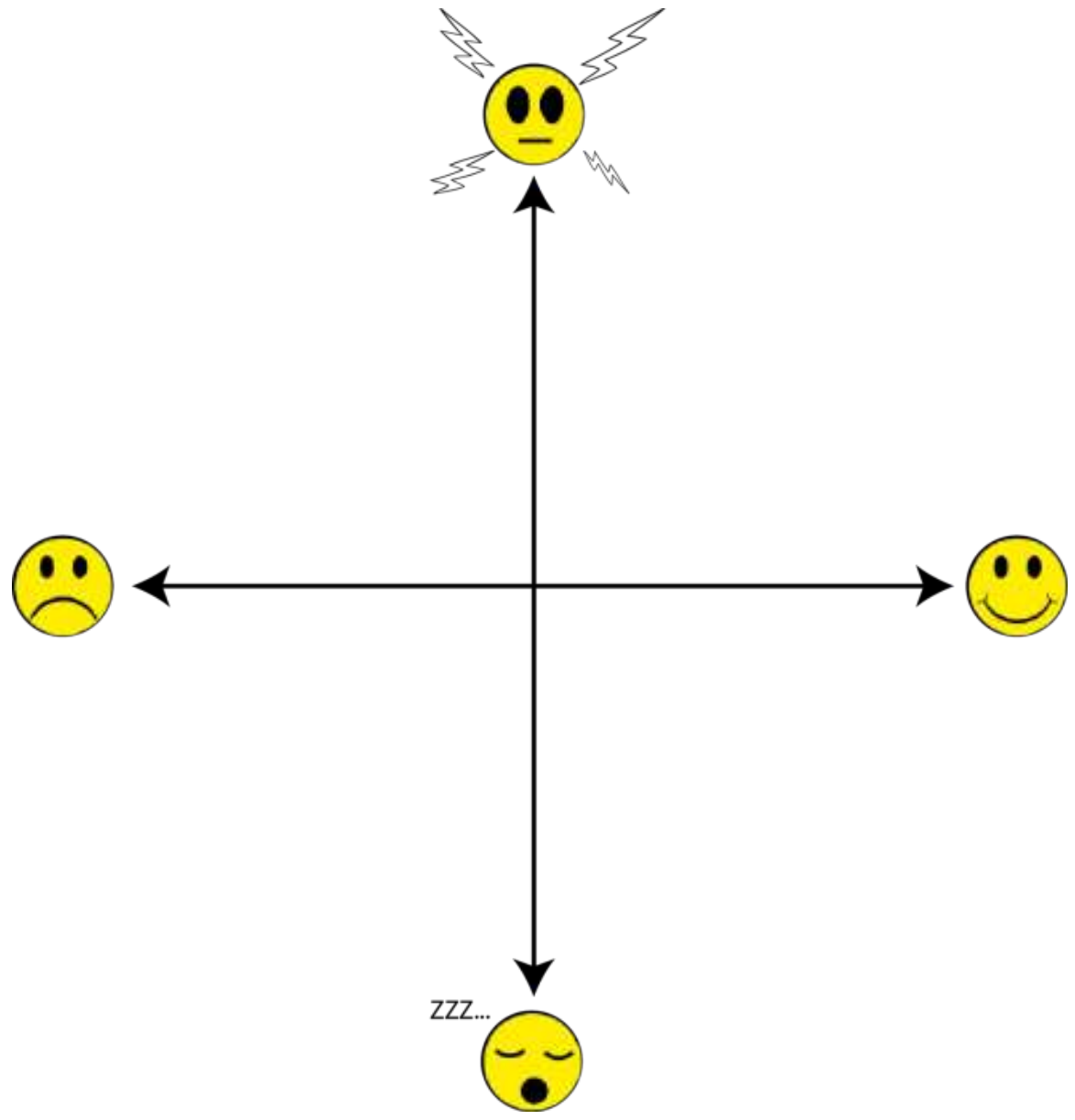

Figure 2. Valence / Arousal space of SYM

For each one, a camera placed behind the visitor and the researcher recorded the interview for conservation purposes or future meta-analysis [26] or in order to aggregate data into a set using, for example, annotation tools such as Advene[3] or Rekall[4].

## 3.3. SYM: Spot Your Mood

SYM [38] aims at proposing a protocol of explicitation support of individual mood in *in situ* experimental conditions. For MoMMa, the software solution consisted in an Android application, running on a tablet, collecting contextualized data to be gathered in a database (for analysis). Tablets were distributed to the co-participants so that they carried them during their visit. At any moment and at any place during this visit, the co-participants could, without any supervision, indicate their mood, their psycho-physiological state.

The indication of this state was done by the user on a diagram of Valence / Arousal as defined by Russell[24]; that is to say, an orthonormal referential where the X axis stands for the hedonistic valence (pleasure / displeasure) and the Y axis stands for the psycho-physiological activation, the arousal (drowsy / excited). *Extrema* of the vectors show smileys representing, in the most explicit non-verbal manner, pleasure (right), drowsiness/sleepiness (bottom), displeasure (left), excitement/high arousal (top) (see Figure 2).

Originally, the first implementation of SYM was using a *node.js* server. The client received a single page with a *SVG* area handling all the interactions through web-sockets. Data was processed and stored in a *MariaDB* database. Even if it could be transformed into an Android application, the initial implementation was a "test of feasibility" design of SYM. To be more reliable, a specific Android application was specially developed during eNTERFACE. It also allowed an easier implementation of a rough indoor localization solution based on WiFi RSSI.

During the development phase of SYM we were able to notice that the totally a-verbal approach could be confusing and a little "user-unfriendly". Originally, when spotting a point on the VA diagram we could not be sure of the correct apprehension and understanding of the tool. Moreover, in some cases, it could be impossible to characterize a unique mood based upon coordinates. For example, "angry" and "afraid" are very different in nature but very close on a VA-space.

It was then decided to design an additional "emotional concepts" database. Considering a lack in terms of available adjective checklists in French we decided to rely on tried and tested ones based on English as a starting point [25][29][15][16][11][6][36][1][2][18][21][32][19][34][31]. Hence we were able to collect 373 different adjectives standing for different emotional concepts. Translating these words with every possible translation and considering every synonym and short sentence gathered on wordreference.com finally resulted in a collection of 2278 different concepts (including polysemic ones).

These concepts linked together nouns, verbs, adverbs and/or short expressions standing for the largest possible "experiencable" mood. Each concept comes with its definition so as to disambiguate in the case of a polysemic one. For example "*bouleversé*" in French (literally "turned upside down") stands for different possible moods from "shattered" to "overwhelmed", we can be "*bouleversé*" by a romantic movie or by the loss of somebody resulting in very different experiences but also resulting in different *loci* on the space.

In order to link similar concepts to each other, a semantic proxemy analysis created a net based on the words leading to emotions. This net was based on the use of Dicosyn [22] in order to create "*intersynonymic*" relationship between the terms weighted as a Jaccard's distance [37]. Then, about 150 words were placed by the team on the diagram in order to produce a temporary tagged space. These words were selected as being the 150 most frequent translations within the corpus of checklists collected.

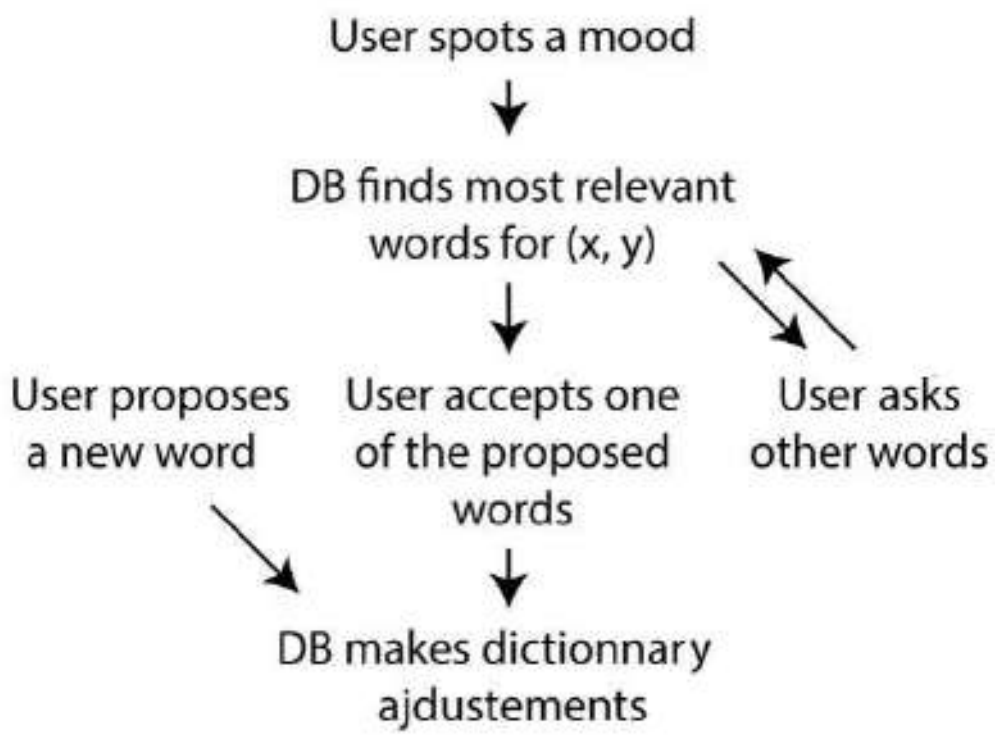

Figure 3. SYM simplified workflow

During the experiment, when the user spotted a mood on the diagram, the coordinates of the point were processed to find the 3 closest emotional concepts in terms of coordinates. The 3 suggestions were then proposed to the user to be accepted or discarded; in this case, 3 other suggestions were proposed each time the 3 previous suggestions were rejected by the user. Each tuple of (x, y, word) was saved for experimentation analysis, and was taken into account to update the positions of the emotional concept on the Valence / Arousal space.

At the very moment the person spotted his/her mood, his/her location was also saved and added to the dataset.

## 4. Analysis

As the experiment took place in a student-city during summer holidays we were afraid not being able to collect enough data. Moreover, the experiment could only be conducted over one week and in a somewhat secluded part of the museum in order not to disturb the public. Due to a lot of technical issues we are going to sum up, the initial experiment had to be lightened.

During the experimentation (one week out of the eNTERFACE month), 12 visitors were equipped with an eye tracker for approximately 20 to 30 minutes to record their subjective visual perspective that is numerically compliant in the heart of a qualitative study [12]. The eye trackers allowed us to go further as compared with previous research[27], to be more precise by increasing the quality of information but also the quantity (we could add to the dataset about the focalization points within the field of view). At that moment, we were able to confirm that the eye trackers are not experienced as foreign nor invasive elements, they do not significantly modify the behaviour of visitors, thus they can be neglected as a significant bias.

Until now, visitors had only been equipped with spy-cameras, then they were asked to show on the screen and comment the items they were looking at during the subjective re-situ interview. During this phase, an additional witness camera, at that moment, was intended to record the interview but also some visual information like the person pointing out a part of the screen. During MOMMA we decided to discard witness camera to avoid some experimental cumbersomeness. In fact, we were able to ascertain that most of the people indicated the same area of the screen, consequently the same area of the video, as the gaze area highlighted by the eye trackers.

As an alternative to the use of the witness camera, we can now advocate the use of screen capture software alongside with an audio recording to reach the same results. This is an important practical contribution provided by the MOMMA experiment in terms of reducing the experimental setup cumbersomeness.

Due to technical issues we were not able to collect correlated mood spotting measurements during the user's route throughout the exhibition. Nevertheless we were able to run tests with available students and observe that not a single person had difficulties in spotting a mood on the Valence-Arousal space. The proposition of words also worked very well, and gave us some feedback on the initial placement of concepts on the diagram. However, the main issue was linked to the usage of the tablet. We had a certain variety of participants, and some of them did not know how to use a tablet, and were unable to get out of sleep mode - and quickly stopped using it for this reason.

Moreover, they felt indisposed by the size, the weight and the brittleness of the device they had to carry all along their visit. This means we will have to find, for future experimentation a more comfortable device to wear for the participant where SYM could be implemented while keeping its presentation and functionalities. As SYM is based upon a client based software, we have been able, so far, to run it on a smartphone without any sizeable adaptation problems. A good perspective to investigate would be to create a graphically smaller version that can be used on small screen devices such as smart watches or smart-phones.

Be that as it may the biggest technical issue we had to cope with concerned the indoor localization we tried to set up. The number of concurrent WiFi networks changed significantly between the installation of the WiFI spots and the actual experimentation. It may be explained by the fact the installation took place on a closing day, when no one was in the building. Also, any smartphone can emit its own WiFi signal to allow net browsing on a laptop, which could happen in the atrium - an area freely open to the public really close to us.

Nevertheless, we managed to draw the route of the visitors not without a lot of noise resulting in a fuzzy drawing. Taking into account the WiFi technique and the smoothing we had to apply, our system was only able to produce a route with a margin of error of around 3 meters. This margin is still too big for further analysis.

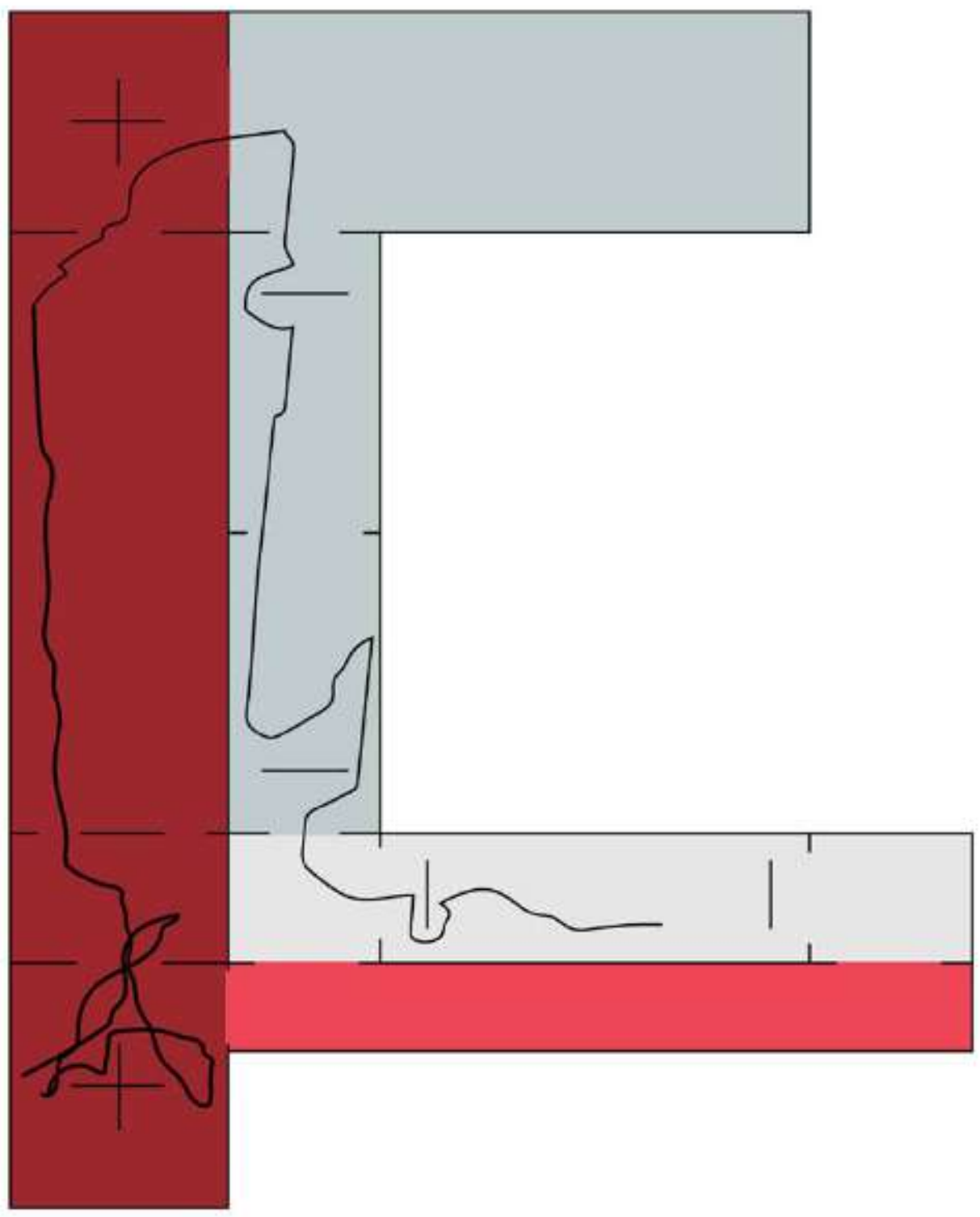

Figure 4. Map of the visit of one interviewee

This lack proved the necessity of getting a reliable localization method, as the mood spotting done through SYM would only be synchronized by the video of the eye tracker - which is a long and painful process. Actually, all along the data gathering process we had to take into consideration the future analysis which needed the data to be synchronized. We were not able, during eNTERFACE workshop to find a suitable technique to do so. Paths we are exploring for future works consist in either the use of the tablet clock or the use of an inaudible high frequency signal to be broadcast by the tablet and recorded by the eye-tracker's built-in microphone.

## 5. Conclusion and future works

Subjective re-situ interviews have been raised to become an invaluable tool for qualitative research about visitor's experience. It reveals that knowledge is not simply transmitted in the sense of accumulating information, but rather constructed by each visitor as an adequate response to his/her expectations through the resources available in the environment and linked to his/her past and knowledge. Hence we cannot neglect such a collection of cues we can sometimes notice in the literature. Furthermore, the addition of eye-tracking tool appears to be insightful, especially when it allows to dispense with a relatively cumbersome or hefty witness camera during the interview procedure. Making the procedure lighter is a great gain when it comes to minimizing experimental biases as it becomes less invasive and persuasive for the interviewee. In this sense, subjective re-situ interviews can be seen as being part of the decision support systems family.

As it is now clear that we need to rely on very private information during the exhibition visit, and as we consider the mood as being one of the most important parameters to explore, SYM appears to be a suitable tool especially when it can be set up alongside localization techniques and on wearable devices.

Despite a certain lack of experimental data, we can however consider that in a pragmatical view, MoMMa can be regarded as a test of feasibility for further improvements. We can also notice that the semantic layer we started to implement could be seen as opening promising avenues transforming SYM into a real verbalization assistance tool. Lighter and quite practical to deploy as compared with subjective re-situ interviews, SYM can be set up as a light diagnostic tool in order to first characterize the "greatest" weaknesses within a given exhibition. It allows researchers to collect qualitative information without the loss of a sizeable quantity of experimenters. Then, it could be possible to deploy a subjective re-situ interview setup in order to inquire deeper into the very heart of the problem. This strategy is the one we choose to develop in our new M4X project.

Acknowledgements. The team would like to thank the *Musée des Beaux-Arts* of Lille for allowing us to stay several days to test the setup and perform the experiments. The team would also like to thank the DeVisu laboratory for lending us the Tobii eyetracking glasses. The team would like to thank the TCTS laboratory for the WiFi hotspots. The team would like to thank Metapraxis for supporting this project and lending us one of the tablets for the experiments. The team would also like to thank Jérôme Hennebert, lecturer at the Lille 3 University, Céline Schall, researcher at *Institut d'Études Romanes, Médias et Arts*, University of Luxembourg, Muriel Meyer-Chemenska, museographer and CEO of Métapraxis, Olivier Aubert, research engineer at the University of Nantes who participated in the experiments and supported this project.

## Team

Daniel Schmitt is Lecturer at Valenciennes University. He holds a PhD from Strasbourg University and he is graduate of the Ecole Nationale Supérieure Louis-Lumière. The 'course of experience' research programme in museums aims to identify, describe, and understand how we construct meaning and knowledge during a museum visit under natural and autonomous conditions (without a guide or teacher). Over a period of more than 20 years he has also managed a large number of museum and exhibition projects while researching interpretative resources for museums in different countries. http://www.univ-valenciennes.fr/DEVISU/membres/schmitt_daniel

Pr. Sylvie Leleu Merviel has been leading both the DeVisu Laboratory, since 1997, and the audiovisual and multimedia department, since 1989, at the University of Valenciennes. Her research covers a very large area in the field of information and communication sciences. Main part of her work is dedicated to evaluation of quality and to sense-making from an anthropocentric and constructivist point of view. As she always tried to open and link research to the industrial world, she enjoys a very good reputation in both media industries (France Television, MediamÃĺtrie, Canal +, etc.) and technical industries at the cutting edge of technology (eg.

ESA). She is chief editor for the 'Ingénierie représentationnelle et construction de sens 'collection, Hermès Lavoisier.

Willy Yvart graduated a Master's Degree in Multimedia, Audiovisual, Information and Communication Sciences from DREAM department at the University of Valenciennes (France) in 2011. Since 2013, he has been a PhD student under the joint supervision of Thierry Dutoit (UMONS, Belgium) and Sylvie Leleu-Merviel (UVHC, France) on the study of semantics metadata in massive music libraries in order to improve indexing and searching techniques. http://www.univ-valenciennes.fr/DEVISU/membres/yvart_willy

Charles-Alexandre Delestage obtained a Master's degree in Audiovisual Communication Management in Valenciennes. He started a PhD in 2014 within DeVisu (UVHC - France). His research is oriented on automation in the audiovisual processes of production and the impact on audiences, focusing on the acceptance of such hedonic content. He is currently working on an automation software for professional broadcast systems as to give researchers in directing algorithms a simple and operational framework. http://www.univ-valenciennes.fr/DEVISU/membres/delestage_charles_alexandre

Muriel Meyer-Chemenska is a museography expert and project manager. She holds a degree in Civilisation from the University of Paris VII and is a member of both the SACD (the French Society of Authors and Dramatic Works) and the SCAM (Civil Society of Multimedia Authors). Since 1990, she has been creating museography programmes, visitor scenarios, designing museography and interpretative resources, assisting clients and exhibition commissioners and providing artistic direction and project management in mediation and visitors experience design for permanent and temporary exhibitions. Muriel Meyer Chemenska is also the director of Metapraxis company dedicated to museography and interpretative resources in all their forms and at all stages of the museography process. It undertakes projects worldwide and for each project, she takes personal responsibility for the management and successful conclusion of all services the agency offers.

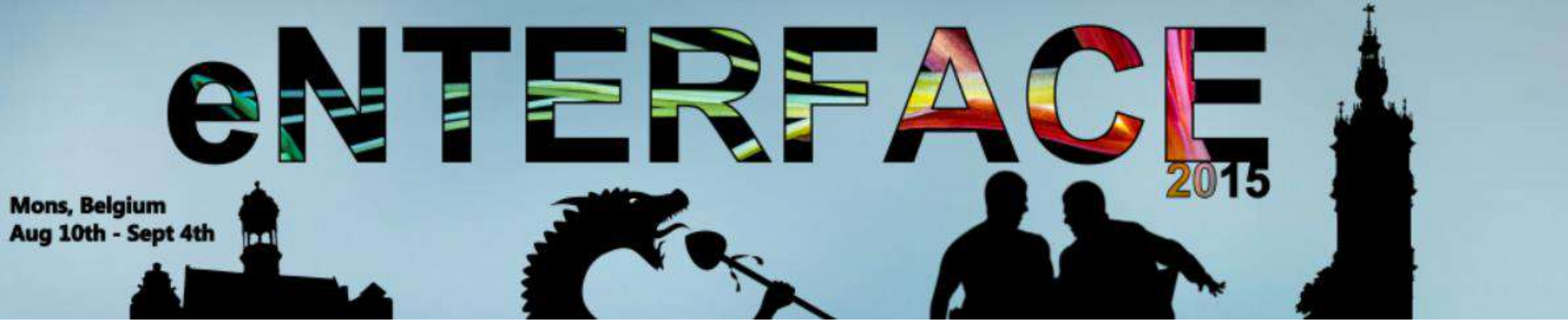

## Project #5:

# VideoSketcher: Innovative Query Modes for Searching Videos through Sketches, Motion and Sound

*Stéphane Dupont, Ozan Can Altiok, Aysegül Bumin, Ceren Dikmen, Ivan Giangreco, Silvan Heller, Emre Külah, Gueorgui Pironkov, Luca Rossetto, Yusuf Sahillioglu, Heiko Schuldt, Omar Seddati, Yusuf Setinkaya, Metin Sezgin, Claudiu Tanase, Emre Toyan, Sean Wood, Doguhan Yeke*

numediart
INSTITUT DE RECHERCHE POUR LES TECHNOLOGIES CREATIVES DE L'UMONS

UMONS
Université de Mons

fnrs
FREEDOM TO RESEARCH

MONS 2015

# VideoSketcher: Innovative Query Modes for Searching Videos through Sketches, Motion and Sound

Stéphane Dupont, Ozan Can Altiok, Ayşegül Bumin, Ceren Dikmen, Ivan Giangreco, Silvan Heller, Emre Külah, Gueorgui Pironkov, Luca Rossetto, Yusuf Sahillioğlu, Heiko Schuldt, Omar Seddati, Yusuf Şetinkaya, Metin Sezgin, Claudiu Tănase, Emre Toyan, Sean Wood, Doğuhan Yeke

## 1 Introduction

This text reports on the results of the work carried on within the VideoSketcher project during the eNTERFACE 2015 summer workshop help in Mons, Belgium. An additional report in video form can also be found on-line[1].

Information retrieval technologies are key enablers of a range of applications in an environment where finding the right information and data becomes critical in many sectors, for efficient decision-making, research, and creative thinking. Multimedia content deserves a particular treatment given its unstructured (non-symbolic) and hidden meaning to the computer, also known as the *semantic gap*. This calls for research on the way multimedia documents can be indexed, queried and stored efficiently. Despite the numerous recent research advances, we are far from there yet, as can be seen through the increasing number of challenging benchmarks and competitions related to *Multimedia Information Retrieval* (MIR). In this context, the study of query modes beyond keywords or text search is of particular interest.

Stéphane Dupont, Gueorgui Pironkov & Omar Seddati
University of Mons, Belgium e-mail: `stephane.dupont@umons.ac.be`

Ozan Can Altiok & Metin Sezgin
Koç University Istanbul, Turkey

Ayşegül Bumin, Ceren Dikmen, Emre Külah, Yusuf Sahillioğlu, Yusuf Şetinkaya, Emre Toyan & Doğuhan Yeke
Middle East Technical University Ankara, Turkey

Ivan Giangreco, Silvan Heller, Luca Rossetto, Heiko Schuldt & Claudiu Tănase
University of Basel, Switzerland

Sean Wood
Sherbrooke University, Canada

[1] VideoSketcher final report video: `http://www.enterface.net/enterface15/wp-content/uploads/2015/10/project7_demo.mp4`

In this paper, we focus on audiovisual databases and develop an integrated prototype enabling to make use of content analysis and recognition through machine learning, multiple query modes (symbolic and non-symbolic enabling to specify the visual as well as audio characteristics of searched video shots), and a scalable database back-end. We evaluate the system in a known-item search task.

### *1.1 Outline*

The report is structured as follows. In Section 2, we provide an overview of previous work in the different technological areas involved in future video search systems. An overview of our approach is then proposed in Section 3. In Section 4, the different components involved in our system are presented, together with the challenges addressed and the new developments made. These components include user interfaces and query modes (Section 4.1), machine learning and computer vision/audition (Section 4.2), and database systems (Section 4.3). In Section 5, we describe the end-user evaluation work we performed. We conclude and discuss avenues for future work in Section 6.

## 2 Related Work

This section presents a brief introductory overview to set the context of this research and the relevant sub-fields. Furthermore, specific references as well as challenges within the identified sub-fields are proposed in the section that will follow later within this report.

Existing approaches to **video retrieval** either focus on audio signals, video frames (images), additional metadata (including subtitles), or a combination of these [24]. QBIC [19] is one of the first systems that successfully combined image and video retrieval by considering color, shape, texture, sketches, and even sample images. The MIRACLE project [25] combines multiple types of video search, such as text searches (transcripts, subtitles, closed captions, and speech recognition), visual information (face clustering and scene change detection), and speaker segmentation. VideoQ [10] addresses movement extraction and supports motion queries by automated video object segmentation and tracking, and use real-time video queries. Other approaches to motion-based video indexing and retrieval are reported in [15, 18, 56, 52]. However, the most visited video search engines on the Internet, Youtube, Google Video, and Yahoo Video still rely on very basic features, mainly text from closed captions, subtitles, or social metadata.

An active area of research in the topic is **machine learning and computer vision (and audition)** where video scenes are automatically analyzed and tagged with information on their semantics [2, 4, 54], including visual concepts as well as human actions and activities displayed in the videos. So far, many classical image-based features have been used and extended [48] to video content, with classification relying on bag-of-features and support vector machines, or other classification approaches. Dense or sparse optical flows have also been used [57]. Since recently, deep artificial neural networks and in particular convolutional ones represent a new state-of-the-art in the area [53, 33]. In general, the use of convolutional neural networks enables classification without requiring classical feature extraction schemes. Concept and action recognition is nevertheless challenging given the large number of spatial and temporal configurations observed in video content. Finally, beside visual information, the audio channels (through voice but also other acoustic sounds and music) may also provide complementary information [3].

Of particular interest is the possibility to rely on **multimodal query modes** beyond keywords and **navigation interfaces** beyond lists of results, as proposed in standard search engines. With this respect, the use of sketches is particularly relevant. In some cases ideas and concepts that are hard to explain can be naturally and concisely described using sketches. In this respect, sketching is a unique modality. Sketches may refer directly to objects or events in the videos, or describe movement. Machine learning algorithms can be used to build sketch recognizers capable of interpreting queries put forward in the form of a drawing. Previous state-of-the-art include [10, 14, 13, 31].

When collections of video data and associated metadata are large, this necessitates efficient and effective **database systems**. De Vries et al. [59] present Mirror, a database that supports content-based multimedia retrieval data and queries. Using Moa, a new relational algebraic framework based on the non-first normal form (NF2), the authors describe the engineering factors for creating the multimedia IR-DBMS. Mirror is implemented on top of the object-relational DBMS Monet. Alvez et al. [1] introduce a system that combines low-level (syntactic) features with semantic features in an object-relational Oracle 11g database. The database is extended by several User Defined Types (UDT) following the MPEG-7 standard descriptors, and operations implemented in PL/SQL, e.g., to evaluate similarity measures. In Moise et al. [36], the authors make use of the map/reduce paradigm for querying large sets of image data in a cloud environment. The authors use extended Cluster Pruning (eCP) for indexing and port it to the map/reduce paradigm on the Hadoop platform. A particular area within databases for video retrieval relates to the vector-space paradigm for video search, where the machine learning and computer vision techniques are used to represent each piece of content as vectors of visual descriptors (rather than a series of keywords or text descriptions), which require efficient and specific **index structures and algorithms** able to deal with such multidimensional data. These will be developed further later on in this report.

# 3 Approach

## *3.1 Architecture*

The proposed system is composed of three main components:

1. A subsystem in which content can be ingested and which triggers the analysis of videos, their segmentation in video shots, and the extraction of features (descriptors) characterizing their audio-visual content. This relies on algorithms for image analysis as well as computer vision and machine learning.
2. The extracted features are then used as the basis for indexing the videos, which necessitated specific approaches as indexing based on such features can generally not rely on indexing schemes evolved for text databases. This is related to the vector space approach used for indexing audio-visual features.
3. A front-end system enabling the specification of multimodal queries including sketches, the adjustment of several parameters affecting the way the index is searched, as well as browsing retrieved results.

Figure 1 shows an overview of the architecture of the system presented in this paper.

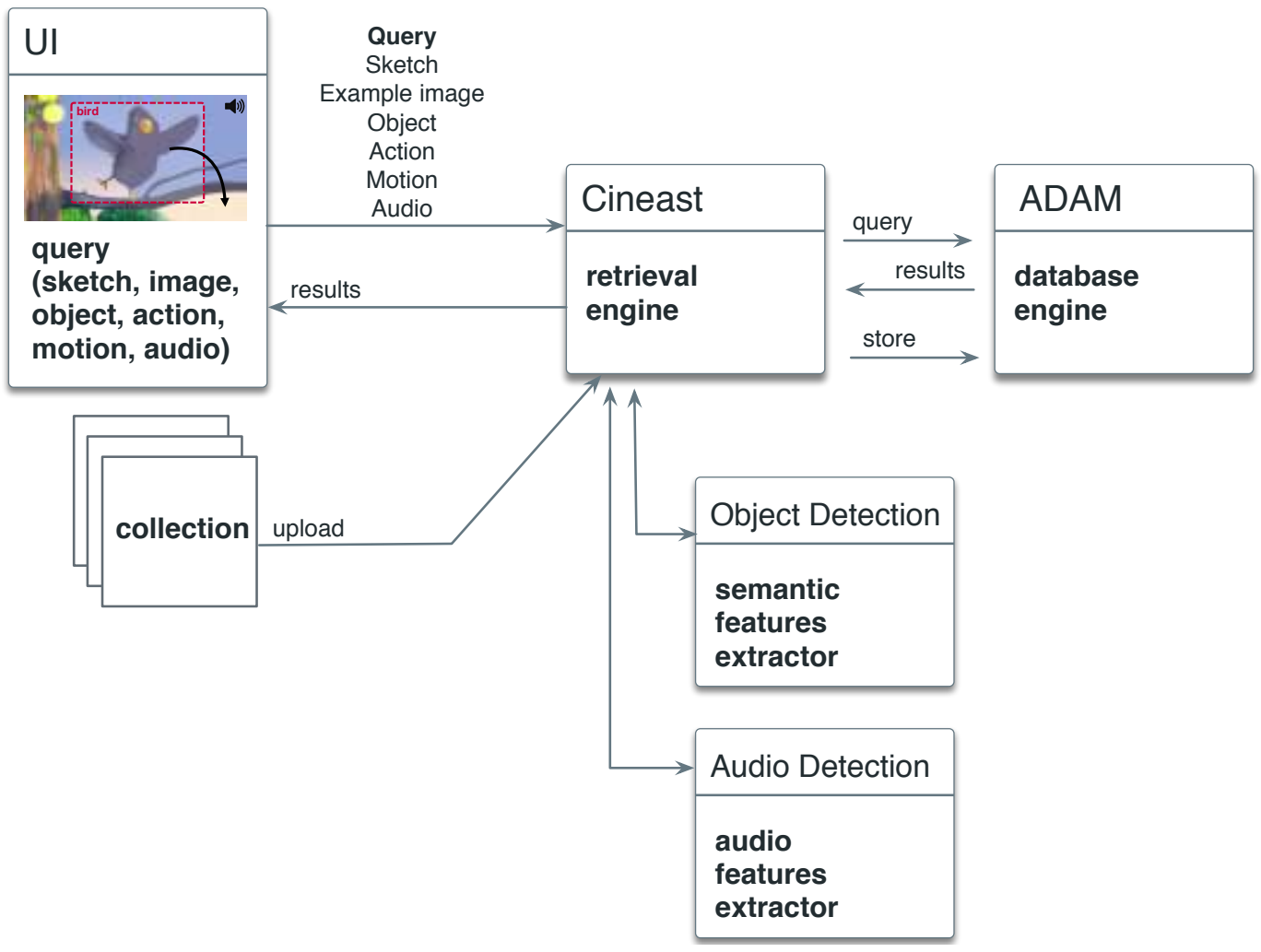

**Fig. 1** Architecture of the proposed system.

## *3.2 Vector Space Retrieval*

The vector space retrieval model [46] is a standard retrieval model used in the field of information retrieval. In this model, every document (and query) is considered to be a multi-dimensional vector. To compute the similarity between a query and a document, a similarity or distance measure is computed between the two vectors.

The vector space model therefore involves, first, the construction of a feature vector that well represents the multimedia object. For this, in a feature extraction/transformation step, the features considered are extracted and a numerical value is stored in the vector representation. Then, to measure the similarity between two vectors, various similarity measures are commonly used. In text retrieval, the cosine measure, which computes the angle between the two vectors, is often used. In multimedia retrieval, the Minkowski norms are often applied to compute a distance measure.

Given a query and a set of feature vectors, in the vector space retrieval model, the nearest neighbors are considered as the most similar documents to the query. A cut-off value $k$ might be defined for the nearest neighbor search, limiting the results only to the $k$ most similar. Note that this approach can enable query modes such as QbE (Query-by-example, where a video shot is used as a query), or QbS (Query-by-sketch, where a sketch of the visual content is used as a query), as long as representations in a common vector space are possible.

Computer vision and audition are fundamental components to work in the vector space model. They enable video understanding by the computer, and hence the automatic extraction of information from the unstructured data represented by the pictures and video shots in the collection. Such information can be semantic (hence enabling keyword or language-based search), or in the form of numerical feature vectors representing the content in a useful way for search (for instance for query-by-example or similarity search). There is hence an interaction between the automatic content analysis approaches through computer vision and audition, and the user interface and query modes that are sought.

Currently, machine learning (unsupervised and supervised) approaches are the workforce of such techniques. They are per se challenging domains, and the research needed to make them useful for generic video search present additional challenges, some of which are introduced in the next sub-section. Machine learning here is not seen as a replacement of image processing algorithms, but rather as a complement. Some image properties, such as color and edge properties, are indeed well represented by more traditional image analysis.

# 4 Components

Subsections 4.1, 4.2, and 4.3, dedicated to the three main components of VideoSketcher, present the results achieved together with some more information about the specific research context and state-of-the-art.

## *4.1 User Interfaces: Query and Result Navigation Interfaces*

### 4.1.1 Image Sketch Queries

In VideoSketcher, three kinds of visual sketches can be used to describe the content for which a user might be looking. Color sketches enable the user to describe the overall composition of the shot to be retrieved, in terms of organization of colors within the picture. Concept sketches enable the user to draw sketches of objects and entities, which can then be used as alternatives to keywords. Finally, motion sketches enable to specify properties of the motion within the video shot.

Sketches can represent objects or entities. The user can draw as sketched representation of the concept to be retrieved, and an automatic sketch recognition system is used to convert the sketch to a vector of concept probabilities. We use a modified version of our convolutional neural network (ConvNet) for sketch recognition described in DeepSketch [49]. We used the TU-Berlin sketch benchmark [17] to train our new sketch recognizer. This benchmark represents the first large-scale dataset of human sketches. It contains 20000 unique sketches for 250 object categories (80 sketches per category). In our new sketch recognizer, we do not use a Softmax layer at the output of the model. As a consequence, the classifier gives independent likelihoods for each category (in a fashion similar to the SVM one-vs-all multi-label classification). All the ConvNets described in this work were built using the publicly available Torch toolbox [12].

As the sketch recognition system has been designed using image recognition based on ConvNets, it may be quite sensitive to resolution and canvas size. As the sketch frames were designed and implemented using relative measurements, at different screens from different resolutions, while sketching, some offsets can occur. This was resolved using absolute measurements for sketch frames and adding a post-processing function to the main script of the UI, which re-computes the coordinates where the mouse is on any of the sketch frames each time the resolution is changed or the page is reopened.

#### 4.1.2 Audio Sketch Queries

The user is able to quickly draw a representation of the visual content of the video shot she is searching for, in form of a sketch. In an analogous manner, one can hence think about "sketching" a representation of the soundtrack of the video shot to be retrieved. Work carried on in this direction is summarized here, where the user performs the audio "sketch" with her voice.

Audio present in videos can be seen as an additional feature for video retrieval. The MAMI database [35] proposed an annotated database for vocal queries, with a collection containing popular music. In our case, videos do not always contain music, therefore we focused on a wider range of sounds and noises. Humming [16] or onomatopoeia [26] were also investigated in the literature, but using only one such theme would limit the user options (lacking whistling, speech, singing, etc.). Another study used the Freesound [20] database for voice query classification [6], however the problem here was not to classify vocal imitations into categories, but rather to map a vocal imitation to similar sounds in the database. To perform this mapping, we investigated a number of audio features including low-level features as well as high-level features learned by a neural network.

We began by investigating the use of low-level features with audio tracks of the Open Short Video Collection 1.0 (OSVC) [41]. Features included pitch density, chroma density, Mel-frequency cepstral coefficients (MFCC), and spectro-temporal modulations. Feature vectors were first computed for each feature type over one second intervals of the entire database. Vocal queries were then processed in a similar fashion, and intervals with the smallest distance between feature vectors returned. On average, nearby vectors in feature space did not yield discernible perceptual similarity. This was likely due to the low-level nature of the features coupled with the limitations of the untrained voice to accurately imitate a variety of sounds. When the query came from the database itself, however, similar examples often corresponded to other intervals from the same video segment. A screen shot of the querying interface is shown in Figure 2. In the top-left section, the user either records or selects a query from an existing audio file. The closest results are subsequently presented in the lower-left section. Visualizations of the features of both the query and its results are displayed in the right half of the interface.

We also explored higher-level feature extraction by training a 2-layer Deep Neural Network (DNN) on a portion of the VocalSketch database [9], more specifically the "single synthesizer" portion consisting of 40 recordings of a single synthesizer with different parameters, in addition to 1036 voice imitations in total. For each sound file, the Mel-frequency cepstrum coefficients (MFCC) were extracted. In order to handle temporal variability, the mean and covariance were then computed, yielding fixed-size features. The DNN was trained using the imitation features as input, and the weights were updated by backpropagating the mean root square error between the recording features and the output layer of the network. Unfortunately, due to the low

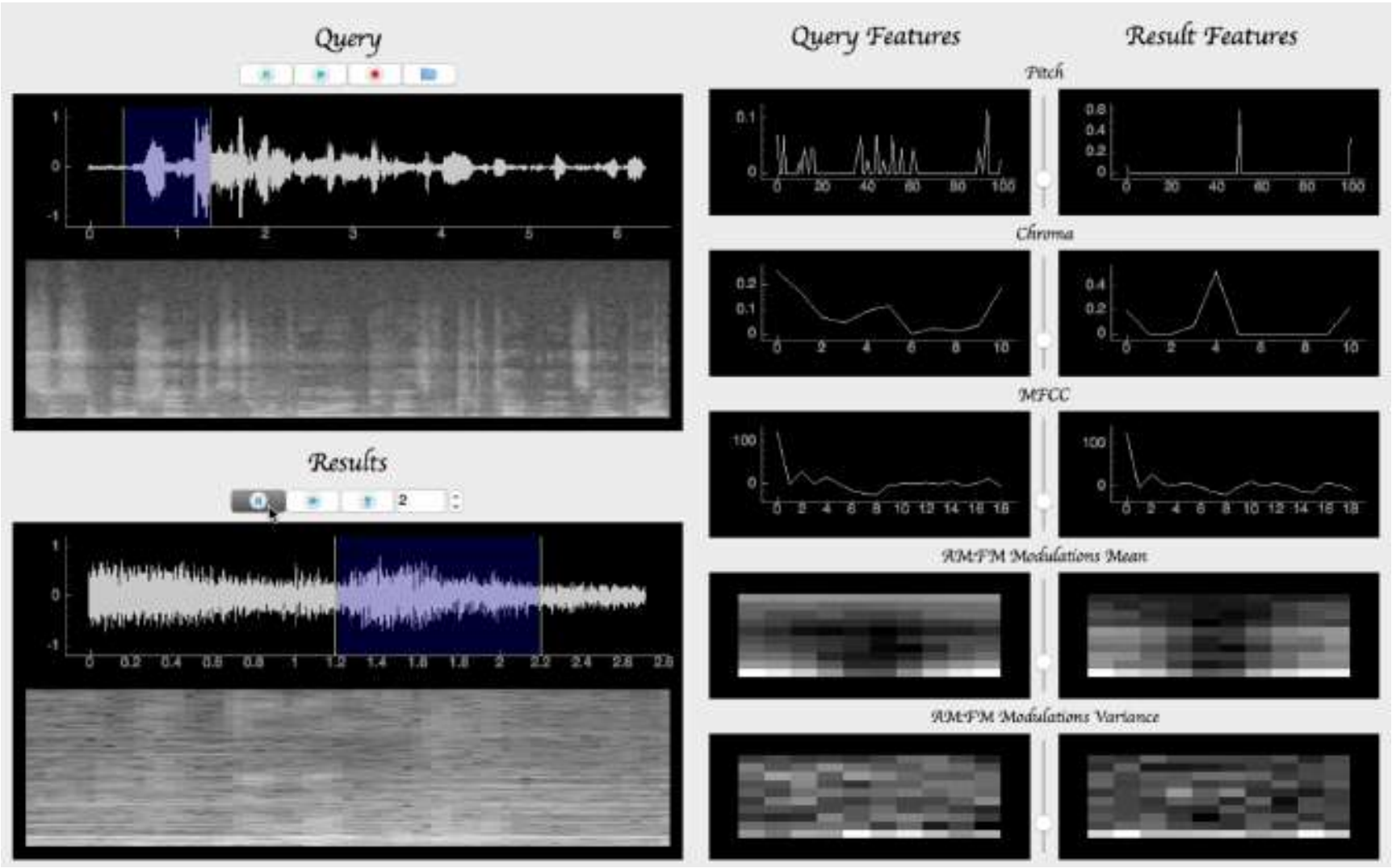

**Fig. 2** Querying interface for low-level features. User queries are depicted in the top-left, proposed results in the bottom-left, while feature values for each are depicted on the right.

number of training samples, the DNN overfitted quickly, leading to poor retrieval.

Improving mapping of vocal imitations to their associated sounds could be done by pre-training the neural network and/or by training on bigger databases. Future work will also investigate fusion of the audio and video features.

#### 4.1.3 User Interfaces and Results Navigation

The UI is composed of two regions. The query panel contains one or more query containers, each with a selectable modality. The user can add and remove query containers with the plus and minus buttons. The user can also select the modality from the menu in the top-left corner. The different modalities have different tools:

1. Color sketch: sketchboard becomes a drawing canvas. The user can select a brush color with a color picker and select a brush size from an on-canvas menu. The color sketch activates the global color, local color and edge features. This is the quintessential Query-by-sketch mode in the traditional sense.
2. Motion sketch: the user can draw motion paths that define general movement of objects or camera motion in the scene.

3. Line/Object sketch: sketches made on this layer are made with a thin black virtual brush. The user should draw objects that he is looking for on this canvas. The resulting sketches are given special treatment of sketch recognition and semantic similarity measure, as described in Sections 4.1.1 and 4.1.4.
4. Audio input: here the user can push a button to record a snippet of sound of maximum 5 seconds, the processed as described in 4.1.2.

The results panel displays search results by showing a scrollable list of thumbnails, each representing a shot. The thumbnails are sorted by relevance and grouped by video, and one extra shot of padding is added before and after the sequence for context. Green bars on the border indicate the score of the shot. Adjusting the sliders in the sliders panel updates the contribution of each feature and causes reordering of the results. Clicking on the thumbnail will create a pop-up with an embedded video player showing the shot in question.

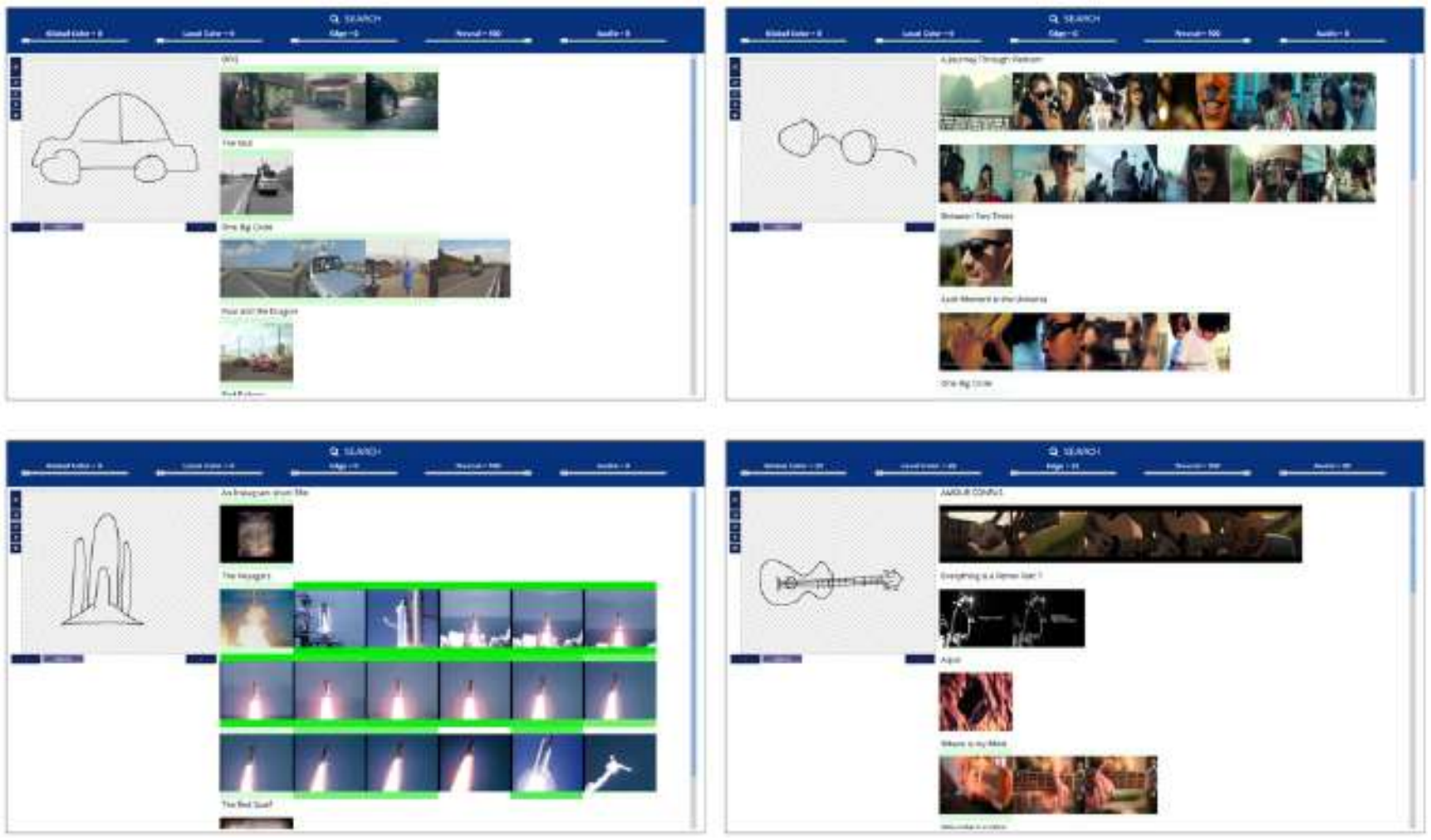

**Fig. 3** UI examples showing query and results for an object sketch.

The challenges encountered while implementing the UI are listed below:

1. Sending Audio to the Back-end: The audio files were getting corrupted during transmission to the back-end. Careful introspection of the problem identified the source of the problem as an interaction between the encoding format, and the browser formatting conventions. In particular, any whitespace in the encoded data stream were being replaced with a plus by the browser. The issue was resolved by modifying the back-end to reverse the process.

2. Layout issues: A number of layout problems were present in the system due to the use of relative measurements. This caused shifts in the coordinates of strokes collected on screens with varying resolutions and sizes. The issue was resolved using absolute measurements for the sketch frames, and recomputing sketch coordinates relative to the updated screen coordinates in case of any subsequent scaling operations.

#### 4.1.4 Query Execution

When a query is processed, many features describing the content work in parallel. They all use vector space retrieval to produce a list of candidate shots that are relevant given their measure of similarity. These candidates are given a similarity score between 0 and 1 each. A late fusion approach is employed on these scores, which combines all these candidate lists by summing over the scores which have been scaled by a weight given to each feature. This is done once per feature category (color, edge, motion, etc.) to produce one list per category using fixed weights and then again to merge these lists into one final list. For the second merging step, the user is able to provide the weights to be used.

The execution of object sketch queries are a special case. Input from the sketchboard is passed directly to the DNN-based sketch classifier described earlier in section 4.1.1. The sketch output is then converted to a 1000-value vector of scores corresponding to the classes of ILSVRC. In the offline phase, the entire collection has been tagged with these 1000 scores using a DNN. Conversion is performed by multiplying the input vector with a similarity matrix. This matrix represents the similarity between each class in the Eitz collection and each synset in ILSVRC and has been computed by manually mapping the Eitz classes to synsets using the NLTK[5] and populating it with the Wu-Palmer distance[64] between synsets. Finally, the newly obtained vector is L1 normalized and a standard nearest-neighbor database query retrieves the closest matching shots. The distances obtained contribute in the final score fusion with the weight defined by the user in the "Neural" slider.

### *4.2 Machine Learning: Analysis, Feature Extraction and Automatic Recognition of Video Content*

During the project, we considered some of these main challenges listed here:

**Selecting useful object categories for video search**: Today, large scale image datasets used as benchmarks for object and environment (context, scene) recognition/classification research are available. But they often contain many categories which are of interest to specific application rather than generic video search as envisioned here. Therefore, categories that better fit the purpose need to be defined.

**Reducing the length of visual feature vectors**: Machine learning approaches and in particular artificial neural networks have a great potential for recognizing the elements of a picture or of a video shot and hence provide additional semantic keywords and meta-data facilitating search. But they also enable to extract feature vectors characterizing the pictures or shots. The can also be stored as meta-data and used in a vector space approach to find similar content (or using a QbE paradigm) through nearest neighbors. Search performance can however be sensitive to the length of these vectors. In order to accelerate the search and reduce the cost, the dimensionality of these vectors need to be reduced.

**Building a better temporal feature extractor**: The extraction of information characterizing the temporal properties within a video is still an emerging research topic. It will enable to better recognize motions, actions, and activities. Here again, data sets used as benchmark for the research may not be diverse enough for a generic video search application. In a previous system, we used a temporal feature extractor trained on the HMDB51 benchmark [34]. Larger datasets with more categories could improve the quality of the feature extractor, as well as the coverage of motions, actions, and activities of interest.

**Integration of temporal information for improved video analysis**: Artificial neural networks, and in particular Convolutional ones (ConvNets) have been shown to be amongst the most promising approaches for video analysis. However, the length of the video sequences that can be present as input to these models need to be fixed, while the time taken for a human action to unfold is obviously variable. This might be a problem for building a good feature extractor. Indeed, in current state-of-the-art, ConvNets are asked to predict the correct action class based on a sequence that could only contain partial information. Time sequence modeling through Hidden Markov Models, Conditional Random Fields, or Recurrent Neural Networks (and predominantly Long Short-Term Memory (LSTM)) hence need to be explored further.

**Use of audio information for improved video analysis**: The audio channels within videos is an additional source of information about the environment as well as human actions performed. Some previous work can be found in the area of feature extraction and machine learning schemes for such computer audition approaches. Integration with visual information remains underexplored though.

**Interaction between query-modes and content analysis**: Computer vision and audition are mainly considered useful at the back-end side, enabling automatic content analysis. They nevertheless also provide novel opportunities at the front-end. Automatic sketch recognition as used here is an example of such an application. Vocal sketches are another example, leveraging audio signals. When such approaches are used for inferring categories from the sketches, the interaction between front-end and back-end can fall back to keywords or semantics. Sketches however offer some novel opportunities as they could also provide information about the attributes (location, pose, size, ...) of objects or actions. In this case, the mapping between front-end sketch analysis and back-end content analysis is an additional challenging issue.

#### 4.2.1 Audio Analysis

Audio analysis has already been introduced previously as a mean to enable sketching through audio. The reader should refer to that section for more information on the audio analysis schemes that can also be used on the video analysis side, in order, for instance, to enable QbE in which the system is able to retrieve video shots which have similar soundtrack (in a vector space sense).

#### 4.2.2 Image Analysis

Three sets of image analysis/recognition approaches have been used here, as described in the following paragraphs.

The input video frames are analyzed using multiple visual features. These include global color information such as average or median color and color histograms of a shot, localized color information from features such as the color layout descriptor (CLD) and edge information obtained from features like the edge histogram descriptor (EHD). A complete list of the lower-level visual features is available [40, 38]. These features are available for matching color-sketch queries to the content (QbS), as well as matching similar video shots (QbE), both in a vector space approach.

Recently, researchers also started using ConvNets features in complement (and sometimes in replacement) of traditional features from the computer vision literature. The availability of large image database for object recognition research, such as ImageNet [45], enable these models to achieve state-of-the-art performance when it comes to features to be used for recognizing objects or scenes. These often outperform standard features such as SIFT, SURF or HoG. Here, we used a ConvNets to extract feature representations also enabling similarity search in a vector space approach, presenting an alternative or complement to the previously proposed color and hedge-based features. We trained a ConvNet on ImageNet using an architecture similar to the one proposed in [65]. Our ConvNet achieve a top-1 state-of-the-art accuracy (using a single center crop) of 59.49% on the validation set (ILSVRC2012). One issue with the selected baseline architecture is that it produces high dimensional feature vectors, making it costly to use for vector-space-based similarity search. In order to reduce the dimensionality of these vectors, we modified the architecture, adding two more layers before the last fully connected layers of our ConvNet (the classification layers). These two layers act as a bottleneck, the first one projects the input into a space of lower dimension, while the second one projects it into a space with the same dimensionality as the original internal representation. After this architectural modification, we only retrained these additional layers as well as the classification layers of our model, keeping the “front-end” layer as they were (no fine tuning). The mean accuracy of our model slightly dropped, from 59.49% to 57.23%. However, this dimensionality reduction technique enabled us to extract feature vectors with a dimension of 128, instead of 4096 previously, reducing the cost of database storage and vector space retrieval computations.

Finally, classifiers were also used to recognize visible object categories, enabling a more semantic search, or even a feature representation based on object probabilities to be used for vector space retrieval. They were also relying on ConvNets. As mentioned previously though, object categories found in the ImageNet benchmark are not all useful for generic search in videos. We hence spent time defining a new list of categories. The preparation of such a new dataset is a complex and costly task. Images corresponding to the different categories are to be found, using other datasets and/or internet crawling. Images coming from internet then need to be filtered manually to avoid mistaken annotations. We prepared a first version of this new dataset and trained a ConvNet on it. This dataset being currently imbalanced, more work will be needed to even the proportion of samples representing the different object categories.

### 4.2.3 Video Analysis

Analysis and recognizing video shots as sequence of frames was also addressed using deep learning approaches relying on ConvNets. Similarly to the use of ConvNets on static images, the approach could be used both for extracting features characterizing the visual input, and for classifying elements of the visual content. The temporal context used by the model is expected to provide complementary information improving the search and classification of human action and activities, or situations where motion is important.

ConvNets applied to images (as described in the previous section) are adapted to capture spatial information. In videos though, changes occur between frames. Different approaches can be used to capture the information conveyed by such changes and motion. One approach is to use 3D-ConvNets [33], which directly take sequences of frames as input. Another approach is to make use of dense optical flow maps as input, instead of pixels, as done in [53]. In this alternative, each map provided as input to the neural network represents an estimation of pixels displacement between two successive frames. This new representation makes the task of temporal information capture easier for the ConvNets, which usually converge faster and better during training.

Here, we focused mostly on the use for feature extraction, to be used in the VideoSketcher system in a vector space retrieval paradigm. We trained a ConvNet on UCF101 [55], a benchmark for action recognition research. This dataset consists of 101 categories and over 12000 videos. The originators propose three splits, we used the first split to train our network from scratch. Our model takes as input a sequence of 10 dense optical flow maps (we calculate these maps using an implementation of TV-L1 algorithm available on OpenCV toolbox [8]). We selected an architecture that produces a short feature vector with 96 values, using a bottleneck approach, similarly to the architecture used for image processing. The short feature vector facilitates further used for video search. As in the previous section, the network is first trained in a supervised manner for classification of the selected categories (here the 101 action categories). Once the training is completed, our ConvNet achieve a state-of-the-art action classification performance of 72.69% mean accuracy for UCF101 split-1. The 96-dimensional feature vector extracted from the bottleneck layer is then used as an additional feature vector that the user can selected while searching content using the full integrated VideoSketcher system.

As stated previously, ConvNet can only process fixed length inputs. For extracting features, we used an analysis period of 10 frames, which means that a feature vector is provided every 400 ms (the video frame rate = 25 frames/second). Video shots are however temporal sequences of variable length, and actually, the length of actions displayed in these shots is also variable and typically longer than the 10 frames passed as input to the ConvNet. Extracting feature vectors that are better representing the visible actions would hence require integration/fusion strategies. The current state-of-the-art when using ConvNets for action classification is to apply a late fusion strategy using the probability vectors extracted by the network at different time instants within the video shot. This is the approach that has been used to obtain the 72.69% accuracy figure reported in the previous paragraph. Here, we also started addressing this issue using time sequence modeling enabled by Hidden Markov Models (HMM) and by Long Short-Term Memory (LSTM) Neural Networks. Our temporal ConvNet is used to extract feature vectors that are used as input to HMMs or LSTMs. Both approaches can provide a solution for the variable length of the action that we try to represent (feature vector extraction) or recognize automatically. In both cases however, this requires careful choices of additional meta-parameters (e.g. the number of states for the HMM, the number of layers and of neurons of the LSTM, etc.). Preliminary results were obtained here on the action classification task, but these were slightly worse than our baseline result. Further studies are ongoing.

Here, as opposed to what has been done for the image analysis part, we did not use the vector of action probabilities as additional semantic feature vector because we did not have the chance to research on defining a relevant set of actions categories (as well as associated data set) for the application to generic video search.

Note that further details on our application of ConvNets to action classification are available [51]. These, different optical flow algorithms and model architectures have been compared.

## *4.3 Database Systems and Video Indexing*

Not only the sheer size of data, but finally also the complexity of data poses a challenge when it comes to storing the data at hand: The feature extraction process (see Section 4.2) produces large amounts of high-dimensional vector data that need to be stored and – at online time – queried in a nearest neighbor fashion, possibly together with Boolean predicates on the corresponding metadata. For example, a user may search for similar images (QbE or more generally similarity search) to the query image that were shot on a specific date (Boolean search). As searching is done at on-line time, it should be handled as quickly as possible to ensure that the user does not have to wait for the results.

Databases traditionally work on data pages; the number of pages to load from the hard disk to respond to a query is the dominating factor influencing the retrieval time (whereas the CPU time for performing the comparison can comparably be neglected). To reduce the query time, in traditional databases, index structures are applied on the structured data that ensure that only the necessary data pages are loaded. For orderable data, B*-trees have become the de facto data structure used in indices reducing the retrieval complexity from $O(n)$ for a sequential scan, where $n$ denotes the number of elements in the database, to $O(logn)$.

B*-trees work very well for linearly orderable structured data and have shown good performance throughout. Vector data, however, is not linearly orderable, but its order is only established with respect to a query (through a nearest neighbor search, for instance). There exists a large corpus on index structures for supporting nearest neighbor retrieval. These include hashing [60]: examples are the Grid File [37] or locality-sensitive hashing (LSH) [27]. Other approaches consider a (balanced) tree-based access structure (e.g., R-Tree [28] and R*Trees, DABS-Tree [7] or M-Tree [11]). The generalized search tree (GiST) [30] subsumes many of the common features of the latter category. The VA-File uses a data structure which is particularly well suited for efficient nearest neighbor search in high-dimensional spaces [62], and currently represents a state-of-the-art approach. These authors have also shown that for increasing dimensionality, all tree-based structures degenerate to an inefficient sequential scan of the data.

There is no approach that fits all needs. And with dataset nowadays increasing in size and dimensionality (different meta-data and representation of different facets of the content), it remains challenging to develop efficient approaches. An approach using multiple index structures has been proposed here. The next Section describes the database back-end platform that has been used and improved for VideoSketcher.

### 4.3.1 Data Storage and Retrieval

The research prototype used and improved here, ADAM [21, 22, 23], focuses on new strategies and methods towards a modern distributed system that stores, organizes and retrieves multimedia data in a large, distributed setting. ADAM follows a modular architecture, based on the nature of the data at hand, the data is placed at different modules. For structured metadata, a traditional, relational database is used. For vector data, ADAM supports a large variety of index structures: locality-sensitive hashing (Indyk et al. [32, 29]), spectral hashing (Weiss et al. [63]), vector approximation-file (VA-File, Weber et al. [62]) that are applied jointly and that allow to increase the performance of the retrieval task. ADAM allows to be distributed over multiple nodes to decrease the retrieval time. The ADAM system will perform a query as follows. First, if Boolean predicates are applied in the query on the structured metadata (e.g., place = “Mons”), ADAM will query the relational database that stores the structured data to retrieve the tuple identifiers that should be further considered. Then, using the built-in index structures, the query vector and the result list from the first step, a nearest neighbor query on the index structures is performed, i.e., potential candidates that will enter the final result list are retrieved whereas results that for sure will not enter the list are pruned. However, it should be noted that the result list may contain both false positives and false negatives (also depending on the index structure used), i.e., it may lack results and it may contain too many results. In the final step, for all elements in the result list, the full feature vector is retrieved and the exact distances are computed, so that results can be re-ranked accordingly.

## 5 End-User Evaluations

A user-centric study has been performed in the last day of the event. The purpose of this event was to test the sketch based video retrieval capabilities of VideoSketcher in the realistic competitive environment inspired by the annual Video Browser Showdown [47]. We also relied on previous work done by us with respect to interactive video search [42]. In the VBS for each visual Known-Item Search task the participants need to interactively find a short video clip (20 secs) in the video collection within a specific time limit.

### 5.1 Video Collection

The evaluation is performed on the Open Short Video Collection 1.0 (OSVC1) dataset [41] which consists of 200 creative commons videos with a high variability with respect to their visual appearance and a duration between 30 seconds and 30 minutes. The total combined duration of the collection is roughly 20 hours. For generating queries, one random 20 second snippet from each video has been automatically extracted. Pictures of the evaluations are provided in Figure 4.

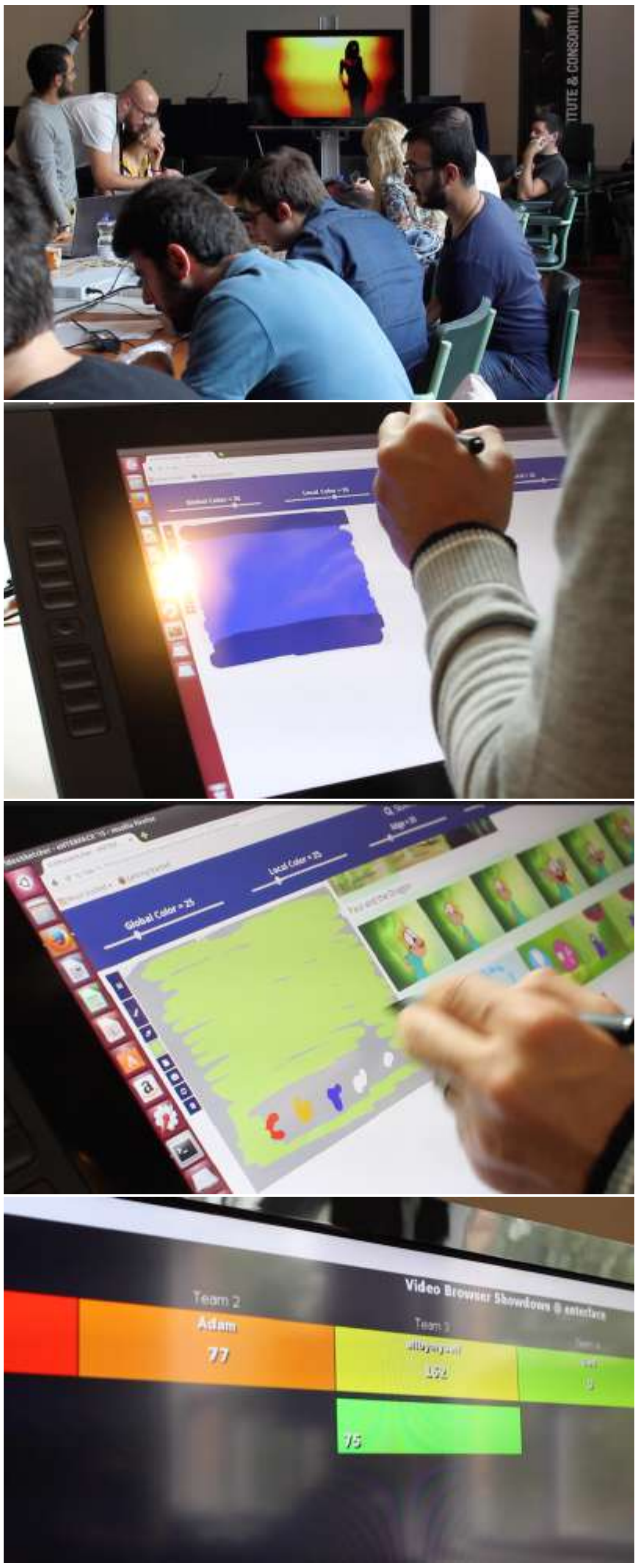

**Fig. 4** Pictures illustrating the setup during the competitive evaluation, the sketch interface and use of a tablet device for sketching and browsing results, and the display of the VBS scoring system back-end.

## *5.2 Setup*

There were 3 teams consisting of 2 or 3 project participants. Each frontend machine was connected to a Wacom tablet to allow stylus sketching. All 3 front-ends queried the same backend system and database.

The event consisted of 3 warm-up/dry run unscored tasks, followed by 10 actual competitive tasks. Tasks were randomly selected from the pool of 200.

Each task began by playing the 20 second snippet on the main screen. As soon as playback ended, participants could begin sketching and browsing for the sequence, and the server would start listening for submissions.

Every task had a deadline of 200 seconds. Each team would be scored with a maximum of 100 points for a task. According to VBS protocol, this maximum score is linearly penalized with time and proportionately penalized for wrong submissions.

## *5.3 Results*

As seen in Table 1, the total number of hits was low: only 4 out of 10 tasks had hits (tasks #4, #5, #7 & #8). However, in 6 out of the 10 tasks, the correct video has actually been found, although not the exact video shot within that video. The main contributing factors for this low success rate are according to us: the short query time of 200 seconds for each task (shorter than in the actual VBS) and the challenging setting of the OSVC collection. Figures 5 & 6 contain a filmstrip of each task and a selection of sketches and submissions made by all 3 teams.

| | #1 | #2 | #3 | #4 | #5 | #6 | #7 | #8 | #9 | #10 | total |
|---|---|---|---|---|---|---|---|---|---|---|---|
| team1 | ~~V~~ | | 6x~~V~~ | 5x~~S~~H | | | 4x~~S~~H | | 5x~~S~~ | 4x~~S~~ | 7x~~V~~ 18x~~S~~ 2xH |
| score | 0 | 0 | 0 | 11 | 0 | 0 | 28 | 0 | 0 | 0 | 39 |
| team2 | ~~V~~ | | ~~V~~ | ~~S~~H | | | 8x~~S~~H | 15x~~V~~ | | | 17x~~V~~ 9x~~S~~ 2xH |
| score | 0 | 0 | 0 | 77 | 0 | 0 | 23 | 0 | 0 | 0 | 100 |
| team3 | | | | ~~SS~~H | ~~SSS~~H | | ~~SSS~~H | ~~SSS~~H | | ~~SSS~~ | 14x~~S~~ 4xH |
| score | 0 | 0 | 0 | 87 | 75 | 0 | 45 | 42 | 0 | 0 | 249 |

**Table 1** Competition results broken by task. ~~V~~ = wrong video submission, ~~S~~ = correct video but wrong shot submission, H = correct submission.

Tasks #1 and #6 were dark sepia and greyscale which made all color features useless, most sketching results returning black frames (which were very frequent in the collection). Tasks #2 and #3 contained a complicated city skyline and a rather bland sequence from an animated cartoon. Both of these are very hard to sketch.

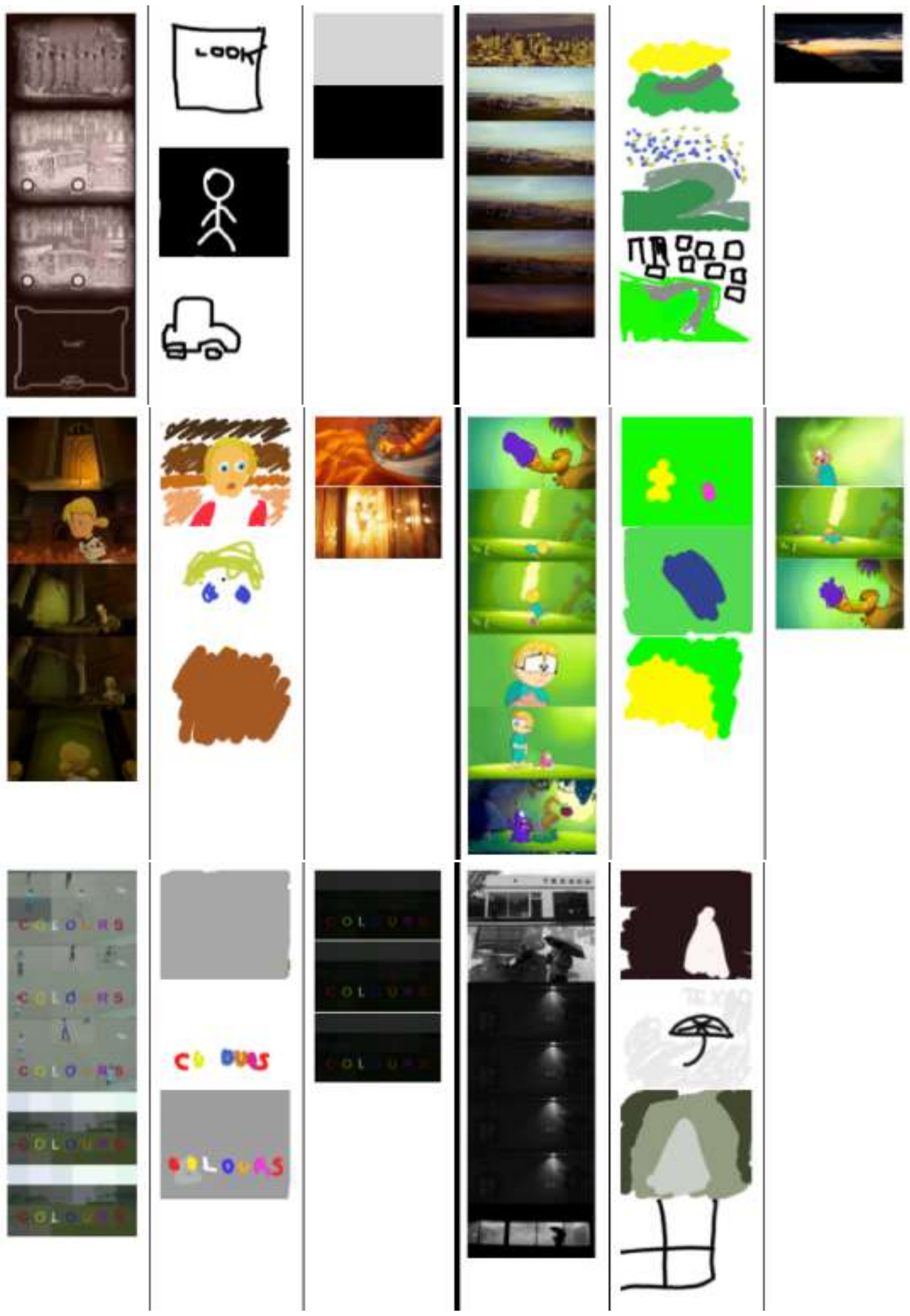

**Fig. 5** Left-to-right, up-to-down tasks #1 through #6. First column shows a filmstrip of the task segment, middle column shows some of the user sketches, rightmost column shows some of the wrong submissions.

Task #4 was found by all 3 teams thanks to its strong color components (green background and purple character). This segment of a cartoon was found even with rather naive sketches, however identifying the exact point in the animation required up to 5 wrong submissions.

Task #5 contained an opening logo with strong colored letters on a grey background which team 3 apparently memorized and was able to correctly sketch and retrieve, but only after 3 false submissions of the same logo from the ending of the video.

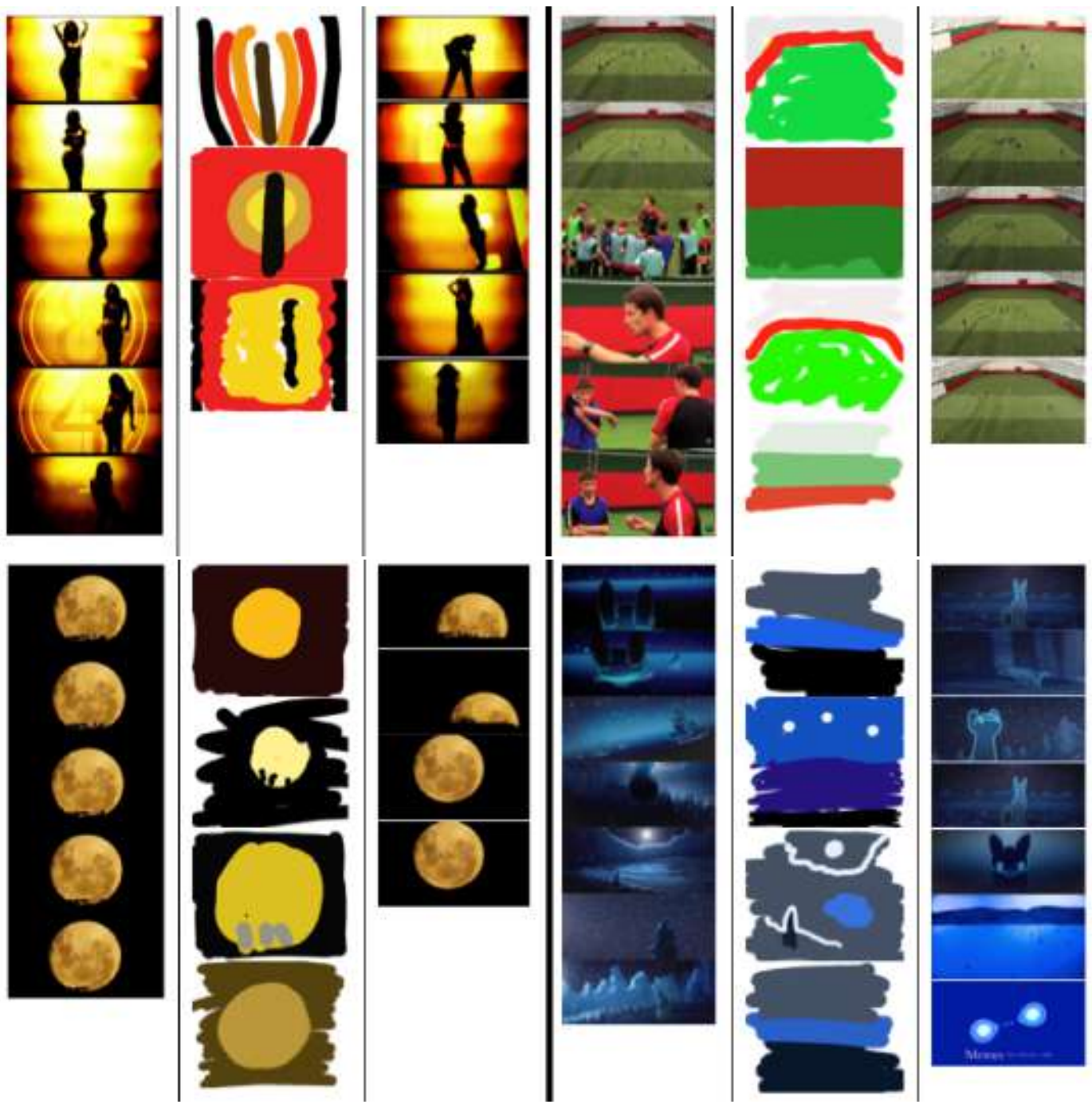

**Fig. 6** Figures 5 continued. Left-to-right, up-to-down tasks #7 through #10. First column shows a filmstrip of the task segment, middle column shows some of the user sketches, rightmost column shows some of the wrong submissions.

Task #7 came from a video that looked the same throughout its entirety, which explains why all 3 teams found the right video, but it took as much as 8 wrong submissions to identify the exact position (nobody managed to find the faded numbers on the background). The same issue can be observed on tasks #8 (a soccer training video mostly filmed from one static camera) and #9 (a music video with a moon setting in the background). Moreover, in task #8 a very similar video (a different lesson in the same soccer training series) was found by team 2, which unsuccessfully tried to browse through, submitting 15 wrong shots.

Task #10 was a blue color dominated cartoon where the video was correctly found by 2 teams. The difficulty of retrieving a snowing scene in a rather complex sequence made it impossible to pinpoint the exact position.

### *5.4 Follow-up work*

The evaluation performed here and the improved system resulting from this work helped us to improve upon our submission to the Video Browser Showdown 2015 [42] and to participate again in 2016 with two submissions [39, 43].

## 6 Discussion and Conclusions

This paper presented the results of the worked carried on during the eNTERFACE 2015 summer workshop held in Mons, Belgium. It gathered contributions from three complementary research domains that can contribute to the future of video search engines: user interfaces enabling innovative approaches for formulating queries through sketches, machine learning enabling advanced computer vision and efficient representations of video content, and database and indexing approaches enabling efficient storage and retrieval. The obtained results have been at the basis of improved components in these three areas.

Ongoing research work and evaluations performed in the area highlight several challenges and avenues for improvement, with novel work that has already started and to be followed further in the future.

On the side of the user interface and multimodal queries, several challenges appear. Sketch-based queries are naturally inclined to be ambiguous as they lack specification in some information channels. In this case, the IR system has to deal with the lack of information in a query, as it cannot deduce whether this information should be absent in the result or whether it has simply not been specified. A novel approach anticipating the intent of the user has been proposed recently [44]. Another issue relates to the fact that users will become more accustomed and expecting real-time feedback when formulating queries, as is already the case with text queries. We proposed solutions for real-time feedback involving partial sketch recognition and matching as well as autocompletion [58, 50]. Regarding the UI itself, the fact that several sketch modes and layers are proposed needs to be considered carefully, through alpha-blending and layer specification.

On the side of machine learning for content analysis, key challenges remain in terms of understanding content and extracting features that can reveal useful for video search. Most current work relies on supervised learning, implying the existence of annotated corpora which are costly to prepare. Semi-supervised, weakly-supervised, or even unsupervised learning need to be pursued further. Initial ideas have been proposed very recently by others [61], and more are poised to come in the future. Active learning is another area worth investigating in this context, with concrete application to video search, while most previous work has been done on less complex data set and use cases. Then, being able to better model and leverage the time sequence rep-resented by video shots, for improving action and activity recognition, seems necessary. We will continue to work with HMMs and LSTMs in order to improve the temporal feature extractor. Further research on combining spatial and temporal visual information is also important [51]. Note also that current research is focused on the challenges of recognizing object and actions, and their spatial location (so-called detection problem). Visual information contains more information, that should enable the recognition of attributes of object and actions. This aspect is likely to become more and more important while the research community will challenge itself with even more complex problems.

When innovating on query modes and video understanding, a fundamental issue lies at the intersection of both, as it becomes necessary to find either a common representation between the query and the content, or else define or learn mappings between two representations. In this work, there are three types of queries. First, semantic image sketch queries are first recognized and the issue can hence be bypassed by relying on a text/keyword representation. For visual (color) sketch queries, we used the same feature set as for the video key-frames analysis, except features computed using machine learning, which have been trained on natural images and won't be appropriate on sketches. This appears to provide positive results, although further research may be required as flat colors when sketching quickly may lead to a significant mismatch between the sketch and the video content feature representations. Finally, for audio sketches, we face the issue of mapping of vocal imitations to their associated sounds. This could be addressed through machine learning, such as pre-training the neural network that has been used and/or by training on bigger databases. Future work is also necessary to investigate fusion of the audio and video features.

## Acknowledgements

This work was partly supported by the Chist-Era project IMOTION with contributions from the Belgian Fonds de la Recherche Scientifique (FNRS, contract no. R.50.02.14.F), the Scientific and Technological Research Council of Turkey (TUBİTAK, grant no. 113E325), and the Swiss National Science Foundation (SNSF, contract no. 20CH21_151571).

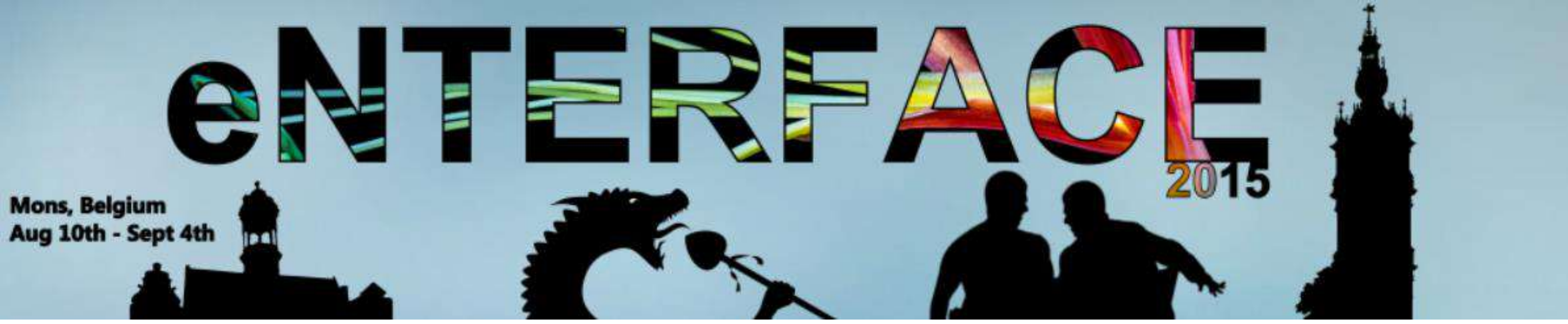

**Project #6:**

# Enter the ROBiGAME: Serious game for stroke patients with upper limbs rehabilitation

*Françcois Rocca, Pierre-Henri De Deken, Alessandra Bandrabur, and Matei Mancas*

# Enter the ROBiGAME: Serious game for stroke patients with upper limbs rehabilitation

François Rocca[1], Pierre-Henri De Deken[1], Alessandra Bandrabur[2], and Matei Mancas[1]

[1] Numediart Institute, University of Mons, Belgium
Place du Parc 20, Mons, Belgium
surname.name@umons.ac.be
[2] IPAL, University Politehnica of Bucharest, Romania

**Abstract.** The objective of the project is to develop an intelligent serious game for rehabilitation of the upper limbs for stroke patients using an interactive rehabilitation robot. The robot screenplay adapts to the patient's functional abilities and the robot mechanical assistance evolves according to patients motivational, motor and cognitive performances when playing a serious video game. This work will be developed on the REAPlan robot providing a distal effector which can mobilize the patient upper limb(s) in a horizontal plane [1] but could also be transferred to other robots or even simpler rehabilitation setups.

## 1 Introduction

Thanks to advances in basic and clinical research, neurological rehabilitation knowledge has greatly developed in recent years [2] and [3]. During the rehabilitation, to learn a task, the exercises in brain-injured patients are improved by the use of rehabilitation robotics that achieves intensity (number of movements by time) higher than the conventional rehabilitation( [4] and [5]). One of the possible optimization of this technique is the implementation of serious games with robots that would combine the benefits of rehabilitation robotics and those from the serious games mainly in terms of patient motivation. Indeed, they have already proved their specific interest in the adult brain-injured patient [6]. The "Enter the ROBIGAME" project can be broken down into four components: medical, robot, video games and analysis of motivation.

This paper is structured as follows. Section 2 provides information about related systems and work, section 3 gives details about the rehabilitation system used in this project. Section 4 gives information about Serious gaming using the game engine Unity. The next section 5 shows how we do user tracking and analysis in three parts: the first one explains the body face tracking, the second one facial action coding system theory and the third explain how we use a machine learning toolkit for expression recognition. Section 6 describes the whole setup in this project and relates the results of the experiment. Finally we conclude in section 7.

## 2 Related works

For the hemiplegia, the most popular and most effective method of rehabilitation is to restore function rather than offsetting the deficit [7]. Different restoration protocols have been developed to relearn both a specific skill than the general activities of daily living [8]. Research in neuroscience shows that relearning protocols result in positive changes in the structure and activity of the brain ([9] and [10]). Different technologies have emerged in recent years to assist doctors and patients during therapy. Some of them are linked to serious games in order to perform not abstract tasks.

### 2.1 Technologies for rehabilitation

While the video game is sometimes seen by the wider public as an entirely fun activity, it has rather great potential for serious activities. Indeed, it can be used as a psychological, cognitive or motor help allowing the player to carry out meaningful tasks. In the medical world, particularly rehabilitation is an area where serious games are becoming more developed ([11] and [12]). For the rehabilitation, different systems use sensors with effectors or not, these systems are generally used with games. Some of them use optical motion system to track the user body (webcam, kinect, etc.), electronic sensors (gyroscope, accelerometer, etc.) or robotic arms. The category of robotic arms for rehabilitation is divided into two families depending on they support or not the entire "joint chain" located between the patient's trunk and hand. Table 1 gives a summary of some of these systems. In the first case we speak of exoskeletons, while in the second we speak of manipulators distal effector. The most famous robot belonging to this second family is the Manus robot developed by Krebs [13] at the Massachusetts Institute of Technology (MIT) and marketed by Interactive Motion Technologies. Robots REAplan and REA2plan developed at Mechatronics Research Centre (CEREM UCL, Belgium) also belong to this category. The latter two robots are those used in this project.

### 2.2 Motivation extraction

When performing an exercise, the ROBiGAME tracker process must be able to capture information related to motivation performance of the patient. The patient's motivation will be assessed from two features extracted from patient's behavior. The first one concerns the orientation of the head [14] and this will be extracted using a 3D camera (Microsoft Kinect sensor). The second one concerns the analysis of the patient's emotional state. From the patient's face that can be automatically detected with the Kinect SDK, it is possible to extract emotional parameters such as "FACS" [15] which are basic units related to the emotions . This would make it possible to know the valence of the face expressions of the patient to determine positive or negative emotions based on expressions and also the degree of energy spent by the patient during the task (neutral, very involved, etc.).

**Table 1.** Summary of some rehabilitation system with serious game for stroke pathologies

| Reference | Function spotted | Technolgies used | Assistance | Type of Serious Game |
|---|---|---|---|---|
| Burke 2009 [16] | Upper limbs motricity | Webcam + Screen | No | Basic Arcade |
| Cameiro 2011 [17] | Upper limbs motricity | Webcam + Data gloves + Screen | No | Basic Arcade |
| Buxbaum 2012 [18] | Hemineglect rehabilitation | Screen | No | Simulation of real task |
| Kim 2011 [19] | Hemineglect rehabilitation | Webcam + Data gloves + Screen | No | Simulation + Arcade |
| Katz 2005 [20] | Hemineglect rehabilitation | Screen | No | Simulation of real task/life |
| Myers 2000 [21] | Hemineglect rehabilitation | VR headset | Yes : Visual +audio | Simulation of real task/life |
| Klinger 2013 [22] | Hemineglect rehabilitation | Webcam +Screen +Keyboard +Gamepad +Mouse +Micro | No | Simulation of real task/life |
| Hasomed [23] | Cognitive Function | Screen +Special Keyboard | Yes: Visual | Object spotting |
| R.O.G.E.R.S. [24] | Cognitive Function | Kinect +Screen | No | Simulation of real task/life |
| Voracyfish [12] | Upper limbs motricity | Kinect +Screen | No | Arcade |

## 3 Reaplan system

The reaplan system is a robotic device for medical assistance which helps in rehabilitation of adults and children affected by stroke. It was developed by Belgian start-up Axinesis (CEREM UCL) to facilitate the recovery of motor function of the upper limbs.

### 3.1 Description and usability

Motor deficiencies (strength, joint mobility, dexterity) will be evaluated clinically according to the validated protocols. An evaluation protocol of these deficiencies by the robot will be developed to provide ROBiGAME quantitative and objective data. Force and position sensors will be used to assess:

- Patient force;
- Comprehensive passive and active range of motion by measuring the largest perimeter the subject can perform with or without assistance;
- Kinematics and kinetics of the arm in standardized movements.

The patient's motor performance will be quantified during rehabilitation exercises using the robot's sensors. The reaplan robot is shown in figure 1, on the right part of this figure you can see the distal effector (in blue), the working plane (in green), the screen (in yellow) and the stop button (in red).

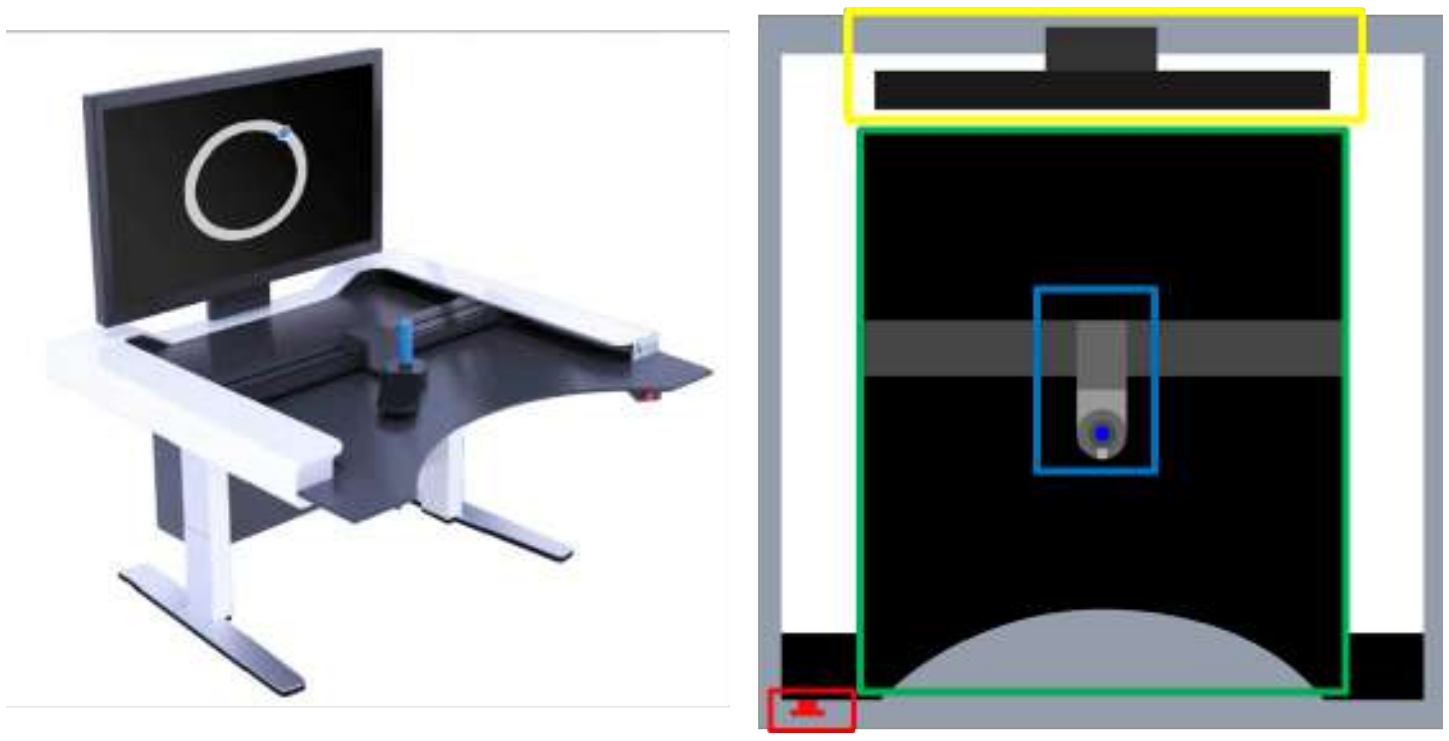

**Fig. 1.** Reaplan systeme with the distal effector in blue, the working surface in green, the screen in yellow and the emmergency button in red.

### 3.2 Communication

From the point of view of the connections, the communication between the REAPlan and the computer is done by USB 2.0 which allows to simulate a port COM. For the software part, an interface created by Axinesis allows to manage the exchange of information and commands between games and robot.

**Existing interface of the REAPlan.** This interface allows the therapists to manage the profiles of the patients. They can also by this way configure the games which will serve during a session of reeducation. During this session, it is easy to verify in live the evolution of the parameters of the patient. At the end of the session, the data and the results are saved for a further analyse and re-used for a later session.

If a new game must be added to the interface, it is necessary to add compiled files in the code of the interface and then to recompile this one in order to test the new game.

**Design without this interface.** To avoid all this procedure of insertion and by the way to improve the development speed, we decided to work without this interface to set up the first games of tests. The Dynamic Link Library file (DLL) which contains all the functions allowing the communication between the interface and the REAPlan was extracted. This file was directly used by the new games.

However the program used for the creation of these games accepts only DLL files with a version of the framework lower or equal to the .net 3.5. This DLL having been realized with the framework .net 4.0, we had to recompile it. This stage also taking some time, we limited the recompilation to the part of code which allows the control of the REAPlan in free mode, that is the mode where the distal effector brings into conflict no forces to the patient. Therefore, the REAPlan will not assist the players in our first game of tests.

## 4 Unity

To create our first games, we used to the Unity framework. This software presents a lot of qualities. It is free in its basic version, numerous tutorials can be found and it has an excellent documentation. Furthermore, an online store allows us to obtain and reuse numerous assets. In the 3.2, we mentioned the use of a DLL, Unity allows the integration of new modules and DLLs. All the programming is realized in c# with the environment of development Monodevelop[25].

### 4.1 Unity description

Unity is a multiplatform game engine and a development platform created by Unity Technologies for designing multiplatform 3D and 2D games and interactive experiences. It is one of the most popular in the gaming industry.

The philosophy in unity is Object-Oriented programming. Any element placed in Unity is a GameObject. A cube with a 3D picture is a GameObject, a camera which gives the point of view of the player is also a GameObject, as well as light sources or sound sources. All the GameObjects have basic properties such as their position, angles of rotation and a size according to axes x, y and z. What differentiates them are the modules which we can bind to these GameObject. For example, the cube will have a module which will allow its visual display. In a game, the various elements can have particular behavior, react to events or simply move. All this is defined in what is called a script. For example, a button will have a module which will bind it to a script which will define what has to occur when the player presses on the button. Finally, the whole level is saved in a file which we call scene. It is therefore possible to have several scenes in a game.

### 4.2 Design of the games

During our first developments, we succeeded in making communicate a Unity-based game with the REAPlan and to build two versions of this game. The first one is a shooting game in 2D with a top view. Not to waste time by recreating from scratch a quite new game and in a purpose of Unity's handling, we decided to rewrite the first game by changing its assets, a part of its code and its parameters. This second game is called "A l'abordage". While the first one was based on spaceships that must destroy asteroids and other enemies, this second one takes place in a maritime environment and implements ships.

**Menu and communications** We gathered both games in a single program with a selection menu which we can see in figure 2. In the starting up of the main program, two scripts are loaded. The first one initializes the communication towards the REAPlan thanks to the dll. The games of the menu are accessible only if the initialization is finished. In case the initialization failed, the communication towards the REAPlan is cancelled and the games become accessible and playable with the control of the mouse. The second script initializes a OSC communication [26] with another computer which takes care to analyze the gaze and motivation of the patient (see section 5.1). These information allow us to edit in real time the level of the difficulty of our game in order to adapt it to the patient's behavior.

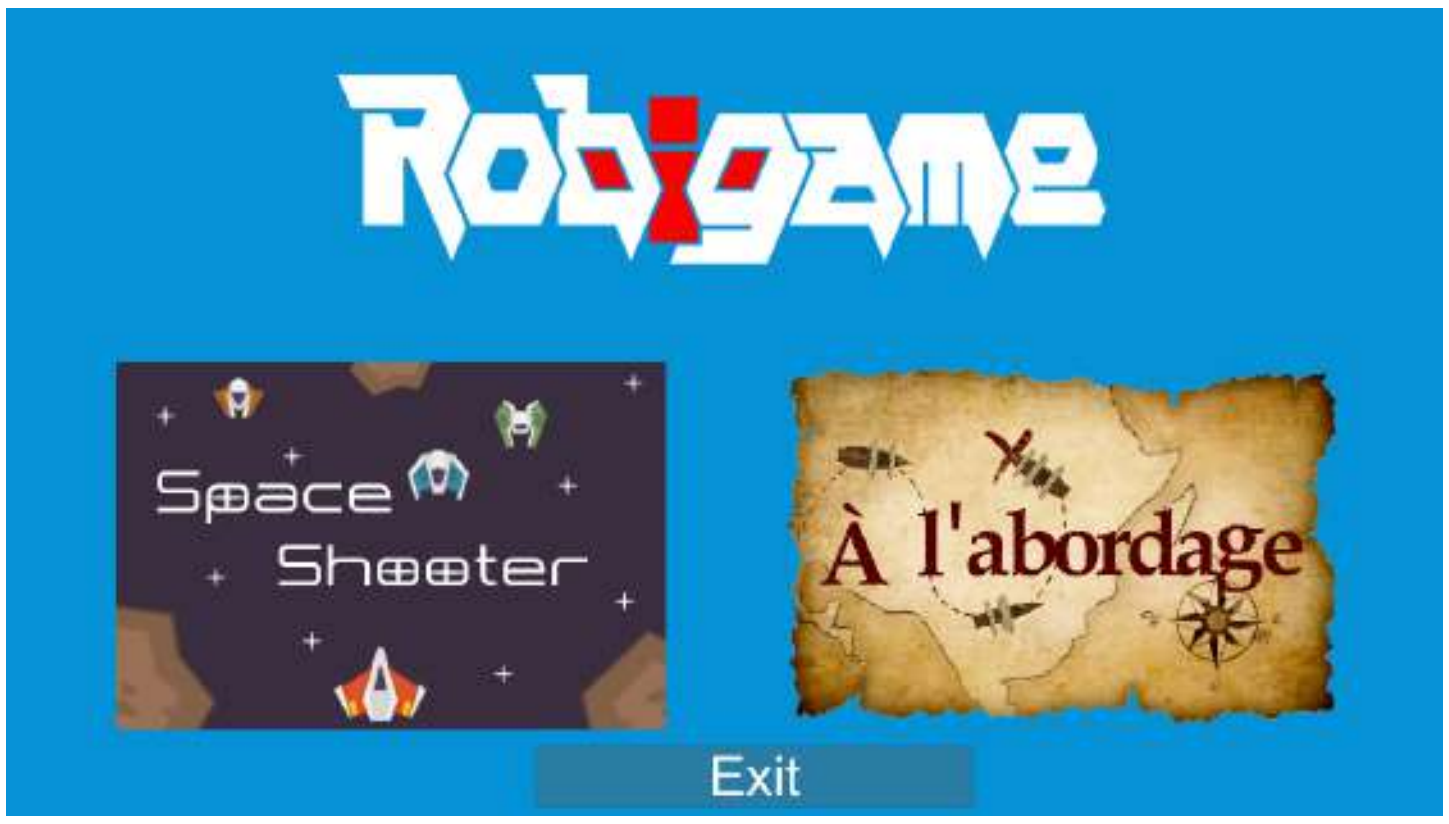

**Fig. 2.** The menu to select the game.

**Games and interface with REAPlan.** Our choice went towards the shooting game in the space for two main reasons. The first one is that the free movement is the only accessible mode for the REAPlan at the moment (see section 3.2), this mode represents the physics in a spatial environment. The second is that the game takes place in a 2D world what corresponds to the movements of control which are allowed by the REAPlan.

To control correctly the game with the REAPlan, a mapping must be realized between the zone of movement of the effector distal (left image of the figure 3) and the play area (right image of the figure 3). We use formule (1) and (2) to reach our purpose.

As it was previously mentioned at the beginning of this section, the second game is a rewriting of the first one with changes of the assets and the parameters. The figure 4 lists the correspondences between both games and specify the parameters used according to the difficulty of the game.

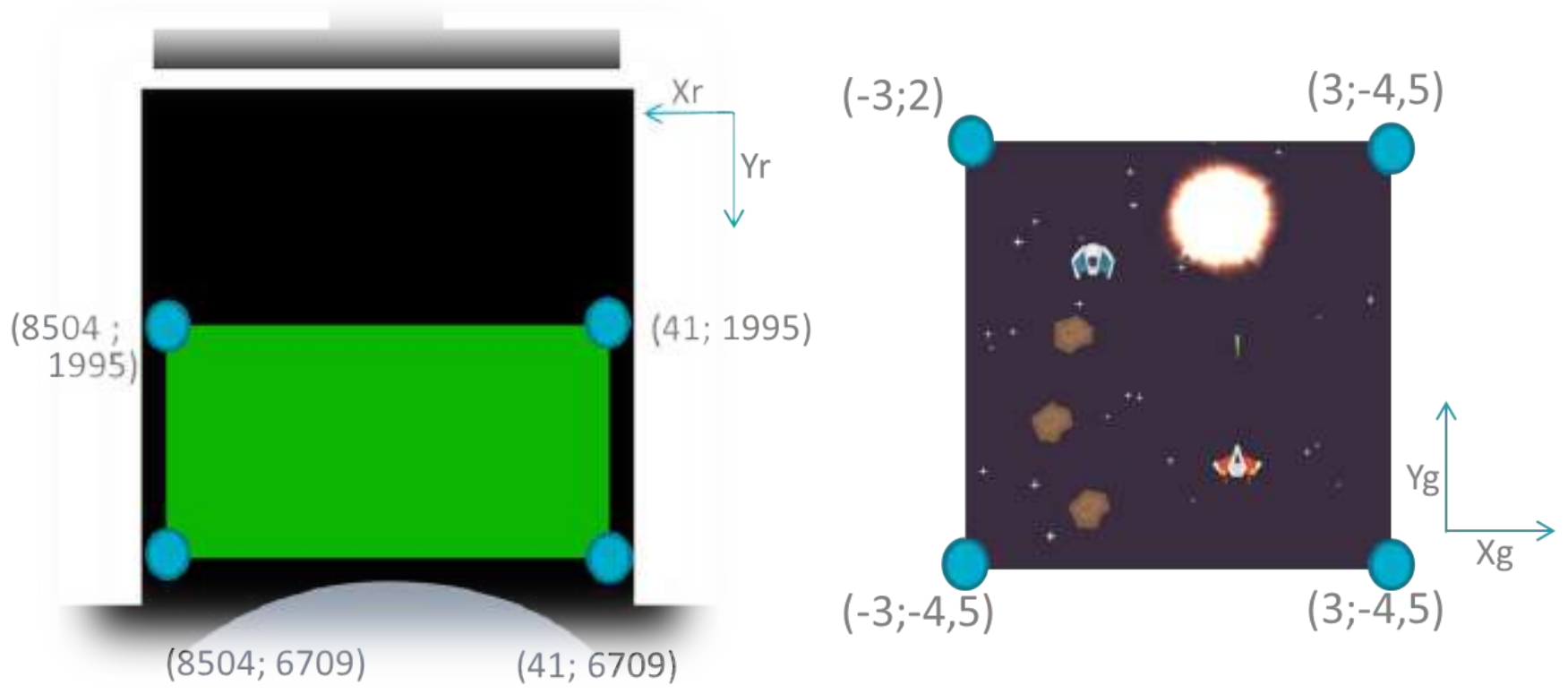

**Fig. 3.** Working plan for rehabilitation to gaming coordinate.

$$Xg = \left[ - \left( \frac{Xr - 41}{8504 - 41} \right) * 6 \right] + 3 \tag{1}$$

$$Yg = \left[ - \left( \frac{Yr - 1995}{6709 - 1995} \right) * 6.5 \right] + 2 \tag{2}$$

## 5 Body and face Analysis

### 5.1 User body and smile detection

For the user body and face tracking, we use the Kinect V2 sensor with Microsoft Kinect SDK. The main use of the Kinect is the user skeleton tracking. Skeletal tracking is able to recognize users sitting on the REAPlan. To be correctly tracked, users need to be in front of the sensor, making sure their head and upper body are visible (see Figure 5). The tracking quality may be affected by the image quality of these input frames (that is, darker or fuzzier frames track worse than brighter or sharp frames). The Kinect sensor contains two CMOS sensors, one for the RGB image (1920 x 1080 pixels) and another for the infrared image (512 x 424 pixels) from which the depth map is calculated. The technology behind the sensor is infrared TOF (Time Of Flight). This sensor measures the time it takes for pulses of laser light to travel from the laser projector to a target surface, and then to come back to an image sensor. Based on this measure, the sensor gives a depth map.

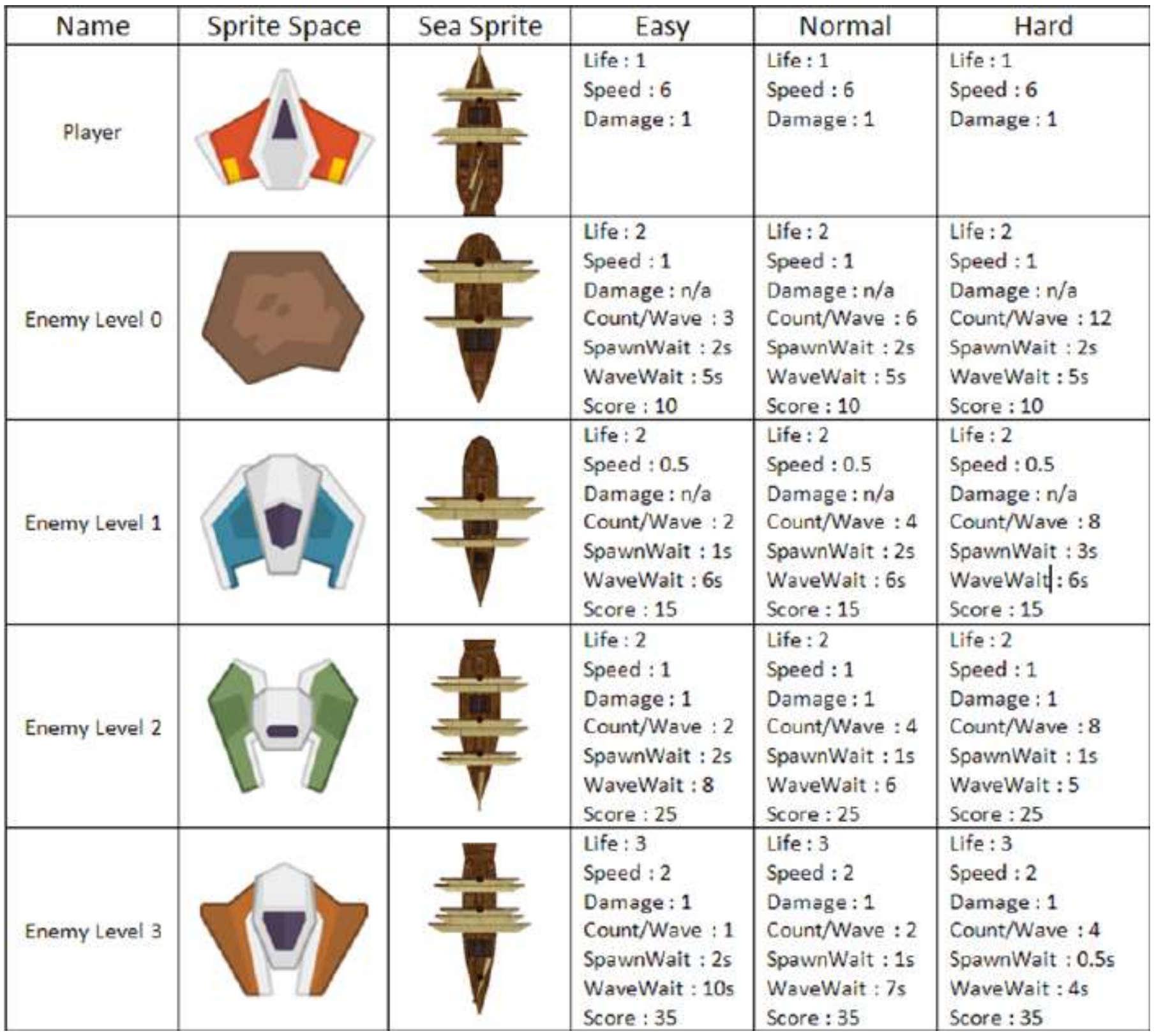

| Name | Sprite Space | Sea Sprite | Easy | Normal | Hard |
|---|---|---|---|---|---|
| Player | | | Life : 1<br>Speed : 6<br>Damage : 1 | Life : 1<br>Speed : 6<br>Damage : 1 | Life : 1<br>Speed : 6<br>Damage : 1 |
| Enemy Level 0 | | | Life : 2<br>Speed : 1<br>Damage : n/a<br>Count/Wave : 3<br>SpawnWait : 2s<br>WaveWait : 5s<br>Score : 10 | Life : 2<br>Speed : 1<br>Damage : n/a<br>Count/Wave : 6<br>SpawnWait : 2s<br>WaveWait : 5s<br>Score : 10 | Life : 2<br>Speed : 1<br>Damage : n/a<br>Count/Wave : 12<br>SpawnWait : 2s<br>WaveWait : 5s<br>Score : 10 |
| Enemy Level 1 | | | Life : 2<br>Speed : 0.5<br>Damage : n/a<br>Count/Wave : 2<br>SpawnWait : 1s<br>WaveWait : 6s<br>Score : 15 | Life : 2<br>Speed : 0.5<br>Damage : n/a<br>Count/Wave : 4<br>SpawnWait : 2s<br>WaveWait : 6s<br>Score : 15 | Life : 2<br>Speed : 0.5<br>Damage : n/a<br>Count/Wave : 8<br>SpawnWait : 3s<br>WaveWait : 6s<br>Score : 15 |
| Enemy Level 2 | | | Life : 2<br>Speed : 1<br>Damage : 1<br>Count/Wave : 2<br>SpawnWait : 2s<br>WaveWait : 8<br>Score : 25 | Life : 2<br>Speed : 1<br>Damage : 1<br>Count/Wave : 4<br>SpawnWait : 1s<br>WaveWait : 6<br>Score : 25 | Life : 2<br>Speed : 1<br>Damage : 1<br>Count/Wave : 8<br>SpawnWait : 1s<br>WaveWait : 5<br>Score : 25 |
| Enemy Level 3 | | | Life : 3<br>Speed : 2<br>Damage : 1<br>Count/Wave : 1<br>SpawnWait : 2s<br>WaveWait : 10s<br>Score : 35 | Life : 3<br>Speed : 2<br>Damage : 1<br>Count/Wave : 2<br>SpawnWait : 1s<br>WaveWait : 7s<br>Score : 35 | Life : 3<br>Speed : 2<br>Damage : 1<br>Count/Wave : 4<br>SpawnWait : 0.5s<br>WaveWait : 4s<br>Score : 35 |

**Fig. 4.** Asset comparisons with games values.

To achieve head pose, at least the upper part of the user's KinectV2 skeleton has to be tracked in order to identify the position of the head. The position of the head is located using the head pivot from the 3D skeleton only on the depth map. The head pose estimation is based on the face tracking and it is achieved on the color images. Consequently, the face tracking is dependent on the light conditions, even if KinectV2 is more stable into dark light conditions than KinectV1. The head pose estimation method returns the Euler rotation angles in degrees for the pitch, roll and yaw as described in Figure 6, and the head position in meters relatively to the sensor being the origin for the coordinates. Based on the head pose estimation, it is possible to know where the user is looking on the screens. It's also possible to know how long they spend watching a part of the screen. Different duration are used to describe the level of attention.

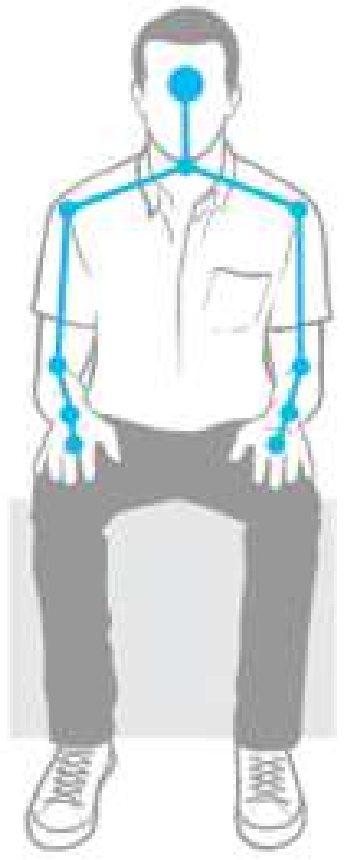

**Fig. 5.** User tracking in seated mode

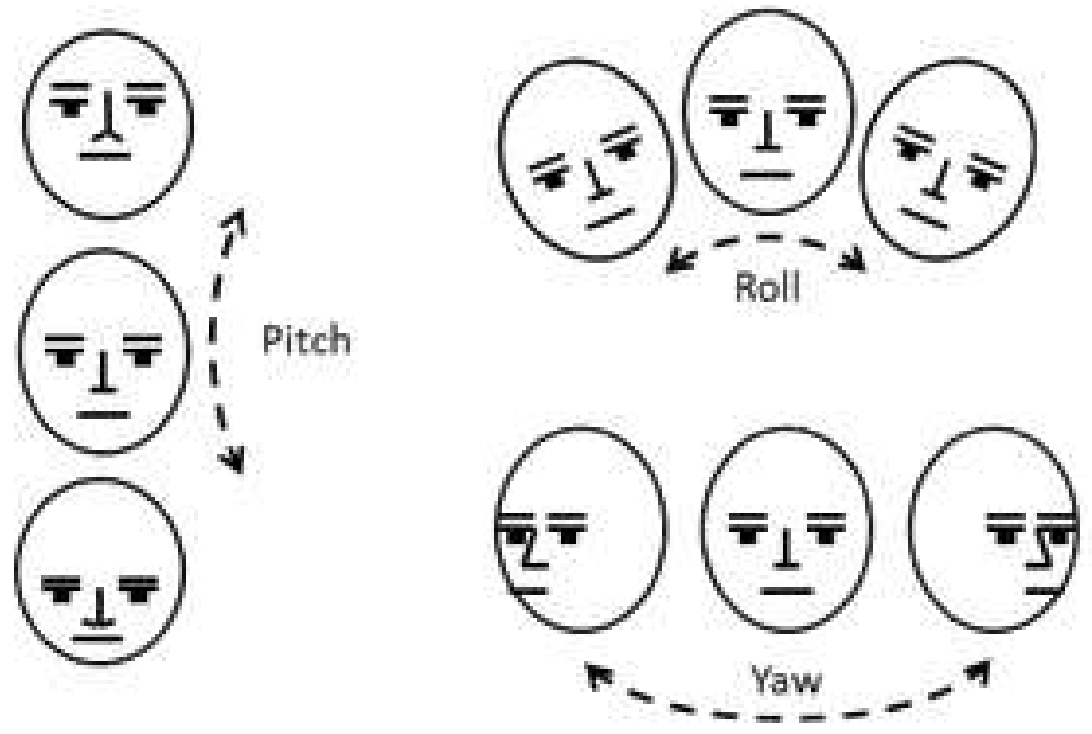

**Fig. 6.** Three different degrees of freedom of the head [27]

Based on the face analyze from the Kinect SDK, we extract: the neutral position of the mouth, brows, eyes, and so on. The Action Units (AU) represent the difference between the actual user face and the neutral face. Each AU is expressed as a weight between -1 and +1. Some basic functions are used to measure the smile (Figure 7) and its possible to determine if user is smiling, is not smiling or is maybe smiling.

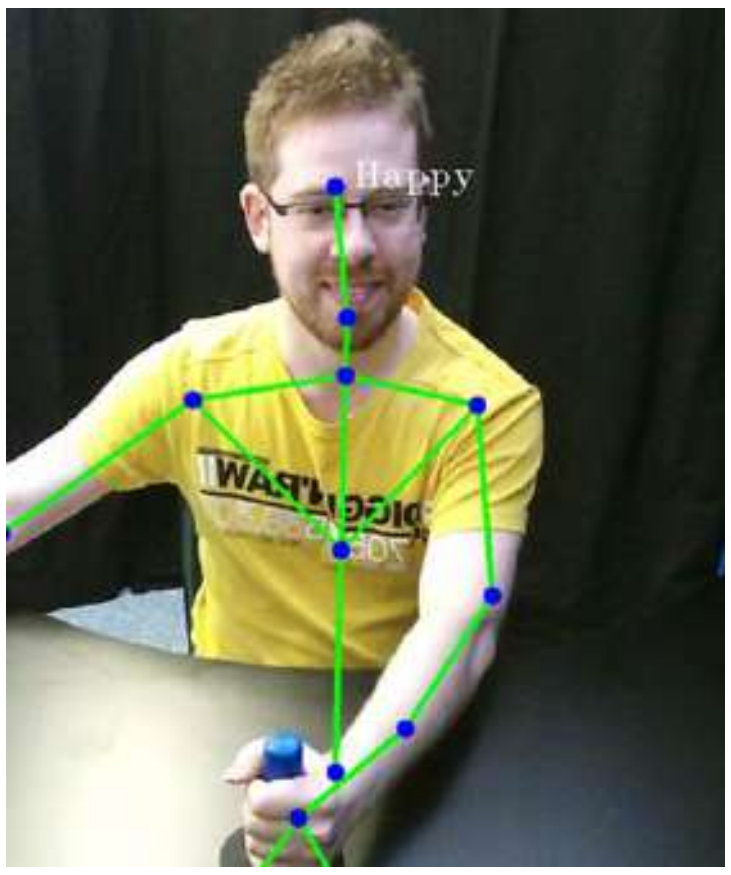

**Fig. 7.** User tracking with an happy face during the therapy.

### 5.2 Facial Action Coding System

The most used system for description of facial expressions was proposed by Ekman and Friesen in [15] and reviewed in [28] and its name is FACS, Facial Action Coding System. The FACS is based on the universality of emotions, namely there are a set of basic emotions which are recognized by people from all cultures. In his first work, Ekman et al [15] proved that there are six basic emotions: happiness, sadness, fear, anger, disgust and surprise. According to FACS, the muscles of a face are able to produce 46 Action Units (AU), where different combinations of AUs form the basic emotions.

An AU is a basic facial action and has three phases onset, apex and offset. This temporal aspect of the AUs can be seen in Figure 8. Facial temporal dynamics are used for recognition of different psychological states as pain [29] or continous affective states [30]. A set of rules that defines AUs and corresponding emotions is shown in Table Table 2 and 3 .

### 5.3 The Wekinator

**Definition.** The Wekinator is a framework for real-time machine-learning, created by Rebecca Fiebrink et al[32]. It is an open-source software application based on Weka framework[33] and it allows one to train and modify machine-learning algorithms in real-time. The application maps the user inputs to specific parameters of synthesis during the training stages.

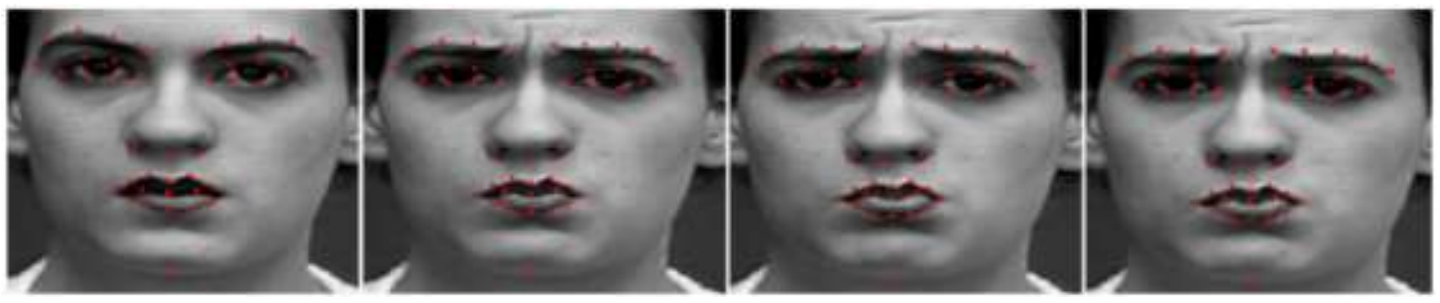

**Fig. 8.** Temporal modification of facial expression from neutral to apex. Image sequence taken from Cohn-Kanade+ database [31]

| AU Code | Description |
|---|---|
| AU1 | Inner Brow Raiser |
| AU2 | Outer Brow Raiser |
| AU4 | Brow Lowerer |
| AU5 | Upper Lid Raiser |
| AU6 | Cheek Raiser |
| AU7 | Lid Tightener |
| AU12 | Lip Corner Puller |
| AU15 | Lip Corner Depressor |
| AU16 | Lower Lip Depressor |
| AU20 | Lip stretcher |
| AU23 | Lip Tightener |
| AU25 | Lips part |
| AU26 | Jaw Drop |

**Table 2.** AUs Description

The datasets consist in features from video, audio or 3D data and they are used to train the learning algorithm responding to this specific inputs. This framework is compatible with Weka, therefore any dataset and classifier can be exported to Weka, and any classifier trained in Weka can be run within Wekinator.

The GUI provides several options for selecting the appropriate learning algorithm that fits a specific problem. Some of these algorithms are SVMs, AdaBoost, decision trees and k-nearest neighbor. The Wekinator provides also an option to modify the training data during the process execution, therefore one can re-train data in real time.

| Emotion | AUs Used |
|---|---|
| Sadness | AU1, AU4, AU15 |
| Happiness | AU12, AU6 |
| Fear | AU1, AU2, AU4, AU5, AU20, AU26 |
| Surprise | AU1, AU2, AU5, AU26 |
| Anger | AU4, AU5, AU7, AU23 |

**Table 3.** Recognizing facial expressions through AUs

The real-time user interaction with a complete learning process, allows a play-along mapping for generating the training dataset while the computer performs the actual task. The Wekinator was originally conceived to create controller mappings for sound synthesis, which provide to the subjective and unique user musical expression. Its purpose was to create an user-centered application, in order to promote real-time exploration of synthesis algorithms into compositional performances.

Rebecca Fiebrink has proven that using a small training dataset, created in a short period of time, could be enough for a personal performance. The training task is a data entry task, but in the same time a musical one as well, because the performer supervises the sounds that he hears during the learning through specific gestures. Thence the user set a musical score, which can be as well a random number, and choose a gesture family to interact with the algorithm.

Thanks to its properties the Wekinator is also suitable for human computer interactions in real-time video applications, interactive games or any other play-along system.

**Algorithm.** The user controller dataset may consist in FFT bin magnitudes or 2D/3D axis positions, that are provided by a joystick, a body tracked with Kinect, a hand tracked by a Leap Motion Controller and so on. The algorithm consists in several real time stages as specifing the input controllers, setting a specific learning method, creating or using an existent training dataset and running the trained to model to perform.

The input dataset and the output response are performed within OSC [34]. The input data is chosen among the built-in feature extractors or one can supply his own specific feature extractors. The former consists in time, spectral, edge detection or color-tracking features, implemented in Processing [35] or ChucK [36]. For training the model, the system has implemented several algorithms with Weka, such as neural networks, decision trees, nearest neighbor algorithm, AdaBoost and support vector machines. Weka is a very well known open-source library written in Java, which provides different classifiers and regressors. Due to this library, the Wekinator is able to furnish as output either a probability distribution, as a likelihood, either a specific class label, as a maximum likelihood. All the communication is achieved through OSC, therefore the output response of the trained model is used in any application that speaks through OSC.

The real time interactive algorithm of this system can be seen in Figure 9. One can set on-the-fly the machine learning methods with inputs, the specific features, choosing the type and the right parameters. In the same process the user builds the training dataset, the system trains the learner and finally the user can evaluate the results.

In the first stage, the user can setup a machine learning method and can also specify the features that will be used for learning. The type of the selected features influences the way of the learned mapper and the discretization over the system.

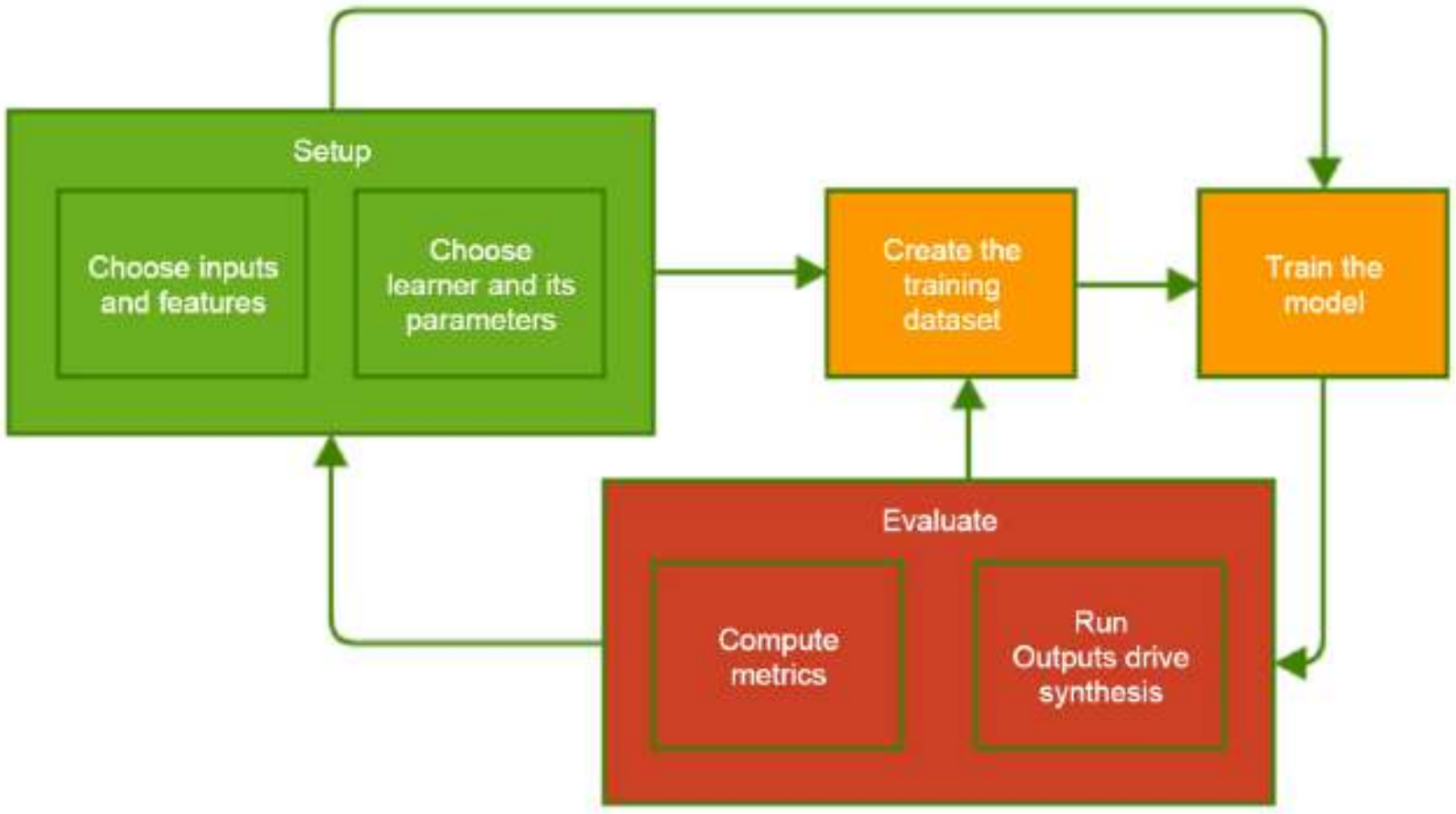

**Fig. 9.** The interactive algorithm of the Wekinator

The training step provides the possibility of choosing the class labels or the regression function. In the learning stage, the Wekinator trains the model with the specific features, extracted in real-time. For example, one can set "1" as the desired output value, while someone raise up several times the left hand, tracked by Kinect. The training process is taking several seconds, and if it is not happening this way, the user can halt the process and tune the model to be faster, by readapting the parameters.

The testing phase consists in a real-time feature extraction, to be fed to the trained model in order to return the specific output values. For example, if one raises up the left hand, the output of the system will be "1" and this value may be used as a shutter for a thirdy part system.

In the evaluation step, if the results are not satisfying, the user is able to increase the size of the training dataset in order to reinforce positive behaviors. Then a new retraining process is needed, to remodel the system.

### 5.4 Application with Wekinator

This system is suitable to our case because it is user orientated, therefore the user can train the model in real time with the specific features, and also because it has a very fast pace of learning and retuning on the order of seconds. Since Robigame is a video game played by patients in real-time, one can very fastly tune the algorithm and start the training for the patient, setting two types of short games, one very easy, which should lead to the lacking of the patient interest and one very hard, which should lead to the frustration of the patient.

The learning is done in a subjective context, since each patient has different ways and different levels of intensity to express their frustration or their motivation. The explicit goal is to create a controllable and suitable map for each user.

The best way to describe the task is by using a regression model, which maps values between 0 and 1. In the training phase, one should start with a value of 0, corresponding to the beginning of the training game and ending with a value of 1, corresponding to the last part of the training game and fitting the apex of the expression.

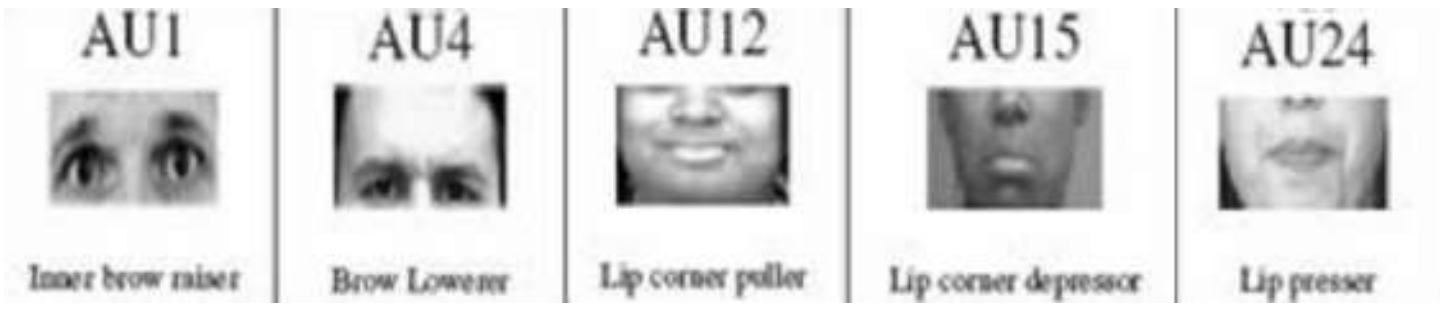

**Fig. 10.** Examples of the required action units extracted from Cohn-Kanade+ database [31]

In order to train the model, the following feature descriptors are used: AU1, AU4, AU20, AU15, AU12. Each of them are described below and can be seen in Figure 10. These algorithm try to track a combination of different emotions as confusion, frustration, anger, agitation, insecurity and sadness, which sum up a negative level in the patient and trigger the decreasing of the game level ([37] and [38]):

– confusion and frustration are described by a high value of AU1 (inner brow raiser)

– anger, agitation and insecurity are described by a high value of AU4 (brow lowerer)

– sadness is described by the following combination
    • low value of AU20 (lip stretcher), high value of AU15 (lip corner depressor) and low value of AU12 (lip corner puller)
    • low value of AU20 (lip stretcher), high value of AU15 (lip corner depressor) and normal value of AU12 (lip corner puller)

## 6 Setup ans Results

In this section we will describe the final setups used in this project and we will give some qualitative results. For lack of time during the project, the previous part about wekinator has not been integrated in the final process but it is interesting to analyze the results without this deeper face analyze. When the patient comes into the field of view of the KinectV2, seated in front of the reaplan, his skeleton is tracked, the head orientation is estimated and the face is analyzed for smiling detection. The Kinect is placed above the screen according result obtained after comparison of the two kinect position for the kinect (figure 11).

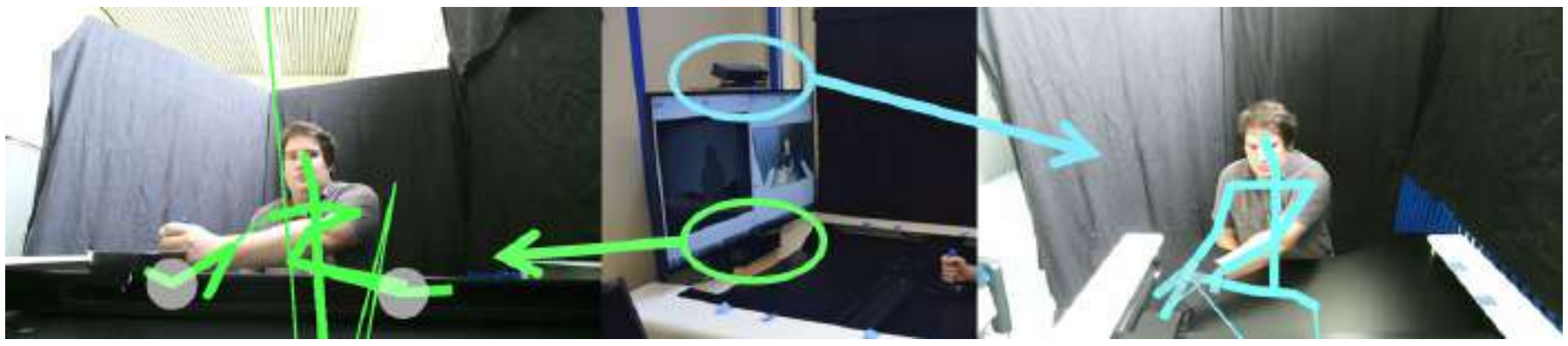

**Fig. 11.** Comparison of the view from the kinect below (green) and above the screen (light blue).

above the screen is the only position to track face correctly and also to track upper limbs. This position is also the one who gives quicker body detection.

When the user body is tracked, a computer is used to perform the kinect process and determines what the user is watching with an accuracy of a few centimeters: Main screen, effector distal plane or elsewhere. These information are completed by the smiling detection and are sent to the game (Figures 7 ). The screen displays the game and the user moves the distal effector to control the game. The difficulty of the game is also displayed on the screen and it changes with the user happiness. The final setup is given on Figure 12.

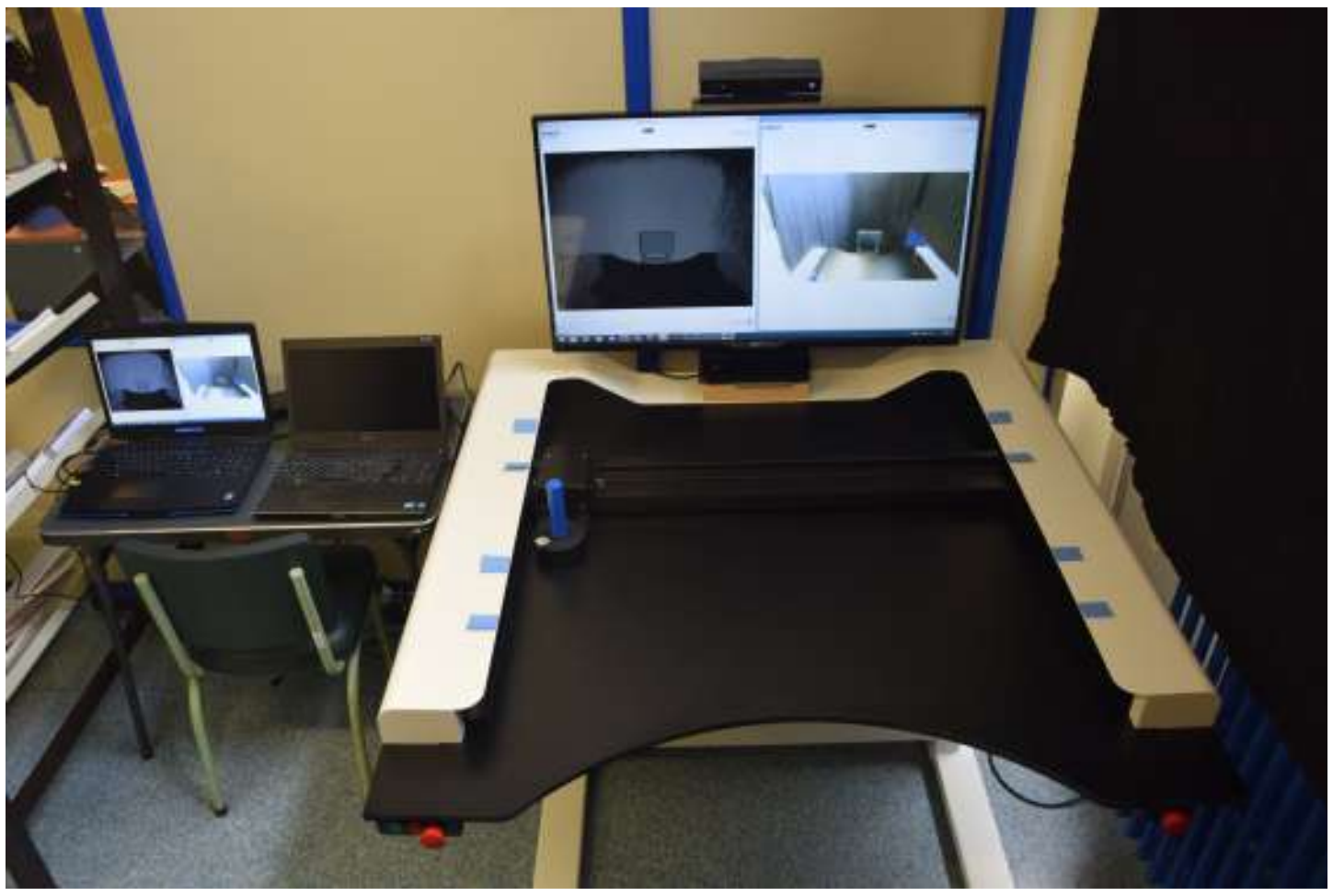

**Fig. 12.** Final setup with computers and REAPlan system .

The results obtained with healthy users are fully satisfactory. Users enjoyed the different games and have not got the feeling the use medical equipment. In addition, users say that the expressive facial analysis creates a new dimension in video games by changing the difficulty of the game. This first version of the robigame setup is functional in real time and is ready to be tested with patients in a therapeutic setting.

## 7 Conclusion and future work

The objective of the main project is to develop an intelligent serious game for rehabilitation of the upper limbs for stroke patients using an interactive rehabilitation robot which screenplay adapts to the patient's functional abilities, and which assistance evolves according to patients motivational, motor and cognitive performances. ROBiGAME will be developed on the REAPlan robot providing a distal effector which can mobilize the patient upper limbs in a horizontal plane but could also be transferred to other robots or even simpler rehabilitation setups. The most important benefit of this work is the integration between Kinect for Windows v2 and REAPlan robot. The application extracts facial features of the user from the data provided by the Kinect. Two games are developed for this purpose and motivation data, computed using the kinect, are used to increase or decrease game difficulty in real-time. In perspective of this work, we need to link face analysis with wekinator to main system. Then we must do quantitative analyses of arm movements. In addition, the analysis of movements must be validated with patients in a therapeutic setting for monitoring patients with real motor difficulties. Finally, a third game is in development and it would be interesting to finish it for testing.

## Acknowledgment

This work is supported by the ROBiGAME Project funded by the Walloon Region from Belgium through the WBHealth Program. This work is also partially funded by the Sectoral Operational Programme Human Resources Development 2007-2013 of the Ministry of European Funds through the Financial Agreement POSDRU/159/1.5/S/132395.

The authors would like to thank our partners of the Robigame project, especially Martin Vanderwegen and Adrien Denis for the handling of the Reaplan.

The authors would like to thank the local organizers of eNTEFACE'15 for their kindness, their local assistance regarding logistics, their nice schedule of social events.The authors would also thanks to Radhwan and Ambroise for their sympathy and for sharing good and bad moments with us during the workshop.

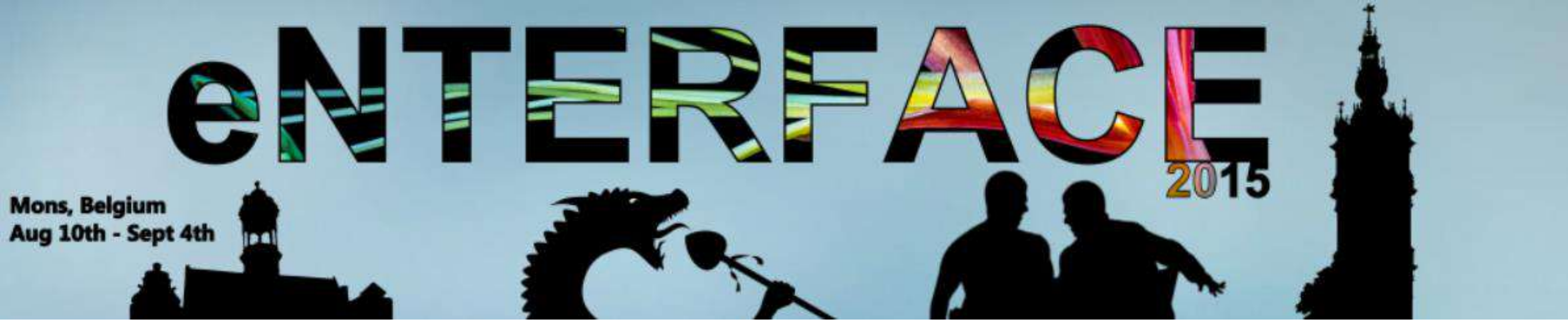

## Project #7:

# BigDatArt: Browsing and using big data in a creative way

*Fabien Grisard, Axel Jean-Caurant, Vincent Courboulay, and Matei Mancas*

numediart
INSTITUT DE RECHERCHE POUR LES TECHNOLOGIES CREATIVES DE L'UMONS

UMONS
Université de Mons

fnrs
FREEDOM TO RESEARCH

MONS 2015

# BigDatArt: Browsing and using big data in a creative way

Fabien Grisard[1], Axel Jean-Caurant[2], Vincent Courboulay[2], and Matei Mancas[1]

[1] Numediart Institute, University of Mons, Belgium
surname.name@umons.ac.be
[2] Université de La Rochelle, France
surname.name@univ-lr.fr

**Abstract.** This paper present a framework to study and artistically illustrate the relationship between multimedia supports. By presenting a global architecture made of several software sub-modules, we introduce new kind of interface to illustrate the concept and complexity relating to big data. The purpose is to provide a playful user interface based on body tracking to let large public browse multimedia content from the web and to create personalized mashups.
A demo video can be downloaded at: http://www.enterface.net/enterface15/wp-content/uploads/2015/09/project9_demo.mp4

## 1 Introduction

The project *BigDatArt* aims in using big data in a creative way so that large public can better understand the concept and complexity of such data. For that aim we will use gaming interactions based on Kinect sensors to enhance collaborative co-creation of mashups of images, videos and text data extracted in real time from the web. The goal is to provide a playful interaction letting large public browse the big data and create mashups from this data in a collaborative way in different cities in Europe. We would like to integrate the project result in different cultural places such as museum or digital spaces.

## 2 System overview

The *BigDatArt* application is composed of three main modules. First, the visualization module is in charge of the display. It is itself decomposed in several sub-modules that will be described in detail in Section 3. Then comes the interaction module. Thanks to the Kinect camera, it is able to recognize several user's gestures that will trigger different responses from the application. Finally, the data module is responsible for gathering data from various sources like Google, Tweeter or FreeSound.

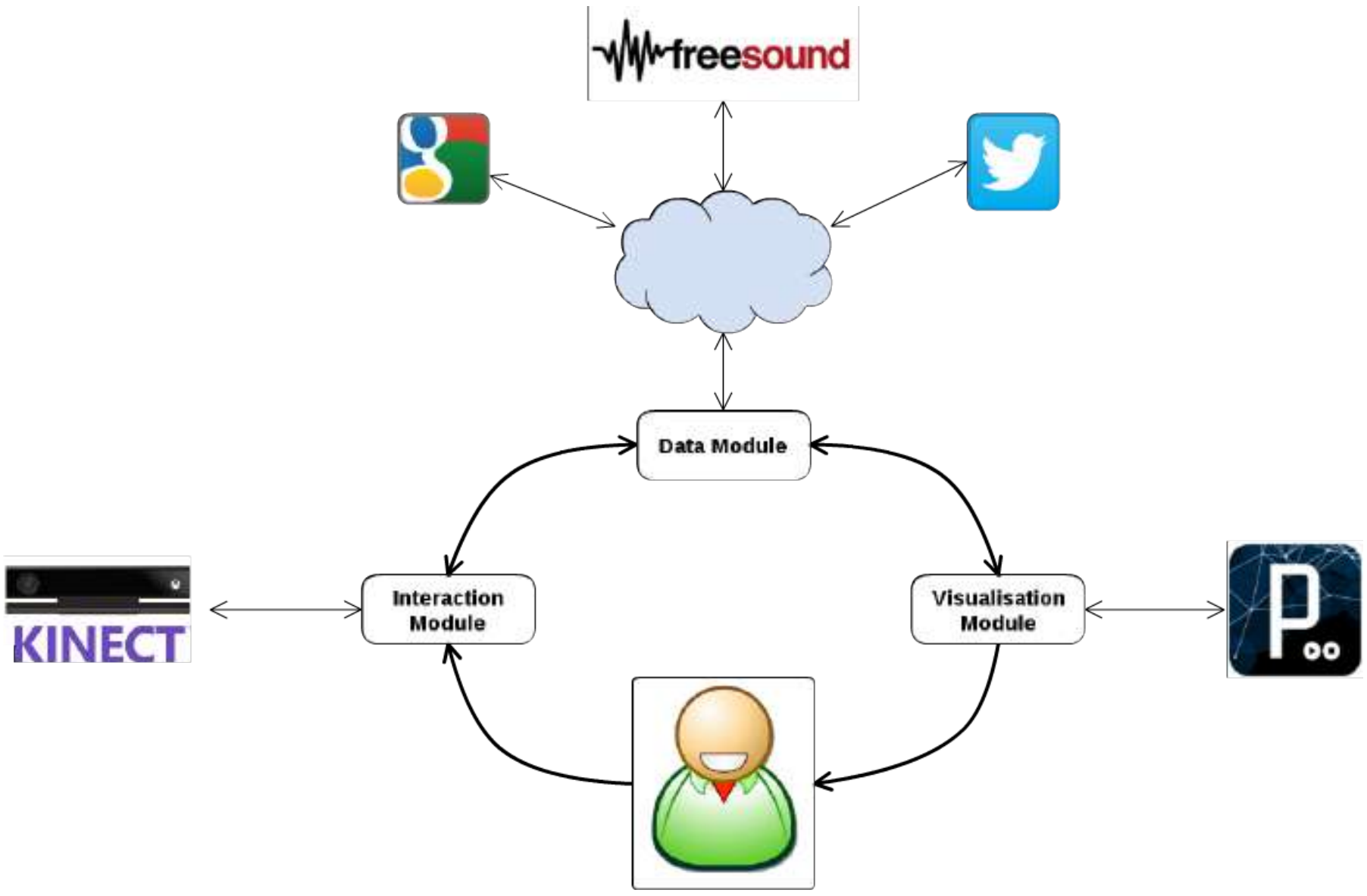

**Fig. 1.** Interaction between modules of the application.

## 3 Software modules

The main program relies on several software components developed in Java, using the Processing framework[5] (v2.2.1).

### 3.1 Main interface

The main interface is composed of four distinct areas. Each of them is autonomous and runs in a separated thread. The main window is in charge of communication between the different modules. Zone **a** is in charge of displaying images and zooming. The word cloud (zone **b**) is created using the WordCram library[9]. To play sounds found on the web (zone **c**), we use the Minim library[3]. Finally we also display tweets in the lower-right area (zone **d**). The display areas are shown on the Figure 2:

**a.** Image area
**b.** Word cloud area
**c.** Sounds waveforms area
**d.** Tweets area

### 3.2 Images selection

Images presented in the application come from various web sites. At first, an image is presented with associated keywords, sounds and tweets. To switch to the next image and keep the application running, it is necessary to select a part of the image, either by letting the user do so or by selecting it automatically. When this is done, we need to query the web for this portion of image. This is done using a java API we developed, responsible of querying Google Search Image. To use this API, the image part first needs to be uploaded in order to be accessible from the web. When this is done, we can feed its URL to the API, allowing it to query Google Image. The results are of two kinds :

– **List of web sites**: when we request Google with an image, several results are available. First, if the image was explicitly recognized and associated with a name, we can recover web sites responding to a text query initiated with the name of the image. Google also returns a list of sites containing the same image. Finally, we can harvest many more sites containing similar images.
– **Similar image**: after the request, we are presented with a list of similar images. We don't have details about the algorithm, but the images tend to be similar in shape, textures and colors.

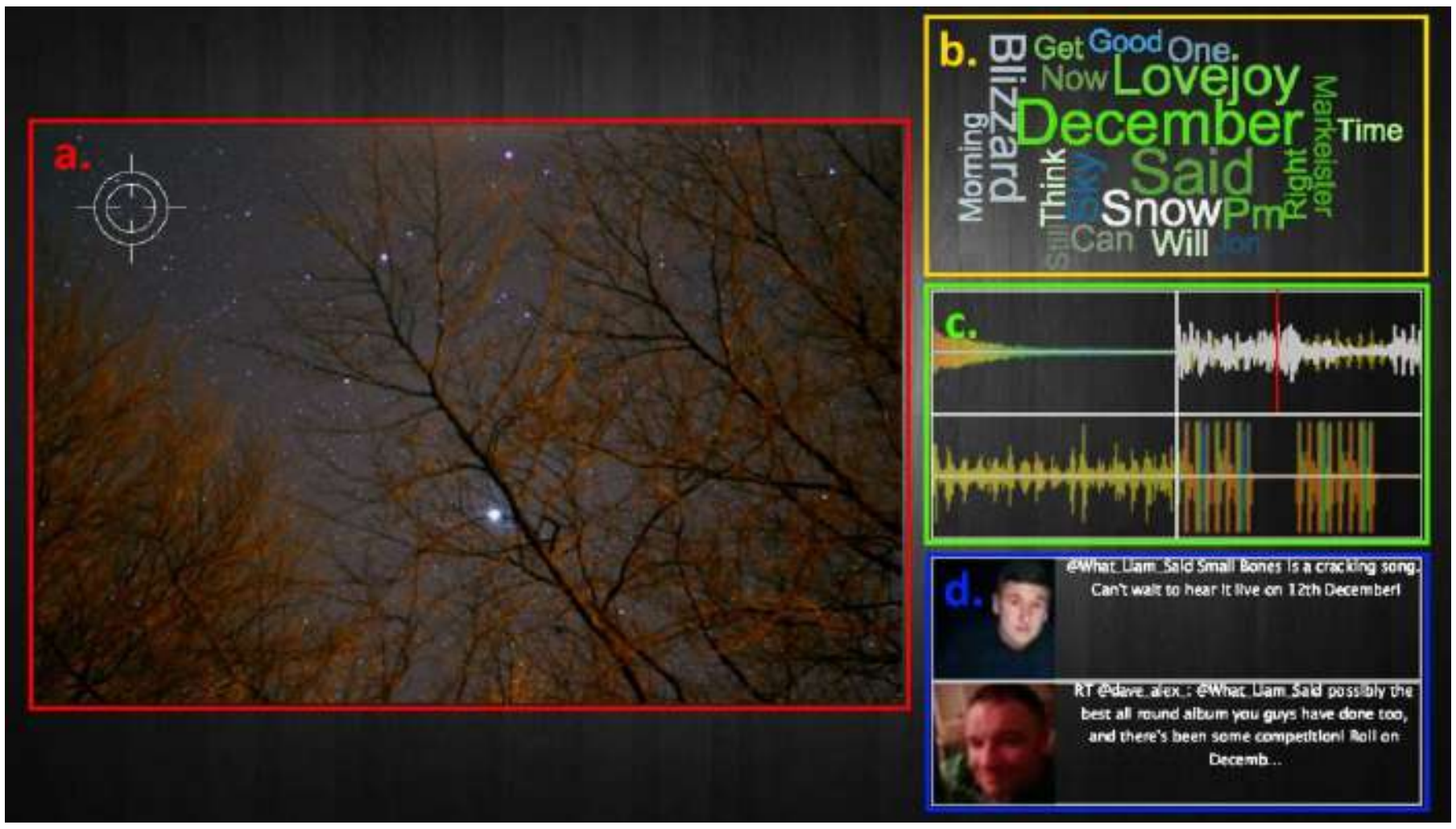

**Fig. 2.** Screenshot of the main interface

We randomly select a similar image to be displayed inside our application. Associated keywords are then extracted from various associated web sites.

### 3.3 Keywords extraction and fusion

Once the next image is selected, we create a list of keywords that describe best the content and the context of this image. This list is needed by the next components to request sounds and text content from social networks. To achieve a correct keywords extraction, our software uses two modules. The first one parses HTML code from the page on which the image has been found and the second one uses a concept recognition API.

**Keywords extraction from HTML code**

When an image is selected, we download it from a website. The source code of the page is retrieved and we then process it with the JSoup Library[12]. The problem with code from web pages is that it contains a lot text in which we do not expect to find some keywords (css, javascript, *etc.*). To cope with this issue, we extract text only from headers and paragraphs (respectively <h> and <p> tags). This allows us to remove some parts of the page like menus or footer. To keep only meaningful words, we use LDA algorithm as proposed by Blei et al.[10]. Then we remove words like "Facebook", "Twitter", "YouTube". They are present on almost every pages and are completely irrelevant to describe the image. Those words are kept on a blacklist we updated regularly during the project develpment. Finally, we end up with a list of keywords and their associated frequencies.

**Keywords extraction from concept recognition**

The Rekognition API[6] is a commercial API to perform concept recognition on natural images. It takes an image URL as an input and returns a JSON file containing the concepts identified, associated with a confidence score between 0 and 1. We create a Java class containing API identifiers and methods to request the API and parse the JSON result. After this step, a new list of keywords is available to describe the image.

**Keywords fusion**

We now have two sources of keywords to describe the image. They are both ordered by relevance (words from the web page are associated with their frequency and concepts with a confidence score). We make the empirically verified assumption that concepts are much more relevant than keywords. To take this into account when merging the two list, the final relevance of concepts is computed by multiplying their confidence score with the best keyword's score.

When the results from both modules are merged into a single list, this list is displayed as a word cloud in the upper right corner of the main interface (Figure 2 b.). It gives the visitor a feedback about the ability of our system to identify the images context and explains the relevance of sounds and text found by the following modules.

### 3.4 Sounds finder

Sounds finder module allows us to perform sound file search on the Freesound website. "Freesound is a collaborative database of Creative Commons Licensed sounds."[1]. We designed a custom API based on URL format request which take the first N words from the keywords list as an input. The API then sends several requests to freesound.org. Each word from the list is translated to a different request, the last one is a AND combination of all the words. For example, with [summer, beach, ball], the API will search for sounds related to summer, beach, ball, summer AND beach AND ball. The results are then sorted by relevance (correspondence between the words from the lists and the words from the sound description) and the duplicates are removed.

### 3.5 Data from social networks

We limited the sources of content to Twitter. An easy-to-use developer API[7] to search for tweets from the last days is freely available and just need the creation of a user account[3]. We used a Java library, called twitter4j[8] and adapted the code from the Bonnamy's tutorial[11] to fit our needs and search for tweets with the keywords list. However, the results were quite deceiving and did not bring any additional value for the visitor. We decided to remove this module from the later versions of BigDatArt.

### 3.6 Mashup Display

Several events can lead the system to switch to mashup display. They are detailed in Section 4. On this display, the system does not search for images any more but presents the ones from the last instance of browsing (Figure 3). As detailled in Section 4.2, the user can move the cursor over the images with the hand. Then, the words associated to the selected image are displayed in a spiral (most relevant words are bigger and on the outer side of the spiral) and the most relevant sound is played.
To quit the mashup display, the user has to move away from the installation and leave it return to autonomous mode.

## 4 System usage modes

BigDatArt is able to run autonomously when no one is present. Otherwise, a user can take control over the system and influence the choice of the content.

### 4.1 Autonomous mode

When no one is in front of the set up, the application runs autonomously. This mode is detailled in the following sections.

[3] https://twitter.com/BigDatArt

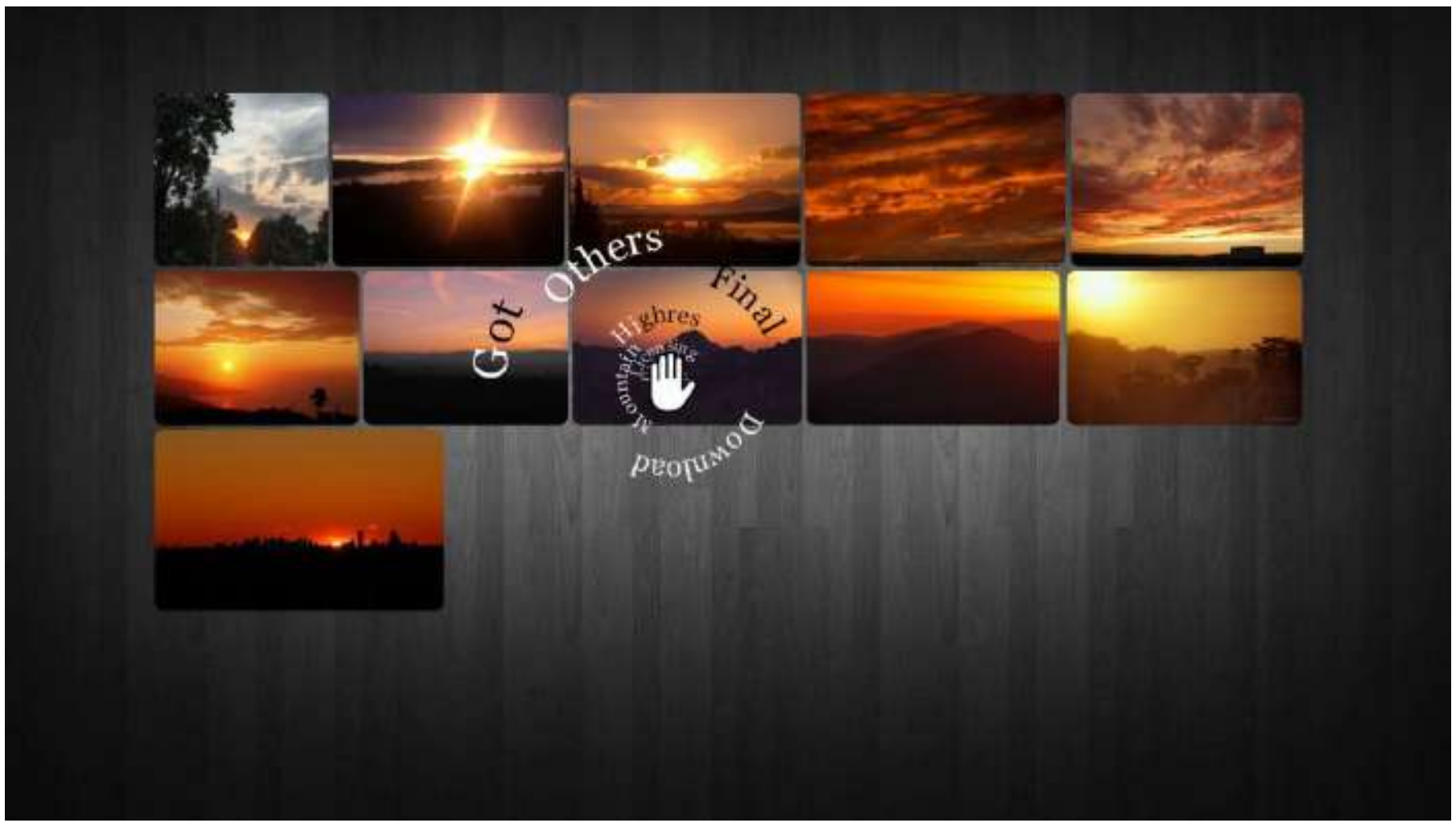

**Fig. 3.** Screen shot of the "mashup" interface

**Saliency-based zooming region selection**

To allow the application to run autonomously when an image is presented, we need to automatically select a part of it in order to continue the execution. We could have make this choice randomly, but the results would not have been very interesting. Instead, we chose to select the most "interesting" part of the image. Because this is highly subjective, we define this as the most salient zone. More explicitely, this is the zone of the image which attracts the most people's attention (*e.g.* a small contrasted object). To find out which zone to select, we use a visual attention model by Perreira Da Silva et al.[14]. It is based on a prey/predator interaction. Different preys focus on different features inside the image (color, orientation, *etc*), and predators focus on preys. The region containing the most predators is considered as the most salient.

Once an image is presented in the application, we start the process of finding the most salient zone, using the model described previously. When the most salient zone is found out, we can start other processes like querying Google and zoom towards this zone.

**Stop condition** The stop condition is fulfilled after 18 iterations or when a user enters the interaction area and takes control over the system as explained in Section 4.2. When a stop condition is reached, the system switches to mashup display (see Section 3.6).

### 4.2 User controlled mode

One of the main objectives of the project was to let the public browse the big data and create interactive mashups playfully. In other words, to propose an accessible user interface. We decided to use a MS Kinect One sensor and the Kinect SDK[2] to track users' skeleton (see Figure 4, b.). We adapted a software written in C++ initially designed to track users' face and identify the direction they are looking at[15] and added our own features. The two parts communicate over OSC, a UDP-based protocol[4]. The Kinect software can be considered as a remote control for the main interface.

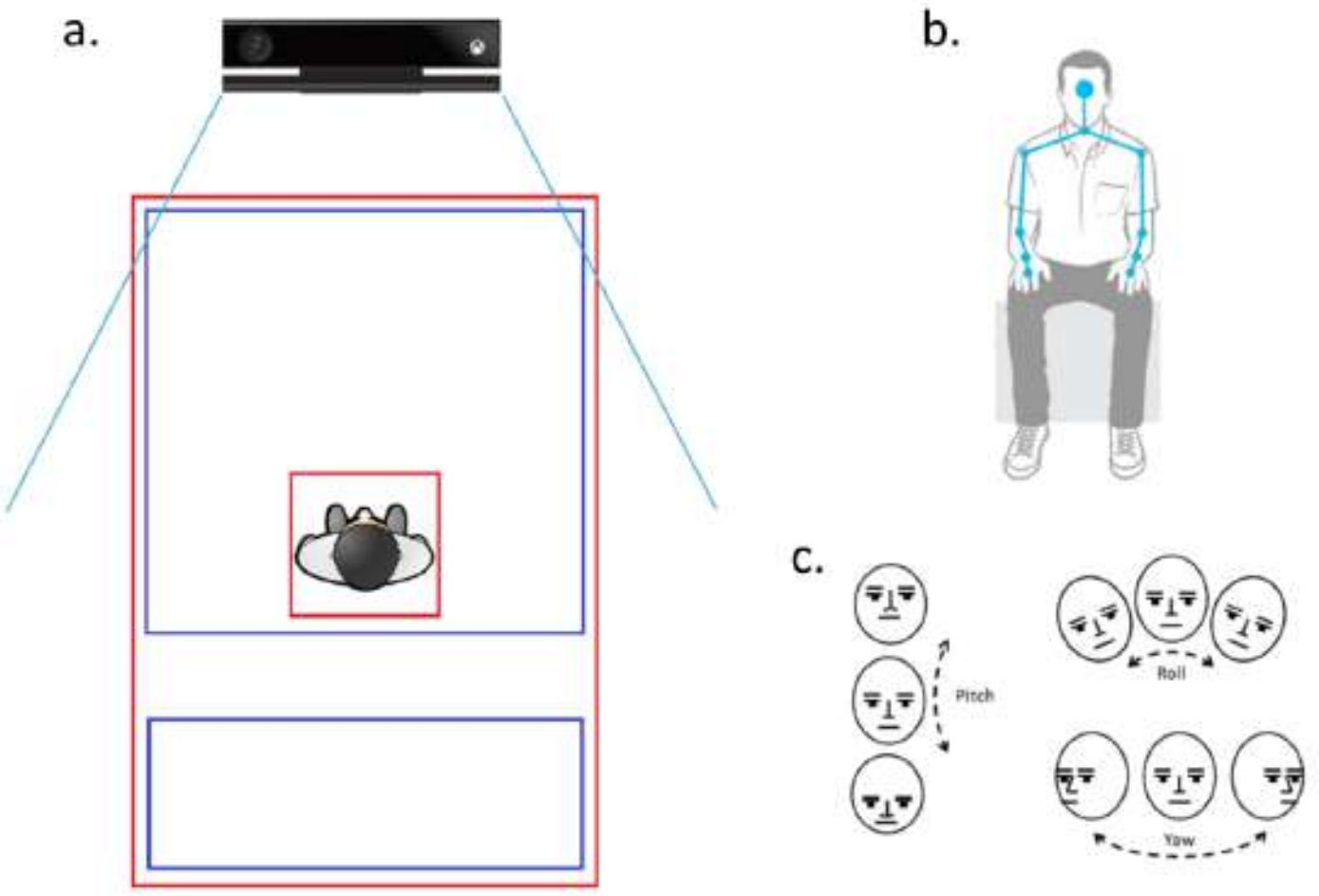

**Fig. 4.** a. Kinect interaction areas, b. Tracked skeleton, c. Euler angles for head tracking

**Interaction areas**

As our system can be used by only one person at a time, we created virtual areas on the ground, as presented on Figure 4. To take control, the visitor has to enter in the small red square. He/she will keep hold on the system as long as he/she stays in the big red rectangle.

We noticed that some images (like logos or plain background) tend to create a loop. The system seems stuck on a subject and only propose images of the same type. To avoid this, we created the two blue areas. When a user has the control, he can move from the biggest zone to the other behind him. This transition will make the system start over with a random image without deleting the queue of content collected until this point.

**Gaze estimation**

As explained in the introduction of this section, we re-used a software designed to identify which part of the screen the visitor is looking at. The method is based on the estimation of his head pose. Accordingly to Murphy-Chutorian[13],

"[...] *Head pose estimation is intrinsically linked with visual gaze estimation* [...]. *By itself, head pose provides a coarse indication of the gaze that can be estimated in situations when the eyes of a person are not visible* [...]." The configuration of our installation (the screen width is close to the visitor-screen distance) increases this effect. The MS Kinect SDK face tracking functions provide Euler angles (Figure 4, c.) and the position of the head in the sensor referential. Knowing the position and size of the screen, we can infer screen coordinates of the user's gaze.
Those coordinates are sent to the main application which displays a target on the interface (Figure 5 a., also visible in the upper left corner of the Figure 2, a.). A 2 seconds countdown fills progressively the target if it is not moving (Figure 5 a. up). When the countdown is finished, the target is locked (Figure 5 a. below), the system selects the area of the image where the target is, zooms on it and searches for a new image similar to the zoomed zone.

**Cursor control**

On a user interface, a cursor is a common and friendly way to select items. We gave the visitors the ability to move a hand shaped cursor on the main interface by moving their hand in front of them. The position of the cursor is calculated relatively to the distance between the hand and the shoulder on the same arm on X and Y axis. The Kinect SDK allows us to know the state of the hand (open or closed, Figure 5, b.). If the cursor is on a clickable item like a sound sample, the user can close the hand and the sound is played.

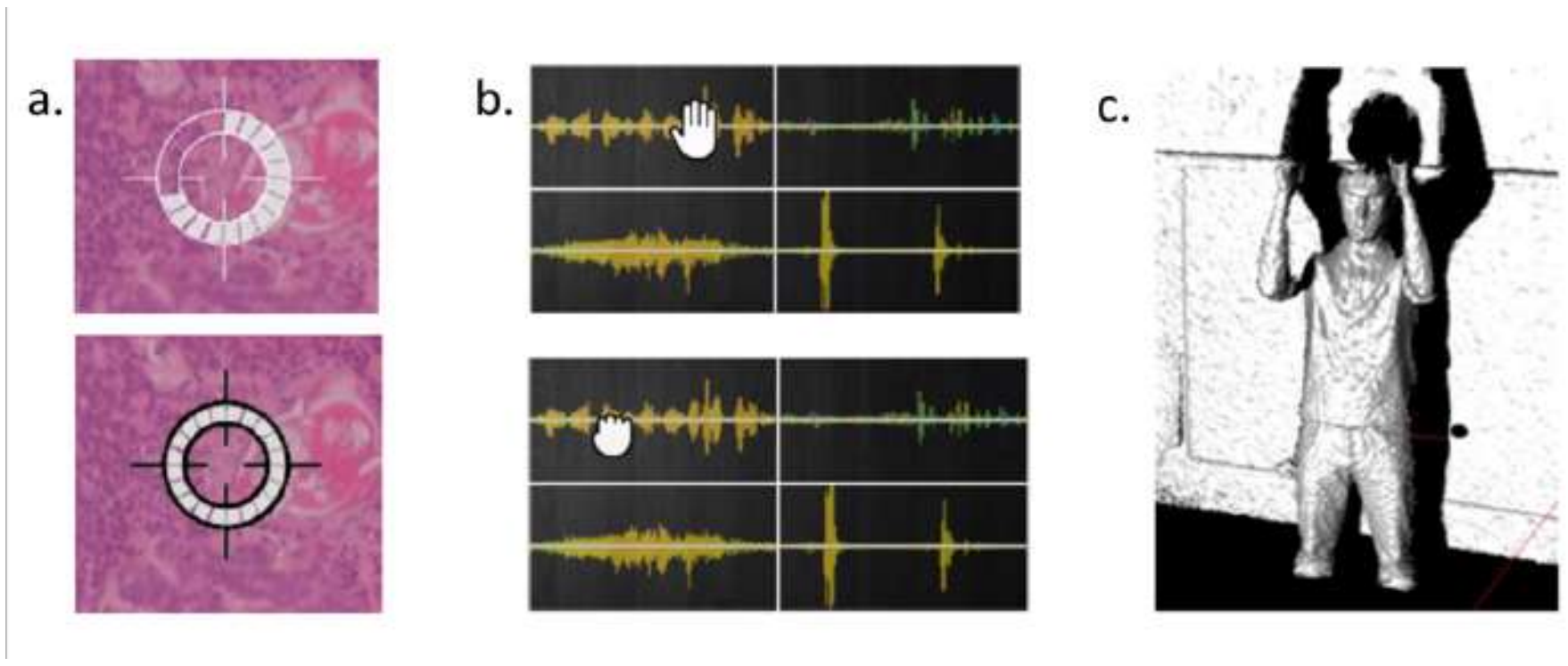

**Fig. 5.** a. Eye target, b. Hand shaped cursor open and closed, c. "Mashup" posture

**Posture recognition**

In user controlled mode, the mashup starts after a fix number of iterations, or the user can launch it by placing his arms parallel in front of him (Figure 5, c.). The posture recognition uses a descriptive method. A vector contains all the descriptors needed to identify this particular posture. For example, the following descriptor checks that the hands and elbows are respectively at the same height with a certain tolerance :

```
abs(rightHand.Y - leftHand.Y) < tolerance &&
   abs(rightElbow.Y - leftElbow.Y) < tolerance
```

Other descriptors check that the arms are not crossed and vertical, and that the hands are higher than the shoulders. If all the descriptors are `true`, the posture is recognized.

## 5 Conclusion and future work

This project proposes new kind of artistic expressions and interrogate us on the information hidden in multimedia sources. It could be improved in several ways. We envisage to modify the saliency algorithm in order to introduce some hazard. Actually we have noticed that the algorithm sometimes loops on a restricted set of images. Yet, the most important evolution concerns the cleaning of the HTML text and the proper extraction of relevant keywords. An algorithm that tackles the problem of salient text could be an efficient solution. At last, we are interested in developping a solution to let people play together through network.

## Acknowledgment

This work is supported by Mons2015, European capital of culture. The authors would like to thank the local organizers of eNTEFACE'15 for their kindness, their local assistance regarding logistics and their nice schedule of social events.

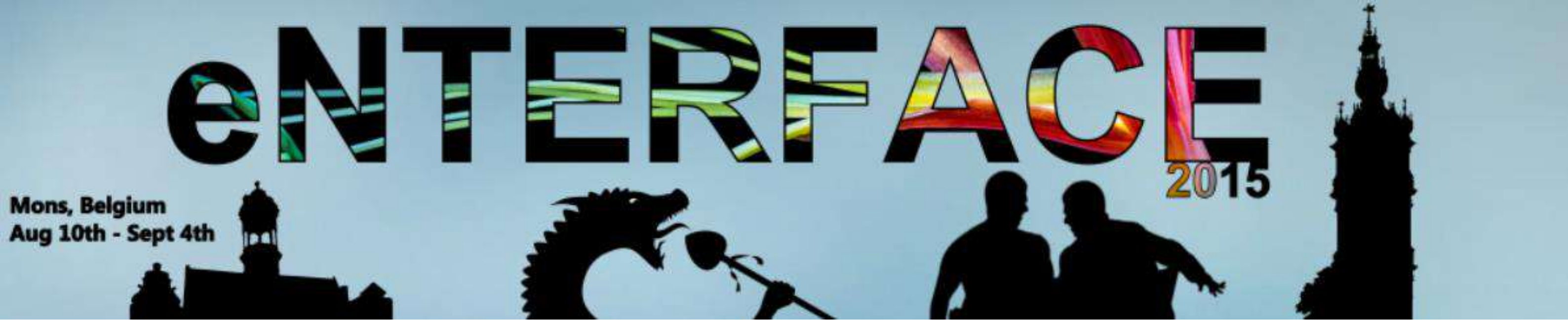

**Project #8:**

## Automatic detection of planar surfaces from uv maps

*Radhwan Ben Madhkour, Ambroise Moreau*

# Automatic detection of planar surfaces from uv maps

Radhwan Ben Madhkour[1], Ambroise Moreau[2]

TCTS Lab, Université de Mons, Belgium
Radhwan.BenMadhkour@umons.ac.be, Ambroise.Moreau@umons.ac.be,
WWW home page: http://tcts.fpms.ac.be/

**Abstract.** In this project, we propose a new method to detect planar surfaces in a scene with a projector-camera pair. First, a structured light method is used to get correspondences between the pixels in the two devices spaces. Then, these correspondences, stored in *uv* maps (i.e 2D-matrices) are analysed to retrieve the different planar surfaces. We based our work on the fact that coordinates of corresponding pixels follow a linear equation when the scanned surface is a plane. Only simple image processing algorithms like Sobel filtering and thresholding are used. The method has been tested on different setups made of four planar surfaces. Our tool could be used to ease the calibration process of a projection-mapping system.

## 1 Introduction

In recent years, projection mapping has become one of the most popular technology for sound and light shows. All around the world, video projectors are used to animate all kind of objects of any size to tell stories and amaze the audience. These light shows usually require an offline calibration step that can be cumbersome to compute the image deformation needed to map the content to the surface. Our goal is to ease the calibration process for certain configurations by automatically detecting planar surfaces in the scene.

Our algorithm relies on a camera-projector system and the correspondences between pixel coordinates in the two devices. In this work, correspondences are obtained through phase shifting, but other structured light methods could be used. Only simple image processing algorithms are applied to the data.

This report is organized as follows. Section 2 briefly presents the structured light method we used. Section 3 describes our solution to detect planar surfaces in the correspondences map, also called *uv* map, along with the processing pipeline. Section 4 shows some results and finally, section 5 concludes on our work.

## 2 Structured light

Structured light refers to active 3D reconstruction methods that rely on the use of a camera-projector pair. Patterns with special properties are projected on the scene by the projector and the camera records the deformations induced by the illuminated objects. The patterns help finding corresponding pixels in the two devices. These correspondences are then used in a triangulation algorithm to build a 3D model.

Structured light methods are classified in two categories depending on the number of patterns used to find correspondences. Time-multiplexing methods project several patterns on the scene to retrieve the *uv* map while spatial or one shot structured light requires only one. For an extensive review of structured light, we refer to the work of Salvi et al [1].

To build the *uv* map, we used the well-known phase shifting algorithm implemented in the CodedLight library [2]. Phase shifting usually projects sinusoidal patterns shifted in one direction and whose phase depends on the pixel position in the projector, as shown in equation 1 [3].

$$I_n^p(x,y) = A\left(x,y\right) + B\left(x,y\right)\ \ cos\left[\phi\left(x,y\right) + \frac{2\pi n}{N}\right] \tag{1}$$

$A(x,y)$ is the background intensity, $B(x,y)$ is the amplitude modulation and $N$ is the number of patterns. Images captured by the camera can be denoted as:

$$I_n^c(x,y) = A\left(x,y\right) + B\left(x,y\right)\ \ cos\left[\phi^c\left(x,y\right)\right] \tag{2}$$

These images are then used to compute the phase value of each pixel:

$$\phi^c(x,y) = atan\left(\frac{\sum_{n=0}^{N} sin\left(\frac{2\pi n}{N}\right) I_n^c\left(x,y\right)}{\sum_{n=0}^{N} cos\left(\frac{2\pi n}{N}\right) I_n^c\left(x,y\right)}\right) \tag{3}$$

The phase given by equation 3 is wrapped, taking value in the $[-\pi,\pi]$ interval. It needs to be unwrapped to get a unique phase value for each pixel so that correspondences between the two devices can be found. Unwrapping algorithms used in the Coded Light library are described in [4].

## 3 Detection of planar surfaces

*uv* maps are 2D matrices whose elements are the coordinates of the corresponding pixel in the other device. One practical way of presenting a *uv* map is by the mean of an RGB image in which the red component is used to store the x-coordinate while the green component stores the y-coordinate. An *uv* map is shown in figure 1a.

Our work is based upon the following observation: when the objects illuminated by the patterns are made of planar surfaces, corresponding pixels coordinates follow a picewise-linear equation. Therefore, it is possible to extract planar surfaces from a *uv* map by observing its derivative. As an example, figure 1b shows the x-coordinates evolution along the horizontal line marked in figure 1a.

The processing pipeline applied to the *uv* map can be divided in three parts:

- the **planar surfaces detection** whose goal is to find planar areas in the scene without telling them apart;
- the **planar surfaces separation** using the first derivative;
- and the **warping computation** to find the transforms that need to be applied to map an image to the planar surfaces.

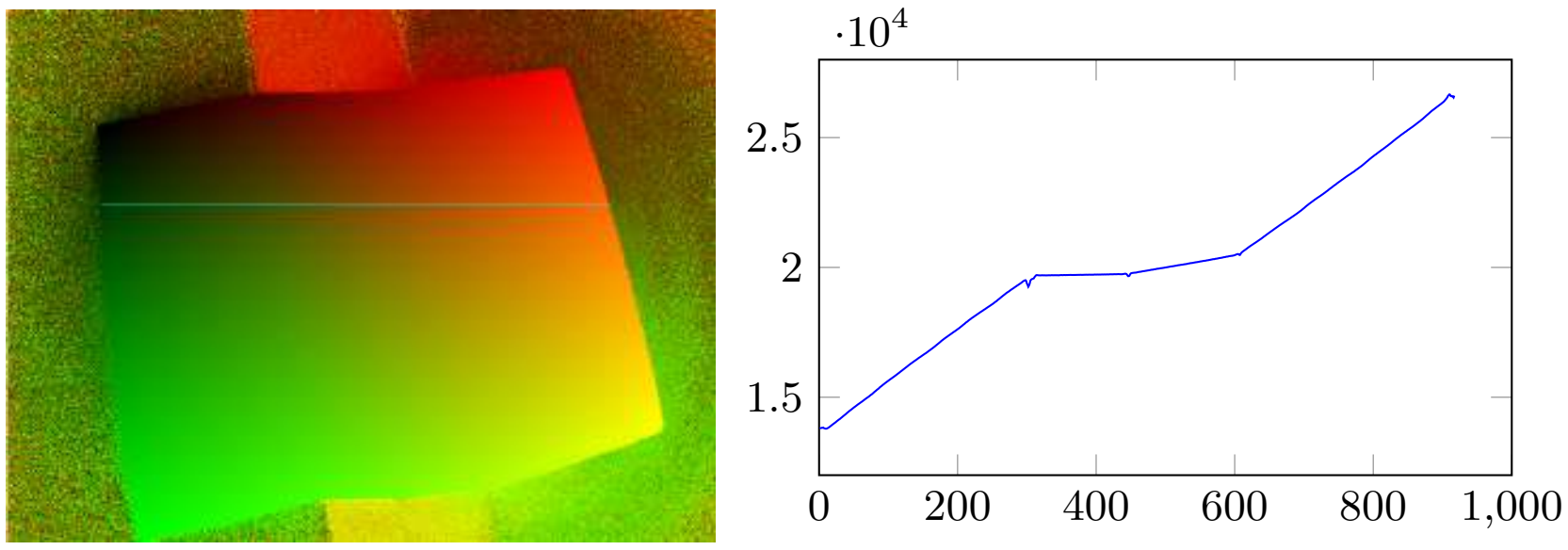

(a) *uv* map (b) Piecewise linear evolution of $x$

Fig. 1: 1a shows an example of *uv* map and 1b shows the evolution of the x-coordinate along the marked line

### 3.1 Planar surfaces detection

Since the coordinates evolution is linear on planar surfaces, the first derivative is constant and the second derivative is equal to zero. Therefore, to find all the planar surfaces in the scene, we need to look for the zones in which the second derivative is null.

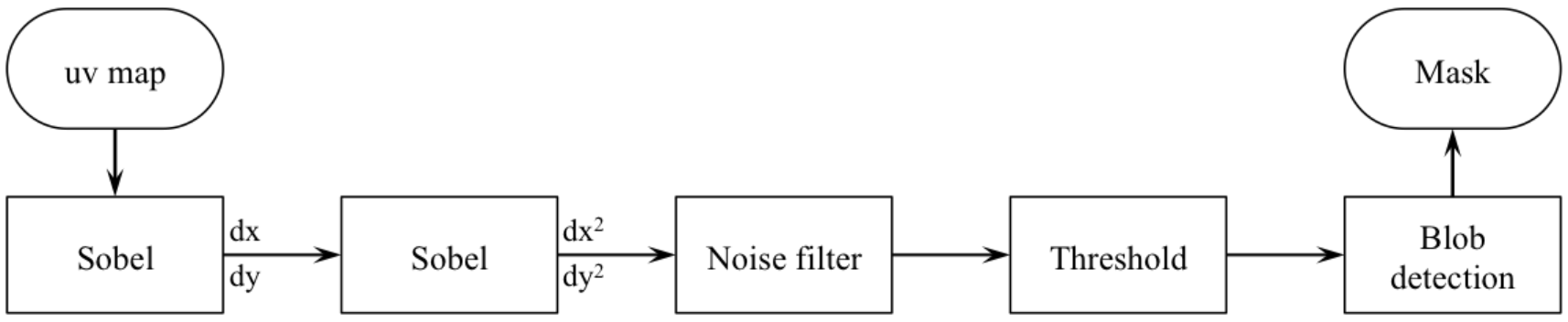

Fig. 2: Detection of planar surfaces with the second derivative

As shown in figure 2, this is done by thresholding the *uv* map second derivatives (along the vertical and horizontal direction) and applying a blob detection algorithm. The output is a binary mask showing all the detected planar surfaces.

### 3.2 Planar surfaces separation

To tell apart the different planar surfaces, we use the first derivatives in both directions (vertical and horizontal) along with the mask obtained in section 3.1. The process is presented in figure 3.

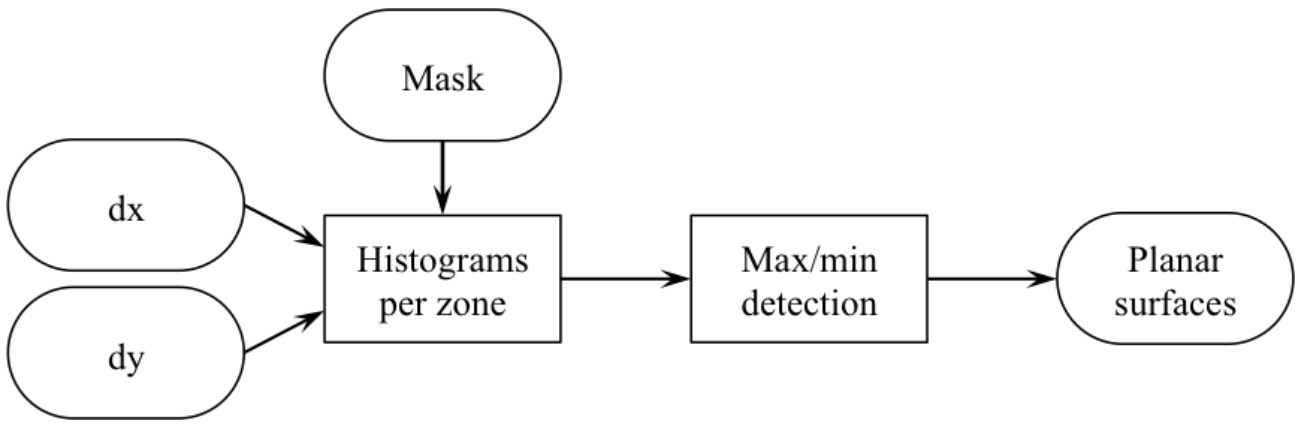

Fig. 3: Separation of planar surfaces

Two histograms are computed for each zone in the mask and these histograms maxima and minima are used to separate the planar surfaces. A spatial approach is used to find extrema. Values under a certain threshold are ignored to avoid false minima.

### 3.3 Warping computation

The goal here is to find the transform that links the projector to a planar surface. For the sake of simplicity, we consider a scene made of a unique plane. When only planes are involved, the transform is a homography. Three different ones can be found:

- $H_{pw}$, the homography linking the projector and the plane in the real world;
- $H_{cw}$, the homography linking the camera and the plane in the real world;
- $H_{cp}$, the homography linking the two devices.

$H_{pw}$ is the one we need to perform the image warping. As shown in figure 4, it is found thanks to $H_{cp}$ and the planar surface corners detected in section 3.2.

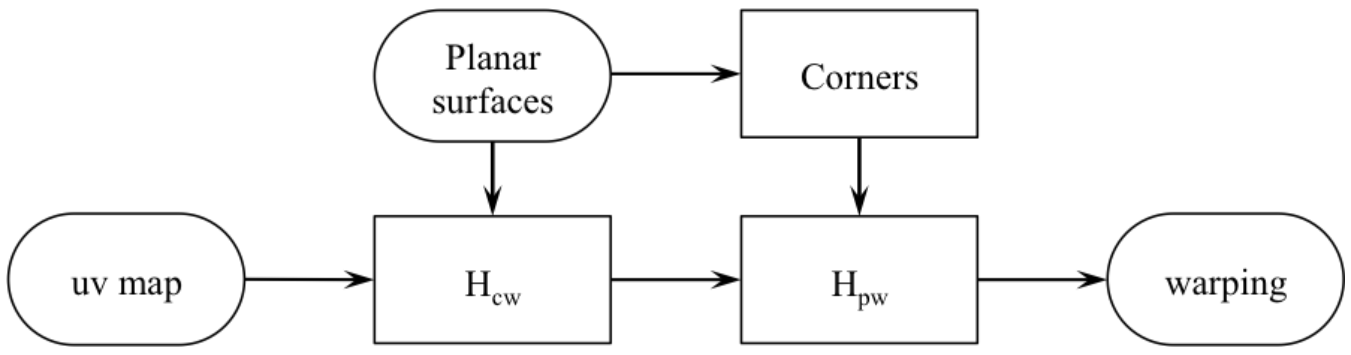

Fig. 4: Separation of planar surfaces

## 4 Results

To assess our algorithm performances, we tested it on different scences made of four wood panels. We used an off-the-shelf DLP projector with a resolution of 800x600 and a Prosilica camera with a resolution of 1280x960. The structured light scan is done with the Coded Light library while the *uv* map analysis is performed in OpenCV [5].

### 4.1 Planar surfaces detection

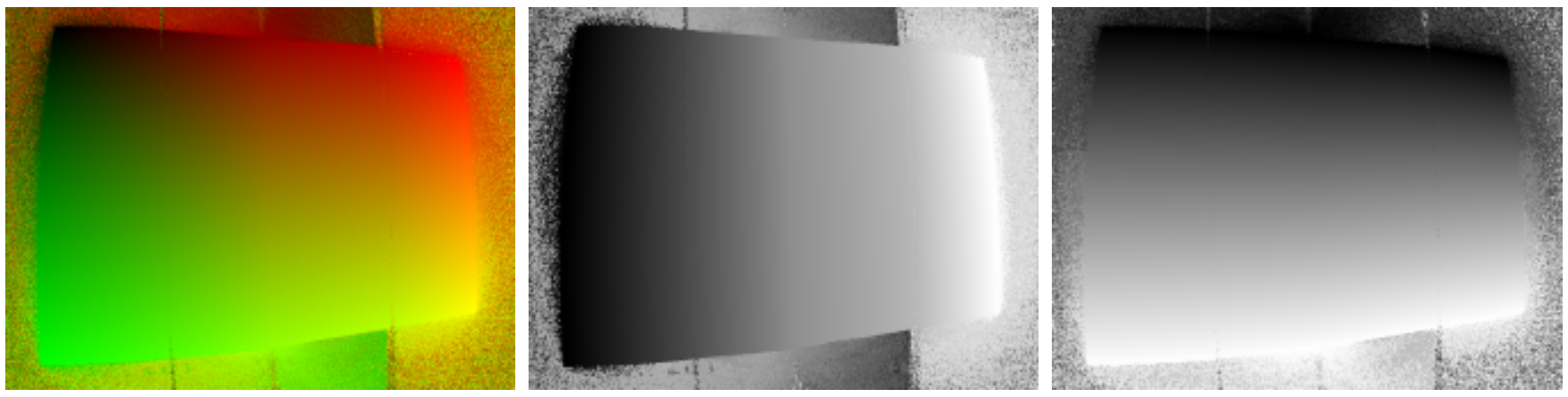

(a) *uv* map (b) x-coordinates (c) y-coordinates

Fig. 5: RGB representation of *uv* map and its components

Figure 5a presents the *uv* map of one our tests. In 5b and 5c we see the coordinates evolution in the horizontal and vertical direction respectively. The latter shows no sign of discontinuity since the four panels are all orthogonal to the ground. Both components partial derivatives are computed in the horizontal direction for 5b and in the vertical direction for 5c. Resulting images are presented in figure 6 where we can see that only the partial derivative with respect to the horizontal direction has different values for the four panels.

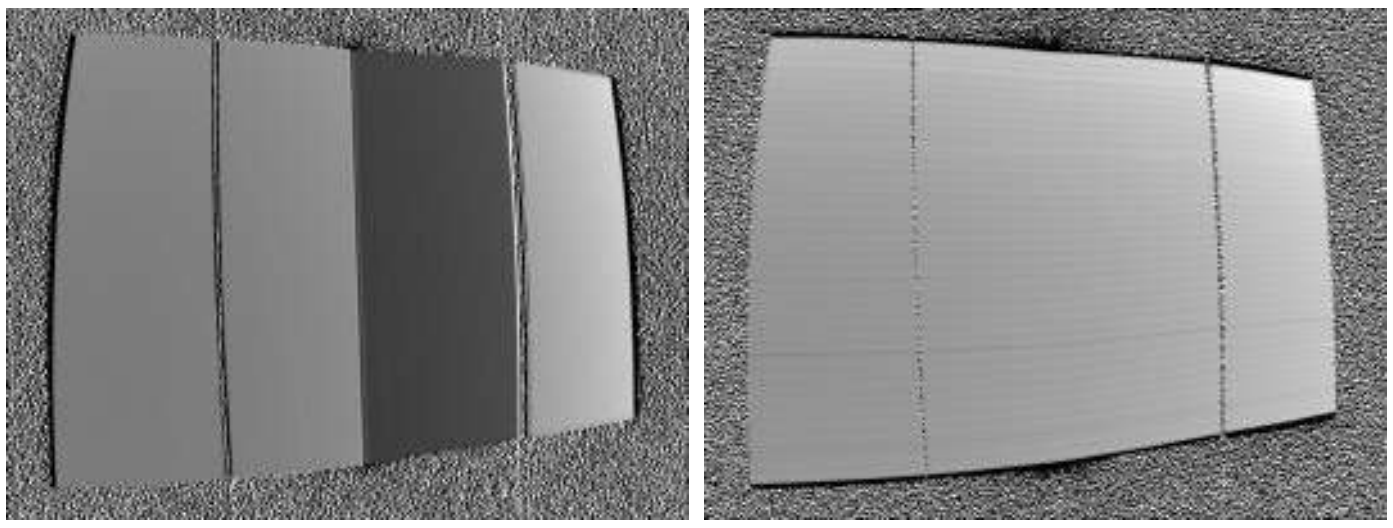

(a) first derivative in the horizontal direction (b) first derivative in the vertical direction

Fig. 6: Partial first derivatives of the *uv* map

Partial second derivatives are computed and combined to extract a binary mask, as shown in figure 7. It appears that second derivatives alone are not enough to separate the different panels since the mask is divided in three zones while our setup has four different panels.

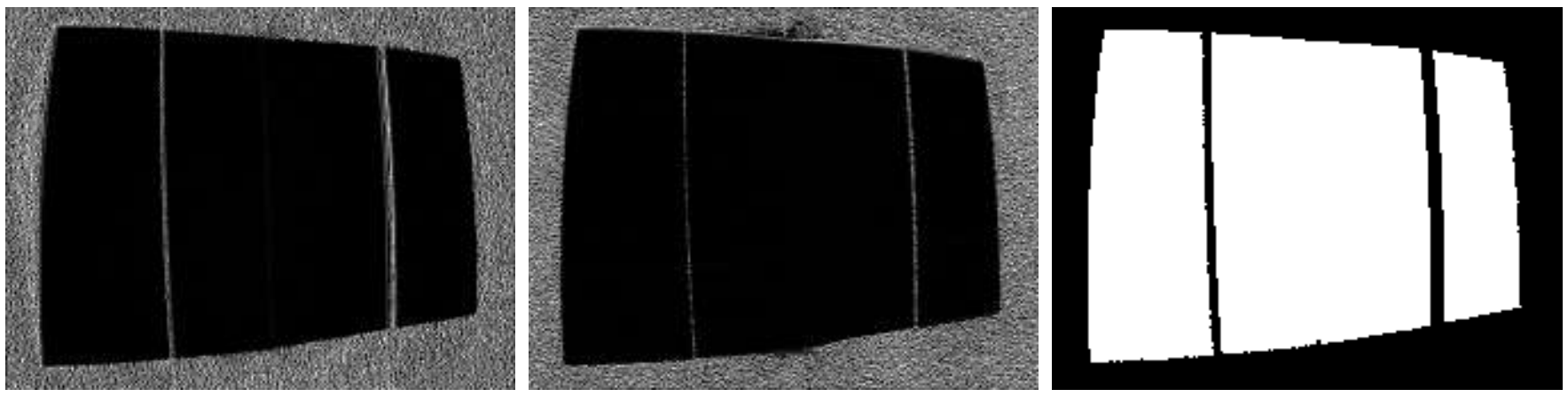

(a) Second derivative in the horizontal direction

(b) Second derivative in the vertical direction

(c) Binary mask of detected planar zones

Fig. 7: Partial second derivatives and detected planar zones

## 4.2 Planar surface separation and image warping

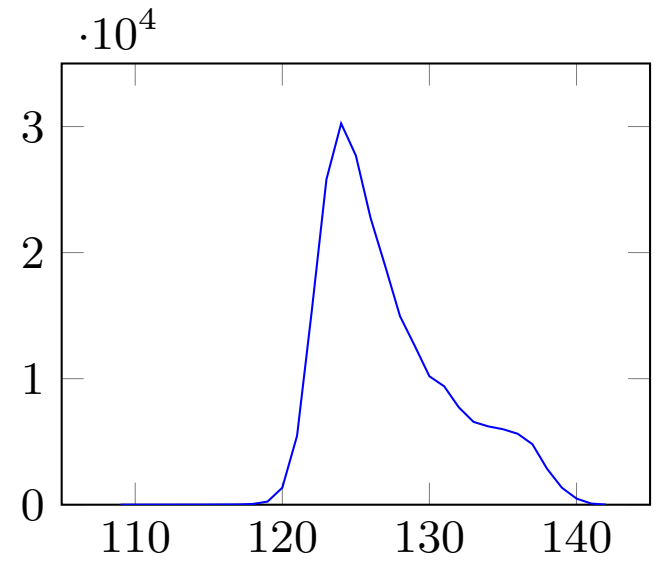

(a) Left zone histogram

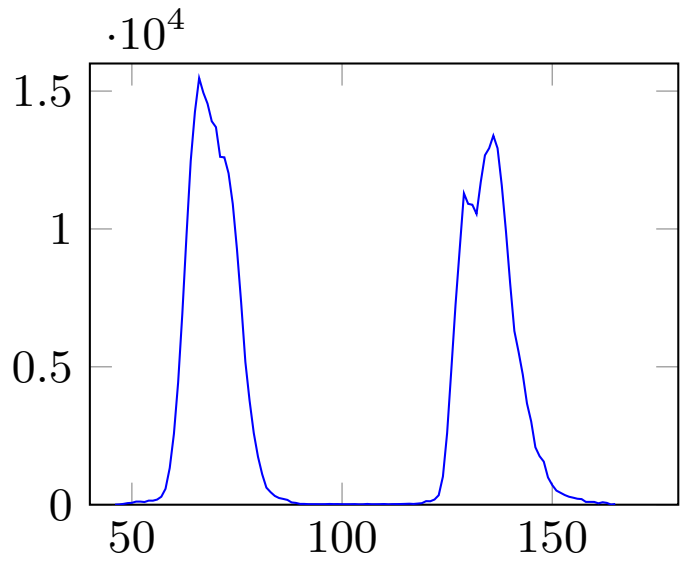

(b) Central zone histogram

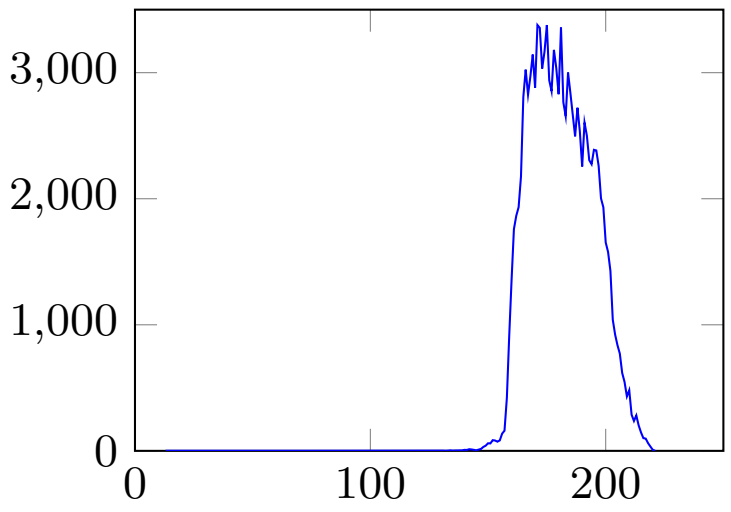

(c) Right zone histogram

Fig. 8: Horizontal derivatives histogram for the three planar zones

Planar surface separation is achieved through a histogram analysis of the first derivatives over the different zones. Histograms for the horizontal derivative are presented in figure 8. To identify the different surfaces, we need to find the different peaks. This is done by looking for extrema. A threshold is used to avoid noisy minima. We used a value of 500, meaning that a derivative value needs to be counted more than 500 times in a zone to be considered as reliable. This analysis performed in both direction is enough to tell the panels apart (figure 9).

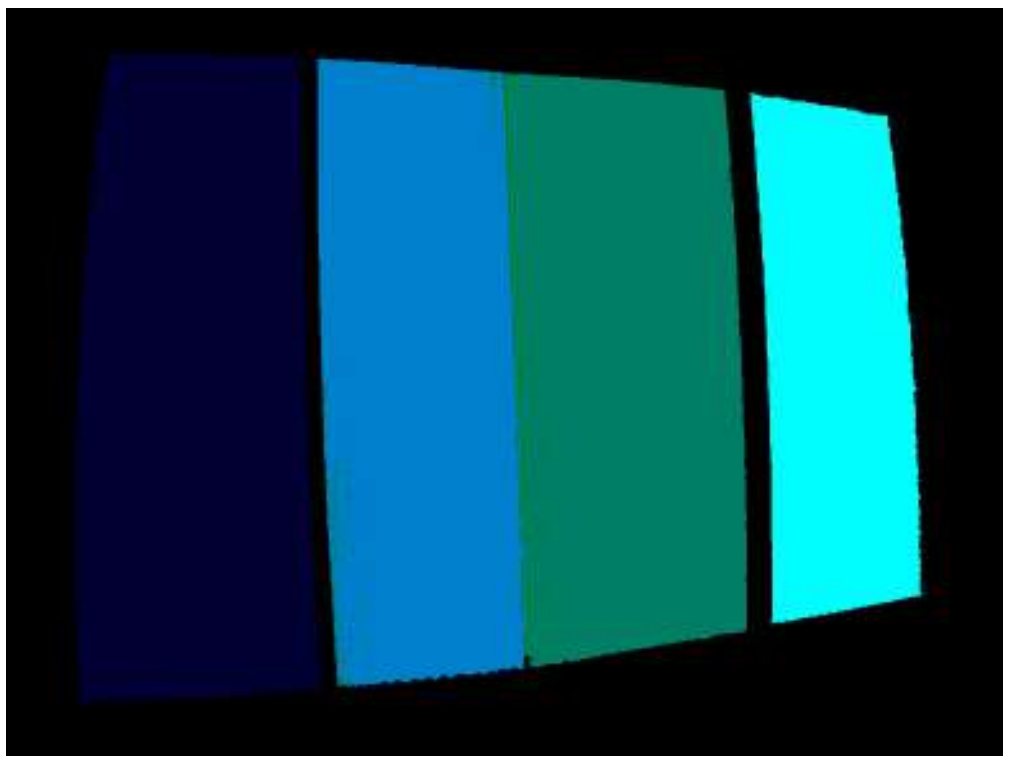

Fig. 9: Detection of planar surfaces with the second derivative

### 4.3 Warping computation

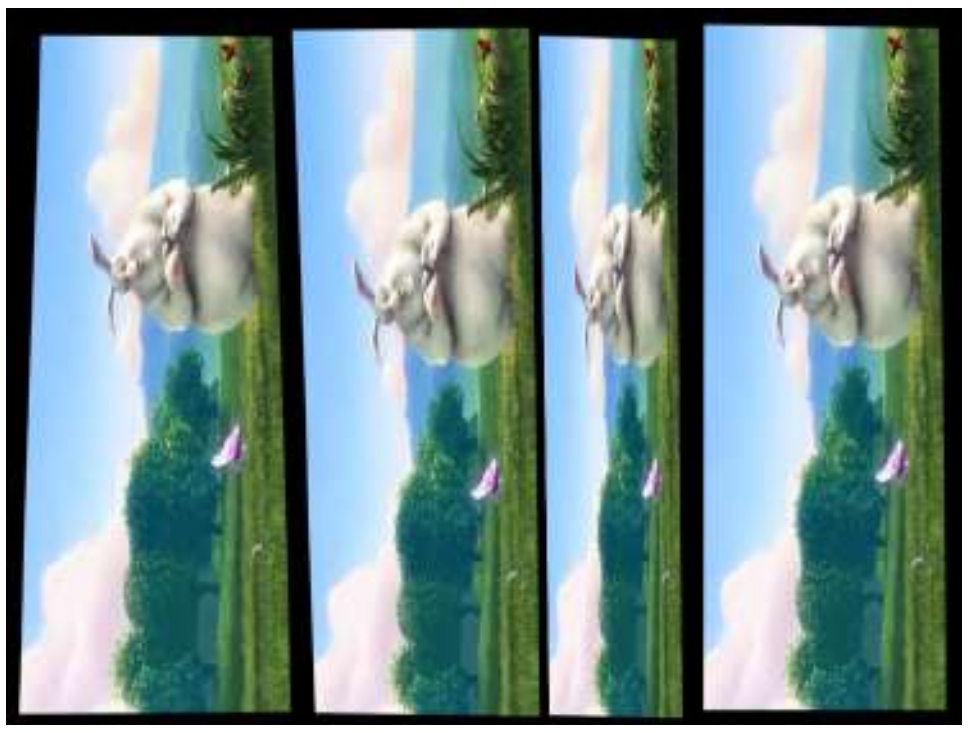

Fig. 10: Image warping to map an image to each surface

All the previous results allow the warping computation. Homographies linking the projector to each planar surfaces are found thanks to the homographies between the two devices and the planar surfaces corners observed from the camera. The resulting warping is presented in figure 10.

## 5 Conclusion

In this project, we proposed a new algorithm to detect planar surfaces in a scene thanks to *uv* maps obtained through a structured light scan. It can be used as an automatic calibration tool for projection mapping on a piecewise planar setup. Only simple image processing algorithms are used to analyse the *uv* map. Partial second derivatives are used to identify planar zones and partial first derivatives tell them apart. This *uv* map analysis could be extended to detect other types of surfaces. It could also be used as a system vision if it were combined with a real-time structured light method.